\documentclass[12pt,twoside]{mitthesis}
\usepackage{lgrind}

\usepackage{url}
\usepackage{slashed}
\usepackage{feynmp-auto}
\usepackage{mathtools}
\usepackage{tikz}
\usetikzlibrary{arrows}
\usepackage{graphicx}
\graphicspath{ {graphics/} }

\newcommand{\fv}[1]{\overleftarrow{#1}}
\DeclareMathOperator{\Tr}{Tr}

\begin{document}

%
%
%
%
%
%
%
%
%
%
%

\title{Measuring the lepton sign asymmetry in elastic electron-proton scattering with OLYMPUS}

\author{Axel Schmidt}
\prevdegrees{B.S., Yale University (2009)}
\department{Department of Physics}

\degree{Doctor of Philosophy in Physics}

\degreemonth{September}
\degreeyear{2016}
\thesisdate{August 26, 2016}


\supervisor{Richard Milner}{Professor}

\chairman{Nergis Mavalvala}{Curtis and Kathleen Marble Professor of Astrophysics\\
  Associate Department Head of Physics}

\maketitle



\cleardoublepage
\setcounter{savepage}{\thepage}
\begin{abstractpage}
%
%
%

OLYMPUS is a particle physics experiment that collected data
in 2012 at DESY, in Hamburg, Germany, on the asymmetry between positron-proton and
electron-proton elastic scattering cross sections. A non-zero asymmetry is
evidence of hard two-photon exchange, which has been hypothesized to cause
the discrepancy in measurements of the proton's electromagnetic form factors.
Alternating electron and positron beams, accelerated to 2 GeV, were directed 
through a windowless, gaseous, hydrogen target, and the scattered lepton and 
recoiling proton were detected in coincidence using a large acceptance magnetic 
spectrometer. Determining the relative integrated luminosity between the electron and 
positron data sets was critical, and a new technique, involving multi-interaction
events, was developed to achieve the desired sub-percent accuracy. 
A detailed Monte Carlo simulation was built in order to reproduce the convolution
of systematic effects at every stage of the experiment. The first stage in the 
simulation was new radiative event generator, which permitted the full simulation of 
the non-trivial radiative corrections to the measurement. 
The analysis of the data and simulation showed that the lepton sign asymmetry rises by 
several percent between a momentum transfer of 0.5~GeV$^2/c^2$ and 2.25~GeV$^2/c^2$. 
This rise as a function of increasing momentum transfer confirms that two photon exchange at least
partially contributes to the proton form factor discrepancy. 

\end{abstractpage}


\cleardoublepage

\section*{Acknowledgments}

The OLYMPUS Experiment was brought to fruition by the collaboration of over
60 scientists from all over the world, from Armenia to Wyoming. The results
I present in this thesis were the product not only of my own efforts but also
the dedication and hard work of my collaborators, for which I am enormously grateful.
The success of OLYMPUS also depended on the tremendous engineering and support 
staff at DESY and at MIT Bates. DESY's MEA~1 and MEA~2 groups provided valuable
assistance in preparing the accelerator, surveying the detectors, and measuring
the magnetic field. I am immensely grateful to Jim Kelsey, Chris Vidal, Peter Binns, 
Brian O'Rourke, Joe Dodge, and Jason Bissouille from MIT Bates for their technical 
expertise and for their help getting OLYMPUS up and running.

I believe that the OLYMPUS collaboration was incredibly fortunate to have 
an all-star roster of post-docs, and I will always be thankful for their 
mentorship. Christian Funke can never receive enough credit for his contributions 
to the OLYMPUS data acquisition system. As a run coordinator, Alexander Winnebeck 
was a super-human force. I often find myself wishing that I could remain in 
graduate school perpetually as an apprentice to J\"{u}rgen Diefenbach. Jan
Bernauer has probably taught me more about how to do physics than anyone
on the planet, and he will always have my gratitude. 

I count myself incredibly fortunate to have had the friendship of Colton O'Connor, 
Lauren Ice, Brian Henderson, and Becky Russell, my classmates and fellow graduate 
students on OLYMPUS. They were an unfailing source of support, they taught me many
amazing things about the world, and they always pushed me to improve. I could not
have wished for better teammates. 

I want to thank Prof. Robert Redwine, OLYMPUS collaborator and member of my
thesis committee, for his sage guidance, insightful questions, and keen editorial
suggestions. I want to thank Doug Hasell for both the amazing technical training 
he gave me, and for his unceasing advocacy on behalf of the OLYMPUS graduate
students. Doug, among others, deserves immense credit for helping to transfer the 
accumulated technical experience from the BLAST experiment to OLYMPUS.
I want to thank my advisor, my chief mentor, and my occasional 
partner in rock and roll, Prof. Richard Milner. In addition to the outstanding
education I received from him as well as years of advice and encouragement,
I am thankful for Richard's pointed and effective questions that cut through
technical minutiae of any discussion to ensure that the physics---what interesting thing
can we learn about the universe---was always in mind. I will be seeking out his
wisdom and judgment for a long time to come. 

I received the help and support from far too many people at MIT to name
them all individually, but several deserve special recognition. Graduate
students who came before me, Asher Kaboth, Richard Ott, and Michael Betancourt
gave me a helping hand on the steep MIT learning curve. I will always be
grateful for the help given to me by Joanne Gregory; she will be dearly missed.
I want to thank Cathy Modica for being a force for good. I want to thank John Hardin for 
his keen insights and for offering his editorial eye. I want to thank
Charles Epstein for stimulating discussions about music as well as physics,
and for being a great bandmate. 

I am grateful to Rick Casten, Volker Werner, Reiner Kr\"{u}cken, and 
Andreas Heinz for taking me on as an undergraduate student and showing
me how much fun experimental physics can be.

I am grateful beyond words to my family---my parents Jean and Martin, my
sister Alissa---not only for their constant love and support, but also for 
nurturing my curiosity about the world. I will always be thankful for the
encouragment I received from my grandfather, Sam Andreades. I have fond 
memories of him enthusiastically answering all of the technical questions I had
saved up specifically for our visits.

Lastly, I consider myself the luckiest man alive to have the love of my 
wife, Ada. For my seven years in graduate school (and several before that), 
she has been an unwavering source of support, an inspiration for new ideas, 
and a caring friend, all while piloting her own career, in the field of
cardiology, into orbit. My achievements would not have been possible
without her but feel all the more rich and meaningful because I share them 
with her.


\pagestyle{plain}
\tableofcontents
\newpage
\listoffigures
\newpage
\listoftables

\chapter{Introduction}

\section{The Proton Form Factor Discrepancy}

More than a decade ago, a new experimental technique for studying the properties of the proton
revealed a glaring inconsistency. The properties of interest were the proton's electromagnetic
form factors, which are a universal way of describing how the proton's electromagnetic charge
and current are distributed within its volume. The proton is a composite particle, made up
of many smaller particles and their associated fields. The proton's positive charge and 
anomalously large magnetic moment are produced by the arrangement of the proton's components.
Understanding the proton's form factors, which convey information about this arrangement,
has long been a goal of the field of particle physics. 

The inconsistency cut between the two experimental techniques for measuring the proton's form
factors. Results obtained with the newer polarization transfer method found different form factors
than older experiments using Rosenbluth separation. New Rosenbluth separation experiments, performed
as a cross-check, confirmed the discrepancy rather than resolving it. Each technique measures
a proton structure that is incompatible with the results of the other.

The leading hypothesis about the cause of this discrepancy is that the results of previous experiments
have been distorted by two-photon exchange. Both polarization transfer and Rosenbluth separation
rely on elastic scattering between electrons and protons, assuming that, in such reactions, exactly
one photon is exchanged between the electron and proton. Reactions in which two (or more)
photons are exchanged are neglected when interpreting the results. If two-photon exchange were
properly accounted for, perhaps the form factor results might be brought into accord.

Since there is no model-independent theoretical technique for calculating the two-photon exchange
contribution, several experiments were conducted to measure it directly. OLYMPUS is one such
experiment, and its design, operation, and results are described in this thesis. 

\subsection{Proton Ftructure and Form Factors}

Conclusive measurements at Stanford in the 1950s determined that the proton is not a point-like particle
\cite{Hofstadter:1955ae,Mcallister:1956ng,Chambers:1956zz,McAllister:1956zz,RevModPhys.30.482,Bumiller:1960zz,Hofstadter:1960zz}.
This discovery was monumental, first for exposing a deeper level in the catalog of nature's
building blocks, and second for spawning a field of research dedicated to understanding the proton's internal workings.
The puzzle of proton structure has several difficulties that define it. Unlike in an atomic nucleus, the proton's 
constituents (quarks and gluons) are light compared to the binding energy of the system; that is, they are highly 
relativistic.\footnote{Gluons, as far as we know, are completely massless. The three valence quarks of the proton have masses
on the order of a few MeV, quite light compared to the proton mass, 938.272~MeV.} And, unlike electrons around an atom, 
the interactions are strong and therefore cannot be understood perturbatively. Calculating the structure of the proton is not 
yet possible using a fundamental theory.  However, there are significant in-roads that can be made with experiments.  

One of the important experimental techniques is elastic electron scattering.  An accelerated electron is a very useful 
probe of the interior of the proton because its interactions are primarily governed by quantum electro-dynamics (QED), a 
theory that is well understood using perturbative techniques.  Elastic electron scattering yields information about the 
electromagnetic charge and current distributions of the proton.  

The common formalism for interpreting elastic scattering experiments makes use of structure functions called form factors.
A proton's electromagnetic interactions are different from a point-like electron because the proton's charge and current
are distributed through the proton's volume. Therefore, it makes sense to modify the proton's Feynman rules in a general
way to account for this difference.
\[
\parbox[][][c]{4cm}{\includegraphics[width=4cm]{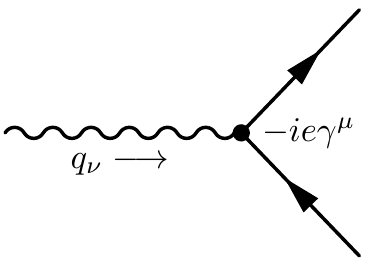}}
\; \mbox{ \Large \( \longrightarrow \) } \;
\parbox[][][c]{4cm}{\includegraphics[width=4cm]{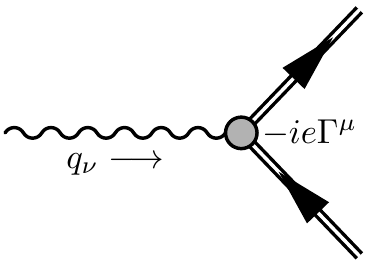}}
\]
\begin{align}
\text{Electron vertex function:~}&\gamma^\mu\\
\text{Proton vertex function:~}&F_{1}(Q^2)\gamma^{\mu} + \frac{i \kappa}{2m_p}F_{2}(Q^2)q_{\nu}\sigma^{\mu\nu} \equiv \Gamma^\mu
\label{eq:proton_vertex_function}
\end{align}
In the above functions, $\gamma$ and $\sigma$ represent Dirac matrices, $q_\nu$ represents the 4-momentum
carried by the electromagnetic field at the vertex, $Q^2 = -q_\nu q^\nu$, $\kappa$ is the anomalous magnetic moment of the proton,
$m_p$ is the proton's mass, and finally $F_1$ and $F_2$ are form factors. The form factors $F_1$ and $F_2$ are functions of $Q^2$.
The form factors specify how the complex charge and current distributions of the proton respond to an 
electromagnetic field. Form factors are the most convenient way for experiments to describe proton structure. 
They should have definite, universal, values that can be measured by different experiments. But they do, in
some sense, obscure what is happening inside the proton at a deeper level. As David Griffiths puts it, using 
form factors as a way to refer to proton structure ``veils our ignorance'' \cite{griffiths2008introduction}.
However, the hope is that someday a complete theoretical description of the proton's internal workings can be 
used to calculate the values of the proton's form factors. And until then, models for the structure of the
proton can be used to predict values for the form factors, which can then be tested against those measured by
experiments.

This expression in equation \ref{eq:proton_vertex_function} is the most 
general electromagnetic vertex function for a spin-$\frac{1}{2}$ particle. Therefore, the same
language can be used to describe the internal structure of the neutron. The differences in electromagnetic
structure of the proton and neutron will be manifest in differences in their respective form factors. For
example, the proton has a total electric charge of $+1$ and, as a consequence, $F_1(Q^2=0) = 1$ for the proton.
However, for the neutron, a neutral particle, $F_1(Q^2=0) = 0$.

\subsection{Experimental Techniques for Measuring Form Factors}

\begin{figure}[htpb]
\centering
\includegraphics{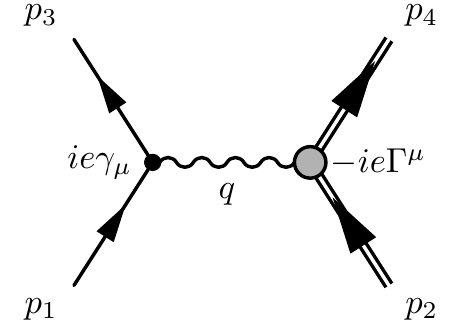}
\caption[Elastic scattering Feynman diagram]{\label{fig:ope_w_labels} There is one Feynman diagram at leading order for elastic 
electron-proton scattering.}
\end{figure}

Using the modified proton vertex function, one can write the matrix element for elastic scattering between an
electron and proton (as seen in the Feynman diagram in figure \ref{fig:ope_w_labels}: 
\begin{align}
\mathcal{M} =& \frac{4\pi\alpha}{Q^2} \bar{u}(p_3) \gamma_\mu u(p_1) \bar{u}(p_4) \Gamma^\mu u(p_2) \\
=& \frac{4\pi\alpha}{Q^2} \bar{u}(p_3) \gamma_\mu u(p_1) \bar{u}(p_4) 
\left[ F_{1}(Q^2)\gamma^{\mu} + \frac{i \kappa}{2m_p}F_{2}(Q^2)q_{\nu}\sigma^{\mu\nu} \right] u(p_2).
\end{align}
Several observables in elastic scattering reactions can be used to extract the values of the form
factors.

\subsubsection{Rosenbluth Separation}
\label{sec:ope}

The basic measurement that can be made using elastic electron scattering is a measurement of the unpolarized differential 
cross-section.  Given the form of the matrix element from above, the resulting lab frame cross section is:
\begin{equation}
\frac{d\sigma}{d\Omega_e} = \frac{d\sigma}{d\Omega_e}_{Mott} \times \left( \frac{(F_{1}+\tau\kappa F_{2})^2}{1+\tau} + 
\tau(F_{1}-\kappa F_{2})^2 \left[\frac{1}{1+\tau} + 2 \tan^2\left(\frac{\theta}{2}\right)\right]\right)
\end{equation}
where \(\tau=\frac{Q^2}{4m_{p}^{2}}\) \cite{Rosenbluth:1950yq}\footnote{I am working in the limit where the
electron has negligible mass, and so the Mott cross section $\frac{d\sigma}{d\Omega_e}_{Mott}$, which describes
the scattering of a spin-$\frac{1}{2}$ particle from a spinless target, can be written as
$\frac{d\sigma}{d\Omega_e}_{Mott} = \frac{\alpha^2 E_3 \cos^2(\theta/2)}{4E_1^3\sin^4(\theta/2)}$.}. Notice that
if \(\kappa F_2\) were taken to zero and \(F_1\) to one, then the correction factor 
would reduce to the expected \((1+2\tau\tan^2(\frac{\theta}{2}))\) for electron scattering from a point-like fermion.  

\begin{figure}[htpb]
\centering
\includegraphics{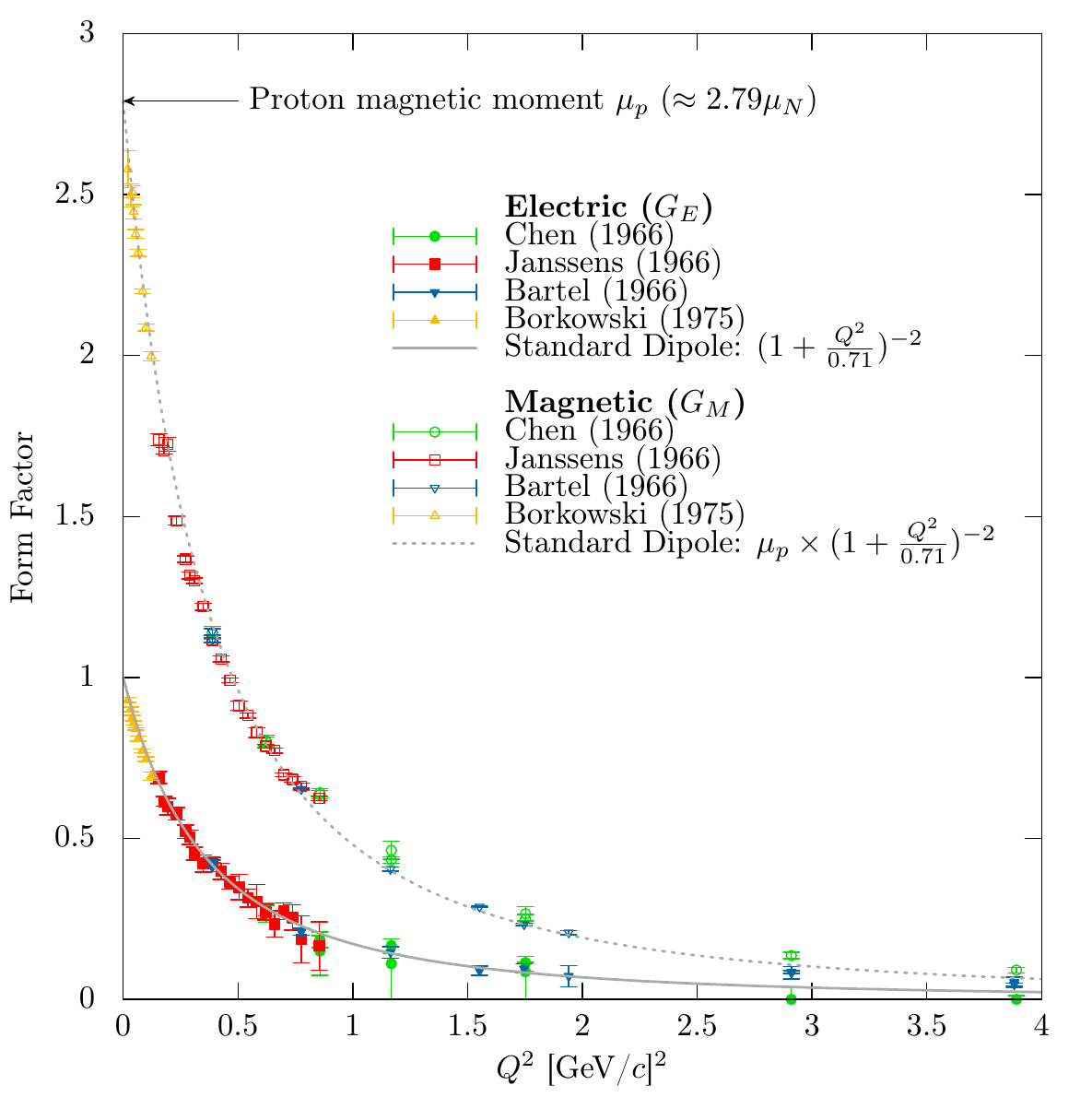}
\caption[Early Rosenbluth separation data]{\label{fig:rosenbluth_data} A representative sample of early Rosenbluth separation data is shown 
\cite{Chen:1966zz,Janssens:1965kd,Bartel:1966zza,Borkowski:1974mb}.
Rosenbluth separation has been used since the 1960s to extract the form factors from cross section data. 
Both the electric and magnetic have an approximately dipole form: $G \propto (1 + \frac{Q^2}{0.71})^{-2}$, 
though the magnetic form factor is scaled higher by a factor of the proton's magnetic moment, $\mu_p$.
This suggests that the proton's charge and current have roughly overlapping spatial distributions. 
Modern data have revealed small deviations between the form factors and the dipole approximation, and 
more accurate, albeit more complicated, approximations have been developed. }
\end{figure}

Ernst, Sachs, and Wali first identified that the linear combinations of form factors \((F_{1}+\tau\kappa F_{2})\) and 
\((F_{1}-\kappa F_{2})\) correspond to the electric and magnetic contributions to the cross-section \cite{Ernst:1960zza}.  
These combinations are now commonly referred to as the Sachs form factors: the former combination being called the electric 
form factor \(G_E\) and the latter the magnetic form factor 
\(G_M\).  With this substitution and with a little algebra, the cross-section can be rewritten:
\begin{equation}
\frac{d\sigma}{d\Omega_e} = \frac{d\sigma}{d\Omega_e}_{Mott} \times \left( \frac{1}{1+\tau} \right) \left(G_{E}^2 + 
\frac{\tau}{\epsilon}G_{M}^{2}\right)
\end{equation}
where \(\epsilon\) is a kinematic variable defined by \(\epsilon^{-1} = 1 + 2 (1+\tau) \tan^2(\frac{\theta}{2})\).  The variables \(\tau\) and 
\(\epsilon\) are a linearly independent pair that completely determine the kinematics of the scattering, just like \(E_{beam}\) and \(\theta\).  
By measuring the differential cross-section at two values of $\epsilon$ for constant \(\tau\), \(G_E\) and \(G_M\) can be separated
\cite{RevModPhys.29.144}.  This
technique is called the ``Rosenbluth separation'' \cite{Andivahis:1994rq}, and was used to make the first measurements of the proton's form
factors.  From these results, we have a good understanding of the shape of the form factors over a wide range of \(Q^2\). A sample of early
form factor data is shown in figure \ref{fig:rosenbluth_data}.

\subsubsection{Methods Using Polarization}

The Rosenbluth separation has several drawbacks.  The measurement of a cross-section will always have limited accuracy.  Large
systematic errors come from being unable to perfectly know the acceptance and efficiency of the detectors and luminosity of the 
experiment.  Furthermore, the kinematic factor \(\tau / \epsilon\) which multiplies the \(G_M\) term grows very large at high momentum 
transfer.  Trying to make a Rosenbluth separation at high \(Q^2\) is nearly impossible because the differential cross-section is 
dominated by the \(G_M\) term.  The uncertainty in the large \(G_M\) term drowns out the small \(G_E\) term.  

With the advent of spin-polarized electron beams, two new techniques became possible.  In both cases, a longitudinally polarized electron
beam is used.  The first option is to use a polarized proton target, and to measure the ratio of the transverse single spin asymmetry to
the longitudinal single spin asymmetry.  The second option is to use an unpolarized proton target, but to measure the polarization of the recoiling
proton, specifically the ratio of transverse polarization to longitudinal polarization.  In both cases, the result yields the ratio of the 
form factors.  
\begin{align}
\nonumber \frac{A_T}{A_L} = \frac{P_T}{P_L} &= - \frac{G_E}{G_M} \times \sqrt{ \frac{2\epsilon}{\tau(1+\epsilon)} } \\
&= - \frac{G_E}{G_M} \times \frac{2M}{E+E^{\prime}} \left[ \tan \left( \frac{\theta}{2} \right) \right]^{-1}
\end{align}

Practically speaking, recoil polarization measurements are used because it is easier to build a polarimeter with high analyzing power than
a target with high polarization, but the two methods are equivalent on theoretical grounds.  Polarization experiments are useful not only 
because their systematic errors are smaller but also
because they yield information on the sign of the form factors, not just their squares as in the Rosenbluth separation.  Recoil 
polarization experiments have been used to accurately measure the proton form factor 
ratio up to a \(Q^2\) of 10 GeV\(^2\)/c\(^2\).  

\subsection{Form Factor Discrepancy}

\begin{figure}[htbp]
\centering
\includegraphics{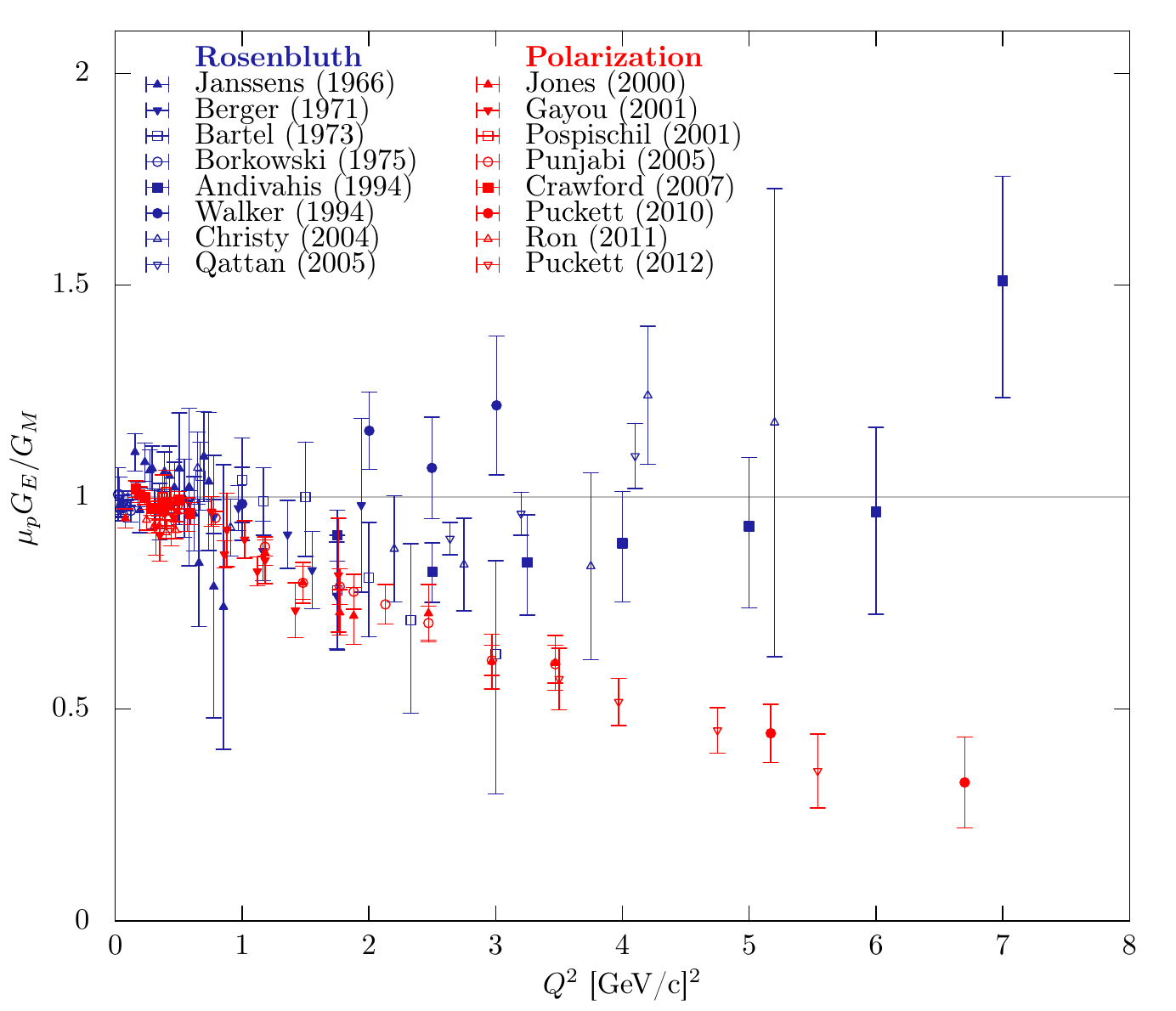}
\caption[Polarization measurements of $\mu_p G_E/G_M$]]{\label{fig:ff_discrepancy} Polarization measurements of $\mu_p G_E/G_M$ (red)
\cite{Janssens:1965kd,Berger:1971kr,Bartel:1973rf,Borkowski:1974mb,Andivahis:1994rq,Walker:1993vj,Christy:2004rc,Qattan:2004ht} 
are inconsistent with Rosenbluth measurements (blue) 
\cite{Jones:1999rz,Gayou:2001qt,Pospischil:2001pp,Punjabi:2005wq,Crawford:2006rz,Puckett:2010ac,Ron:2011rd,Puckett:2011xg}.}
\end{figure}

Contrary to expectations, measurements of the ratio $\mu_p G_E/G_M$ from polarization 
experiments do not agree with their Rosenbluth counterparts. Figure \ref{fig:ff_discrepancy}
shows that the recent measurements of the proton form factor ratio are inconsistent
between the two experimental methods. Unpolarized Rosenbluth experiments have
measured a form factor ratio that is consistent with unity and is independent of $Q^2$.
Polarization experiments consistently measure a ratio that falls with increasing $Q^2$. 

After the emergence of the form factor discrepancy, suspicion was first cast toward
the Rosenbluth measurements. In high-$Q^2$ measurements the $G_M$ term in the Rosenbluth formula
dominates the cross section, obscuring $G_E$ and leading to large uncertainties. If the
uncertainties in old Rosenbluth separation experiments were underestimated, it might be 
possible that the Rosenbluth and polarization results were consistent after all. However, 
a new set of Rosenbluth experiments, conducted in the early 2000s, showed results that
were consistent with the old Rosenbluth data and inconsistent with the polarization
data \cite{Christy:2004rc,Qattan:2004ht}. Furthermore, the latest polarization experiments
reconfirmed the discrepancy \cite{Ron:2011rd,Puckett:2011xg,Meziane:2010xc}.

The proton form factor discrepancy clouds our understanding of proton structure. Without
its resolution, it is not clear which experimental method, if either, is measuring the true
form factors. The effort to find a resolution has prompted a new campaign with both
theoretical and experimental efforts, which will be presented in the following chapter.
Thus far, the most promising hypothesis is that the discrepancy is caused by hard two-photon
exchange. 

\section{Two-Photon Exchange}

\subsection{Theoretical Challenges}

\label{ssec:theory_challenges}

The electron-proton cross section presented in section \ref{sec:ope} assumes that the
electron exchanges only one virtual photon with the target proton. This assumption makes
the calculation far more tractable, and has a tenable justification. Adding an additional
photon suppresses the matrix element by a factor of $\alpha$, the fine structure constant.
Since $\alpha \approx 1/137$, one can argue that the effect of two-photon exchange should be
a sub-percent level correction to the one-photon exchange cross section. 

However, a full calculation of the two-photon exchange contribution in a model-independent 
fashion has not yet been achieved. All calculations so far make some approximation or have
some model dependence. One approach is to assume that one of the exchanged virtual photons
is soft, i.e.\ it carries negligible 4-momentum. With this approximation, the one-photon
exchange matrix element can be factored out, leaving behind a correction factor. Typically,
when experimental data are analyzed, this soft two-photon correction is applied (as part of
a standard radiative correction procedure), and any remaining hard two-photon contribution 
is assumed to be negligible. For this reason, the hypothesized cause of the discrepancy is
the hard two-photon exchange, which was not accounted for. Any distinction between hard
and soft two-photon exchange is artificial and arbitrary, and is only a result of the 
approximation that has typically been made during the analysis of previous experiments.

There have been numerous theoretical attempts to calculate the hard two-photon contribution
to the electron-proton cross section, and these will be presented in section \ref{sec:theory}.
To summarize, there is not a theoretical consensus as to the expected magnitude of the 
effect. Polarization measurements are probably less susceptible to two-photon exchange,
since a ratio of cross sections are being measured; a small correction to both the 
numerator and denominator is less likely to sway the ratio. Rosenbluth separations, however,
which require fitting absolute cross section measurements, are more likely to be affected
should hard two-photon exchange contribute significantly to the total cross section.

\subsection{Experimental Techniques}

In the absence of a theoretical consensus, experiments to determine the hard two-photon
exchange contribution are needed. Three new experiments recently collected data, and
one of those, OLYMPUS, is the subject of this thesis. 

The most direct way to measure the contribution from hard two-photon exchange is to
look for a lepton sign asymmetry in elastic electron-proton scattering. Any difference
between positron-proton and electron-proton elastic scattering cross sections is
evidence of two-photon exchange. Furthermore, for two-photon exchange to resolve
the form factor discrepancy, the asymmetry must have a specific kinematic dependence,
favoring positrons as $Q^2$ increases and as $\epsilon$ decreases. 

All three contemporary experiments are aiming to quantify the lepton sign asymmetry
over a range of kinematic space, though the specific techniques used are all slightly
different. If the three experiments have consistent findings, the fact that they all 
used different techniques will generate confidence in the results. To measure a lepton 
sign asymmetry, one needs beams of accelerated electrons and positrons and a proton
target. One needs a detection apparatus that can distinguish elastic scattering
events from inelastic background. And one needs to monitor the relative luminosity
between the electron scattering and positron scattering data sets. 

\section{The OLYMPUS Experiment}

OLYMPUS is a collaboration of over 60 scientists, from 13 institutions, in six countries.
The OLYMPUS experiment took place at the Deutches Elektronen Synchrotron (DESY), a particle
accelerator complex in Hamburg, Germany, completing data collection in early 2013. 
OLYMPUS had a very tight timeline; funding was secured in early 2010, and the start of
data collection was in February 2012. In that time window, there was an intense effort to
complete all of the necessary preparations. Among the long list of accomplishments were the design and construction
of a new windowless hydrogen gas target, the assembly and installation of the detector
apparatus, and the writing of custom software needed to operate the equipment and collect
the data. 

The data collection took place in two running periods, the first in February 2012,
and the second in the fall of 2012. Following the end of data collection, the collaboration
undertook a survey of all of the detector positions, made a map of the detector's 
magnetic field, and then began the process of analyzing the data. The first results
from the OLYMPUS experiment are included in this thesis, and will be submitted
for publication soon. 

In this section, I will introduce the reader to the various aspects of the OLYMPUS experiment,
previewing what will be covered in detail in this thesis.

\subsection{How OLYMPUS Worked}

To make a lepton-sign asymmetry measurement, OLYMPUS used alternating electron and positron
beams from the DORIS storage ring at DESY. The beams were directed on a stationary proton
target, consisting of a tube of hydrogen gas, that was internal to the vacuum of the ring.
Elastic scattering events were recorded by detecting the scattered lepton and the recoiling
proton in coincidence in a toroidal magnetic spectrometer. The electron beam and positron
beam were alternated, approximately once per day, to accumulate an electron data set and a
positron data set. The relative luminosity of the two data sets, needed for normalization,
was monitored by three independent systems. 

\subsection{OLYMPUS Spectrometer}

The main detection apparatus in OLYMPUS was a toroidal magnetic spectrometer. 
One of the reasons that a tight timeline was possible was that the spectrometer
was repurposed from the BLAST experiment spectrometer \cite{Hasell:2009zza}. The BLAST
experiment had a successful run at the MIT Bates Linear Accelerator Center from 2001--2005. 
OLYMPUS saved a great deal of time, money, and effort, by reusing a proven detector system.
Some refurbishment and upgrades were undertaken, so, with those in mind, I will in this
thesis refer to the former BLAST spectrometer as the OLYMPUS spectrometer. 

The spectrometer magnet, which spanned nearly 4~m wide by 4~m long, by 4~m high, was made
up of eight electromagnetic coils, which were arranged around the beamline like the segments
of an orange. The two segments in the horizontal plane were instrumented with particle 
detectors. Drift chambers measured the trajectories of outgoing particles, while walls of
scintillating plastic were used to measure the time-of-flight. The particle trajectories
curved slightly in the field of the magnet, and the particle's momentum was inferred from
the amount of curvature. OLYMPUS made a coincidence
measurement, detecting the scattered lepton in one sector and the recoiling proton in
the other. The spectrometer was left-right symmetric; leptons and protons could be detected
in either sector. This symmetry was employed in the data analysis to control systematic
uncertainties. 

\subsection{Luminosity Monitoring}

The relative luminosity between the electron and positron data sets was crucial
in making an asymmetry measurement, and so several new luminosity monitors were
built specifically for OLYMPUS. Luminosity could be monitored online by integrating
the beam current and the approximate target density, but a percent-level asymmetry
measurement required more accurate systems. A pair of tracking telescopes was installed
along the $12^\circ$ scattering angle in order to measure the rate of forward 
elastic scattering. In addition to the $12^\circ$ telescopes, a pair of electromagnetic
calorimeters were positioned at approximately $1.29^\circ$ to measure the rates of
symmetric scattering from atomic electrons in the target. Both systems achieved
systematic uncertainties of less than 0.5\% in measuring the relative luminosities
between running modes. 

\subsection{Analysis Approach}

Discriminating elastic scattering events from background was not a significant challenge.
By making a coincidence measurement, and measuring the momenta, angles, and flight time of the 
both the lepton and proton, the kinematics of elastic scattering events were heavily over-constrained. 
Kinematic cuts were employed to separate most of the background from signal, and the remaining
background was estimated and subtracted. 

The analysis was neither as simple nor as straightforward as calculating the asymmetry
in the detected elastic scattering events. The asymmetry caused by hard two-photon exchange 
needed to be extracted from asymmetry caused by other sources. 
As mentioned in section \ref{ssec:theory_challenges}, a set of standard
radiative corrections needed to be applied to remove contributions from soft two-photon exchange.
These corrections were convolved with properties of the
detector, such as resolutions, acceptances, and efficiencies. Our solution for
accounting for these convolutions was to simulate everything. We simulated our
radiative corrections, using a new radiative event generator, which includes 
corrections for soft two-photon exchange. We simulated the propagation of these events 
in the geometry of our detector and produced a full simulated data set. We analyzed 
the real and simulated data sets with the exact same software. The asymmetry we 
present is corrected for the asymmetry we found in simulation.

\section{In This Thesis}

This thesis is divided into chapters, which can be read as self-contained
descriptions of some aspect of the experiment. In the section, I'll give a 
brief preview of each chapter.

In chapter \ref{chap:background}, I lay out the scientific motivation for OLYMPUS.
I present the various theoretical methods for calculating the two-photon exchange
contribution. I also show how the previous asymmetry data are inadequate for 
determining if two-photon exchange is responsible for the form factor discrepancy.
Lastly, I briefly compare OLYMPUS and the two other contemporary asymmetry experiments.

In chapter \ref{chap:apparatus}, I give an overview of the OLYMPUS apparatus. Detailed
descriptions of the apparatus as well as various component detectors can be found in a
variety of other published sources, and I do not aim to exceed any of them. However, I 
do want to give the reader a chance to get familiar with all of the different mechanical
aspects of the experiment while this thesis is open. I try to list, wherever possible,
the sources that go into greater detail, so that the curious reader can find them.

In chapter \ref{chap:field}, I present the survey of the OLYMPUS magnet, which was
undertaken after data collection was complete. I try to give a sense of how we planned
and conducted the magnetic field measurements, and how we analyzed the data. This 
chapter goes into more detail than our recently published article on the same subject
\cite{Bernauer:2016hpu}.

In chapter \ref{chap:rc}, I present the radiative corrections for OLYMPUS. I try to 
argue why radiative corrections need to be made by experiments in order for results
to be interpreted. Next, I explain what specific corrections need to be made for OLYMPUS.
Then I go into detail about the new radiative event generator that we wrote specifically
for the OLYMPUS analysis. I conclude with a discussion of our attempts to validate 
the generator.

In chapter \ref{chap:lumi}, I present an alternate method for extracting luminosity
from the symmetric M\o ller/Bhabha calorimeter (SyMB) data. This alternate method 
makes use of multi-interaction events (MIEs) and effectively trades precision for a
more accurate relative luminosity determination. The MIE analysis was initially pursued 
as a cross check of the main SyMB analysis, but, over time, proved to be a more 
robust approach. In this chapter, I show
a derivation of the method, estimate its systematic uncertainties, and compare
the results to those from the $12^\circ$ tracking telescopes.

In chapter \ref{chap:recon}, I discuss the problem of track reconstruction in the 
OLYMPUS spectrometer. I explain how the drift chamber data must be processed prior
to reconstruction. Next, I address how we identify track candidates using pattern matching,
and then fit the trajectories of track candidates using the Elastic Arms Algorithm. 
I conclude with a discussion of the problem of determining the time-to-distance
relationship for the sense wires in the drift chambers. 

In chapter \ref{chap:analysis}, I lay out the OLYMPUS analysis framework. I first
examine the strategy we used in analyzing the data, including our use of simulation at
every stage. I then present the analysis chain, and cover each step taken in processing
the data. Finally, I show the approach I used in selecting elastic events and estimating
backgrounds.

In chapter \ref{chap:results}, I discuss the results of the OLYMPUS experiment. I do a
preliminary estimate of the systematic uncertainties. I also compare the results to 
various predictions and to the results of the other contemporary experiments. These
results are some of the first to be publicly released from OLYMPUS, and the reader
should understand that they are in preliminary form. That is, they will be superseded
by analyses which combine my work with the work of my colleagues and make final estimates
of the systematic uncertainties. 

\section{A Remark on Collaboration in This Work}

The success of the OLYMPUS experiment was only possible through the hard work of over 60 scientists. 
There was no shortage of problems to be solved, and the original scientific work I contributed 
could easily fill a thesis. However, this kind of science does not readily lend itself to the 
classification of ``her work'', ``his work'', ``my work'', and so on. Almost everything that 
I accomplished in bringing OLYMPUS to fruition was done either in collaboration with a partner,
or by working in a small team, or, at the very least, through solo effort that was guided by
frequent discussions with one or more of my colleagues. The topics I choose to cover in this 
thesis are those where I feel I made the most significant contributions. Because OLYMPUS is the work of 
of ideas and contributions of many people, I will often use the pronoun ``we'', when describing
some of my accomplishments. I cannot help it. It really was ``we.'' There are, however, some cases
where I touch on areas of expertise of my colleagues. In those cases, I try 
to recognize my colleagues' work explicitly, often to point the reader to a different thesis, 
where more detail can be found. 

\chapter{Scientific Motivation}
\label{chap:background}

\section{Calculations of Hard Two-Photon Exchange}
\label{sec:theory}

Hard two-photon exchange was first suggested as a possible solution to 
the proton form factor ratio problem in a pair of papers in Physical 
Review Letters Volume 91, Number 14, in the year 2003. The letter by
P.~A.~M.~Guichon and M.~Vanderhaeghen~\cite{Guichon:2003qm} suggests
a phenomenological approach to describing the effects of hard two-photon
exchange, while the following letter by P.~G.~Blunden, W.~Melnitchouk, 
and J.~A.~Tjon~\cite{Blunden:2003sp} uses a model-dependent hadronic 
calculation of the hard two-photon contribution. Both letters point 
out that a precise comparison of $e^+p$ and $e^-p$ elastic cross 
sections is the needed experiment to quantify the two-photon exchange 
effect.

Guichon and Vanderhaeghen make several points in their work that are
relevant to the theoretical discussion. The first is that there is no
known model independent way to calculate hard two-photon exchange. Any
calculation of hard two-photon exchange will rely upon assumptions which
are difficult to validate. Secondly, any model for the two-photon exchange
effect must also preserve the linearity of Rosenbluth separations. The 
Rosenbluth separation technique relies on fitting a line to reduced cross 
section measurements plotted versus $\epsilon$ (at fixed $Q^2$). This 
linear fitting procedure works over a large range of kinematics, so any
model for two-photon exchange which would spoil this linearity can be 
rejected out of hand. Broadly speaking, this constraint implies that 
any hard two-photon exchange contribution must be roughly linear in 
$\epsilon$ as well. Thirdly, they point out that even a small 
contribution from two-photon exchange can have a large effect on the 
Rosenbluth form factors, without distorting the polarization result.

In their phenomenological approach, Guichon and Vanderhaeghen demonstrate
that the inclusion of hard two-photon exchange results in a third form
factor, which they label $F_3$. All three form factors become functions
of both $Q^2$ and $\epsilon$, but, because Rosenbluth data appear linear
in $\epsilon$, any $\epsilon$ dependence must be weak. By neglecting any
$\epsilon$ dependence in $G_E$ and $G_M$, they are able to numerically 
extract the third form factor from fits to existing data. The results 
imply that a hard two-photon contribution must be on the order of a 
few percent of the total cross section, and must increase with $Q^2$,
since the form factor discrepancy is greater at higher momentum transfer.

In their letter, Blunden, Melnitchouk, and Tjon attempt to calculate the
two-photon exchange diagram using a hadronic model. The two-photon exchange
diagrams (the box, and crossed-box diagrams) are difficult to calculate
because they have off-shell proton vertices and an off-shell hadronic
propagator. In a hadronic model, one attempts to clear these hurdles by 
assuming that on-shell vertices are reasonably accurate, and that a 
propagator can be constructed as a coherent
sum of propagators of all of the hadronic states available to the 
proton. The proton can propagate as a proton, or as an excited
$\Delta^+$, or in any of the other excited states of the proton. In practice
one has to terminate the sum, and in this paper only the proton propagator
term is used. The results show that with only the proton term included, 
the two-photon exchange is sufficient to explain about half of the 
form factor discrepancy. The authors hope that by including the $\Delta^+$
term and further hadronic terms, the discrepancy can be fully explained.

Following these letters, there has been a flurry of papers attempting
to quantify the magnitude of the two-photon exchange contribution. Without
a model-independent theoretical framework, theorists have followed several
different model-dependent approaches, each with advantages and drawbacks. 
The approaches can be grouped in a few major categories based on the models
used: phenomenological models~\cite{Guichon:2003qm,Rekalo:2003xa,TomasiGustafsson:2004ms,
Tvaskis:2005ex,Chen:2007ac,Arrington:2007ux,Borisyuk:2007re,Guttmann:2010au,Borisyuk:2010ep}, 
hadronic models~\cite{Blunden:2003sp,Blunden:2005ew,Kondratyuk:2005kk,Kondratyuk:2007hc},
partonic models~\cite{Chen:2004tw,Afanasev:2005mp,Borisyuk:2008db,
Kivel:2009eg,Kivel:2012vs}, and models using dispersion relations
\cite{Gorchtein:2006mq,Borisyuk:2006fh,Borisyuk:2010cv,Borisyuk:2012he,Borisyuk:2013hja},
though there are other calculations that defy these classifications
\cite{Jain:2006mu}.

\subsection{Phenomenological Approaches}

\label{ssec:phenom}

Following the initial phenomenological work of Guichon and Vanderhaeghen
\cite{Guichon:2003qm}, there have been numerous attempts to extract 
the two-photon exchange amplitude from existing data. Several works
have argued that two-photon exchange can be ruled out from the lack
of non-linearities in existing Rosenbluth data 
\cite{Rekalo:2003xa,TomasiGustafsson:2004ms,Tvaskis:2005ex}.
Other studies have extracted values of two-photon exchange that
are consistent with a resolution of the form factor discrepancy
\cite{Arrington:2007ux}, or at least qualitatively consistent 
with other two-photon calculations~\cite{Borisyuk:2007re}, though
the results are dependent on the functional form chosen to parameterize
the two-photon exchange contribution. 

An alternative approach is to attribute the cause of the discrepancy 
entirely to hard two-photon exchange so as to constrain $F_3$ and the
asymmetry between positron-proton and electron-proton cross sections.
This technique serves as a very useful benchmark for interpreting
lepton-sign asymmetry measurements. These phenomenological predictions
indicate where measurements would have to fall to fully confirm the
hypothesis that two-photon exchange causes the proton form factor
discrepancy. Phenomenological predictions~\cite{Chen:2007ac,Guttmann:2010au,Borisyuk:2010ep,Bernauer:2013tpr} 
that include the latest polarization transfer data~\cite{Meziane:2010xc}
predict asymmetries on the order of a few percent. That suggests that
OLYMPUS aim for uncertainties at the sub-percent level to definitively 
confirm the two-photon exchange hypothesis.

\subsection{Hadronic Calculations}

The hadronic calculation in the letter by Blunden et~al.~\cite{Blunden:2003sp}
was followed by a more detailed paper by the same group~\cite{Blunden:2005ew}.
The push was then made to include further hadronic intermediate states other
than the proton. The $\Delta^+$ state was included in a calculation and 
surprisingly was found to reduce the contribution from two-photon exchange
\cite{Kondratyuk:2005kk}. The addition of many more higher intermediate states
was found to have diminishing effect~\cite{Kondratyuk:2007hc}. Even by including
many intermediate states, hadronic calculations still suffer, however, because 
unknown off-shell form factors are approximated, typically with their on-shell
values. 

\subsection{Partonic Calculations}

In a partonic calculation, the two photon exchange diagram is broken down
into interactions between the photons and individual partons within the proton.
The calculation requires a parton distribution function as input. The weakness
of this method is that parton distribution data will need to be extrapolated
in a model-dependent way to elastic kinematics. Nevertheless, Chen et~al. 
performed a partonic calculation using Generalized Parton Distributions (GPD)
taken from Deep-Inelastic Scattering (DIS) data, which they describe in an
initial letter~\cite{Chen:2004tw} as well as in a more detailed paper
\cite{Afanasev:2005mp}. The results of the calculations suggest that hard 
two-photon exchange can ``substantially reconcile''~\cite{Afanasev:2005mp} 
the form factor discrepancy. Partonic calculations can be improved by 
including QCD corrections, by using perturbative QCD (pQCD), although the
reliability of these corrections is limited to the kinematic region in which 
the partons inside the proton are weakly interacting (high energies and 
large scattering angles.) These pQCD calculations typically use proton
Distribution Amplitudes (DAs) as their input. Dmitry Borisyuk and Alexander 
Kobushkin produced a partonic calculation that can be extrapolated towards 
hadronic results at low $Q^2$~\cite{Borisyuk:2008db}. Nikolai Kivel, working 
with Vanderhaeghen, performed a subsequent partonic calculation using two 
different DA models, showing that hard two-photon exchange can substatially 
alter the slope of a Rosenbluth plot without introducing nonlinearities 
\cite{Kivel:2009eg}. Kivel and Vanderhaeghen also introduced a formalism based 
on Soft Collinear Effective Theory that makes more modest claims~\cite{Kivel:2012vs}.

\subsection{Dispersion Relations}

Dispersion relations are useful tools for removing the dependence of 
off-shell form factors from hadronic calculations. By making use of
the unitarity and analyticity of the scattering amplitudes, one can
relate the real and imaginary parts of on-shell form factors, circumventing
the need to rely on approximations of off-shell form factors. Mikhail 
Gorchtein performed a calculation at high energies using dispersion relations
and found that two-photon effects should play a role at the few percent
level at forward angles~\cite{Gorchtein:2006mq}. Borisyuk and Kobushkin 
made significant progress using dispersive techniques. Their first 
calculation showed agreement with earlier hadronic calculations 
\cite{Borisyuk:2006fh}. A subsequent calculation was able to fully
remove any off-shell form factors~\cite{Borisyuk:2010cv}. They also
added contributions from the $\Delta^+$ intermediate state~\cite{Borisyuk:2012he} 
as well as the P33 partial wave~\cite{Borisyuk:2013hja}. The general 
tenor of their results shows a hard two-photon correction that helps 
resolve the discrepancy.

\subsection{Criticism of the Two-Photon Exchange Hypothesis}

There are calculations that contest that hard two-photon exchange
is responsible for the form factor ratio discrepancy. E. Tomasi-Gustafsson 
et~al.\ have long argued that two-photon exchange would necessarily 
introduce non-linearities in Rosenbluth data~\cite{Rekalo:2003xa, 
TomasiGustafsson:2004ms, Bystritskiy:2006ju, TomasiGustafsson:2006pa}.
They suggest instead that the discrepancy might be caused by inadequate
radiative corrections, specifically the treatment of bremsstrahlung.
In references~\cite{Bystritskiy:2006ju, TomasiGustafsson:2006pa}, they
develop a formalism calculating multi-photon bremsstrahlung that treats
the electron vertex as a composite object QED object (made up of soft photons
and $e^+e^-$ pairs). The momentum distribution of the hard electron 
within the composite vertex can be described by structure functions
governed by the Altarelli-Parisi equations (typically used for 
partons within hadrons). By implementing a radiative correction based
on this formalism, Tomasi-Gustaffson~et~al.\ show a modification to
the Rosenbluth data that can fully explain the form factor discrepancy
without any need for hard TPE.

\subsection{Outlook}

There is not a theoretical consensus on the magnitude of the hard
TPE effect. A wide variety of calculations and phenomenological
extractions support the hypothesis that hard TPE contibutes 
significantly to the elastic $ep$ cross section and is responsible
for the form factor discrepancy. However these calculations all
have model-dependencies and the phenomenological extractions all
contain assumptions and approximations. Furthermore, this opinion
is not unanimous. A minority view, supported by calculations with
the electron structure function formalism, is that hard TPE is
negligible and that instead our focus should be improving radiative
correction procedures. 

This state of discord is a clear call for experimental verification.
Experiments that measure the size of the hard TPE contribution
can validate or disprove the claim that it bears responsibility for the 
form factor discrepancy. 

\section{Measurements of Hard Two-Photon Exchange}

\label{sec:tpe_exp}

\subsection{Lepton-Sign Asymmetry}

The most direct technique for measuring hard two-photon exchange is 
the detection of an asymmetry between the cross sections of $e^-p$ and
$e^+p$ elastic scattering reactions.\footnote{This is not the only
technique. A measurement of a scattering-plane-normal single-spin asymmetry
is also way to detect two-photon exchange, since such asymmetries are
zero in the one-photon exchange approximation. However, these asymmetries
are proportional to the imaginary part of the two-photon exchange amplitude. 
The imaginary part of the two-photon amplitude only enters the $ep$ cross 
section at the three-photon exchange level or higher. Thus, the results
of the single-spin asymmetry technique do not shed any light on the 
form factor discrepancy and are irrelevant to this work. For a detailed
treatment, see reference \cite{Carlson:2007sp}.} In the one-photon 
exchange approximation, there is no asymmetry between the two lepton
charge signs. The next-to-leading term, the interference between 
the one- and two-photon exchange amplitudes, is odd in the sign of the
lepton charge. Any lepton-sign asymmetry is proportional to this
interference amplitude, seen in equation \ref{eq:asym}.
\begin{align}\label{eq:asym}
\mathcal{M} =& \mathcal{M}_{1\gamma} + \mathcal{M}_{2\gamma} + \mathcal{M}_{3_\gamma} + \cdots \mathcal{O}(\alpha^4) \\
\left|\mathcal{M}\right|^2 =& \left|M_{1\gamma}\right|^2 + 2\text{Re}\left(\mathcal{M}_{1\gamma}^\dagger \mathcal{M}_{2\gamma} \right)
 + \cdots \mathcal{O}(\alpha^4)
\end{align}
\begin{equation}
A_{2\gamma} \equiv \frac{\sigma_{e^+p} - \sigma_{e^-p}}{\sigma_{e^+p} + \sigma_{e^-p}} 
\approx \frac{2\text{Re}\left(\mathcal{M}_{1\gamma}^\dagger \mathcal{M}_{2\gamma} \right)}{\left|M_{1\gamma}\right|^2}
 + \cdots \mathcal{O}(\alpha^4)
\end{equation}
Absolute cross section measurements have accuracy that is limited by
knowledge of the detector acceptance. By measuring an asymmetry, the
only relative cross section measurements are needed.

A non-zero asymmetry indicates two-photon exchange, but the kinematic
dependence of the asymmetry can determine whether or not it is responsible
for the proton form factor discrepancy. To bring Rosenbluth measurements
into agreement with polarization measurements, the asymmetry must have
a negative slope with respect to $\epsilon$. As $\epsilon$ decreases, the
asymmetry should grow larger, i.e., the positron cross section should
exceed the electron cross section. Furthermore, since the form factor
discrepancy grows at large $Q^2$, we would expect that the asymmetry
also grows with respect to $Q^2$. Phenomenological predictions (mentioned
in subsection \ref{ssec:phenom}) vary in their estimates, but suggest
that an asymmetry that grows to several percent at large angles and
low $\epsilon$ would be sufficient to resolve the form factor
discrepancy.

\subsection{Radiative Corrections}

Extracting the two-photon exchange amplitude from a measured lepton sign
asymmetry is complicated by the problem of radiative corrections. The standard elastic scattering
radiative correction procedures all include corrections for soft two-photon
exchange, two-photon exchange calculated in the limit that one of the exchanged
photons carries infinitesimal momentum, i.e., is soft. Since all previous
elastic scattering measurements have accounted for soft two-photon exchange,
the form factor discrepancy could only be caused by two-photon exchange,
in which both photons carry significant momentum. This additional two-photon exchange 
is given the appellation: ``hard two-photon exchange.'' In the standard radiative corrections procedures,
hard two-photon exchange is assumed to be negligible. 

Measurements of the lepton-sign asymmetry are seeking to quantify hard
two-photon exchange. This means that these measurements must correct their
results for the effect of soft two-photon exchange, as well as the other
radiative effects that enter at order $\alpha^3$. Isolating hard two-photon 
exchange essentially amounts to the problem of disentangling one radiative 
correction from all of the others. It is a problem that requires careful
attention, and, for the OLYMPUS experiment, was a primary concern in the design of our 
simulation and analysis. The specifcs of the OLYMPUS radiative corrections
procedure are described in detail in chapter \ref{chap:rc}.

\subsection{Previous Experiments}

\begin{figure}[htbp]
\centering
\includegraphics{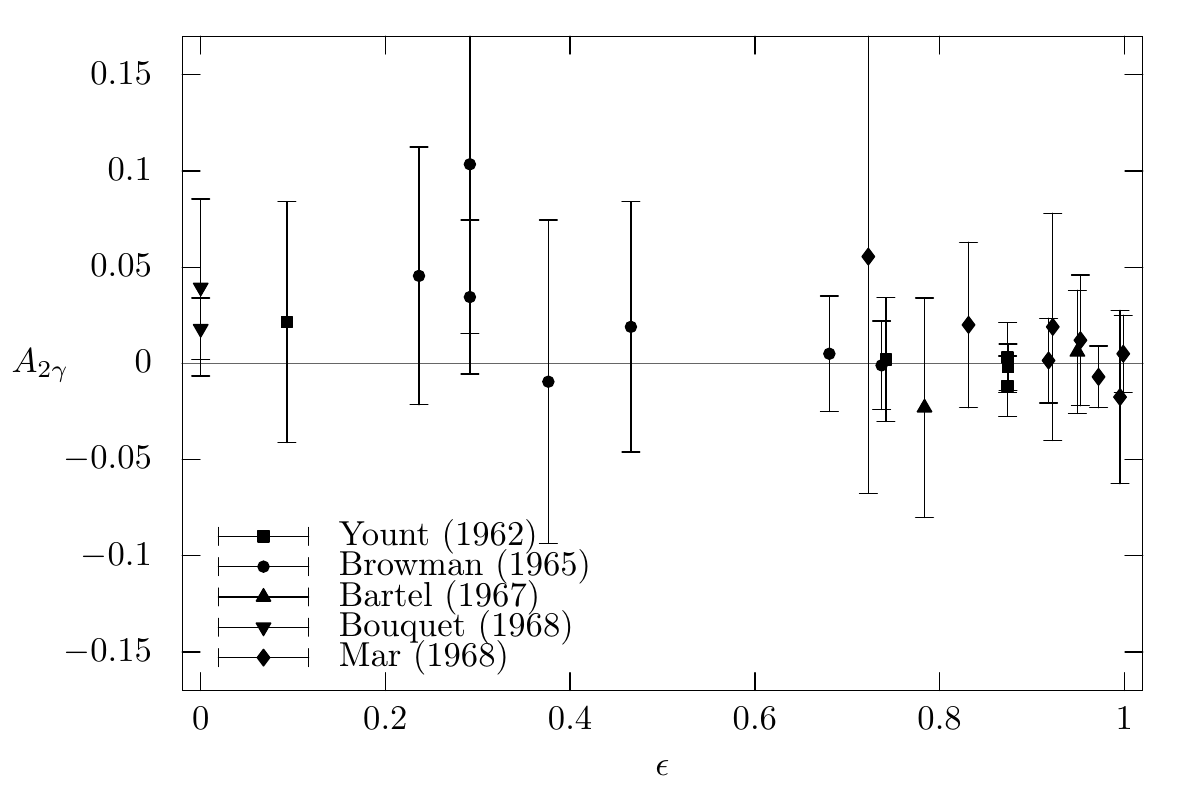}
\includegraphics{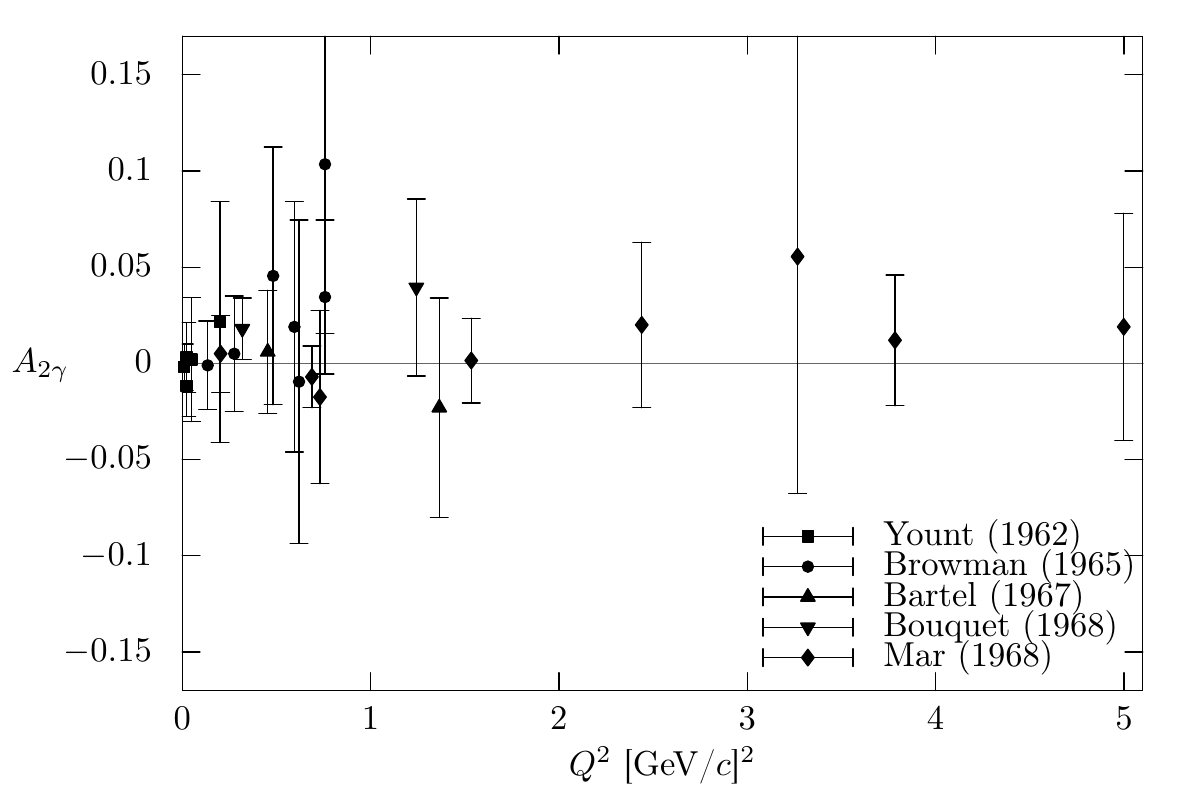}
\caption[Previous asymmetry data, projected by $\epsilon$ and $Q^2$]
{\label{fig:1d}Previous asymmetry data, all from the 1960s, lack the precision
to make any significant claims about two-photon exchange.}
\end{figure}

\begin{figure}[htbp]
\centering
\includegraphics[clip=true,trim=0 3cm 0 2cm]{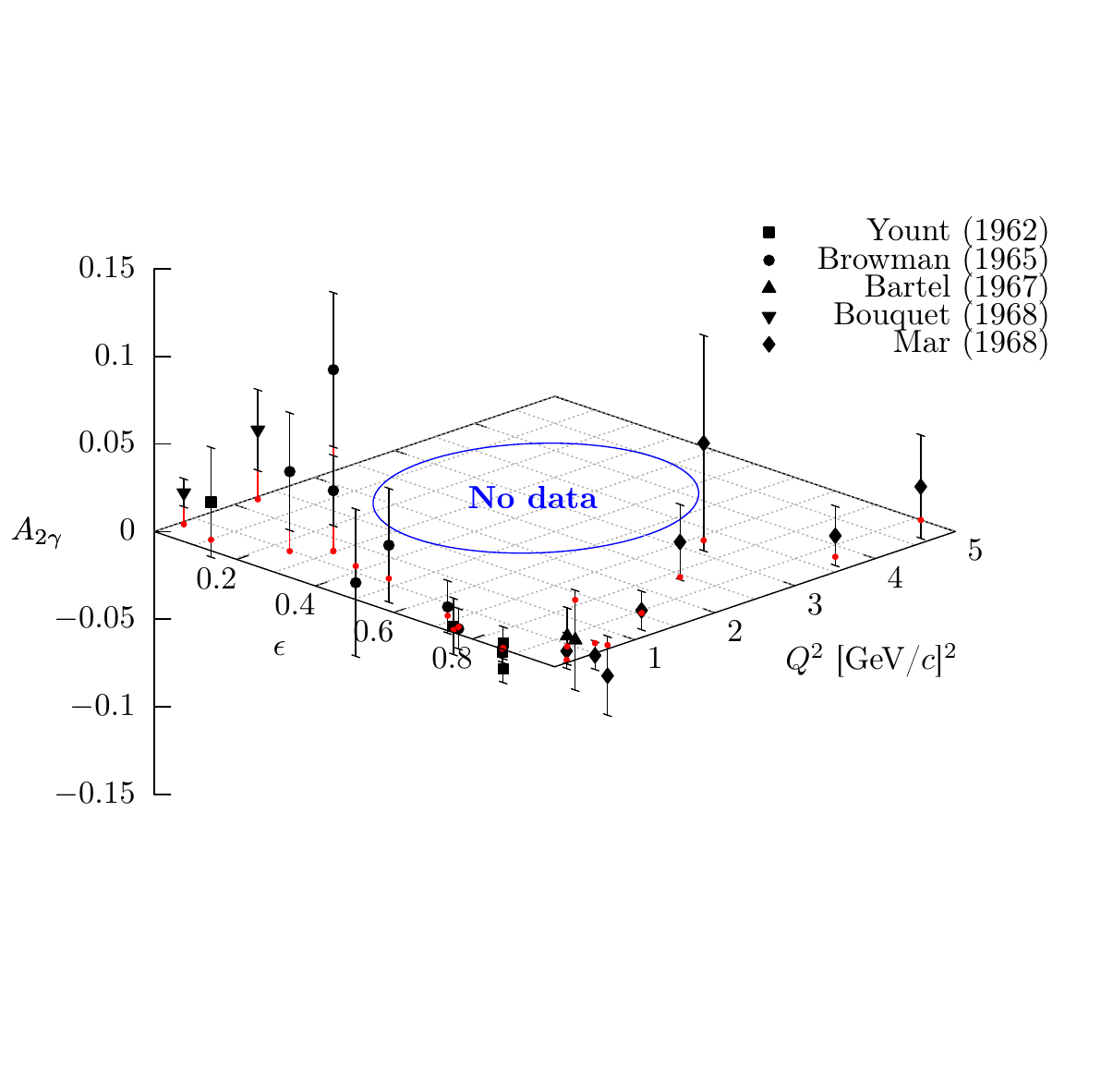}
\caption[Three-dimensional representation of previous asymmetry data]
{\label{fig:2d}Previous asymmetry measurements were either made at high $\epsilon$ where
the asymmetry approaches zero, or at low $Q^2$ where the form factor discrepancy 
is small. The interesting kinematic region, at low $\epsilon$ and high $Q^2$ has
gone unexplored.}
\end{figure}

A handful of experiments comparing the cross sections of positron-proton to electron-proton
elastic scattering were conducted in the 1960s \cite{yount:1962aa,browman:1965zz,anderson:1966zzf,bartel:1967aa,
cassiday:1967aa, bouquet:1968aa, mar:1968qd}, and their data are shown in the plots in figures \ref{fig:1d}
and \ref{fig:2d}. Some of these results were published in the form of a cross section 
ratio:
\[
R_{2\gamma} = \frac{\sigma_{e^+p}}{\sigma_{e^-p}} \approx 1 + 2A_{2\gamma},
\]
but for the sake of consistency, I present all of the results re-worked as asymmetries. 

In figure \ref{fig:1d}, the data are plotted with respect to $\epsilon$ (upper plot), 
and with respect to $Q^2$ (lower plot). Most of the measurements have uncertainties
on the order of 5--10\%, making them unable to distinguish between the null hypothesis
(hard two-photon exchange is negligible, $A_{2\gamma}=0$) and the alternative hypothesis (hard
two-photon exchange resolves the discrepancy: $A_{2\gamma}$ rises several percent with 
decreasing $\epsilon$ and increasing $Q^2$). If anything is to be concluded experimentally
about hard two-photon exchange, better data are needed.

A further weakness of these data can be clearly seen by plotting the data over the 
two-dimension $\epsilon, Q^2$ plane, shown in figure \ref{fig:2d}. The specific value
of $Q^2$ and $\epsilon$ for each data point is marked with a red dot on the plane. 
The asymmetry value of each point is shown with a black point using the vertical axis.
The data largely fall into two groups. Measurements were either made at high values
of $\epsilon$, where the asymmetry is theoretically constrained to be close to zero,
or the data were measured at small values of $Q^2$, where the form factor discrepancy
is small. The kinematic region relevant for two photon exchange---where $\epsilon$ is 
low and $Q^2$ is high---is not covered. New experiments are needed to push into this
region.

\subsection{Contemporary Experiments}

\label{ssec:new_exp}

\begin{figure}[htpb]
\centering
\includegraphics{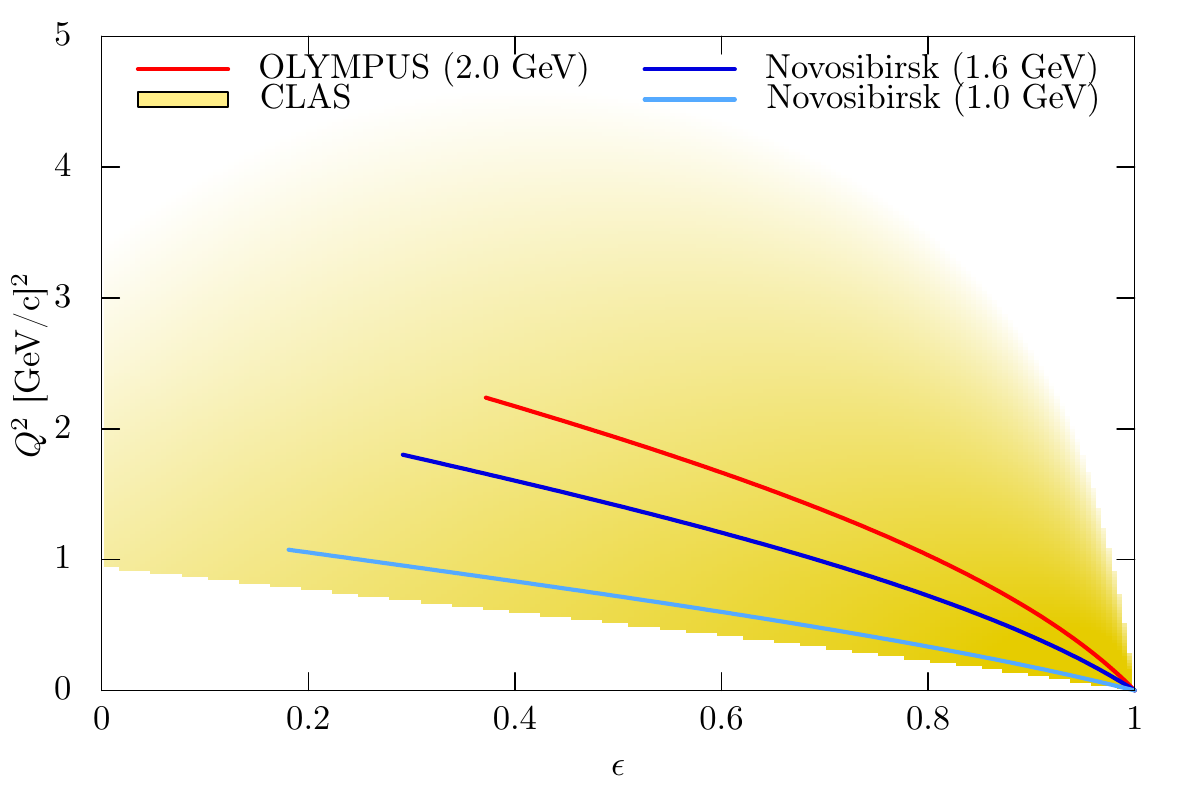}
\caption[Kinematic reach of new asymmetry measurements]{\label{fig:reach} The three contemporary asymmetry experiments push into
the high $Q^2$, low $\epsilon$ kinematic region. The CLAS experiment does not have a
fixed beam energy, so its kinematic reach extends over a large area on the kinematic
plane. The shading approximately represents the log of the event rate in CLAS.}
\end{figure}

The emergence of the form factor discrepancy, as well as the inadequacy of previous
data, prompted the design of three new experiments to measure the lepton sign asymmetry.
In addition to OLYMPUS were experiments that collected data in Novosibirsk, Russia, and
at Jefferson Lab, in the United States. Both will be described in this section. All three 
experiments push into the low $\epsilon$, high $Q^2$ kinematic region, though all have 
slightly different coverage, as seen in figure \ref{fig:reach}. The experiments also use 
different techniques, reducing the susceptibility to a single common systematic error. 
The results of all three experiments will be discussed in chapter \ref{chap:results}.

\subsubsection{Experiment at VEPP-3, Novosibirsk}

The Novosibirsk experiment collected data in two runs in 2009 and 2012 at the
VEPP-3 storage ring in Novosibirsk, Russia \cite{Rachek:2014fam}. 
Alternating electron and positron beams were directed 
through an internal gaseous hydrogen target. Above and below the target
were wide-angle non-magnetic spectrometers, consisting of drift chambers
for tracking, scintillators for triggering, and NaI crystals for energy
measurements. The lepton and proton were detected in coincidence. Combining
both runs, the dataset contains almost 1~fb$^{-1}$ of integrated luminosity.

The strength of this experiment was its non-magnetic design, ensuring
both species of lepton had identical acceptances. This came at the cost
of having no momentum analysis for background rejection. 
A weakness of this experiment was that it depended on forward-angle
electron-proton scattering for its luminosity monitoring. Since
its luminosity monitoring was susceptible to two-photon exchange
effects, the Novosibirsk results quote the asymmetry as being
relative to that of the ``luminosity normalization point'' (LNP).
The LNP is at high $\epsilon$ where the lepton sign asymmetry 
should be close to zero. However, this normalization introduces
some degree of uncertainty that makes their results less able
to reject hypotheses.

The results from both runs were published in 2014 \cite{Rachek:2014fam}. 

\subsubsection{CLAS Two-Photon Experiment, Jefferson Lab}

The CLAS two-photon experiment at Jefferson Lab (E-07-005) used a much more exotic technique. 
To avoid the problem of luminosity normalization, they scattered a beam of $e^+e^-$ 
pairs on a liquid hydrogen target and collected electron scattering data
and positron scattering data at the same time. To produce the beam, they
directed the 6~GeV electron beam from the CEBAF accelerator against
a gold foil, producing a secondary beam of bremsstrahlung photons.
This photon beam was directed against a second gold foil to produce
a tertiary beam of $e^+e^-$ pairs. A tungsten shield blocked any
remaining photons, while the electrons and positrons were steered
around it by means of a magnetic chicane. The fluxes of electrons and
positrons on the target were matched by periodically flipping the 
chicane polarity.

The experiment used the CEBAF Large Acceptance Spectrometer (CLAS) in
Jefferson Lab's Hall B for its detector. CLAS is a toroidal magnetic
spectrometer that has nearly full azimuthal coverage. CLAS has drift
chambers for particle tracking inside a layer of scintillator, which
measures time-of-flight. A forward electron calorimeter was used for
triggering, but not for energy measurements in this experiment.
Leptons and protons were detected in coincidence.

A side effect of the pair-produced beam was that the incoming lepton energy
was not fixed, but rather varied between several hundred MeV and up to 
5.5~GeV. The incoming lepton energy was reconstructed kinematically in
each event. Whereas OLYMPUS and the Novosibirsk experiment sample lines 
of fixed beam energy in the $\epsilon,Q^2$ plane, the CLAS experiment samples an
area over that plane. Furthermore, the sign of the lepton in any given event was 
not predetermined. The identification of positrons versus electrons was made 
using the curvature direction of the tracks. 

The strength of this experimental approach is the novel method for
handling the problem of luminosity normalization. The weaknesses are
the low luminosity afforded by a tertiary beam, and from the systematics
associated with the various magnetic fields used in their apparatus.
The CLAS two-photon collaboration has not yet reported estimates of the
luminosity they collected, but one can estimate, based on the sizes
of their statistical uncertainties on their results, that they collected
on the order of hundreds of inverse picobarns. Large bins were needed
to accumulate enough statistics. Magnetic fields are always a potential 
source of asymmetry between electrons and positrons, and to limit any 
false asymmetry, CLAS relied on a complicated protocol
for flipping the magnet polarities. Not all magnet polarities could 
be changed: a mini-torus used for sweeping background from M\o ller and
Bhabha scattering went unflipped. Furthermore, not all of the polarity
configurations produced usable data.
 
The CLAS two-photon experiment collected data in a proof-of-concept run in 2006
\cite{Moteabbed:2013isu}, followed by a full production run at the end of
2010 and beginning of 2011. Preliminary results have been published \cite{Adikaram:2014ykv},
and full results have been submitted for publication \cite{Rimal:2016toz}.

\chapter{Apparatus}

\label{chap:apparatus}

\section{Introduction}

The apparatus of the OLYMPUS experiment was a large-acceptance magnetic
spectrometer. Much of this spectrometer was previously used in the
BLAST experiment, at MIT-Bates. Some components of BLAST were not
needed for OLYMPUS (neutron detectors, for example), and some new
components were built specifically for OLYMPUS (most notably, the 
target and luminosity monitors). OLYMPUS was conducted at the DESY laboratory,
in Hamburg, Germany, which meant that the BLAST components had to
be shipped from Massachusetts to Hamburg. Shipments arrived in
Germany in 2010, and the spectrometer was assembled in a staging
area near the storage ring. In July, 2011, the spectrometer was
rolled on rails into place, and the ring beam pipe was reconnected. 
Two periods of dedicated data taking took place. The first was a 
month of running over February of 2012. The second period was 
three months of running in the fall of 2012. Cosmic ray data
was taken for the first month of 2013. Then the detector and
magnetic field was surveyed. In the spring of 2013, the 
detector was disassembled, and many components were recycled into
new experiments.

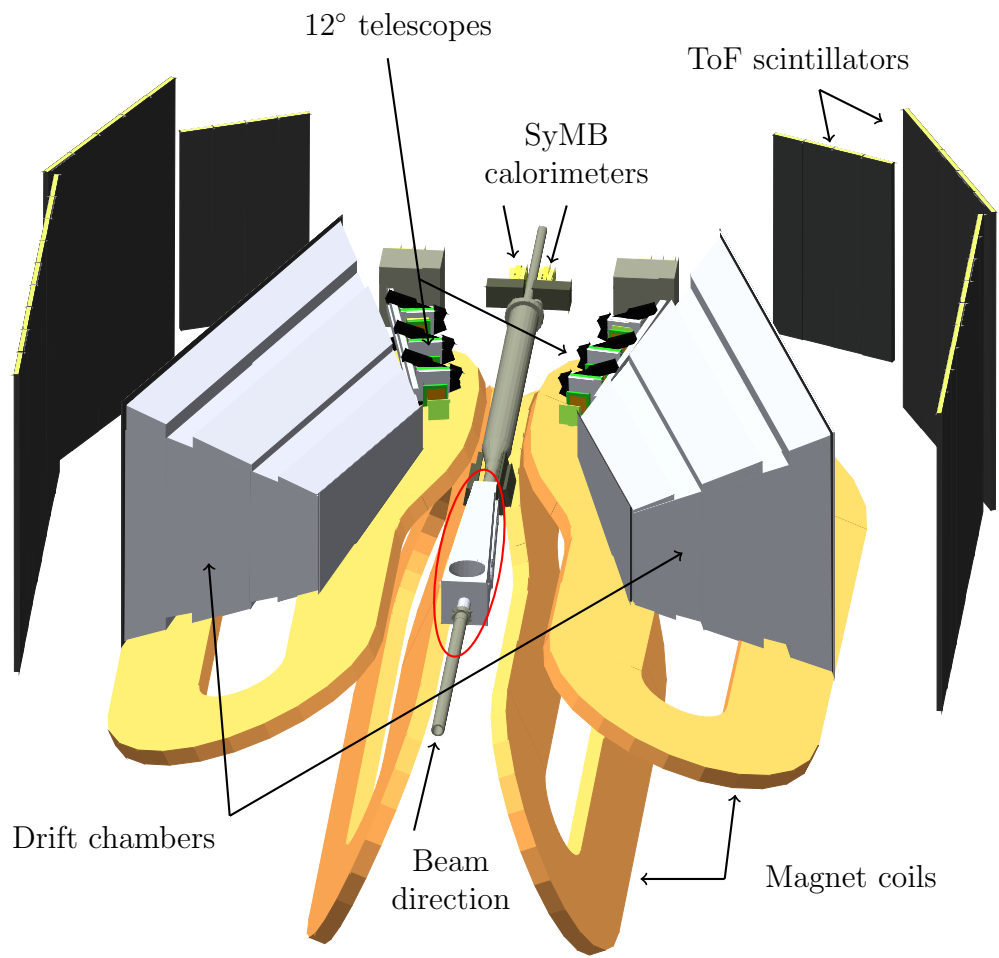
\begin{figure}
  \centering
  \begin{tikzpicture}[scale=1.4]
    \node[anchor=south west,inner sep=0] (image) at (0,0)
         {
           \includegraphics[clip=true,trim=0cm 4cm 0cm 6cm,width=14cm]{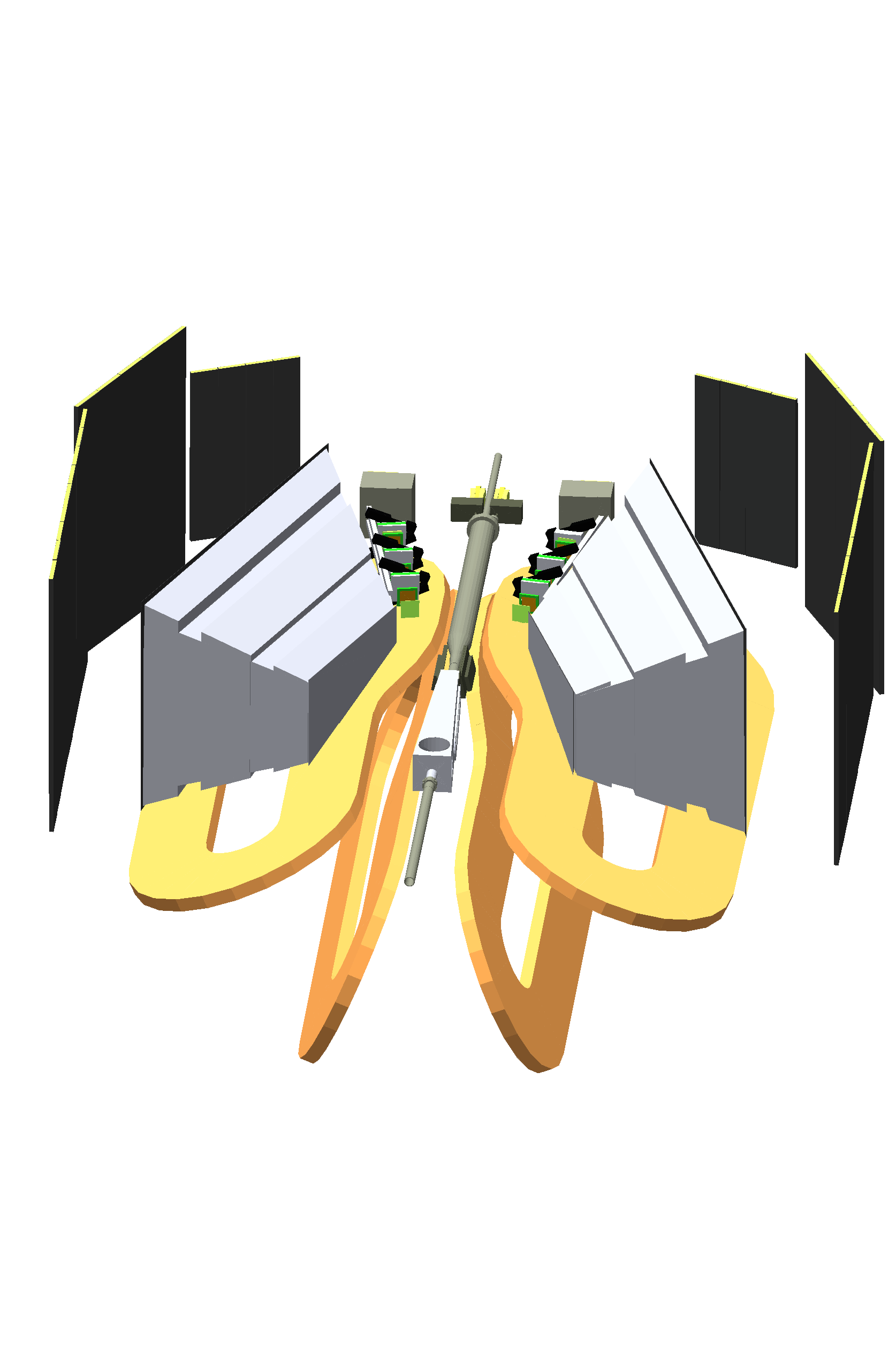}
         };
         \node[align=center,black] at (4.7cm,1.5cm) {Beam\\direction};
         \node[align=center,black] at (8.5cm,1.5cm) {Magnet coils};
         \node[align=center,black] at (1.5cm,1.9cm) {Drift chambers};
         \node[align=center,black] at (8.0cm,9.3cm) {ToF scintillators};
         \node[align=center,black] at (4.2cm,9.6cm) {$12^\circ$ telescopes};
         \node[align=center,black] at (5.8cm,8.4cm) {SyMB\\calorimeters};

         \draw[arrows=->,draw=black,thick] (4.35cm,1.96cm) -- (4.55cm,2.8cm) node[midway,above]{};

         \draw[arrows=->,draw=black,thick] (5.2cm,7.9cm) -- (5.3cm,7.4cm) node[midway,above]{};
         \draw[arrows=->,draw=black,thick] (5.8cm,7.9cm) -- (5.65cm,7.35cm) node[midway,above]{};

         \draw[arrows=->,draw=black,thick] (8.2cm,9.0cm) -- (8.3cm,8.5cm) node[midway,above]{};
         \draw[arrows=->,draw=black,thick] (8.2cm,9.0cm) -- (8.8cm,8.7cm) node[midway,above]{};

         \draw[arrows=->,draw=black,thick] (2.6cm,2.1cm) -- (2.4cm,4.5cm) node[midway,above]{};
         \draw[arrows=->,draw=black,thick] (2.6cm,2.1cm) -- (6.9cm,4.6cm) node[midway,above]{};

         \draw[arrows=->,draw=black,thick] (7.3cm,1.5cm) -- (6.5cm,1.5cm) node[midway,above]{};
         \draw[arrows=->,draw=black,thick] (7.3cm,1.5cm) -- (7.4cm,2.3cm) node[midway,above]{};

         \draw[arrows=->,draw=black,thick] (4.1cm,9.3cm) -- (4.49cm,6.57cm) node[midway,above]{};
         \draw[arrows=->,draw=black,thick] (4.4cm,7.2cm) -- (5.8cm,6.5cm) node[midway,above]{};

         \draw[draw=red,thick,rotate around={-10:(4.9cm,4.7cm)}] (4.91cm,4.5cm) ellipse (0.3cm and .9cm);

  \end{tikzpicture}
\caption[The OLYMPUS spectrometer]{\label{fig:dawn_labels} This illustration shows the detectors that make up the OLYMPUS 
spectrometer. The top four magnet coils are not shown. The scattering chamber is marked by the red oval.}
\end{figure}

This chapter will describe the various components of the OLYMPUS
apparatus as they were used in data collection. An illustration
of the spectrometer with the top four magnet coils removed can
be found in figure \ref{fig:dawn_labels}. Since much of the
apparatus was designed for the BLAST experiment, very little of 
this chapter contains original design work on my part, although
I will make a note where it occurs. Instead, this chapter presents
the germane information for understanding how the detectors 
functioned in order to make the lepton sign asymmetry measurement. 
More detailed information on the apparatus can be found in a couple 
of published sources. The BLAST technical design report \cite{BLAST:TDR} 
contains a number of specific technical details, but one should 
exercise caution because it presents a proposal for BLAST, not what 
was actually built. A more concise article describes BLAST as it was 
realized \cite{Hasell:2009zza}. Valuable technical details can be 
found in a several of the early PhD theses of BLAST students
\cite{BLAST:crawford, BLAST:maschinot, BLAST:zhang}. 
Finally, at the time of this writing, there have been several articles 
published with concise technical descriptions of the OLYMPUS apparatus
\cite{Milner:2013daa} and several of its components \cite{Bernauer:2014pva, Benito:2016cmp, Bernauer:2016hpu}.

\section{Accelerator}

\begin{figure}[htpb]
\centering
\includegraphics{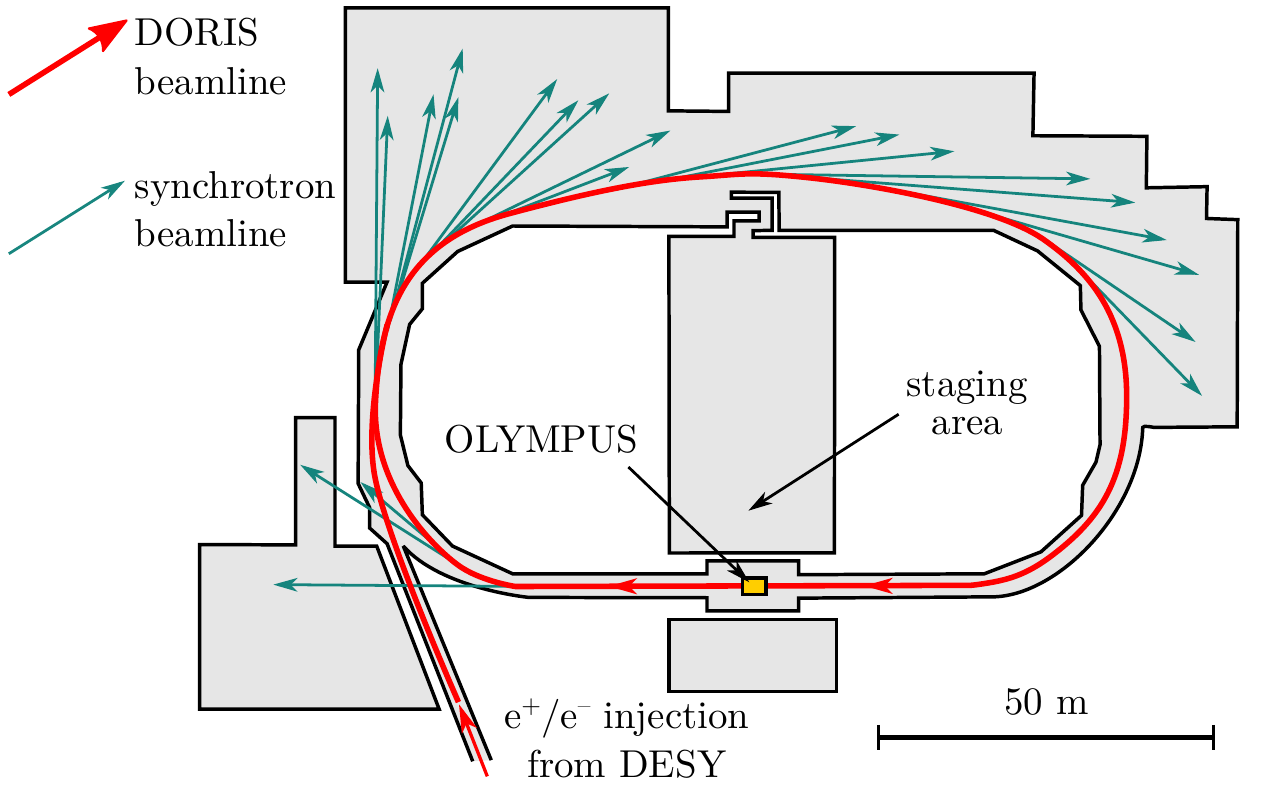}
\caption[The DORIS storage ring]{\label{fig:doris} OLYMPUS was assembled in a staging
area, and then rolled into place in the south hall of the DORIS
ring.}
\end{figure}

The OLYMPUS experiment was conducted at the DORIS storage ring at
DESY, in Hamburg, Germany. The layout of the DORIS ring is shown in
figure \ref{fig:doris}. DORIS is a German acronym for 
\emph{DOppel-RIng-Speicher}, meaning double-ring store. As its
name implies, DORIS was once an $e^+e^-$ collider, but it was
reconfigured several times during its lifetime. In the time leading 
up to OLYMPUS, DORIS operated primarily as a synchrotron light source. 
Wiggler magnets and numerous photon beamlines had been installed in 
curved sections of the ring. In typical running a 4.5~GeV, 140~mA,
positron beam was stored in five bunches, providing a steady source 
of synchrotron X-rays in the photon beam lines. OLYMPUS, by contrast,
ran with alternating electron and positron beams of 2.01~GeV and
40--65~mA stored in ten bunches. Running with a greater number of 
bunches was preferable since it reduced the maximum instantaneous
current and reduced the number of random coincidences per 
unit of luminosity. The current was chosen as a trade-off between
luminosity and background rates. A table of the properties of the
DORIS ring for OLYMPUS running is shown in table \ref{table:doris}.

\begin{table}
\centering
\begin{tabular}{ | l | l | }
\hline
Parameter & Value \\
\hline
\hline
Ring circumference & 289.193 m \\
Revolution frequency & 1.036652 kHz \\
Beam energy & 2.01 GeV \\
Beam current & $\approx$ 65 mA \\
Bending radius & 12.18 m \\
Horizontal emittance & 200 nm$\cdot$rad\\
Vertical emittance & 5 nm$\cdot$rad\\
RF frequency & 499.6665 MHz \\
Harmonic number & 482 \\
Bunch spacing & 100, 100, 100, 100, 96 ns\\
\hline
\end{tabular}
\caption[DORIS beam parameters]{\label{table:doris} These were the parameters for the DORIS ring during OLYMPUS running.}
\end{table}

The DORIS ring could be configured to store either electrons or
positrons, and switching between these configurations took 
approximately one hour. OLYMPUS took data alternating between
electron and positron beams, switching between the beam species
daily. Counter-intuitively, positron running tended to be more 
stable. The positive beam repelled positive ions from residual
gas in the beam pipe, whereas the electron beam attracted these
ions and scattered from them at higher rates. For this reason
DORIS operated with positrons when running as a synchrotron
light source. Probably the higher reliability for positron running
was in part also due to the accelerator operators being more
practiced at correcting instabilities in positron mode.
Whatever the cause, electron running had perceptibly more frequent 
high-voltage trips and beam losses. 

During the injection of particles into DORIS, the OLYMPUS 
detectors saw noticably higher rates. For this reason, during 
injection, a signal was sent to the OLYMPUS data acquisition
system (DAQ) to inhibit data taking. Though there were occasional
high voltage trips, for the most part the detectors could remain 
in their powered state through injection. During the February 
running period, the ring was filled to 65~mA, and then data 
were collected as the current decayed to 40~mA, at which point
the ring was refilled. In the fall run, DORIS was operated
in ``top-up'' mode, in which the ring was filled to 65~mA,
and then replenished every few minutes to maintain as constant
a beam current as possible.

\section{Target}

A new target system and accompanying beam pipe were designed and
built specifically for OLYMPUS and have been described in a recent
article \cite{Bernauer:2014pva}. Much of the scientific program at
BLAST required spin-polarized targets, so cryogenics, polarimetry,
and spin transport were the primary technical concerns. OLYMPUS had
no need of spin polarization, but did have to contend with the 
high bunch-charge in the DORIS bunches, so wake-field suppression
and thermal conductivity were significant factors in the target
design.

\begin{figure}[htpb]
\centering
\includegraphics{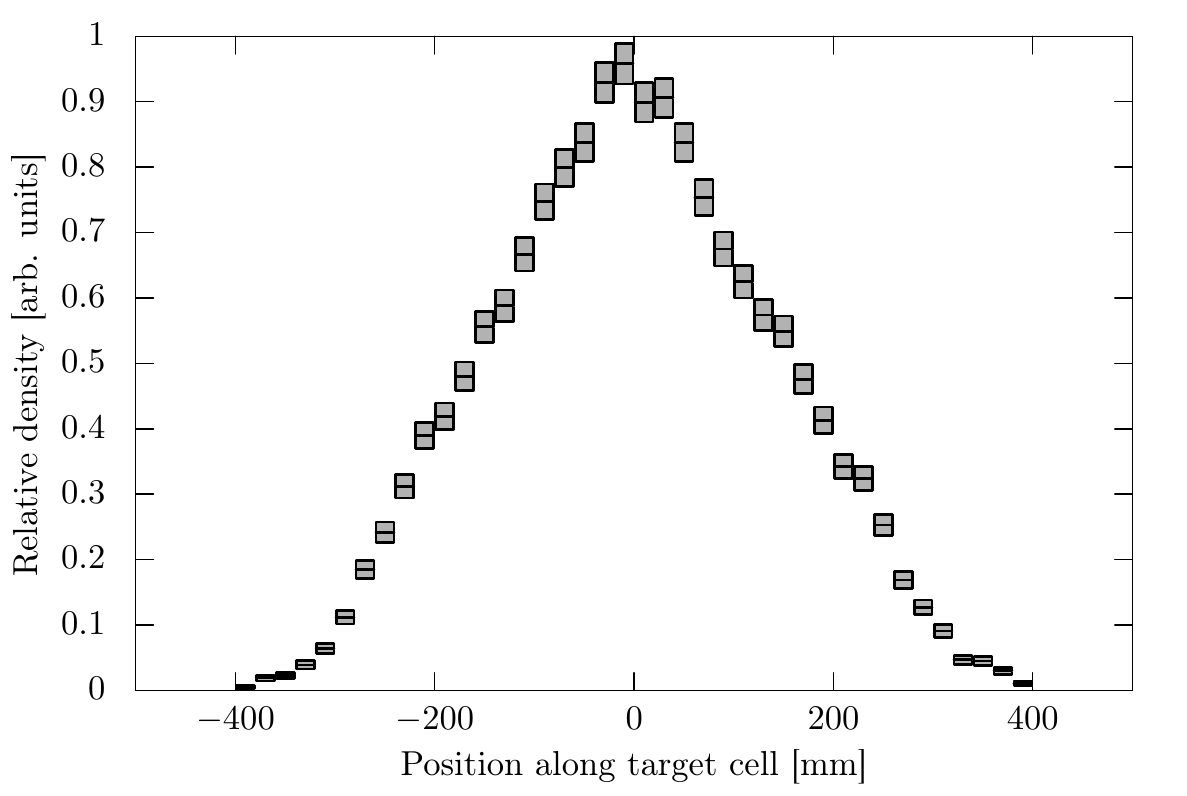}
\caption[Target density distribution]{\label{fig:target_triangle} The density profile along the
length of the target, as reconstructed from tracks in data, shows
the characteristic triangular distribution.}
\end{figure}

The OLYMPUS target was windowless and internal to the DORIS beamline
vacuum. The target itself was a thin aluminum cell, 60~cm long, open at both
ends, held inside the beamline vacuum. The beam was steered to pass
down the axis of the cell. Hydrogen gas was fed into the 
cell through a small opening at its midpoint. The gas diffused to the
ends of the cell, after which it was removed from the beamline by a 
system of six turbomolecular vacuum pumps. This windowless design
produced a characteristic triangular density profile, shown in figure
\ref{fig:target_triangle}.

\begin{figure}[htpb]
\centering
\includegraphics[width=12cm]{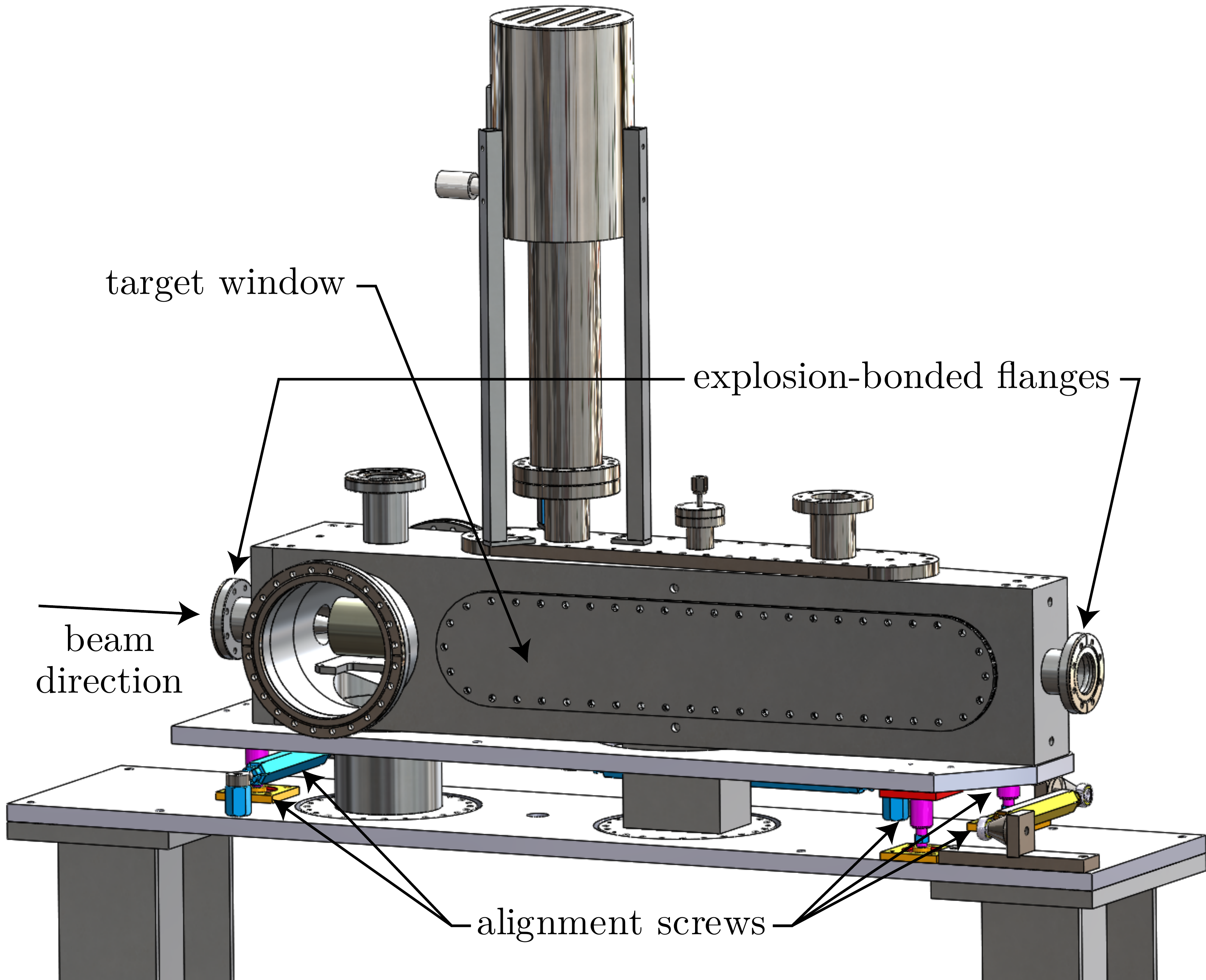}\\
\includegraphics[width=12cm]{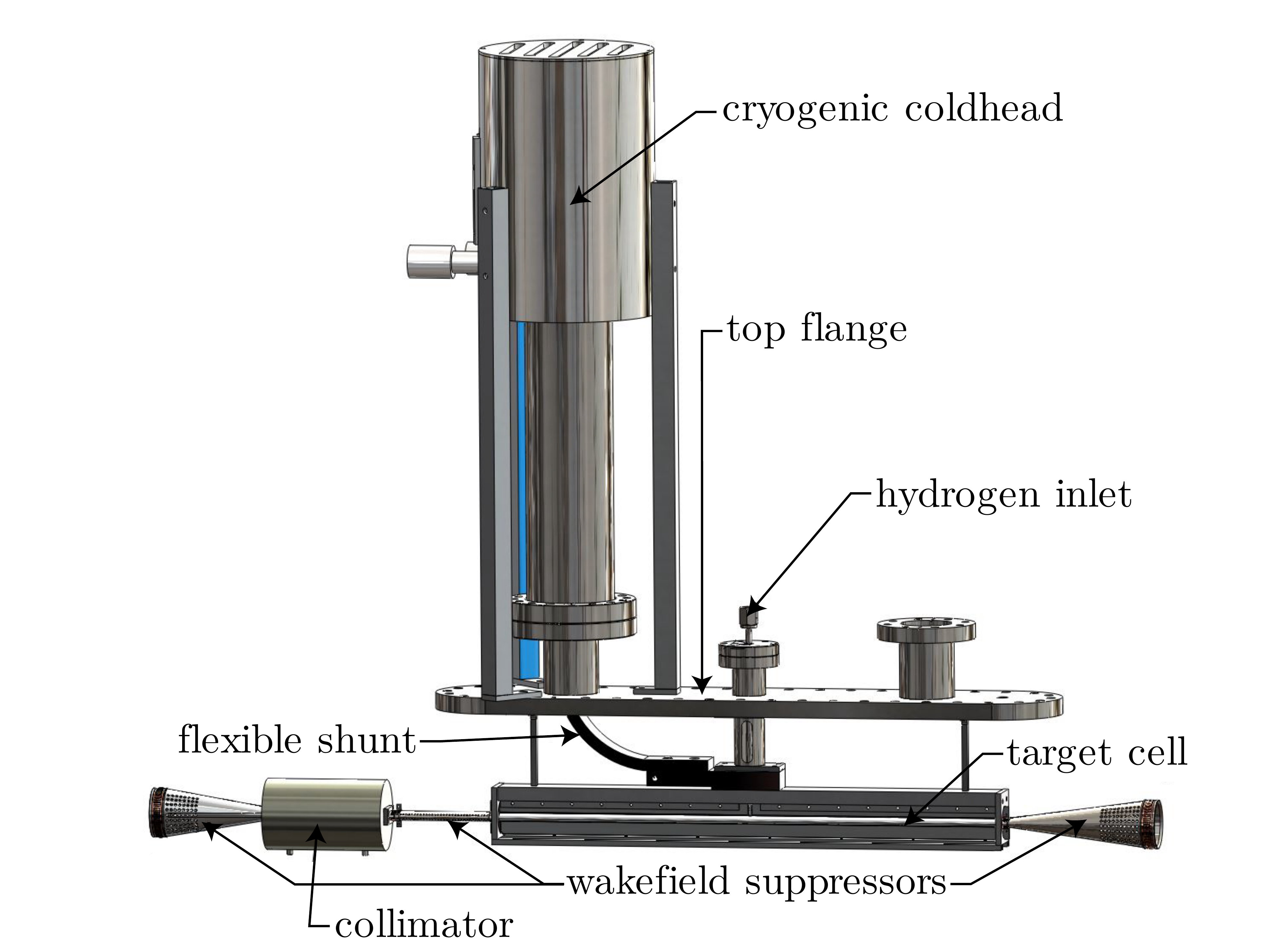}
\caption[Target schematic]{\label{fig:target_chamber} The target cell, collimator, and wakefield
suppressors were placed in the vacuum of the accelerator, inside a specially
built scattering chamber.}
\end{figure}

The target was held inside a specially built scattering chamber,
which maintained the integrity of the beamline vacuum, but also
had thin aluminum windows on either side, to allow scattered 
particles to pass through without having to traverse a lot of
material. An illustration of the scattering chamber and its contents
is shown in figure \ref{fig:target_chamber}. The front and back faces had flanges to connect with
the up- and down-stream beam pipes. The bottom face had ports
for the vacuum pumps, and the top faces had ports for the
hydrogen gas, for various sensors, and for a cryogenic coldhead. 
The coldhead was used to cool the target cell to cryogenic
temperatures (in practice, approximately 70~K), and in turn
cool the hydrogen gas inside, increasing its density. OLYMPUS
aimed to achieve a target density of $3\times 10^{15}$~atoms/cm$^2$
without significantly degrading the beam.

A tungsten collimator, 15~cm long and 10~cm in diameter, was
positioned just upstream of the target cell in order to
protect the thin cell from beam-halo particles and from
synchrotron radiation. The target aperture was slightly 
smaller than the target cell, and was slightly flared
downstream to reduce showering and small-angle scattering. 
The outer dimensions were chosen based on the results of 
simulated beam-halo showers.

The collimator and target cell were connected to each
other and to the upstream and downstream beampipes 
with conductive metal tubes to suppress wake-fields. 
Since any sharp transitions between conducting surfaces
could be a source of induced wake-fields, these wake-field
suppressors were designed to provide continuous electrical
conductivity through the entire length of the scattering
chamber. The first wake-field suppressor was shaped like a 
cone, and joined the upstream beamline port of the scattering 
chamber to the upstream face of the collimator. The second
was an elliptical tube that connected the downstream aperture
of the collimator to the upstream end of the target cell.
The third connected the downstream end of the target to the 
downstream beamline port of the scattering chamber. The wakefield
suppressors had small holes to allow hydrogen gas to pass through
them, allowing more efficient vacuum pumping.

The three meters of beamline downstream of the target were
also specially designed for OLYMPUS. In addition to large
diameter ports for turbomolecular vacuum pumps, the beamline
was widened to transmit symmetric M\o ller and Bhabha leptons
to the SyMB calorimeters. In front of the calorimeter apertures,
the wide beamline had a cap with three ports. The center port
connected to the narrow downstream beamline to circulate the 
stored beam. The left and right port had thin aluminum windows
to allow symmetric M\o ller and Bhabha electrons to exit 
the vacuum pipe and enter the calorimeters. This part of the beamline
is relevant to the discussion of the luminosity analysis presented
in chapter \ref{chap:lumi}, and is shown there as figure
\ref{fig:bl_schematic}.

The target system had a test run at DORIS in early 2011. 
The cryogenic and vacuum systems performed well under 
beam conditions that were to be used for OLYMPUS running.
After the test concluded, the target was left in place, 
without hydrogen gas, for DORIS synchrotron operations. The
target could not withstand the higher current synchrotron
conditions and had to be removed. It was found that different
parts of the target cell frame were made of different types
of aluminum, and expanded differently under thermal stress. 
This caused the cell and frame to deform, which broke the 
electrical contact of the wake-field suppressors. Subsequent
wake-field heating damaged the target cell, and a new cell
had to be produced. In this second iteration, 6063 aluminum
was used throughout, and the wakefield suppressors were 
redesigned. The wakefield suppressor connectors were made
more robust, and the holes in the wakefield suppressors,
necessary for removing hydrogen gas from the target cell,
were reduced in number and moved to surfaces that were
farther from the beam.

After the conclusion of the February running period,
a gap was discovered between the tube that supplied the
hydrogen gas to the cell, and the cell inlet itself. Analysis
of the luminosity monitor data indicated that only 
approximately one eighth of the hydrogen was entering
the cell; the rest entered the scattering chamber and
was removed by the vacuum system. Consequently, the 
February running period only had one eighth of the expected
luminosity. In the summer of 2012, the inlet joint was
redesigned to eliminate this gap, and a new target was
produced for the subsequent Fall running period. Luminosity
monitor data indicated that the Fall target had 
the desired density and that no gas was leaking from
the inlet.

\section{Magnet}

\label{sec:app:magnet}

\begin{figure}[htpb]
\centering
\includegraphics[clip=true,trim=0cm 4cm 0cm 6cm,width=12cm]{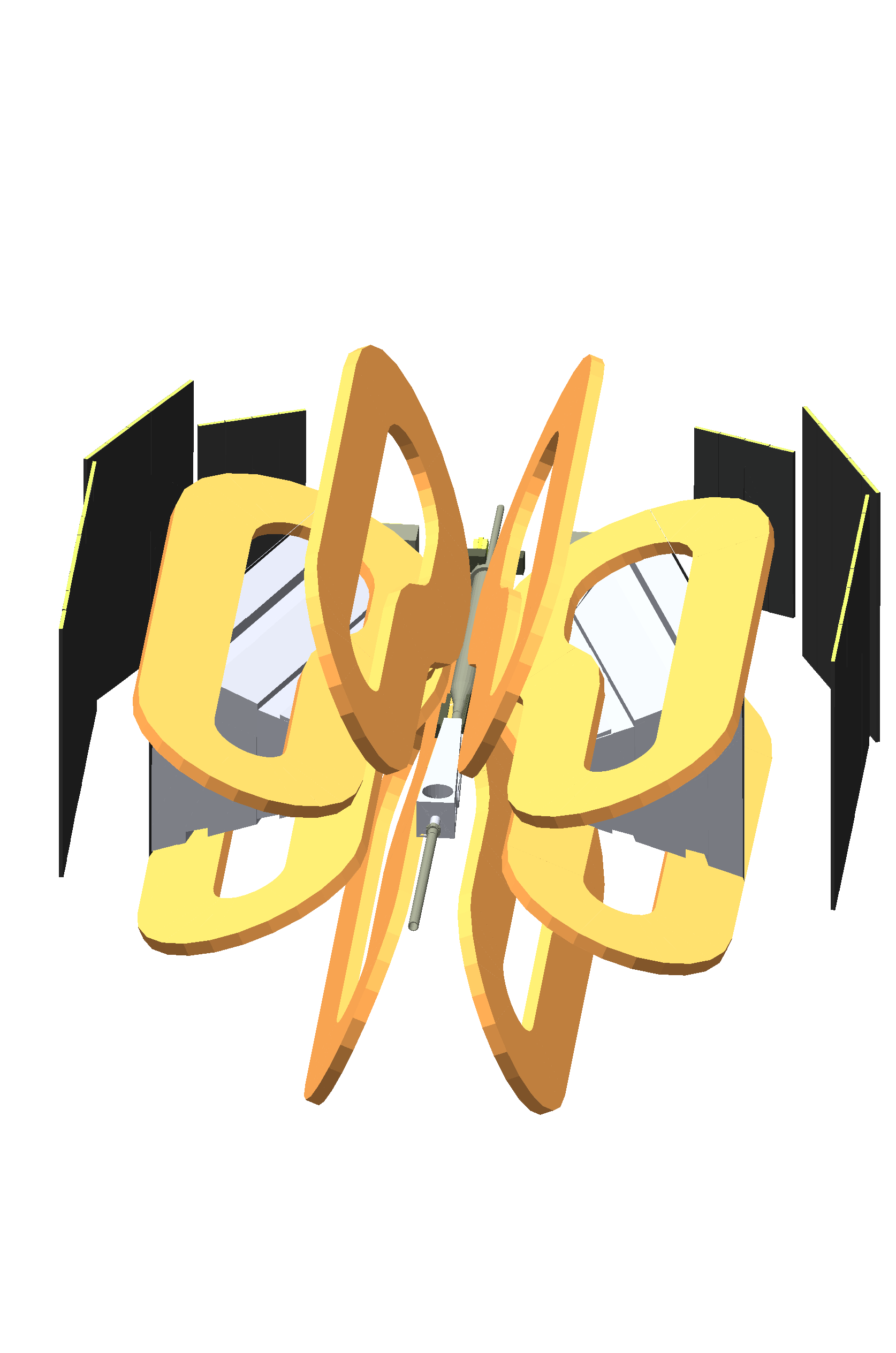}
\caption[The OLYMPUS spectrometer, showing all eight magnet coils]
{\label{fig:dawn_full} The eight coils of the OLYMPUS magnet divided the volume
around the beamline into eight sectors, shaped like wedges in an orange.}
\end{figure}

The magnetic field of the OLYMPUS spectrometer was created by eight electromagnetic
coils, arranged like the segments of an orange around the beamline. These coils produced
a toroidal magnetic field in the region of the tracking detectors. By measuring the 
curvature of their tracks, the momentum of charged particles could be inferred. The
OLYMPUS magnet was previously used as the BLAST spectrometer magnet, where it was
previously described \cite{Dow:2009zz}.

\begin{figure}[htpb]
\centering
\includegraphics[width=12cm]{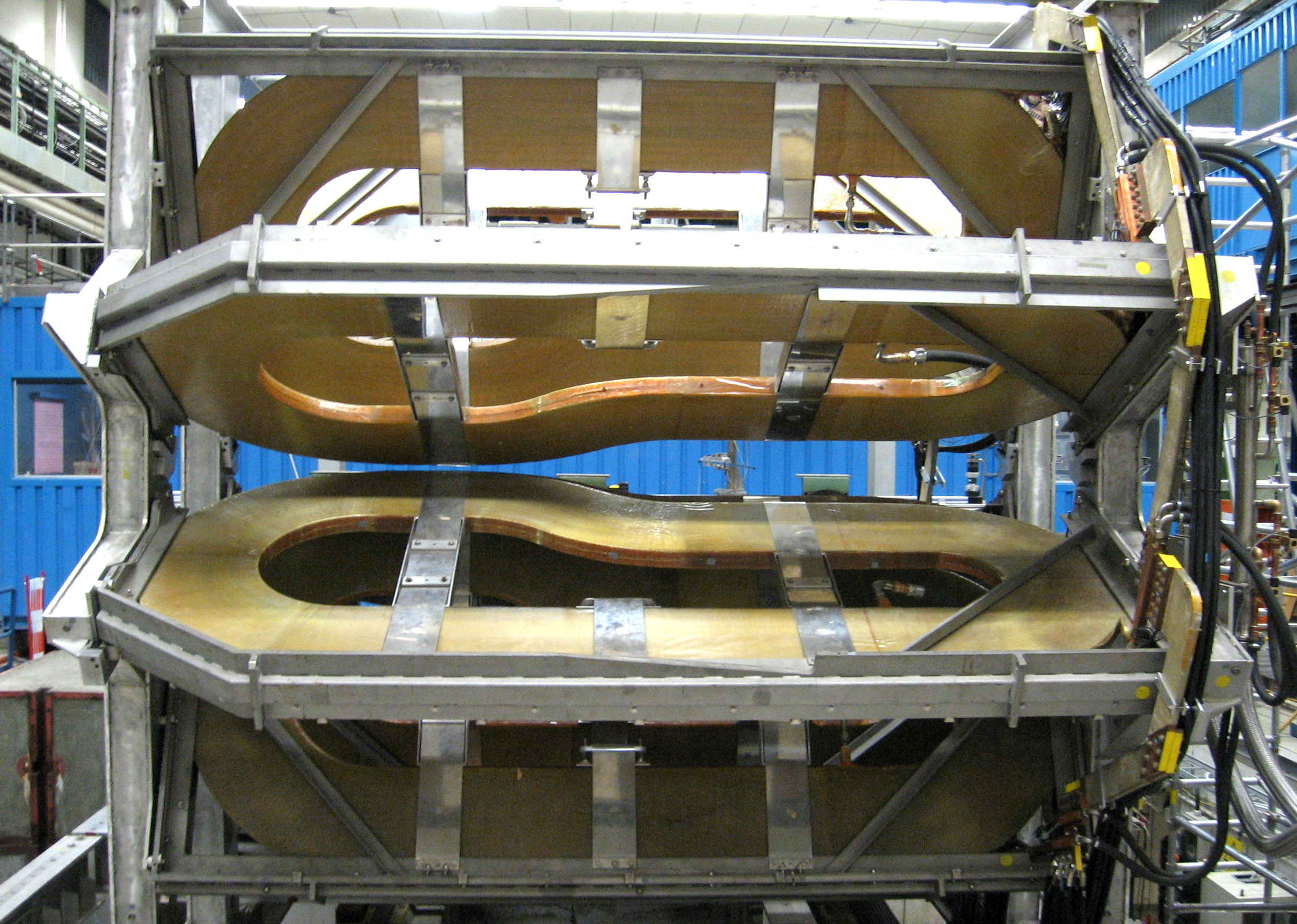}
\caption[Photograph of the magnet coils]{\label{fig:magnet_side_photo} This photograph shows the magnet coils
from the side, prior to the installation of any of the detectors.}
\end{figure}

The OLYMPUS coils were made of hollow copper bars, wound 26 times
in two layers of thirteen windings. Each bar was wrapped in fiberglass
and then the entire coil was potted in a rigid resin to make the 
assembly strong enough to resist the strain experienced from magnetic 
forces. The coils were conventionally conducting, so resistive heat 
needed to be removed. This was accomplished by flowing cooling water 
through the hollow recess of each bar. The coils had an irregular
shape; the coils were narrower upstream to accommodate the scattering
chamber, and then expanded downstream to produce a higher field
at forward scattering angles. A photograph of the coils, prior to
the installation of the OLYMPUS detectors, is shown in figure
\ref{fig:magnet_side_photo}.

The coils were arranged to produce a toroidal field. The beam passed
through the central axis of the toroid, with the target slightly
upstream of the center. The field was azimuthally symmetric, which
mirrored the azimuthal symmetry of the cross section. The 8 coils
divided the region around the target into eight wedge-shaped sectors,
each roughly 45$^\circ$ wide. Two of these sectors, the one left and
the one right of the target in the horizontal plane, were instrumented 
with detectors. 

At BLAST, the magnet was operated with a current of 6730~A, producing
a peak field of approximately 0.4~T. At OLYMPUS, the magnet was operated at 
5000~A, producing a peak field of 0.3~T. This naively corresponds to 
a radius of curvature of 11~m for electrons with 1 GeV/$c$ of momentum. 
Given a tracking plane spacing $L\approx 0.5$~m, for perpendicular tracks the
sagitta $s$ is given by $s \approx L^2/2R$, which comes to about one
centimeter. Thus, we can expect sagittae on the order of a few millimeters
for higher momentum tracks, and sagittae on the order of a few centimeters
for lower momentum tracks. 

The magnet could be operated with two different polarities, which were referred
to as positive and negative, although they represent, respectively, the 
magnetic field lines pointing counter-clockwise and clockwise around the 
beamline when facing downstream. It was originally intended that the 
magnet polarity be flipped every few hours to reduce systematics stemming
from electrons and positrons having different trajectories. In the 
high-luminosity environment of the fall run, negative polarity running
had to be abandoned. With negative polarity, low-energy M\o ller electrons
are bent by the magnetic field into the drift chambers, swamping them 
with noise and current. The schematic in figure \ref{fig:main_spec}
shows the track curvature produced with positive polarity.

At the conclusion of data taking, there was an intense effort to survey
the magnetic field in the volume relevant for track reconstruction. 
Those efforts were described in a recent paper \cite{Bernauer:2016hpu},
and are the subject of chapter \ref{chap:field} in this thesis.

\section{Main Spectrometer}

\begin{figure}[htpb]
\centering
\includegraphics[width=14cm]{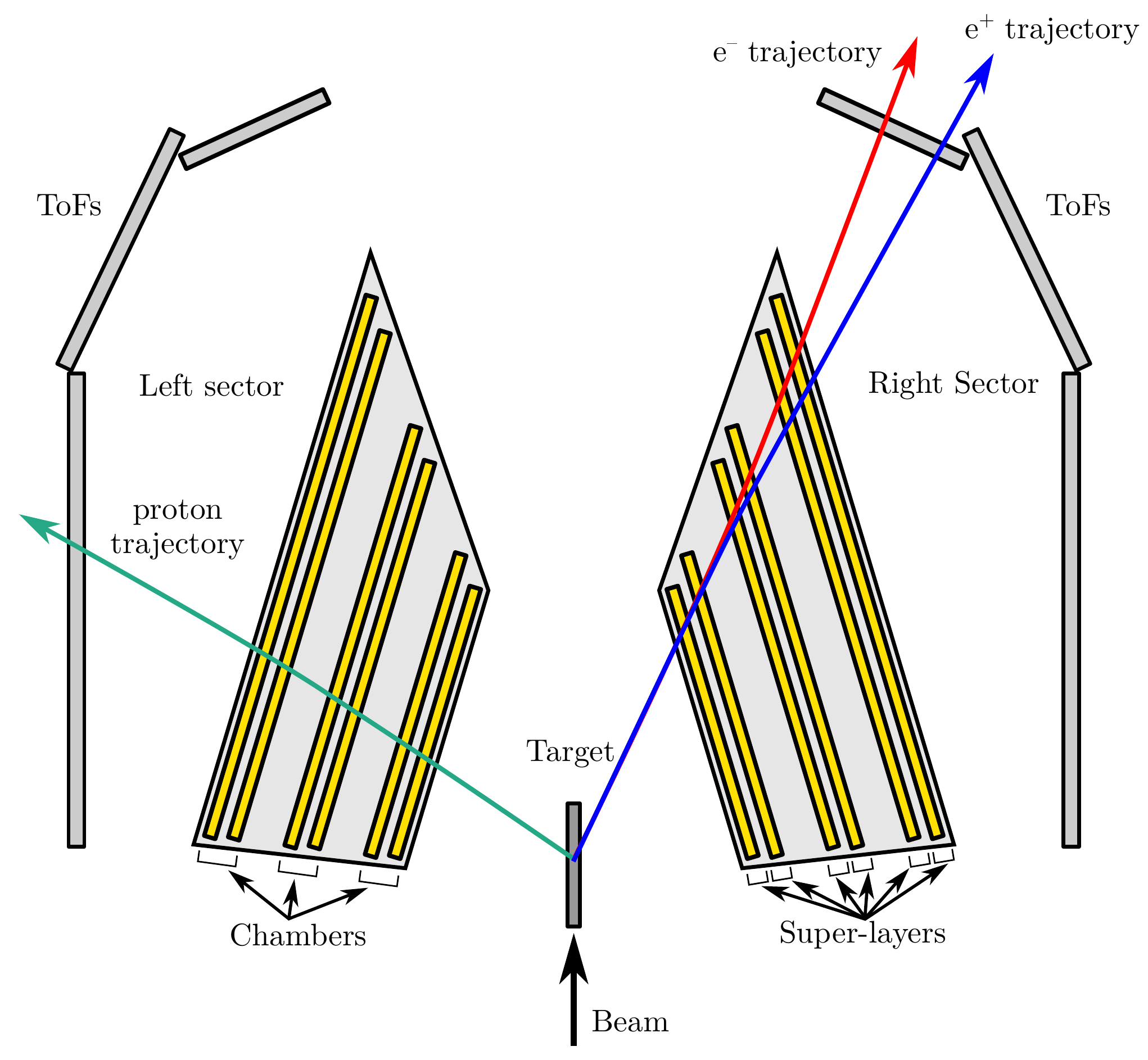}
\caption[Schematic showing an example elastic event]{\label{fig:main_spec} This schematic shows an example elastic
scattering event for a lepton scattering to the right at an angle of $25^\circ$. 
Both the $e^-$ and $e^+$ trajectories are drawn in order to show how the magnetic
field produces a slight curvature in opposite directions for the two lepton signs. }
\end{figure}

The left and right segments around the target were instrumented with 
detectors to track scattered leptons and recoiling protons. This system
of detectors will be referred to as the ``Main Spectrometer'' to distinguish
it from the 12$^\circ$ luminosity monitors, which also form a spectrometer
of sorts. The main spectrometer detectors came from the BLAST spectrometer.
Figure \ref{fig:main_spec} shows a schematic of an example elastic scattering
event in the main spectrometer.

Moving from the target outwards, the first detectors encountered by outgoing
particles were a series of drift chambers. These detectors made accurate 
determinations of the passing particle's trajectory. This was used to
determine the particle's scattering angle, and, using the curvature 
caused by the magnetic field, the particle's momentum. Beyond the drift chambers 
were the time-of-flight scintillators (ToFs): panels
of scintillating plastic which provided accurate timing information. 
Since these signals could be read out quickly, the ToFs were used to 
trigger the data acquisition system. 

Both of these systems will be described in more detail in the following
subsections.

\subsection{Drift Chambers}

\label{ssec:drift_chambers}

The drift chambers were the tracking detectors in the OLYMPUS spectrometer.
Drift chambers work by collecting the electrons from ionization of a charged
particle passing through a gaseous medium. The passing charged particle
will ionize gas atoms along its trajectory. The newly liberated electrons
are pulled, using an electric field, to high-voltage wires strung throughout 
the medium. Close to the wires, the electric field creates an amplification
cascade so that a measureable current can be detected on the wire. By measuring
the time it takes for the ionization electrons to drift to different sense
wires, the trajectory of the particle can be inferred. 

Before describing the operational details of the OLYMPUS chambers, I will first
describe the layout and introduce the vocabulary that we used to refer to the
different parts of the chambers. Some of this vocabulary is shown in figure \ref{fig:main_spec}.
As mentioned earlier, the OLYMPUS spectrometer had two sectors, one to the left 
and the other to the right of the target. Each sector had its own drift chamber 
frame: an aluminum trapezoid about the size of a large bathtub. Each frame contained 
three drift chambers within a single gas volume. The three chambers---the inner,
middle, and outer---had wires that were strung nearly vertically from the top and
bottom faces of the frame. The inner chamber wires were the shortest, while the 
outer chamber wires, which were strung at the widest part of the trapezoid, were
the longest. Each sector had nearly 5000 wires, of which 477 were read-out.

\begin{figure}[htpb]
\centering
\includegraphics[width=\textwidth]{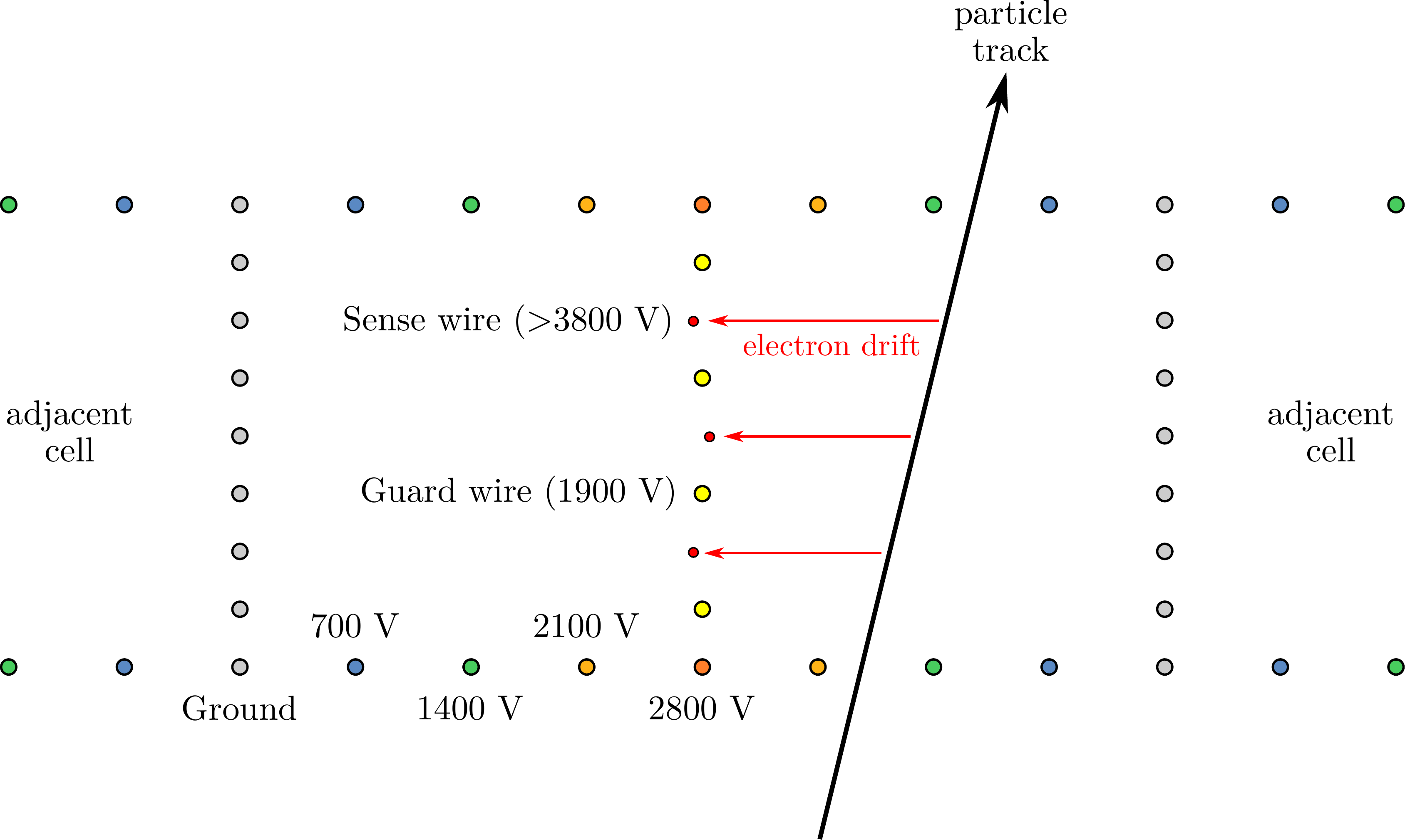}
\caption[Wires in a drift cell]{\label{fig:cell_cartoon} This illustration shows a cross section of a drift cell
within a super-layer, looking parallel to the wires, whose cross sections appear as circles. The colors
represent the voltage of each wire. The highest voltage wires, the sense wires, attract
the ionization electrons produced by nearby tracks. The size of the wires are not to 
scale.}
\end{figure}

\begin{figure}[htpb]
\centering
\includegraphics{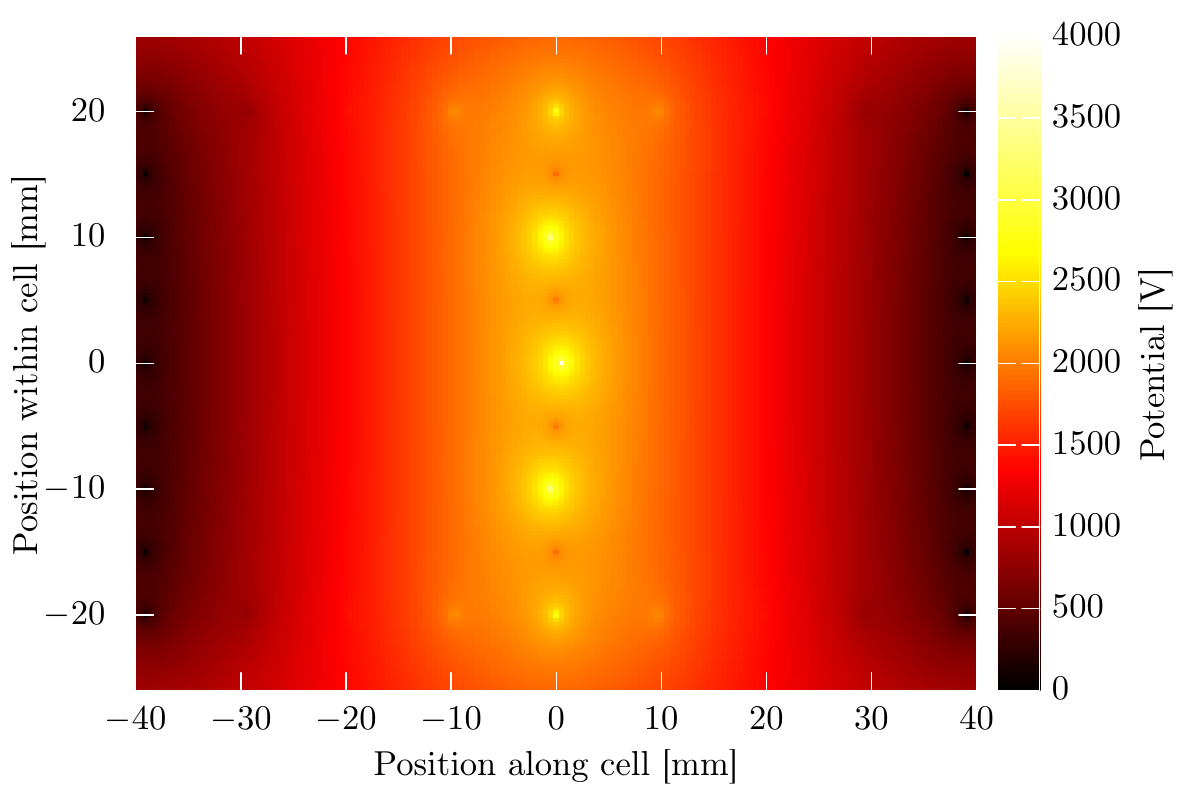}
\caption[Drift cell electrical potential]{\label{fig:drift_cell_potential} The potential is highest at the center
of the drift cell, and decreases uniformly to the cell edges.}
\end{figure}

Each chamber was made up of two super-layers, which in turn were made up of rows
of a repeated element called a drift cell. A schematic of a drift cell within a 
super-layer is shown in figure \ref{fig:cell_cartoon}. Each drift cell was 78~mm long
by 40~mm deep, and made up of several dozen parallel wires. The wires 
were held at specific voltages in order to create a potential gradient across
the cell. The wires forming the boundary of the cell were held at ground potential.
The center of each cell was held at high voltage. Ionization electrons produced
in the cell would drift toward the high voltage wires at the center of the cell.
A calculation of the potential within a drift cell is shown in figure \ref{fig:drift_cell_potential}.
The number of drift cells in each super layer is shown in table \ref{table:cells_in_sl}.

\begin{table}
\centering
\begin{tabular}{| l | c | c |}
\hline
Chamber & Cells in inner super-layer & Cells in outer super-layer \\
\hline
\hline
Inner & 18 & 19 \\
Middle & 26 & 27 \\
Outer & 34 & 35 \\
\hline
\end{tabular}
\caption[Cells in a super-layer]{\label{table:cells_in_sl} The trapezoidal shape of the chambers
necessitated more cells per super-layer moving outwards from the target.}
\end{table}

Only three of the wires at the center of the cell were read out, called the 
sense wires. These wires were smaller (only 25~$\mu$m in diameter, compared
to 100--150~$\mu$m for the other wires) in order to create larger electric
fields and a greater amplification in their vicinity. The sense wires carried
the highest potential in the drift cell; they were held typically between 
3800~V and 4000~V, balancing signal efficiency with the possibility of sparking.

The arrival times of the ionization electrons at the three sense wires provide
information about the track position, but there is an ambiguity about the 
direction of drift. Given an arrival time, the track could either have originated
to the left of the sense wire, or to the right. To resolve these ambiguities,
the sense wires are staggered by $\pm 0.5$~mm relative to the high voltage plane.
With staggered wires, the ambiguity can be resolved because the correct side
will be a better fit to the three arrival times. 

The arrival times provide information about the track position perpendicular to 
the wire, but no information about the position of the track in the direction 
parallel to the wire. For this reason the wires of each super-layer were rotated 
$\pm 5^\circ$ relative to vertical. Adjacent super-layers had a $10^\circ$ stereo 
angle between them, enough to localize the position of a track in two dimensions.

\subsubsection{Refurbishing the Chambers at DESY}

After the conclusion of the BLAST experiment, the drift chambers sat idle for
several years and, by the time of OLYMPUS, were in need of refurbishment.
It was deemed too great a risk to ship them while fully strung. The wires 
collectively (nearly 5000 per sector) exert significant tension on the frames. 
It was feared that if many wires broke during shipping, this might initiate a 
catastrophic failure that could damage the frames. The chambers were unstrung 
at MIT Bates Laboratory, and then shipped to DESY in the spring of 2010. Over
the summer of 2010, both sectors were completely restrung by me and several
others in a clean room at DESY. In January of 2011, new electrical connections 
were soldered to the newly strung wires. In the spring of 2011, the chambers
were taken to the DORIS staging area and installed on the OLYMPUS subdetector
frames. 

In addition to new wires, new high-voltage distribution boards were designed
for OLYMPUS. The first iteration was beset by sparking problems: traces carrying
several thousands volts passed too closely to ground lines. Despite efforts to 
insulate the boards with insulating acrylic, spray-on plastic, and RTV silicone,
a second iteration of boards was needed. These were rapidly produced at the end
of 2011, and installed just before the start of the February run in 2012. 

\subsubsection{Read-Out Electronics}

Several steps of electronic processing are needed to transform current pulses,
produced by ionization electrons arriving at a sense wire, into a time signal.
The first step is to transform the current pulse into a voltage pulse, which
is done by a capacitive circuit on the high-voltage distribution board. Common-mode
noise between the sense wire and the adjacent wires (called guard wires) is also
removed at this stage. The voltage pulse is amplified by a fast amplifier then 
sent to a discriminator to be turned into a digital pulse. The digital pulse is
sent to a time-to-digital converter (TDC), operated in common stop mode. The 
wire chamber signal provides the start, and the stop is provided by a trigger
signal that has been delayed by a fixed amount of time that is longer than the 
maximum drift time in the chambers. This arrangement produces the counter-intuitive
situation where longer drift times are represented by smaller output times on the TDC. 
The time of zero-drift was calibrated for each wire.

\subsubsection{Time-to-Distance Functions}

Analyzing drift chamber data requires converting times from the TDCs to distances
between a track and the sense wires, a problem which we gave the name ``time-to-distance''
or ``TTD''. The track reconstruction works best when its internal time-to-distance function is
as close as possible to the inverse of the true ``distance-to-time'' function for
the chambers. Developing a good time-to-distance function for the track reconstruction
was one of the major analysis hurdles that had to be cleared. To give the reader
a sense of scale for the subsequent discussion, an ionization electron could drift
the 39~mm from the ground plane to the high-voltage plane in 1--2~$\mu$s. 

The drift cells were designed to make the time-to-distance relationship as linear
as possible. The electric field is roughly constant in the cell volume, and points
from the ground plane to the high-voltage plane. But the field close to the sense
wires is much higher than in the bulk, and so, in the region of small times, the 
function is non-linear. As well, in the region close to the ground plane, ionization
electrons feel very little field, and can drift very slowly for hundreds of nanoseconds,
before they enter the high-field region and get pulled toward the high voltage plane.
This produces a plateau in the time-to-distance function for large times.

To further complicate matters, the time-to-distance function depends on the incidence
angles of the track, as well as the local magnetic field. The Lorentz force of a magnetic 
field has the effect of changing the drift direction of the ionization electrons. 
Rather than drifting along the electric field lines, the electrons will drift at an angle,
called the ``Lorentz angle''. Since the toroidal field is non-uniform, every cell has a 
slightly different Lorentz angle, and requires a slightly different time-to-distance
function.

After trying many different approaches, we settled on a simple parameterization that
captured the linear-region, the non-linear behavior for very short and very long times,
the dependence on incident angles, and the Lorentz angle. We then iteratively fit
the parameters for each wire to trajectories found by the track reconstruction software. 
After several iterations, a set of best-fit parameters was found.

\subsubsection{Gas Mixture}

The specific gas mixture used in a drift chamber can greatly affect the chamber
performance. We made several mistakes with regards to our gas mixture, and I want
to discuss those here in the hope that the reader may avoid similar mistakes. 
First, I remind the reader that typically a mixture of two gases is used; a
drift gas which makes up the majority of the mixture, with a small amount of 
a quench gas. The drift gas should have a low affinity for electrons. The quench
gas should have many rotational and vibrational modes in order to absorb photons
released during amplification, so as to prevent re-ionization. 

At BLAST, a mixture of 82.3\% helium (drift gas) and 17.7\% isobutane (quench gas)
was used \cite{BLAST:maschinot}. Using isobutane at DESY would have required the hassle
of additional safety protocols because of its flammability, so the gas mixture at OLYMPUS was 
switched to 90\% argon (drift gas) and 10\% carbon dioxide (quench gas). Both
were inert and could be used safely. However, the drift velocity of electrons
in the two mixtures were significantly different. Whereas at BLAST, electrons
drifted at approximately 20~mm/$\mu$s \cite{BLAST:crawford}, at OLYMPUS, the drift 
velocity was closer to 30~mm/$\mu$s. This substantially reduced the position
resolution of the drift chambers and even made it difficult to resolve the 
$\pm 0.5$~mm staggering of the sense wires. Since the staggering was necessary
to resolve left/right ambiguities in tracks, track reconstruction became
an enormously difficult enterprise.

This was not our only lapse in judgement when it came to the gas mixture. At the
beginning of the fall run, one of the problems with the drift chambers was ``ringing'':
TDC times that appeared to be the result of noise, but were also correlated
with the trigger, and spaced an even 100~ns apart. To combat this ringing, we added
a small amount of ethanol vapor to the gas mixture. Ethanol is a very effective
quench gas, and so it was hoped that a small amount of ethanol might eliminate 
the ringing. However, the addition of ethanol ended up causing more problems
than it solved.

First, it is possible that the gas mixture had nothing to do with the ringing noise. 
The source of the ringing was never conclusively determined, but with the benefit 
of hindsight, I suspect that the majority of this noise was caused by background 
particles produced by the beam (or beam halo) that caused additional ionization in 
the drift chambers. That would explain the time correlation with the trigger (which 
was itself, correlated in time with the beam bunches), and the 100~ns time spacing 
(which is the same as the DORIS bunch spacing). It also explains why this noise was less of a problem in
February run. At that time, due to the leak in the target inlet, the target density was lower.
This would have reduced the beam halo from beam-target interactions. If this hypothesis
is correct, the gas mixture was unrelated to the ringing, and the ethanol might have
been a needless complication.

\begin{figure}[htpb]
\centering
\includegraphics{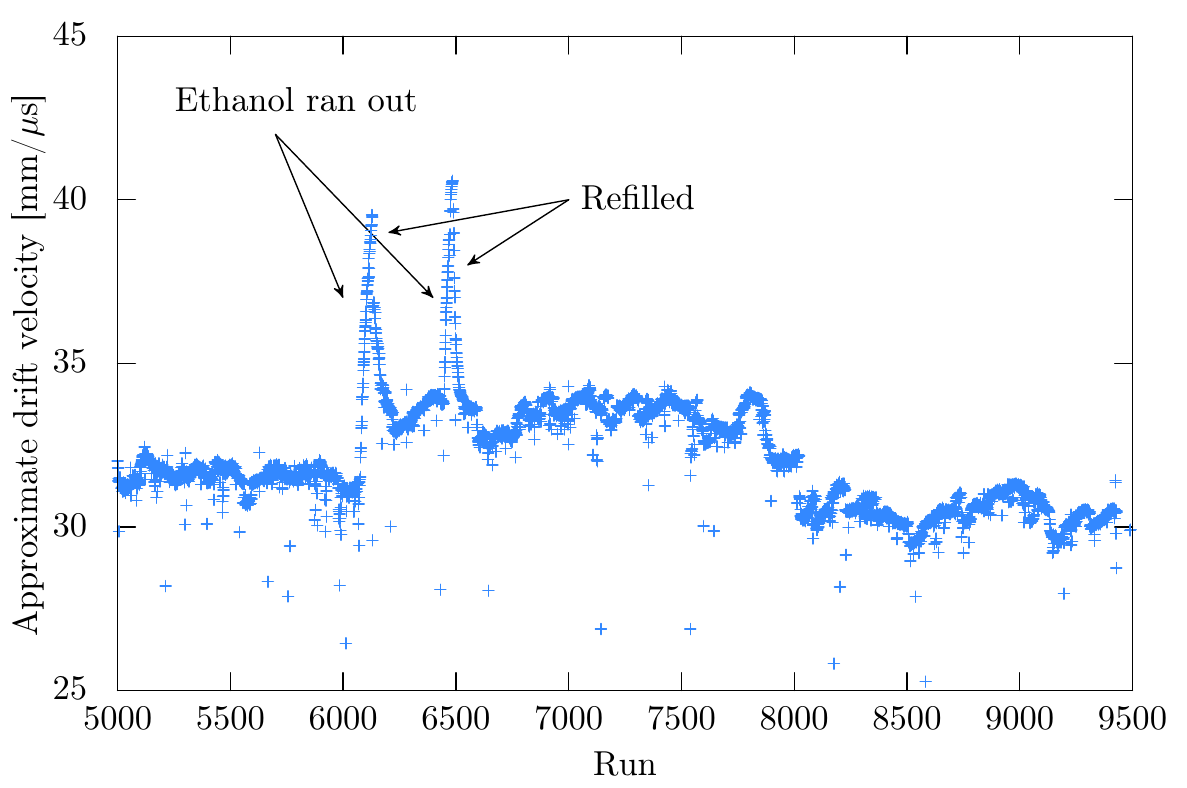}
\caption[Variations in drift speed in the drift chambers]
{\label{fig:ethanol_church_length} The ethanol concentration was not stable,
causing the drift velocity to vary over time. The change in drift velocity was so great
that it alerted us on two occasions that the ethanol bubbler had become empty.}
\end{figure}

Second, in our hurry to add the ethanol, we did not come up with a mechanism to 
accurately control the amount of ethanol we added. We passed the dry gas mixture 
through a liquid ethanol bubbler, which we put in a small refrigerator. The 
ethanol added to the gas was controlled only by the temperature of the refrigerator.
It was clear that the refrigerator temperature varied, we could see changes
over time in the drift velocity, shown in figure \ref{fig:ethanol_church_length}.
In fact, we could even tell, from the drift velocity alone, when the bubbler ran out of ethanol. 
The consequence of this temperature variation was that we had to produce different
time-to-distance functions for different periods of time during data taking. 
And since we were never quite sure of the exact proportion of ethanol vapor, we had
enormous trouble trying to calculate the drift properties of the gas.

Third, we subsequently learned that we had misunderstood how ethanol is commonly
used in drift chamber noise reduction. According to \underline{Particle Detection with Drift Chambers}
by Blum, Riegler, and Rolandi \cite{blum:chambers}, ethanol is often temporarily
introduced to chambers that use hydrocarbons (like propane or butane) as their quench gas.
In these gas mixtures, the hydrocarbon molecules will adhere to the wires
over time, forming long chains or ``whiskers''. As these whiskers get longer,
they can be source of noise and dark current. Introducing ethanol vapor can
help break up these whiskers, restoring the chambers to optimal conditions.
However, Blum et al.\ suggest that ethanol should be periodically added during
breaks in operation and then removed, not introduced as a permanent fraction
of the gas mixture. In the case of our drift chambers, we were never at risk
of whisker growth, given that our quench gas was carbon dioxide, and because
we left the ethanol in the chambers during data taking, it substantially
altered the drift chamber performance.

\subsection{Time-of-Flight Scintillators}

\begin{figure}[htpb]
\centering
\includegraphics[width=12cm]{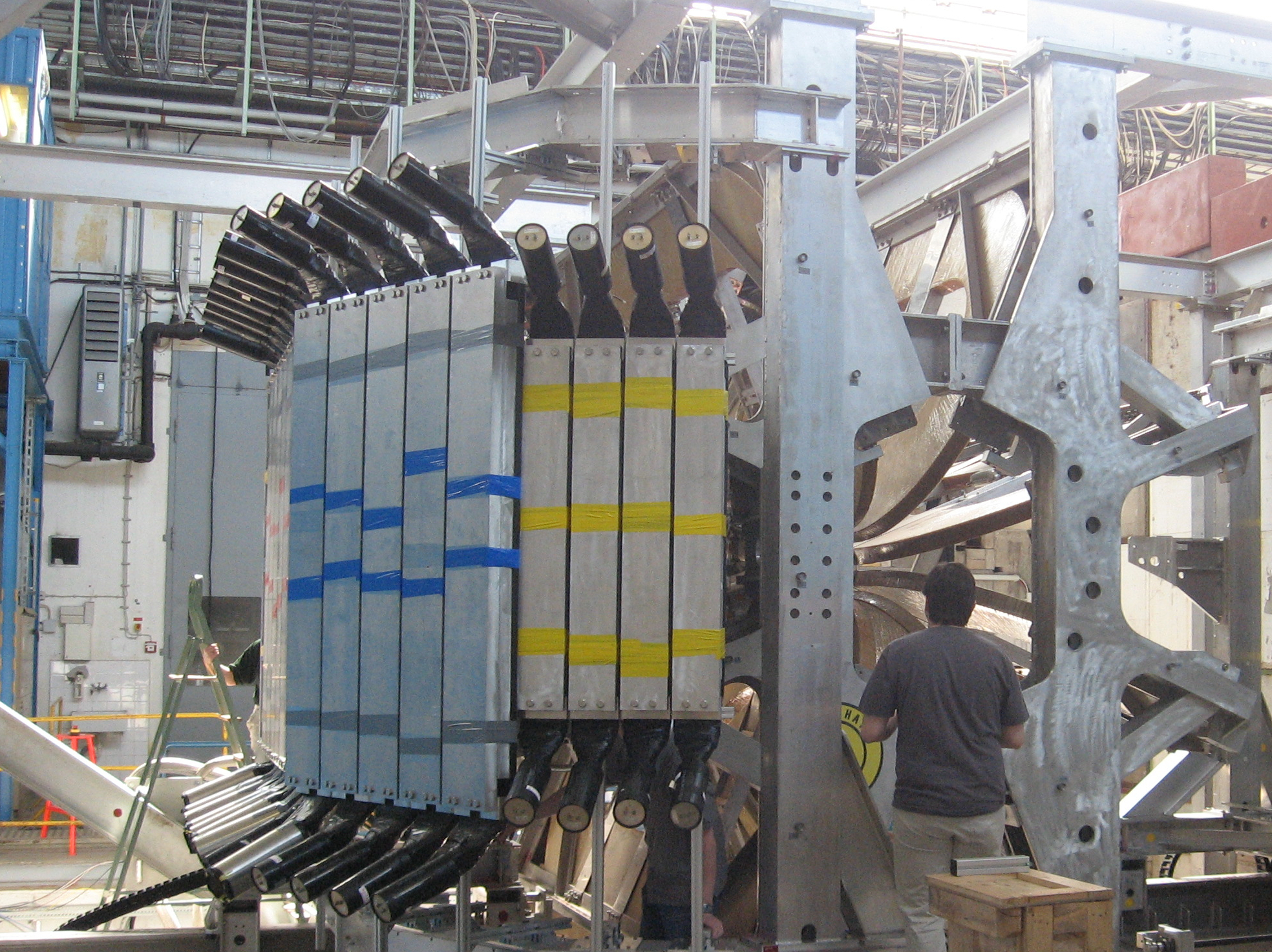}
\caption[Photograph of the time-of-flight scintillators]
{\label{fig:tof_photo} This photograph shows the right sector ToFs when the 
spectrometer was being assembled in the staging area.}
\end{figure}

The Time-of-Flight (ToF) scintillators were bars of scintillating
plastic that were arrayed in panels outside of the drift chambers.
The scintillation light was detected by photo-multiplier tubes 
(PMTs) attached to the top and bottom ends of each bar. Because
the signal and readout from the ToFs was fast, the ToFs were used
for timing and triggering. Each sector had three panels for the downstream, middle, and
upstream regions of the acceptance. The downstream bars were 
slightly smaller: 120~cm tall, 17.8~cm wide, and 2.54~cm deep.
The other bars were 180~cm tall, 26~cm wide, and 2.54~cm deep. 
Each sector had a total of 18 bars. Figure \ref{fig:tof_photo}
shows the ToF bars on the right sector.

Each scintillator was wrapped in reflective foil to keep light
from escaping. A thin sheet of lead was placed on the face of
the bar that faced the target to block low-energy electrons. 
Plastic light guides were glued to the top and bottom faces.
PMTs were positioned flush to faces of the light guide, and 
coupled with optical grease. The entire assembly was wrapped
in black plastic to form a light-tight seal. The PMTs were 
also wrapped with a layer of Mu-metal to shield them from the
magnetic field.

The output from each PMT was split and sent to both an 
amplitude-to-digital converter (ADC) to record the signal amplitude
and to a TDC to record the signal time. The signal amplitude
contained information about the light produced by the passing
particle, and thus the energy loss, $dE/dx$, though this had
to be unfolded from the amount of attenuation that occurred
between the track and the PMTs. The mean of the signal times
from the two PMTs could be used to determine the time of 
flight for the particle. The time difference between the two
PMTs indicated the track position along the bar. The analysis
of the ToF data was performed by my colleagues Lauren Ice and
Rebecca Russell, whose recent thesis describes the analysis
in more detail \cite{russell:thesis}.

\subsection{Trigger}

The main spectrometer was triggered using a combination of ToF
and drift chamber information. The level 1 trigger, which needed
to be fast, made use only of ToF information. In each sector, the
top and bottom PMTs of a single bar needed to have a signal in
coincidence. Furthermore the left sector bar and right sector bar
needed to be a specific combination that was possible for an 
elastic $ep$ scattering event. The background rates of the most 
downstream ToF bars were very high, and by requiring a valid
kinematic combination, a background signal in a left downstream
bar and a background signal in a right downstream bar would no
longer create a trigger. For scale, the trigger rate was on
the order of several hundred Hz, while we estimated the rate
of elastic $ep$ events to be on the order of about 10--20 Hz. 

In the Fall run, the luminosity increased, and so a second level
trigger was developed, which used information from the drift
chambers. For the second level trigger to fire, each sector
had to have at least one signal from a wire in either the 
middle or outer chambers. Even this minimal requirement
reduced the trigger rate to acceptable levels.

In addition to the main trigger, a number of more permissive
test triggers were included, but heavily pre-scaled. These were
useful in studying aspects of the detector performance, for example,
the ToF efficiencies \cite{russell:thesis}.

\section{Luminosity Monitors}

\begin{figure}[htpb]
\centering
\includegraphics{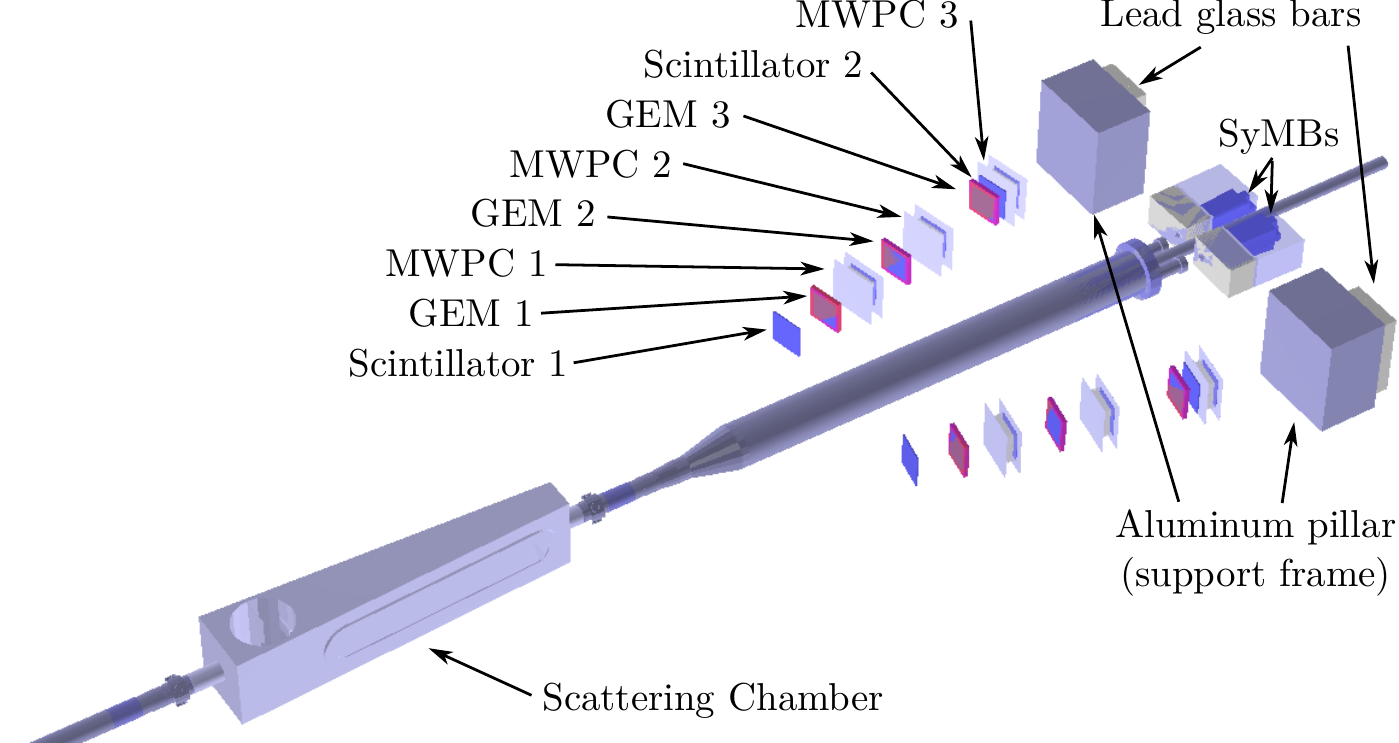}
\caption[Luminosity monitors]{\label{fig:telescopes} The $12^\circ$ telescopes and the symmetric
M\o ller/Bhabha calorimeters were positioned downstream of the scattering chamber.}
\end{figure}

In order for the data from electron running to be compared to that from positron
running, the relative luminosity of the two data sets was needed. A rough estimate
of luminosity could be made online using the beam current, the target flow rate
and target temperature, but for a percent level asymmetry measurement, highly-accurate
luminosity monitors were needed. Two independent systems, the $12^\circ$ telescopes
and symmetric M\o ller/Bhabha calorimeters (SyMBs), were specially built for 
OLYMPUS and will be described in this section. A schematic showing their position
relative to the downstream beamline can be found in figure \ref{fig:telescopes}. 
A description of the luminosity analysis can be found in chapter \ref{chap:lumi}. 

\subsection{$12^\circ$ Tracking Telescopes}

The $12^\circ$ tracking telescopes were a set of tracking detectors positioned
along the $12^\circ$ scattering angle from the target. Two telescopes were built,
and installed in each sector, mounted to the downstream face of the drift chamber
frame. Figure \ref{fig:telescopes} shows the different detector elements in the
telescopes and their positioning relative to the beamline. Each telescope had two 
planes of scintillator, three gas electron multiplier (GEMs)
detectors, and three multi-wire proportional chambers (MWPCs). The scintillators
were read out using silicon photomultipliers (SiPMs) and were used to trigger
the telescopes. The GEMs and MWPCs were the tracking detectors; they could
record the position of a passing charged particle in two dimensions. The position
resolution of the MWPCs was on the same scale as the 1~mm spacing between the 
wires. The GEM detectors had a much finer position resolution of approximately 100~$\mu$m.
There was substantial magnetic field in the volume of the telescopes, so a momentum
measurement of passing particles could be made.

The $12^\circ$ monitored luminosity by measuring the rate of elastic $ep$ scattering
in which the lepton passed through the telescope, with an approximately $12^\circ$ 
scattering angle. The recoiling proton would pass through the opposite sector of the 
main spectrometer. The system was triggered by the coincidence of signals in both
telescope scintillators and in both PMTs of an opposite sector ToF bar. For most
(about 90\%) leptons that passed through the telescopes, the proton also passed
through the drift chambers, though this was not required for the trigger.
The telescope trigger efficiency was tested using a pre-scaled secondary trigger 
which only required a signal in bars of scintillating Pb glass, placed at the
downstream end of the telescopes. The only convenient way to mount them in position 
was to strap them to aluminum pillars that served as part of the support frame for the toroid
magnet. This meant that passing particles lost some of their energy traversing the
pillar, but this was not a problem for a simple trigger test. 

An analysis of the $12^\circ$ telescope data was performed by my colleague Brian Henderson,
and the reader is encouraged to see his thesis for details \cite{henderson:thesis}. The
results are also summarized in section \ref{ssec:12deg}. 

\subsection{Symmetric M\o ller-Bhabha Calorimeters}

The symmetric M\o ller-Bhabha calorimeters (SyMBs) were a pair of lead fluoride (PbF$_2$) Cherenkov
calorimeters, that were positioned 3~m downstream from the target, very close to the beamline,
as can be seen in figure \ref{fig:telescopes}. An article describing the design and operation of the 
calorimeters was recently published \cite{Benito:2016cmp}.  The calorimeters monitored luminosity
by measuring the rate of scattering from atomic electrons in the hydrogen gas of the target---M\o ller
scattering during electron running, and Bhabha scattering during positron running. The calorimeters
were positioned at the symmetric scattering angle, in which both final state leptons emerge at 
the same angle, one entering the left calorimeter, the other entering the right. This angle is 
approximately $1.29^\circ$ for a beam energy of 2.01~GeV. 

Each calorimeter was made up of nine lead-fluoride crystals, arranged $3\times 3$, each connected to its own PMT. SyMBs sat 
in a region of weak magnetic field, and so the entire apparatus---crystals and PMTs---were placed in
a mu-metal box, shielding the PMTs from the field. The calorimeter aperture was defined by a collimator:
a block of lead with a cylindrical hole. The collimator was positioned in front of the central crystal
in the calorimeter. The central crystal contained the bulk of the electromagnetic shower---the crystals
were approximately 25~mm in width, and the Moli\`{e}re radius for lead fluoride is 21.24~mm.

The SyMBs had to withstand high rates, on the order of hundreds of kHz. Rather than being read out by the 
main OLYMPUS data acquisition system (DAQ) every time the SyMBs recorded a signal, the SyMBs had their 
own data acquisition system. The energy deposited in each calorimeter was recorded in several fast histogramming
cards, which could be operated with essentially zero dead time. The histogramming cards could be filled faster 
than the DORIS bunch frequency, i.e., the SyMBs could distinguish and record signals from
adjacent bunches. The histograms were periodically written to the main OLYMPUS DAQ data stream. This meant that
event-by-event information was not recorded; only the aggregation of events between readouts was recorded.

The analysis of the SyMB data was undertaken by my colleague Colton O'Connor, and the interested reader should
consult his thesis for the most detailed description of that work \cite{oconnor:thesis}. Unfortunately, a 
luminosity extraction directly from M\o ller/Bhabha scattering rates ended up not being possible, due to 
large systematic effects that which were not fully considered when the SyMBs were designed. Fortuitously,
a secondary analysis, based on the rates of multi-interaction events, turned out to be a much better way
to extract luminosity from the data. This analysis is the subject of chapter \ref{chap:lumi}.

\section{Survey}

\label{sec:survey}

Many of the detectors described in this chapter were used to make precise measurements of the
position of particle tracks. These position measurements were only as accurate as our understanding
of where the detectors were physically positioned in space. For this reason, at the end of data
taking, a survey of all of the detectors was undertaken. This work was largely performed by the 
DESY survey group MEA 2. 

The survey group used theodolites with laser range-finders to measure the three-dimensional position 
of special reflective targets. These survey targets were affixed 
to the walls of the experimental hall (to provide a global reference frame), to parts of the OLYMPUS frame
(to provide local reference systems), and to known fixed positions on all of the OLYMPUS subdetectors. 
The first stage of the survey was conducted in the experimental hall, with the OLYMPUS detector unmoved.
Once the position of the OLYMPUS frame within the hall was established, the two sectors of the main spectrometer
(and including the $12^\circ$ telescopes, which were mounted to the drift chambers) 
were disconnected, and carried by crane into the staging area. The second stage of the survey was to
measure the positions of the drift chambers and ToF scintillators with respect to the subdetector frames.
With the ToFs and drift chambers out of the way, a third stage of surveying was possible back inside
the experimental hall. There was now an unobstructed view of the scattering chamber, the SyMBs and elements 
of the beamline. After this was complete, the magnetic field was also surveyed, and this will be described
in detail in chapter \ref{chap:field}.

The analysis of the survey data was performed by my colleague Jan Bernauer. Jan had to integrate over 5000
position measurements from dozens of reference frames to produce the most likely three-dimensional coordinates
for all of the survey targets. He then had to combine these coordinates to estimate the positions and orientations
of all of the elements in the OLYMPUS apparatus. This was information used by Colton O'Connor to make a digital map
of the apparatus geometry, written in the mark-up language GDML \cite{gdml:note}. This map was used at many stages in the analysis:
by the simulation's event generator to place events relative to the beam position monitors, by the simulation's 
propagator when simulating trajectories, and in the track reconstruction. 

\chapter{Magnetic Field Survey}

\label{chap:field}

\section{Introduction}

The OLYMPUS spectrometer is a magnetic spectrometer; it uses a magnetic
field to transform a charged particle's momentum information into position
information, which can be measured accurately. For this system to work, 
the magnetic vector field needs to be known throughout the spectrometer 
volume. Without a magnetic field map, particle tracks cannot be simulated 
and tracking software cannot determine a track's momentum. Furthermore,
the magnetic field bends electrons and positrons in different directions.
Inaccuracy in a magnetic field map can lead to systematic differences
between the two lepton species.

The magnetic field could, in principle, be calculated from current in 
the toroidal magnet and from the shape of the coils. In practice, this was 
not feasible because of uncertainty in the true positions of the coils.
The coils deformed slightly under the strain of magnetic forces when the 
magnet was powered. Furthermore, the current-carrying copper bars in each
coil, described in section \ref{sec:app:magnet}, were potted together
with epoxy during construction. Even if we had a perfect survey of the 
exterior of the coils, we would not know, with high accuracy, where the 
bars physically sat within the epoxy potting. 

A further risk with trying to calculate the field comes from the fact
that our field is non-uniform. There are areas of low-field (around
the beamline) and high-field (around the drift chambers) and large
gradients between them. Slight inaccuracies in the assumed positions
of the coils can thrown off the field dramatically, especially in 
the high-gradient regions.

Rather than calculating the magnetic field, we attempted to measure it.
Using a Hall probe positioned by a system of mechanical translation 
tables, we measured the magnetic field at over 36,000 positions in and
around the spectrometer. We analyzed the resulting data to produce a
software map that could estimate the field as a function of position,
esentially interpolating between measurement positions. This map is 
used in analysis (specifically by the Geant4 path-swimming routines)
to solve the trajectories of charged particles moving through the
spectrometer.

A recently-published paper has a concise description of the OLYMPUS 
magnetic field measurement and data analysis \cite{Bernauer:2016hpu}. 
In this chapter, I'll try to provide a more detailed account of our efforts and our decision
making. I'll start with an explanation of our measurements, including
a description of the apparatus and our procedure. 
Next I'll describe how we analyzed the data from these measurements
to produce a magnetic field map. Then I'll show how we used spline
interpolation on a precomputed grid to provide the speed up necessary
for large-scale simulations. I'll conclude with some reflections about 
what would have made our magnetic field measurement better.

\section{Measurements}

\subsection{Overview}

The measurements of the magnetic field were made using a three-dimensional
Hall probe, which provided the full magnetic field vector at the probe's position.
The probe was moved through the spectrometer volume by a system of translation 
tables. Early on in the measurement process, it was discovered that the tables
did not provided the desired millimeter-level accuracy in their movements. To 
correct for this, we used theodolites to determine the probe position. 

The measurements were conducted in the spring of 2013, at the conclusion of the OLYMPUS 
detector survey. The detectors had been removed from the magnet, giving the Hall
probe unobstructed access to the spectrometer volume. The apparatus was first
assembled on the left side of the magnet to survey the left sector. Then it was
taken apart and reassembled on the right side of the magnet for the right sector
survey. The process was completed in about four weeks, and over 36,000 field
measurements were made. 

\subsection{Apparatus}

\begin{figure}[htpb]
\centering
\includegraphics[width=12cm]{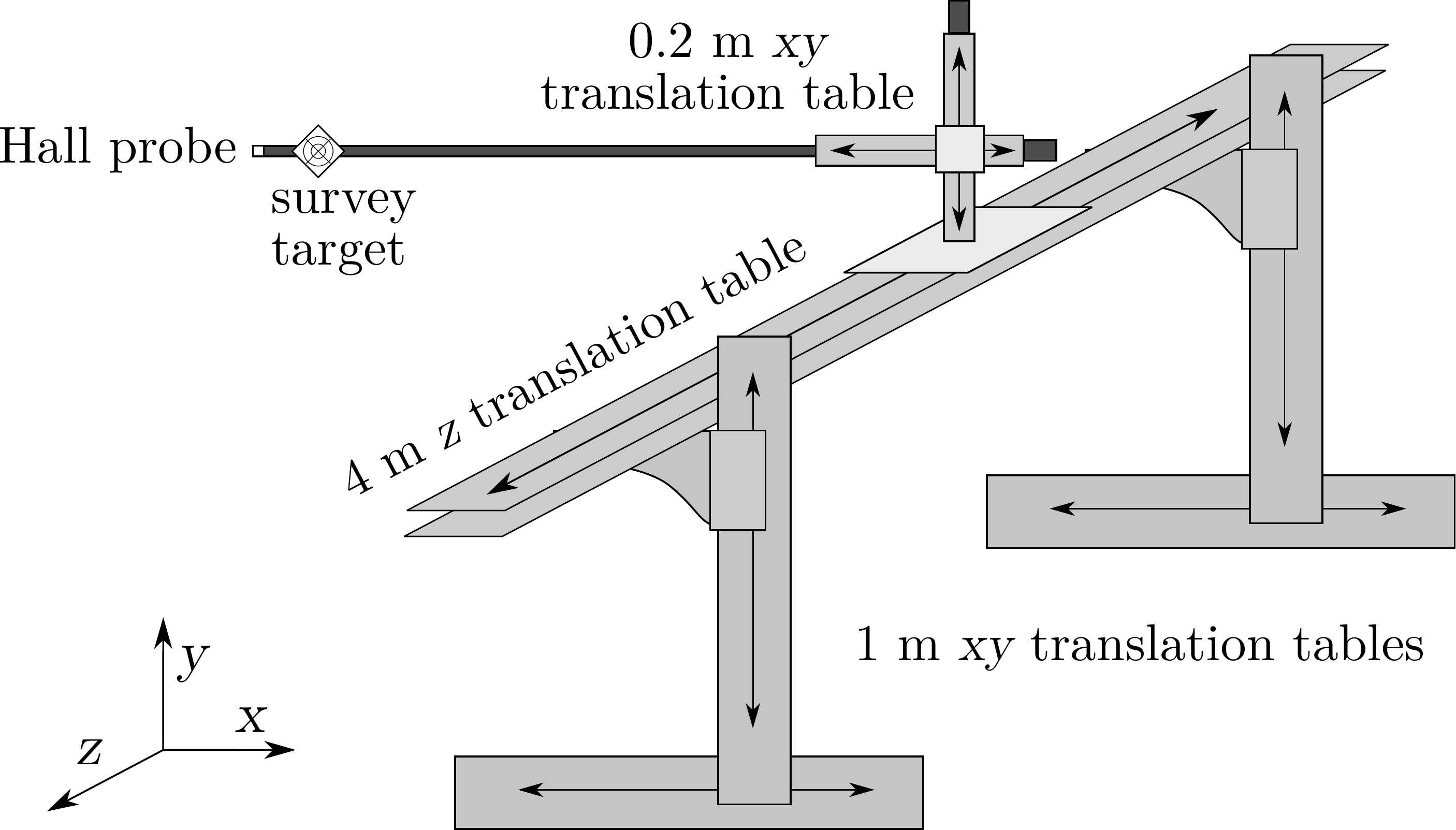}
\caption[Magnetic field survey apparatus]{\label{fig:magnet_survey_schematic} 
The magnetic field measurement apparatus used a system of translation
tables to move the Hall probe to various positions within the spectrometer.
This schematic shows the system when set up to measure the left sector.
}
\end{figure}

\begin{figure}[htb]
\centering
\includegraphics[width=12cm]{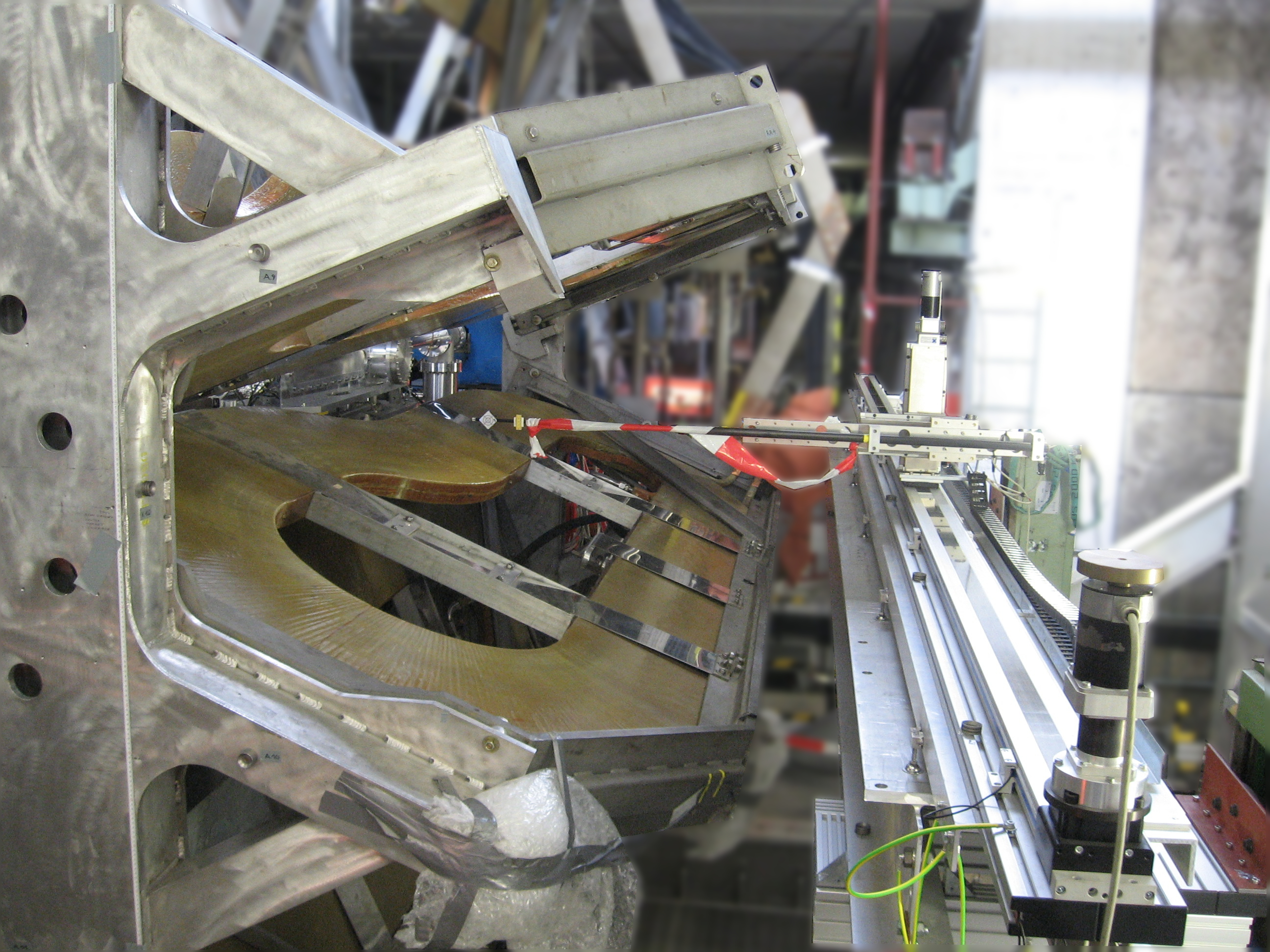}
\caption[Photograph of the magnetic field survey apparatus]{\label{fig:magnet_survey_photo}
This photograph shows a magnetic field scan in progress on the left
sector. The probe is just to the left of the diamond-shaped survey target.
}
\end{figure}

The magnetic field measurements were made using a 3D Hall probe, moved by a system
of translation tables. This set-up, a schematic of which is shown in figure
\ref{fig:magnet_survey_schematic}, was originally built for a survey of the 
undulator magnets of DESY's free-electron laser FLASH \cite{Grimm:2007zz}. 
The base of the apparatus was a pair of translation stands with 1~m range
in both the $x$ and $y$ directions. These stands were moved synchronously
and operated as a single table. The stands supported a 6~m aluminum I-beam,
onto which was bolted a three-dimensional translation table. The 3D table
had a 4~m range in $z$, and $0.2$~m ranges in $x$ and $y$. 

The motors of the translation tables could not operate within a strong magnetic
field, so the apparatus was assembled to the side of the magnet, and the probe
was attached to the end of a carbon fiber rod, which extended into the magnet
volume. The rod was held parallel to the $x$ direction. By using several 
different rods of different lengths, some additional range $x$ was achieved. 
The rod was attached to the tables by means of specially designed set of brackets. 
Depending on how the brackets were attached, the rod could be extended by 0.2~m
in $y$ or $x$, giving additional position range.

\begin{figure}[htpb]
\centering
\includegraphics{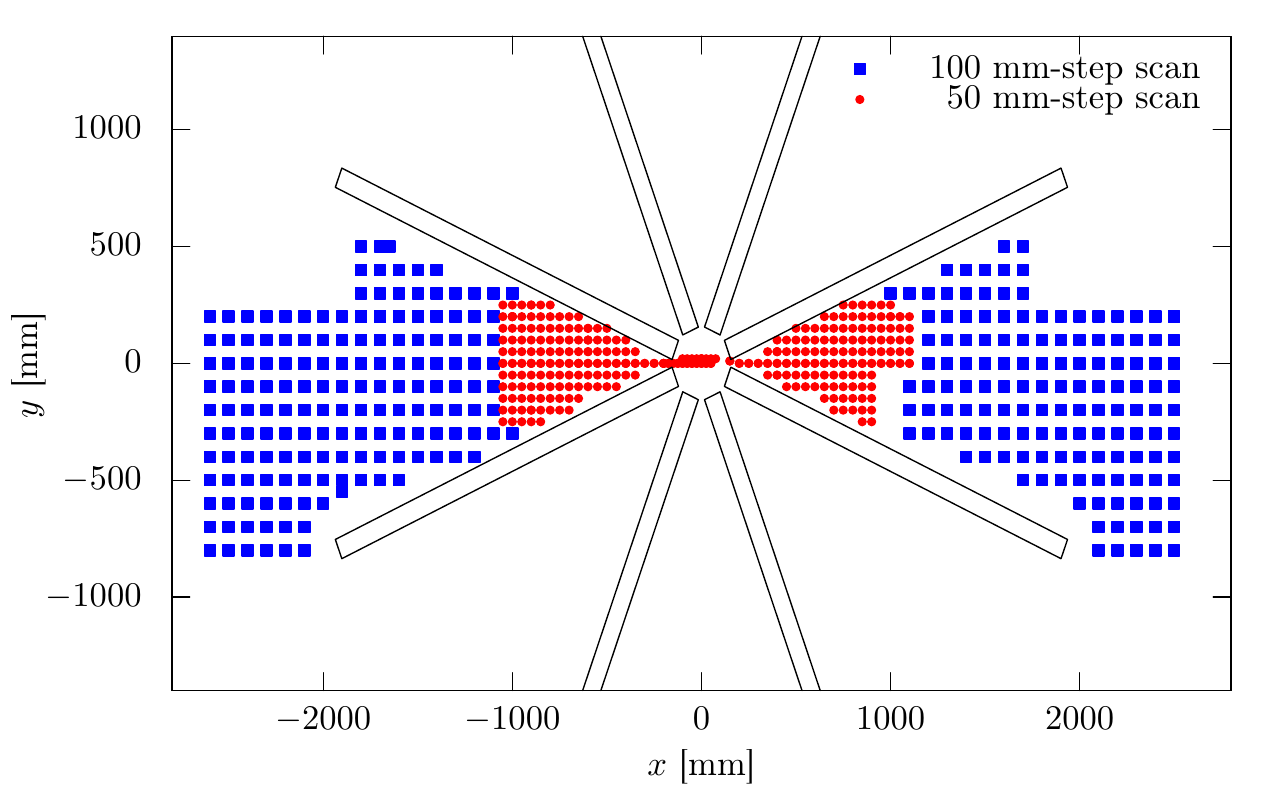}
\caption[Positions of the magnetic field scans]{\label{fig:magnet_scans}Scans were made with greater density in the 
inner region of the magnet where the field gradients are highest.}
\end{figure}

The apparatus was designed to perform scans along the $z$ direction at a fixed
$x$ and $y$. At the beginning of each scan, the probe would be moved to the 
desired $x$ and $y$ using a combination of the 1~m translation stands and 
the three-dimensional table. Then the probe would be stepped along in the
$z$ direction: the probe would move to a new position, wait one second for
vibrations in the rod to dampen, measure, and then move to a new $z$ position.
Scans were taken at 50~mm spacing in the high-gradient region of the spectrometer,
and 100~mm spacing further away. In all, 703 scans were made, the nominal
positions of which are shown in figure \ref{fig:magnet_scans}.

\subsection{Surveying}

During the set-up of the translation tables, the DESY Survey Group used a theodolite
with a laser range-finder to measure the positions of survey targets on the I-beam
to align the the tables with the axes of the OLYMPUS coordinate system. The
Survey Group also measured the position of a target on a $z$-axis stepping motor
over the course of a scan. The results indicated that the target did not follow a
straight line, but in fact followed a trajectory which wiggled several millimeters
up and down over the range in $z$. The shape of the trajectory was found to
be repeatable, and so could be corrected, so long as the probe's start and end position
were known. We realized that we needed to survey the probe position at the start and end 
of each scan. 

\begin{figure}[htpb]
\centering
\includegraphics[width=12cm]{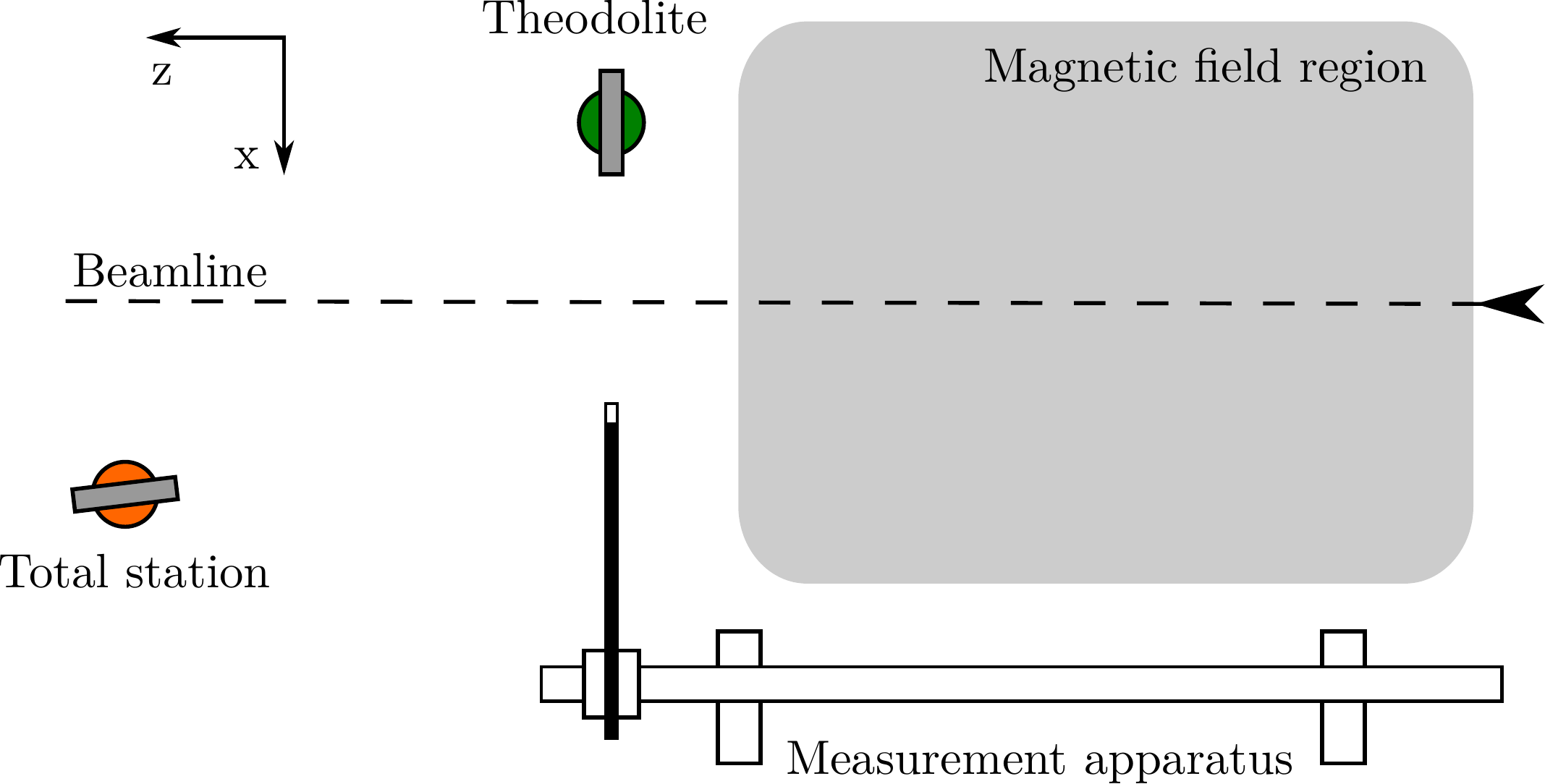}
\caption[Orientation of the survey equipment]{\label{fig:ts_position} This schematic shows the orientation of
the total station and theodolite relative to the magnet and measuring apparatus during
measurements of the left sector, looking down on the experimental hall from above 
(not exactly to scale). The apparatus position is shown for the start of a scan. During 
a scan, the probe would be moved across the magnetic field region (from left to right in this image).}
\end{figure}

The DESY survey group provided a Leica Wild T3000 Total Station, and a Kern E2 Theodolite 
to allow OLYMPUS personnel to start and end positions in between measurement scans. 
Both devices can measure the polar and azimuthal angles to a point in space fixed
by a telescope and cross-hairs. The total station can additionally measure the distance 
to a set of special reflective targets using a laser range-finder. We positioned the
total station in the DORIS tunnel just downstream of OLYMPUS, while the theodolite
was positioned on the opposite sector to have a face-on view of the probe at the
start of each scan. A schematic of this positioning is shown in figure \ref{fig:ts_position}.

To use the laser-range finding feature of the total station, we needed to attach
reflective targets to the measurement rod. The best system we could devise was to
use cable ties to tightly fasten a loop in the target's plastic backing around the
rod. We aimed to keep the target cross hairs 50~mm from the probe position, which 
we checked by ruler. When using the longest rod, the target was sometimes blocked
from view by one of the magnet's supporting K-beams. For this rod, we attached
a second reflective target and used the total station to calibrate its position
relative to the first target.

This survey system depended on our knowing the positions of the theodolite and
total station. We calibrated these positions daily using the survey targets
that had already been fixed to the OLYMPUS support frame and the walls of the
experimental hall. These positions of these targets had previously been ascertained by 
survey. By remeasuring them, the total station and theodolite positions could be fixed.
Occasionally, the survey apparatus would get bumped out of carelessness, but
it was easy to recalibrate the position by remeasuring the survey targets.

\subsection{Procedure}

During data taking, a crew of three people operated equipment to make measurements.
One person controlled the translation tables from a computer station near the side
wall, away from the fringe field of the magnet. A second person operated the 
theodolite to survey the probe tip start position. The third person used the total 
station to survey the reflective targets at each scan's start and end
position.

All of the scans for a given rod were performed at once, to minimize the number of times
that a rod had to be exchanged. First, the medium rod measurements were taken
(for the region 950~mm $< x <$ 1700~mm). Next, the short rod was installed for the
region 1800~mm $< x <$ 2500~mm. Then the long rod was installed for the
region 150~mm $< x <$ 900~mm. The brackets attaching the long rod were then
adjusted to extend the long rod to cover the M\o ller region ($|x|<100$~mm). Finally,
the vertical extension was added and the medium rod reattached to cover the area above
the beamline, near the upper edge of the tracking volume. After these scans were completed on the left
sector, the apparatus was taken apart and reassembled to measure the right sector.

A daily calibration procedure was designed and followed during data taking. Each morning,
the theodolite and total station were each used to measure targets on the walls and frame
to fix the theodolite and total station positions. The positions of the 1~m translation stands 
were checked with rulers against the positions specified by their electronic controllers.
The zero-point on the Hall probe was reset using a Mu-metal cylinder. Finally, a test run
was taken with the magnet turned off. 

\subsubsection{Changing Rods}

A set of procedures was followed in order to change the rod. The work of mounting the
probe in the rod, and attaching the rod with brackets to the translation tables was performed
by a technician in the DESY group MEA 1. No easy way was found to align the probe in the rod,
so the alignment was checked by using the theodolite, and then adjusted by hand. A reflective
survey target was attached to the rod and positioned using a ruler. Finally, a survey scan was
taken. The probe was moved and the target resurveyed at every step point in a scan. These data
were used to correct for deviations in the probe position relative to its nominal position
specified by the translation tables. 

\section{Analysis of the Survey Data}

\subsection{Determing the Probe Position}
\label{ssec:probe_pos}

The first task for analyzing the magnetic field scans was the determination of the spatial
position and orientation of the probe at each measurement point. The scanning procedure
was designed to map out a grid in space, the translation tables deviated from their nominal
positions by several millimeters, necessitating the survey procedure. The integration of the
survey data was divided into three principal tasks:

\begin{itemize}
\item Determining the positions of the targets at the beginning and end of each scan
\item Determining the probe positions relative to these targets
\item Modeling the trajectory of the probe along the length of the scans.
\end{itemize}

Each of these topics is covered in the following sections.

\subsubsection{Determining Target Positions}

For each scan, the scanning rod was surveyed using total station and theodolite.
The two devices are similar; they both have a telescope with crosshairs
and can report the polar and azimuthal angles of the crosshairs.
The total station can also measure the distance to a reflective target using
a laser. This additional distance measurement, combined with the measurement of the
two angles, uniquely constrains the target point in three dimensional space, making
the total station far more useful than the theodolite. Though the theodolite was 
extremely useful for making sure that the probe was aligned during the installation of
each new rod, we never ended up using its survey data, and for the remainder of this
chapter, when I mention survey data, I'll be referring only to that of the total station.

The first task in analyzing the survey data was calibrating the total station's position
and orientation. This was determined from the daily calibration measurements of targets
around the hall and on the OLYMPUS frame. This gave us a way to translate the total stations
measurements into three dimensional positions in the OLYMPUS coordinate system. We found
that the total station position varied only slightly day to day, with the exception of 
a few large displacements, which we attribute to times when the total station was
moved by accident.

\begin{figure}[htpb]
\centering
\includegraphics{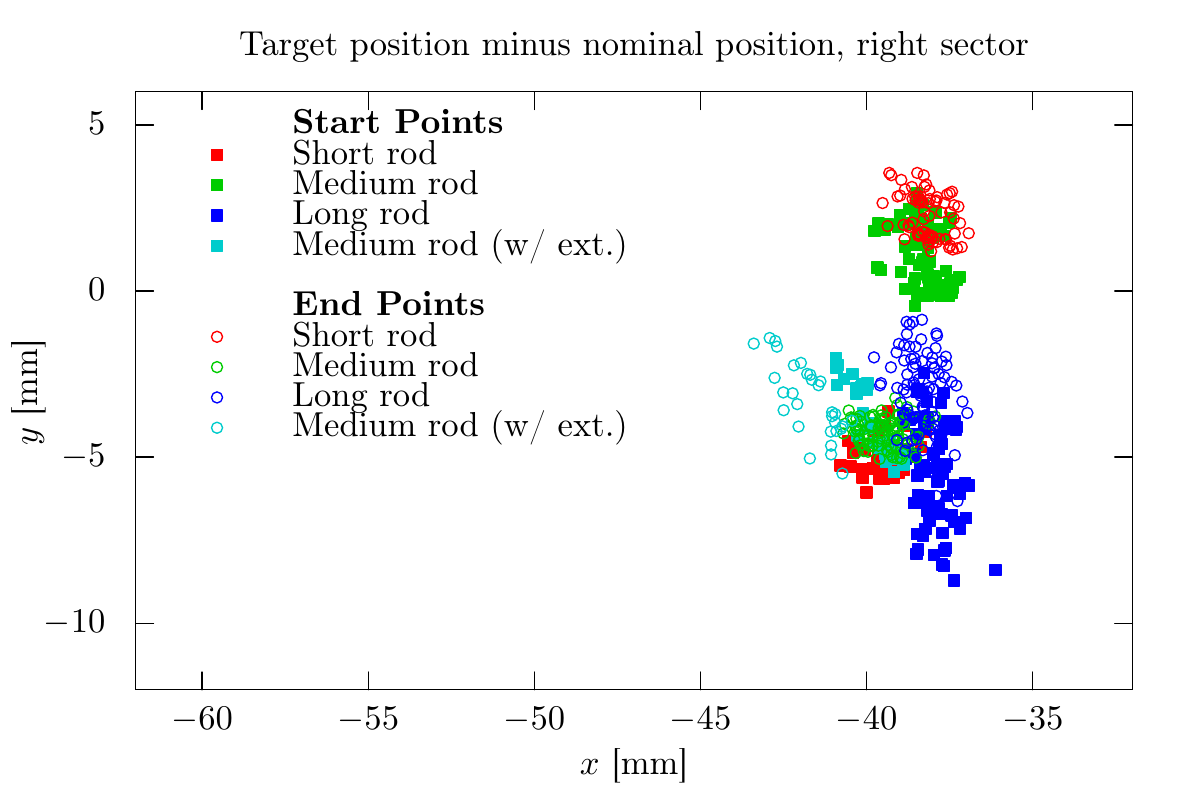}
\caption[Scan start and end positions]{\label{fig:magnet_start_end} The start and end positions were not exactly 
at the nominal scan position due, demonstrating how valuable having a survey was. 
The positions are all well-clustered, indicating that we have fixed all of the
(major) transcription errors. }
\end{figure}

The start and end positions of each scan were then mapped into OLYMPUS coordinates.
On examination of this data, it was found that a handful of points had anomalous 
positions due to transcription errors. The total station measurements were recorded
by pen and paper, and if digits were transposed or mis-written, this, thankfully,
produced a noticable abberation in position. We endeavored to correct as many of
the transcription errors as we could find, approximately a few dozen over 700 scans. 
To search for these errors, the position data were plotted in several ways to look for
inconsistencies. One way was to plot the difference between the target position
and the nominal scan position. If the position data were correct, then they would appear
in this plot to be clustered around a constant value: the difference between the target
position and the probe position. Positions that were very far from the cluster,
were likely to have transcription errors. Another useful plot was the
difference between the start and end positions of a scan. In the $x$ and $y$ directions,
this difference clustered around zero. Outliers were likely to have errors.
After examining every measurement, all of the positions were well-clustered to within 
a few millimeters or better, as can be seen in figure \ref{fig:magnet_start_end}.

One observation that was made during this process was that during the left sector measurements,
the $Z$ axis of the translation table came out of alignment with the OLYMPUS $Z$ axis.
This occurred between the measurement of the end point of scan 93 and the start point of scan 94.
Between these scans, the large $XY$ tables were adjusted, so it is possible that during this
movement, the tables made contact with part of the magnet. This misalignment posed no
problems because the survey data allowed the probe positions for scans after 94 to be corrected.

\subsubsection{Determining the Probe Positions}

The next task was to determine the position of the Hall probe relative to the positions
of the total station targets. The reflective targets were fixed to the measurement
rods by hand using cable ties, but their positions were not precisely
known. A set of offsets was needed to establish the position of the Hall probe relative
to the positions of the targets on each rod.

Reflective targets were fixed to the rod using cable ties, and positioned 50~mm from
the probe using a ruler. We attempted to use the total station to check the angles
between the probe and target but could not fix the probe position in the cross hairs
since we were looking side-on. Therefore we assumed that the probe was always 50~mm
away from the target in $x$, and 0~mm in $y$. The deviation $z$ was directly related
to the thickness of each rod, which we checked using a micrometer with calipers.

For some scans the target close to the probe was not visible, and a second target further
up the rod was needed. We used the total station to fix the distance between the two targets
in three dimensions when both targets were visible. This allowed us to produce an offset
for use when the near target was not visible.

\subsubsection{Scan Trajectories}

Early on, during the set up of the measurement apparatus, it was noticed that the translation
tables did not maintain a consistent vertical position over the course of the full range in the
$Z$ direction. Instead, the table height fluctuated several millimeters over the 
course of a scan. Corrections to the scan positions were required to accurately
reflect the true trajectory of the Hall probe over the course of a scan.

\begin{figure}[htpb]
\centering
\includegraphics{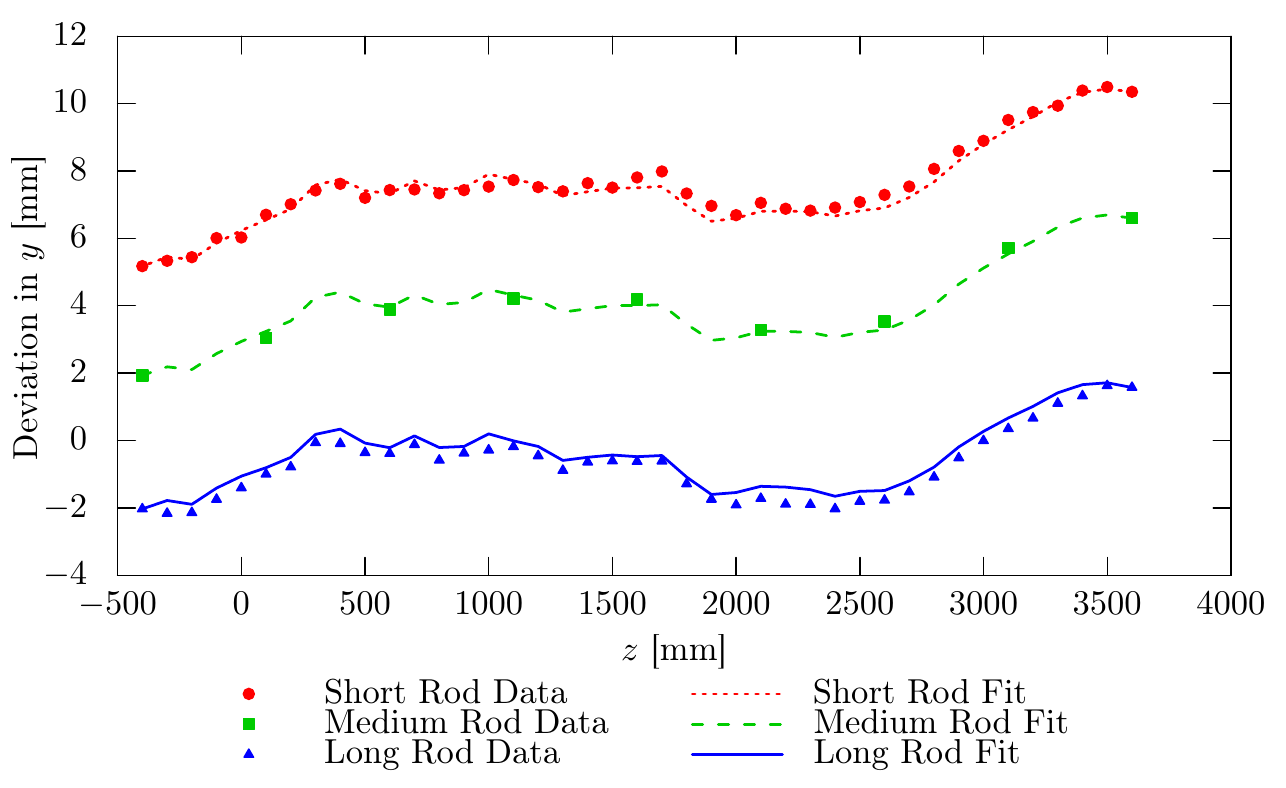}
\caption[Position deviations along a scan]{\label{fig:magnet_zscan_traj} The deviations in scan position could be corrected
using calibration data fit with a simple parameterization.}
\end{figure}

To measure these trajectories, a test scan was made for each rod, in which the rod was
stepped in increments of $Z$ and the target positions were measured with the total station.
From these measurements, it was found that the overall shape of these fluctuations was
consistent for each set-up of the apparatus, i.e., all scans on the left sector had similarly
shaped trajectories, but these were different from the shape of the trajectories on the right
sector. It was also found that the magnitude of these fluctuations depended predictably
on the length of the rod being used. In fact it was sufficient to parameterize the trajectories
of the Hall probe in the following way:
\begin{equation}
x_{r,s}(z) = x_s(z) + L_r \cos\left(\theta_s (z)\right) + \text{linear term},
\end{equation}
\begin{equation}
\label{eq:fitY}
y_{r,s}(z) = y_s(z) + L_r \sin \left(\theta_s (z)\right)  + \text{linear term}.
\end{equation}
Index r represents the rod being used, while index s represents the sector (left or right). The
trajectory could be expressed in terms of three functions $x_s(z)$, $y_s(z)$, and $\theta_s (z)$
that were consistent across each sector, as well as the rod length $L_r$ which was measured.
The three functions were extracted using a fit to the calibration data, and the residuals
were found to be sufficiently small. A linear term was used to match the trajectory to each 
scan's start and end points. Fits to trajectories on the left sector from three rods are shown
in figure \ref{fig:magnet_zscan_traj}.

\subsubsection{Results}

\begin{figure}[htpb]
\centering
\includegraphics{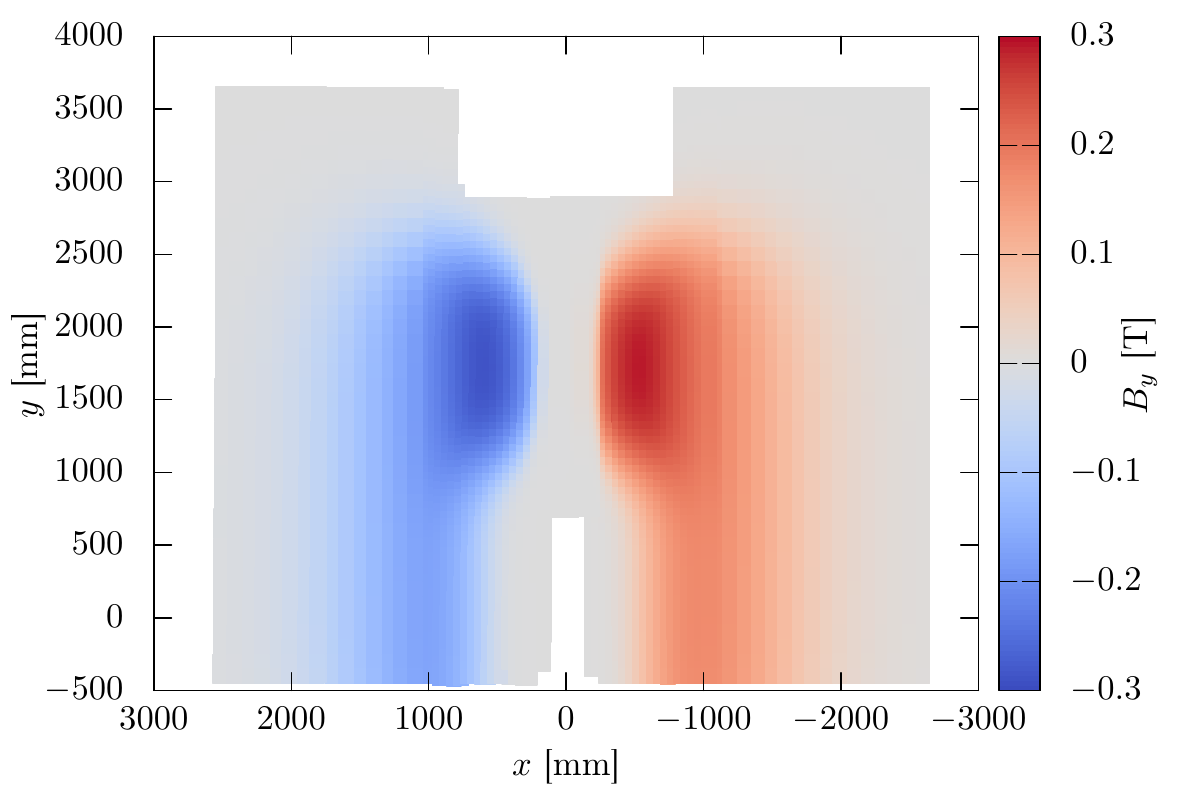}
\caption[Magnetic field measurements in the $y=0$ plane]
{\label{fig:field_as_meas} This figure shows our measurement of the magnetic field in the $y$ direction
along the $y=0$ plane.}
\end{figure}

By determining the start and end points of each scan, mapping the probe position to the target
position, and correcting the trajectories between endpoints, we have mapped the magnetic field 
data into OLYMPUS coordinates. The field in the $y=0$ plane, which is nominally the center of the
tracking acceptance, is shown in figure \ref{fig:field_as_meas}. We can see that the field points
downwards in the left sector ($x > 0$) and upwards in the right sector ($x<0$).

\subsection{Generating a Field Map}

\subsubsection{Overview}

The result of the work in section \ref{ssec:probe_pos} was to convert the magnet field
measurements into a list of positions and corresponding field vectors. The challenge of
this section is to turn that data into a field map; a value of the magnetic field as a
function of position. This requires determining the magnetic field between nearby 
measurement points (interpolation) and at points that are outside the spatial extent 
of the measurements (extrapolation). Interpolation in three dimensions is straight-forward
if the measurement points fall on a regular grid. Because of the position deviations of
our apparatus, our data do not. Direct interpolation on an irregular grid is a much more
difficult proposition and is one that we sought to avoid.

Instead, we chose to build a model of the magnetic field, parameterized by the positions
and orientations of the magnet coils, and to fit these parameters to our measured data.
This approach carried two advantages. The first was that after fitting, we could then 
use the model to calculate the field at any arbitrary point in space. The second was that
any errant data points, for which we assigned an incorrect position or recorded the wrong field,
would have their influence in the fit dampened by all of the other measurements.
The challenge of this approach was to create a model that could both accurately reproduce 
the measured data, while also being well constrained by it. We settled on a 35-parameter 
model, in which the parameters represented position offsets, rotation offsets, and scaling
parameters.

\subsubsection{Magnet Model}

We chose to model the OLYMPUS magnet with a collection of line segment-shaped currents
that traced the position of the current-carrying copper bars within each coil. In reality,
current is carried through the volume of each copper bar, but by approximating the bars 
by filaments, we avoided having to model the current distribution over the cross-sectional
extent of each bar. The field produced by a line-segment of current is also easy to calculate.
By integrating the Biot-Savart Law, we find that for a segment centered at point $\vec{c}$,
carrying current $I$, and with a length vector $\vec{L}$, the field at position $\vec{p}$ is 
given by:
\begin{equation}
\vec{B} = \frac{\mu_0 I}{4\pi \beta^2} \left[
\frac{\left(\alpha + |\vec{L}|/2\right)}{\left|\vec{c} - \vec{p} + \vec{L}/2 \right|} -
\frac{\left(\alpha - |\vec{L}|/2\right)}{\left|\vec{c} - \vec{p} - \vec{L}/2 \right|} \right]
\cdot \left( \vec{c} - \vec{p} \right) \times \vec{L},
\label{eq:biotsavart}
\end{equation}
\begin{equation}
\alpha \equiv \hat{L} \cdot \left( \vec{c} - \vec{p} \right),
\end{equation}
\begin{equation}
\beta^2 \equiv \left( \vec{c} - \vec{p} \right)^2 - \alpha^2 .
\end{equation}

The nominal OLYMPUS coil shape is irregular, but can be broken down into straight segments
and circular arc segments. The straight segments were trivially represented by filaments.
Rather than calculating the field produced by arcs, which requires special functions called
Appell functions, we chose to approximate the arc-shaped segments with collections of line-segments.
The proper placement of these line segments is a non-trivial exercise. If the start and end points 
lie on the path of the arc, then the area of the arc is diminished and the dipole moment decreased. 
The segments can be spread out to preserve the area of the arc, but this means that the end points 
first and last segment will no longer lie on the end points of the arc. If the arc is a portion of 
a current loop, then the loop will not be continuous at the start and end points of the arc.
To balance these concerns, it was decided to position the segments so that the first
and last segment began and ended, respectively, on the endpoints of the arc, preserving
continuity of any current loops. The other segments' endpoints would be expanded
outward to a larger radius $R'$ to maintain the area of the arc. The expression for $R'$, derived 
from geometry, is
\begin{equation}
R' = R \left( \sqrt{\frac{1}{(N-2)^2} + \frac{\theta_a}{(N-2)\sin(\theta_a / N)}} - \frac{1}{N-2}\right),
\end{equation}
where N is the number of segments used to approximate the arc, and $\theta_a$ is the
angle subtended by the arc.

In our model, each coil was identical in shape and composition. Each coil was represented
by 26 current loops, arranged in two rows of 13, in the nominal positions of the copper
bars within the coils. The two $180^\circ$ arcs of each loop were approximated by 180 line
segments, while the two $38.58^\circ$ arcs were approximated by 40 segments.

\subsubsection{Fitting the Magnetic Field Data}

To fit our model, we had to choose a set of free parameters to vary. When we began, all we 
knew were the nominal positions of each coil. We did not know what freedom needed to be 
introduced to the model to accurately describe our magnetic field data. Did the coil positions
need to vary only in radius from the beamline, or in orientation as well? Was the toroid
axis where we thought it was? Were the coils the correct size?

Our approach was to try many different combinations of free parameters and then to simultaneously
monitor the fit residuals and the parameter values themselves. Our fitting criterion was that 
we minimize the sum of the squared residuals: $\sum |\vec{B}_\text{meas.} - \vec{B}_\text{model}|^2$.
Fits that converged on minima where the parameters were wildly different than their expected values 
indicated that some  parameters in our model were not well-constrained by the data, and should not be 
given freedom. If adding a free parameter reduced fit residuals without causing any parameters to go wild, 
we took that as a sign that more freedom could be added. 

One natural parameter set consisted of the three positions and three rotations for all eight coils.
However the final positions for the top two and bottom two coils were wildly different
than what one would expect. We have no measurements close to those coils, and so they
had no firm constraints on their positions and rotations. Instead, it was more productive 
to give the four coils surrounding the tracking volume (numbered 0, 3, 4, and 7) more freedom,
while restricting the top and bottom pairs of coils (numbered 1, 2, 5, and 6).

We settled on a parameter set with 35 free parameters. The four coils adjacent to the
tracking volume were given 24 parameters to fully describe their positions and orientations.
The remaining four coils were collectively given eight: three positions to specify the toroid
origin, three rotations to specify the toroid axis, one to specify the distance of the coils
from the axis, and, lastly, one parameter to describe the amount of rotation in the plane of
the coil.\footnote{The in-plane rotation was indicated by a study in which we used the total
station to see how much the magnet moves due to magnetic forces. This study is described
in Appendix \ref{app:magnet}.} Of the remaining three parameters, two were scaling factors
in the plane of the coils, allowing all eight coils to collectively shrink or expand in two
dimensions. The final parameter was the current in the magnet, which the fit predicted to
within 0.2\%. The values of these parameters, after the fit converged, are shown in table
\ref{tab:params}. For reference, if the magnet occupied its nominal position, the parameters
would have all been 0, with the exception of the scaling factors $S_x$ and $S_y$, which would be 
1, and the current, which would be 5000~A. 

\begin{table}[htbp]
\centering
\begin{tabular}{|c c | c c | c c | c c |}
\hline
Par. & Value & Par. & Value & Par. & Value & Par. & Value\\
\hline \hline
$\Delta \rho_0$ & 26.58 mm & $\Delta \rho_3$ & 24.64 mm & $\Delta \rho_4$ & -4.962 mm& $\Delta \rho_7$ & -1.494 mm\\
$\Delta \phi_0$ & 0.01$^\circ$& $\Delta \phi_3$ & -0.006$^\circ$ & $\Delta \phi_4$ & -0.006$^\circ$ & $\Delta \phi_7$ & 0.002$^\circ$\\
$\Delta z_0$ & 8.214 mm & $\Delta z_3$ & 1.429 mm & $\Delta z_4$ & 0.332 mm & $\Delta z_7$ & 22.74 mm\\
$\Delta \theta_{x0}$ & -0.040$^\circ$ & $\Delta \theta_{x3}$ & -0.065$^\circ$ & $\Delta \theta_{x4}$ & 0.12$^\circ$&$\Delta \theta_{x7}$ & 0.547$^\circ$\\
$\Delta \theta_{y0}$ & -0.939$^\circ$ & $\Delta \theta_{y3}$ & 1.141$^\circ$ & $\Delta \theta_{y4}$ & 1.1$^\circ$&$\Delta \theta_{y7}$ & -0.891$^\circ$\\
$\Delta \theta_{z0}$ & 0.402$^\circ$ & $\Delta \theta_{z3}$ & 0.635$^\circ$& $\Delta \theta_{z4}$ & -0.846$^\circ$ &$\Delta \theta_{z7}$ & -0.492$^\circ$\\
\hline
$x$ & 3.357 mm & $\alpha_x$ & 1.368$^\circ$ & $S_x$ & 0.986 & $\theta_{\text{coil}}$ & -0.05$^\circ$ \\
$y$ & -21.7 mm & $\alpha_y$ & 0.017$^\circ$ & $S_y$ & 0.998 & $\Delta \rho$ & 12.35 mm \\
$z$ & 4.633 mm & $\alpha_z$ & -0.163$^\circ$ & $I$ & 5007 A & & \\
\hline
\end{tabular}
\caption[Best fit parameters for the magnetic field model]
{The final fit had 35 free parameters. Coils 0, 3, 4, and 7 were given full freedom because they surrounded the tracking volume
and were well constrained by the measurements.}
\label{tab:params}
\end{table}

\subsubsection{Results}

\begin{figure}[htpb]
\centering
\includegraphics{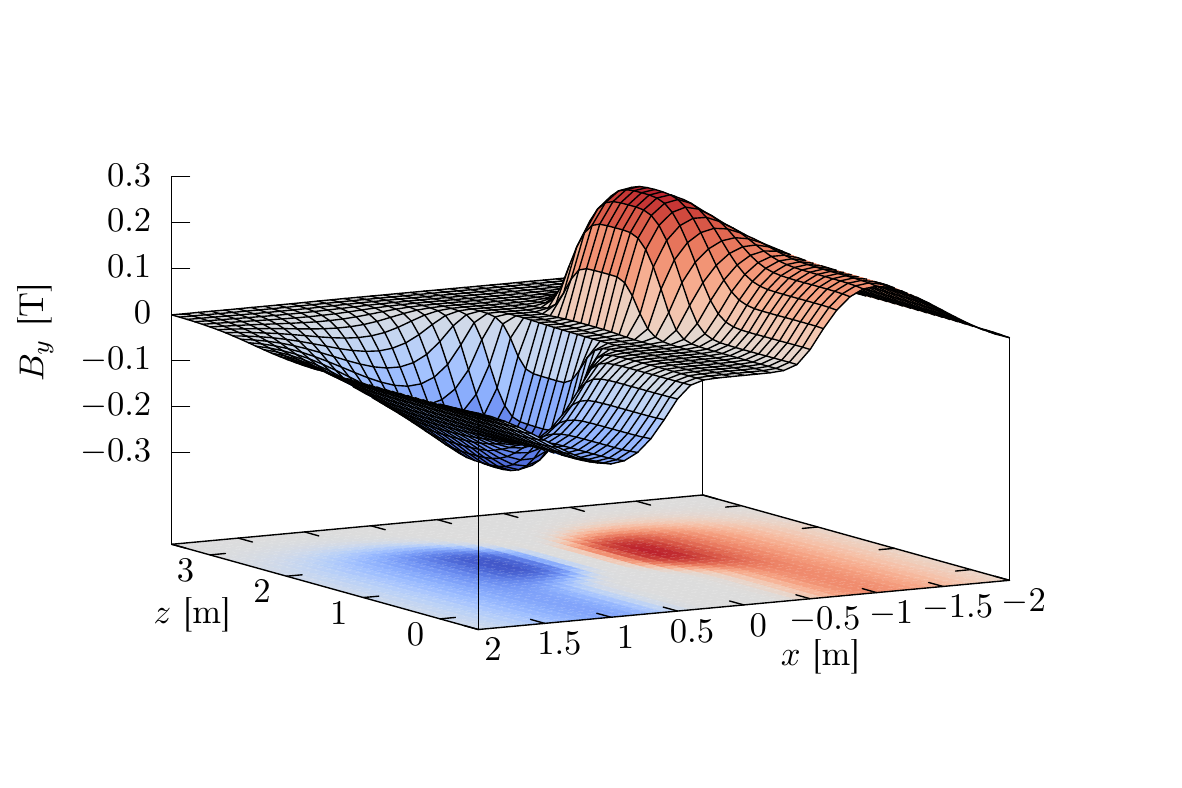}
\caption[Magnetic field model in the $y=0$ plane]{\label{fig:field_in_plane} The magnetic field map is shown in the $y=0$ plane.}
\end{figure}

\begin{figure}[htpb]
\centering
\includegraphics{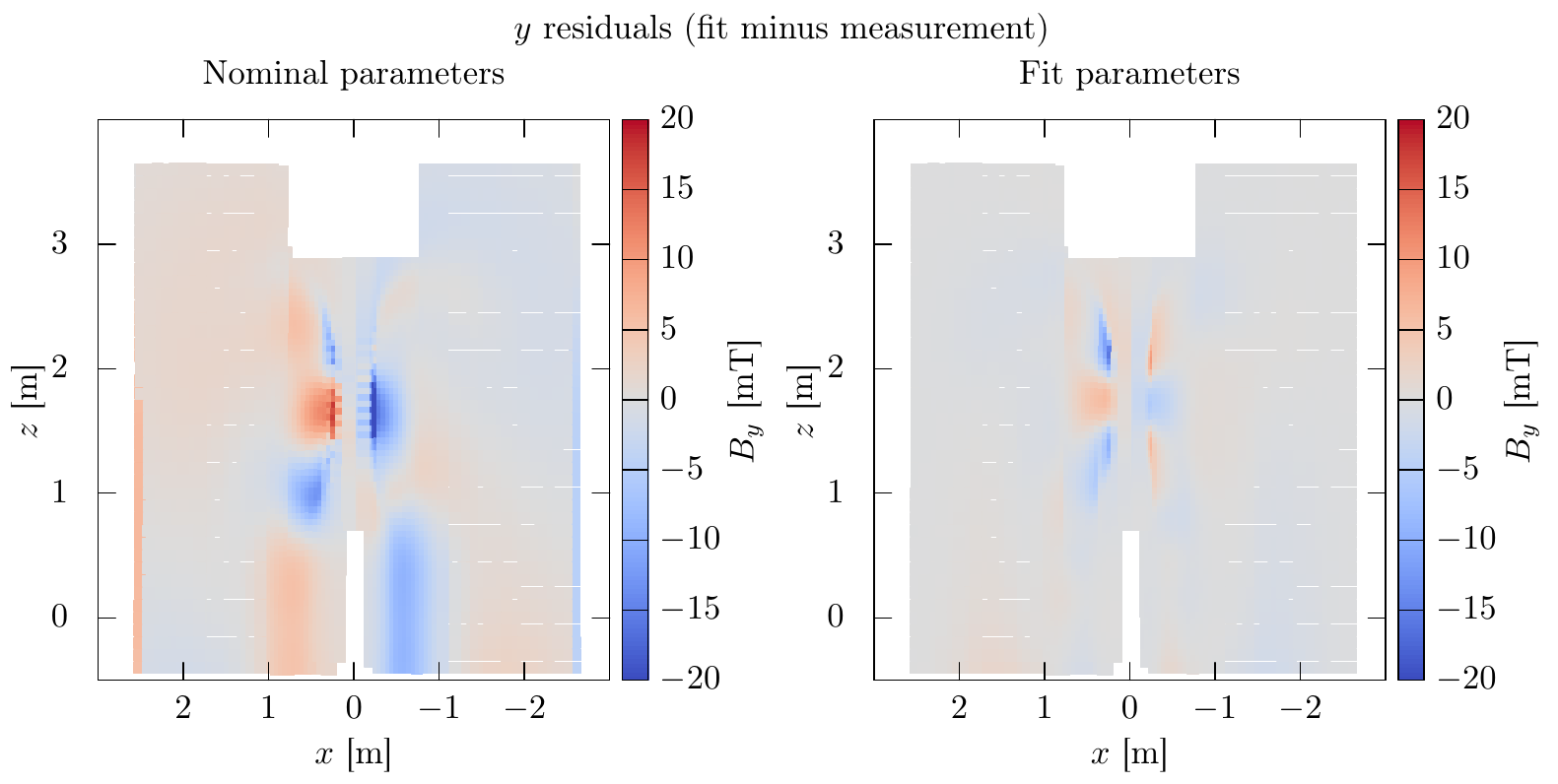}
\caption[Magnetic field residuals before and after fitting]
{\label{fig:magnet_res_y} The nominal field was not too far off from reality, but with slight adjustments
in the positions of the coils, the fitted model did a better job of matching the field measurements.}
\end{figure}

Figure \ref{fig:field_in_plane} shows the magnetic field in the $y$ direction, as calculated from 
the field model after fitting the parameters to data. Qualitatively, there are no significant changes
to the shape of the field compared with our expectations. More can be learned from looking
at the residuals of the fit, shown for the $y$ direction in figure \ref{fig:magnet_res_y}. 

The fit improved the model's match to our measurements from an r.m.s.\ residual 
$\sqrt{\frac{1}{3N}\sum |\vec{B}_\text{meas.} - \vec{B}_\text{model}|^2}$ (where $N$ is
the number of measurement points) from 3.198 mT to 1.873 mT. These residuals do
not represent random Gaussian errors, as can be seen in figure \ref{fig:magnet_res_y}.
There are specific regions that are not modelled well. This can be caused both by
inadequacy of the model and by errors in the survey. As examples, in the first case, 
perhaps the coil arcs are not exactly circular or the conducting bars are unevenly 
spaced. In the second case, we may have assigned the wrong positions to measurement
points because of mis-measuring the probe to reflective target distance. In the first
case, the model is to blame. In the second case the model helps by constraining what
would otherwise be inaccurate measurements. 

\section{Using the Field Map in Software}

\subsection{Spline Interpolation}

\label{ssec:magnet_spline_interp}

To calculate the magnetic field at a given position, the magnetic field model of the previous section 
works by applying equation \ref{eq:biotsavart} to each current segment in the ensemble that makes
up all eight coils. This method is fast enough for some applications (such as fitting free parameters,
or examining the shape of the field), but is far too slow to be used for simulation of tracks or
for track reconstruction. Both of these applications require hundreds if not thousands of field
queries per track, and the rates need to scale in order to track the billions of events in the
OLYMPUS data set. A significant speed up was needed. 

Our solution was to use the magnetic field model to pre-compute field vectors (and derivatives, via finite differences)
on a grid over the spectrometer volume. A fast field calculation could be performed by using
three dimensional spline interpolation between the nearest grid points. Spline interpolation
was preferred to linear interpolation so that the first derivatives of the field would be
continuous and non-zero, helping the simulation software query the field appropriately. 

Interpolation would have been possible directly on the measurement data. There were three reasons
for not pursuing this approach. Primarily, the measurement data only covered a limited amount of space,
and extrapolation was needed as well as interpolation. However in addition to this, the algorithms for
interpolation on a non-uniform grid are much slower than those on a uniform grid. Lastly, the
presence of a fitting step in between the measurements and interpolation served to smooth
the data and remove errors that might be associated with faulty survey measurements.

The grid was chosen to range from -2.5~m to 2.5~m in $x$, from -1~m to 1~m in $y$, and from
-0.5~m to 3.5~m in $z$ with 50~mm spacing. The grid had a total of 418,241 points. The choice
of grid had to balance point density, which would reduce the errors from interpolation with
the size of the grid, which the simulation software would need to store in memory.

The interpolation scheme was based on an algorithm by Lekien and Marsden \cite{NME:NME1296}
that uses eight values per field direction $i$, per grid point $P$: 
\[
C_{i,P} = \left\{ B_i, \frac{ \partial B_i }{\partial x }, \frac{ \partial B_i }{\partial y },
\frac{ \partial^2 B_i }{\partial x \partial y }, \frac{ \partial B_i }{\partial z },
\frac{ \partial^2 B_i }{\partial x \partial z },\frac{ \partial^2 B_i }{\partial y \partial z },
\frac{ \partial^3 B_i }{\partial x \partial y \partial z} \right\}.
\]
My colleague Brian Henderson developed several improvements to the algorithm \cite{henderson:thesis}.
The first was to refactor the spline basis functions to reduce the memory needs, or alternatively
reduce the number of multiplications needed per query. To see how this works, consider that a spline 
can be written in a basis such that: 
\begin{equation}
B_i(x,y,z) = \sum_{l,m,n=0}^3 a_{i,lmn} x_f^ly_f^mz_f^n \:\:\:\: i\in \left\{ x,y,z \right\},
\end{equation}
where the coefficients $a_{i,lmn}$ are polynomial functions of the field and derivatives at the
eight surrounding grid points. The calculation of $a_{i,lmn}$ requires three $64\times 64$ matrix
multiplications, which can either be computed on the fly, or precomputed and stored. The first 
option is slower; the second option requires significant memory, which scales with the volume of
the grid. Brian refactored the splines into basis functions:
\begin{align}
f_0(x) =& (x - 1 )^2 (2x + 1)\\
f_1(x) =& x (x - 1)^2\\
f_2(x) =& x^2 (3-2x)\\
f_3(x) =& x (x - 1),
\end{align}
so that the spline could be written in the form:
\begin{equation}
B_i(x,y,z) = \sum_{l,m,n=0}^3 b_{i,lmn} f_l(x) f_m(y) f_n(z) a_{i,lmn} \:\:\:\: i\in \left\{ x,y,z \right\},
\end{equation}
where the coefficients $b_{i,lmn}$ are the values $C_{i,P}$ for the eight surrounding grid points.
If we denote these eight points as a vector $P_j$:
\[
 P_j = \{xyz, Xyz, xYz, XYz, xyZ, XyZ, xYZ, XYZ\},
\]
where miniscule denotes a grid point with a smaller value of the coordinate than the
queried point, and majiscule denotes a grid point with a larger value of the coordinate than 
the queried point, then
\begin{align}
b_{i,lmn} =& C^k_{i,P_j},\\
j =& 4\left(\frac{n}{2}\%2\right) + 2\left(\frac{m}{2}\%2\right) + \left(\frac{l}{2}\% 2\right),\\
k=&4(n\%2) + 2(m\%2) + (l\%).
\end{align}
The memory costs of precomputation are eliminated, and the matrix multiplications are avoided. 

The second improvement was to store the values $C_{i,P}$ in an array in memory so that the coefficients
for each of the eight grid boxes were grouped into sets of continuous blocks of memory. This 
permitted Single-Instruction Multiple Data (SIMD) computing, which could improve the speed
of queries for some computer architectures. 

\subsection{Field in the Beamline Region}

\begin{figure}[htpb]
\centering
\includegraphics{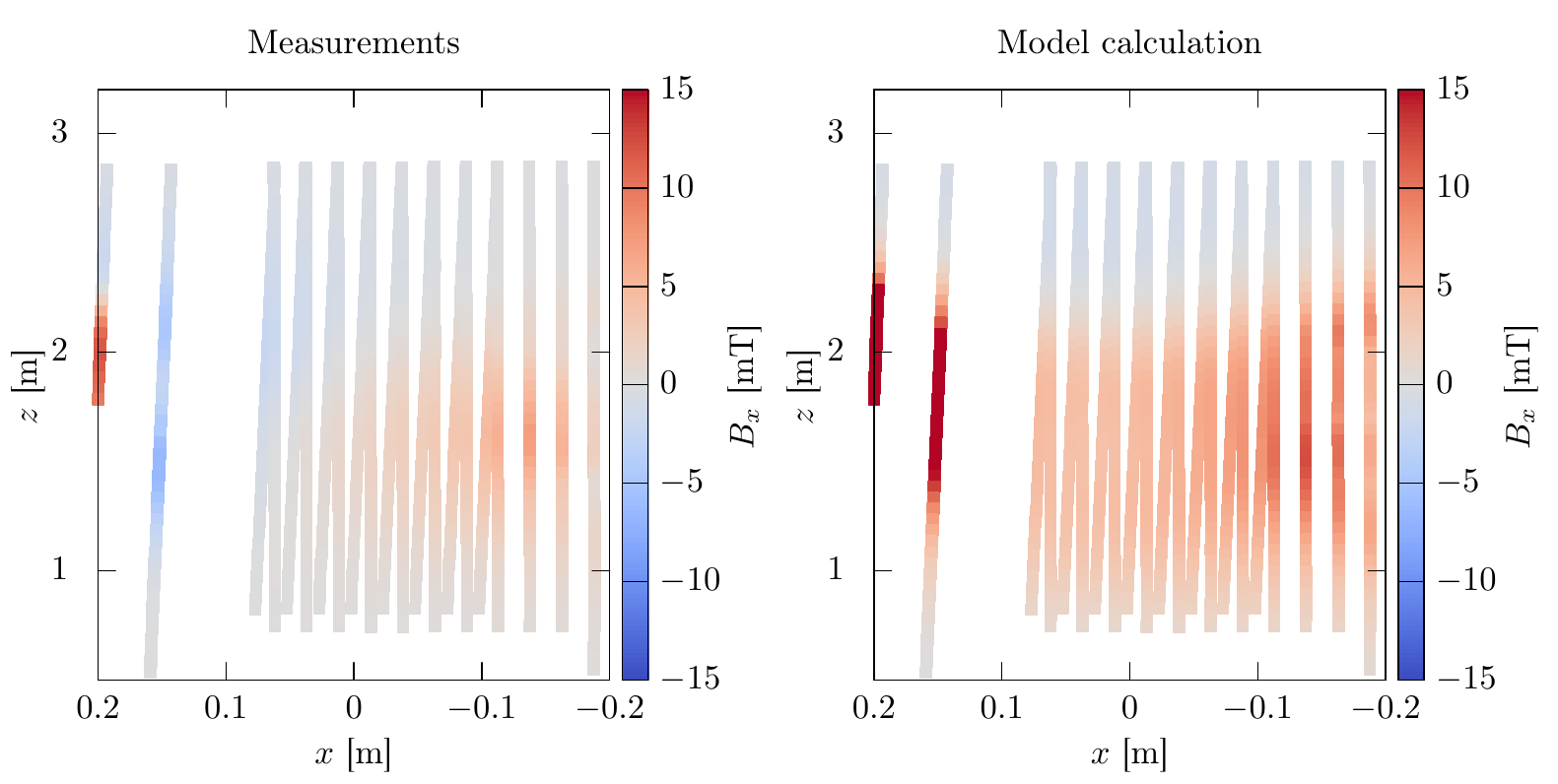}
\caption[Residuals in the beamline region]{\label{fig:magnet_bl_x} The magnet model did not do a good job in the beamline region,
as seen here for the $x$ component of the field.}
\end{figure}

The field model was unable to reproduce features in the magnetic field in the beamline region.
This can be seen in figure \ref{fig:magnet_bl_x}.
This was a difficult region to model correctly because it is physically close to the edges of
all eight coils, and any changes in position in any coil can introduce gradients in the beamline
region. It is also a region where not many measurements were made. It was difficult to stick the
probe in between the coils, and measurements could only be taken from the left or right, not from 
above or below. Still, the measurements indicated some clear features that the model could not
reproduce, which provoked concern for the simulation of M\o ller/Bhabha events which pass 
through the beamline region. We decided that simulations in this region (defined by $|x|<100$~mm,
$|y|<50$~mm, and $500 < z < 3000$~mm) would use a different interpolation scheme. 

\begin{figure}[htpb]
\centering
\includegraphics{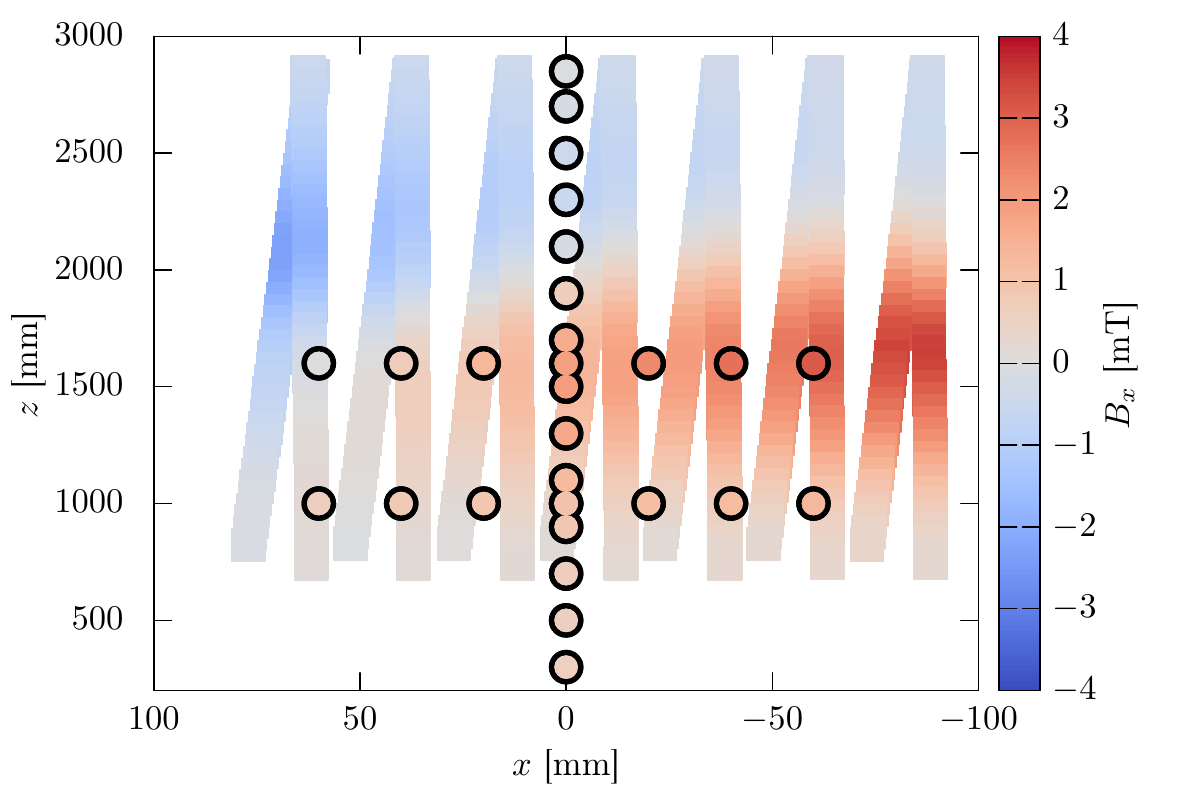}
\caption[Comparison 0f 2011 and 2013 measurements in the beamline region]
{\label{fig:magnet_bl_2011} The 2011 measurements of the beamline region (shown in circles) match all the
features from the main 2013 measurements (shown as bands). }
\end{figure}

Our first cross check was the data from field measurements taken in 2011 of the field in the beamline
region. These measurements were used to align the coils, since the beamline field is zero if the toroid
is perfectly aligned. These measurements were made with a completely different apparatus and survey
technique. The results, shown in figure \ref{fig:magnet_bl_2011}, confirmed that the features we measured
in 2013 are correct, and not the result of spurious measurements.

We decided to interpolate directly on the beamline field measurements. Since the measurements were all
made at $y=0$, we made the approximation that the field did not vary in $y$. This allowed us to use a 
two-dimensional Lagrange polynomial interpolation over $x$ and $z$ \cite{Bevington:1305448}. 

In the region inside the target chamber (defined by $|x|<100$~mm, $|y|<50$~mm, and $|z|<500$~mm),
we had no 2013 measurements, since the chamber was still in place and obstructed the probe. But we 
did have 2011 measurements, which became a lot more trustworthy when we showed that they matched the 
2013 measurements in the beamline region. The 2011 measurements in the target chamber region were taken
along the $x=0, y=0$ line. We assumed that, in this region, there was no variation in $x$ or $y$, and
we modelled the $z$ variation by fitting a quadratic function to each component from the 2011 measurements.

\subsection{Field Near the Coils}

Our spline interpolation procedure ran into trouble around grid points that were very close to any of
the line-segments of current in the model. One effect of modeling the conductors as line segments is that
the field diverges at the segments. The field and derivatives get very large close to the segments.
Ordinarily, tracks never get that close to the magnet coils. The coils are positioned at $\phi = \pm 22.5^\circ$,
while the tracking acceptance covers the region $|\phi|<15^\circ$. However, when interpolating on a
grid, some of the grid points will inevitably be close to segments, and any grid point with enormous
derivatives will spoil the interpolation over any of the boxes with which it is associated. The worst
areas were a few hundred millimeters from the beamline where the tracking acceptance passes closest
to the inner edges of the coils. 

Our solution was to recalculate the field and derivatives for grid points with $|\phi| > 15^\circ$ using
a different technique, making use of the approximate azimuthal symmetry of the toroid. For such a grid point,
we would first calculate the magnetic field and derivatives at a point with the radius from the beamline, and
same $z$, but for $\phi$ equal to the nearest $\pm 15^\circ$ plane. We would set all azimuthal derivatives to
zero (assuming perfect azimuthal symmetry), and then rotate the vectors back to the original grid point.
For example, if a grid box were to fall at $r=1$~m, $z=2$~m, and $\phi=20^\circ$, we would first calculate the field and derivatives 
$\{ \frac{\partial \vec{B}}{\partial r}, \frac{\partial \vec{B}}{\partial z}, \frac{\partial^2\vec{B}}{\partial r \partial z} \}$ 
at the point where $r=1$~m, $z=2$~m, and $\phi=15^\circ$, assuming all other derivatives were 0. We would then 
rotate all the vectors by $5^\circ$ so as to properly correspond to the position where $\phi=20^\circ$.

We used spline interpolation on this new grid exactly as before. We found that the simulated trajectories
at $\phi=\pm 15^\circ$ were essentially the same as those simulated with the non-interpolated (and very slow)
Biot-Savart field model.

\section{Improvements for Future Field Measurements}

Though the OLYMPUS magnetic field measurements turned out successfully, if I were involved in
a similar project in the future, there a few things I would do differently. I share these ideas
in case they can help the reader save time and work on any future field measurements.

\subsection{Reliability of the Translation Tables}

The largest fraction of total effort went into surveying and interpreting survey data. This could
have been avoided if we have used translation tables that could deliver the desired accuracy. The
problem, as far as we can tell, was that the long 4~m table rested on an I-beam which was slightly 
deformed, giving the entire table a degree of deformation. Because of the table position and probe
position now deviated by several millimeters, we had to introduce a survey procedure. Having a theodolite
operator and total station operator effectively tripled the man-hours needed for the magnetic field
measurements. And furthermore a great deal of analysis time was sunk into making sense out of the
survey data. I now appreciate that work done ahead of time to test the accuracy of the translation
tables would have been recovered many times over if we had avoided this pitfall.

\subsection{Uniform vs.\ Non-Uniform Measurement Spacing}

Since the translation tables had deviations, our measurements were no longer in a regular grid.
Interpolation on regular grids is simple, while interpolation on irregular grids is an ordeal.
If our tables had delivered uniformly spaced measurements, direct interpolation would have been
an option for us.

If uniformly-spaced measurements aren't feasible, then it makes no difference if the measurements
are spaced nearly evenly or completely randomly. It struck us that a Hall probe mounted to a small
helicopter drone, combined with a automatic laser tracker could have given us about the same performance
as our immense translation table apparatus. A drone could fly randomly back and forth through the magnet, and
as long as the laser tracker could record the position at each measurement point, it would be quick work to 
survey the magnetic field. 

\subsection{Better Survey Marks}

It felt a little bit silly to use a high-precision laser range-finding total station to measure
reflective targets that were positioned using a ruler. There is a always weakest link, and in our
survey, it was the calibration of the probe to target distance. Had we not been under time pressure
to complete the measurements, I think we could have devised a modification to the carbon fiber rods
that would have fixed the survey target in relation to the probe. At the very least we might have
been able to check the distance with a micrometer and calipers instead of a ruler.

\chapter{Radiative Corrections}
\label{chap:rc}

\section{Introduction}

``Radiative corrections'' is the general moniker for the procedure of accounting
for higher order terms in a perturbative expansion. Often, radiative corrections
depend on the properties of an experiment's apparatus, e.g., the resolution,
acceptance, or threshold of a detector. In these cases, it is crucial for the 
experimenters to apply radiative corrections to their data because, without them, 
the results of the experiment cannot be interpreted. 

\begin{figure}[htpb]
  \centering
  \includegraphics{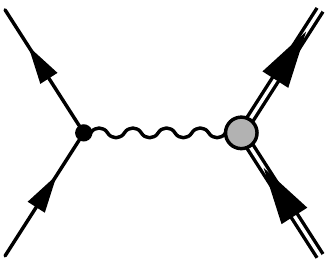}
\caption[The leading order one-photon exchange diagram]
{\label{fig:ope}There is a single leading order diagram: one-photon exchange.}
\end{figure}

\begin{figure}[htpb]
  \centering
  \includegraphics{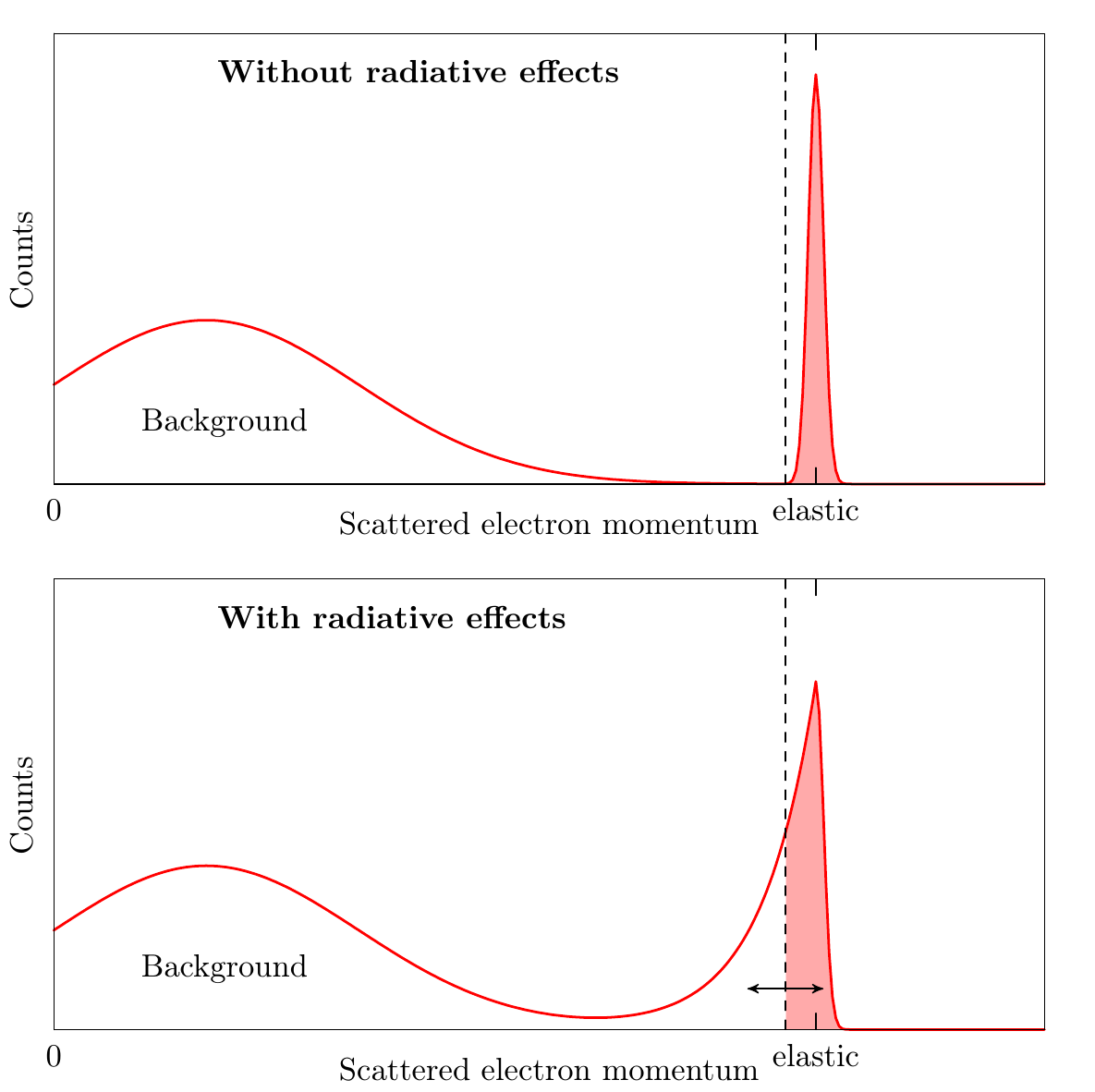}
\caption[Example spectrum showing the effect of radiative tails]
{\label{fig:rc_example} This figure shows how radiative effects complicate
the analysis of an inclusive elastic electron scattering experiment. Both plots show
sketches of what momentum spectra might look like in a spectrometer positioned at fixed
scattering angle. The top plot represents an idealized case, where radiative effects
do not exist; it is simple to separate elastic events from background with a simple
cut on momentum. The bottom plot represents a more realistic case, where the elastic
peak has a long radiative tail. The choice of cut position can greatly change how many
elastic events are recorded. Radiative corrections, which model the shape of the tail,
are needed to remove the dependence on cut position.}
\end{figure}

As an example, let's consider radiative corrections for the case of an unpolarized
elastic electron-proton cross section measurement. The leading order diagram
is the one-photon exchange diagram, shown in figure \ref{fig:ope}. By attributing the 
entire cross section to one-photon exchange, one can extract the form factors $G_E$ and $G_M$ that describe
the proton vertex. Let's imagine that this example experiment is an inclusive
measurement, that is, we will position a spectrometer at scattering angle $\theta$
to detect the electron and the proton will go undetected. The electron spectrometer
will tell us the momentum of the detected electron. To measure the cross section, 
we will count the number of electrons detected by our spectrometer with the proper 
elastic momentum, and divide by the integrated luminosity. We will disregard electrons 
with lower momenta because they are the result of inelastic processes (and of course,
it is not kinematically possible for electrons to emerge with greater than the elastic 
momentum). The energy spectrum from the spectrometer might look like the top plot in
figure \ref{fig:rc_example}. 

This approach is complicated by the emission of soft bremsstrahlung photons. These photons,
which go undetected, remove a small amount of energy from the scattered lepton. Instead
of seeing a narrow peak in our momentum spectrum at the elastic momentum, we will see a long
tail in the distribution towards lower momenta, sketched in the bottom plot of figure
\ref{fig:rc_example}. When we analyze our data, we must make a distinction between elastic and 
inelastic electrons, but where we make this distinction clearly affects our result. If we 
require more stringent limits on electron momentum, our extracted cross section will be smaller. 
If we consider more of the tail to be elastic, then our extracted cross section will be larger. The 
result is meaningless, because it can be tuned.

\begin{figure}[htpb]
  \centering
  \includegraphics{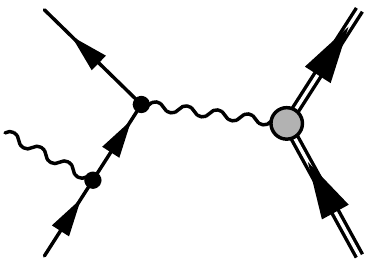}
  \includegraphics{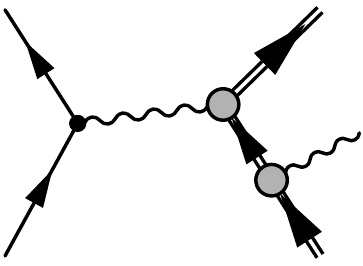}
  \includegraphics{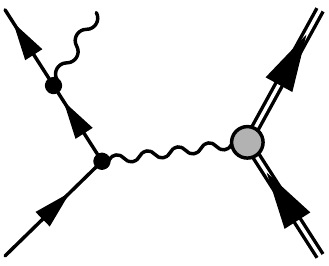}
  \includegraphics{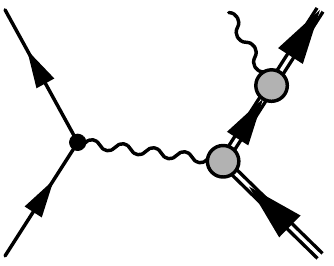}
  \caption[The $\mathcal{O}(e^3)$ bremsstrahlung (inelastic) diagrams]
          {\label{fig:brems} There are four $\mathcal{O}(e^3)$ bremsstrahlung (inelastic) diagrams.}
\end{figure}

\begin{figure}[htpb]
  \centering
  \includegraphics{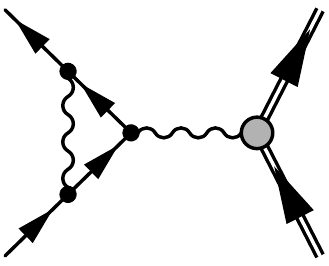}
  \includegraphics{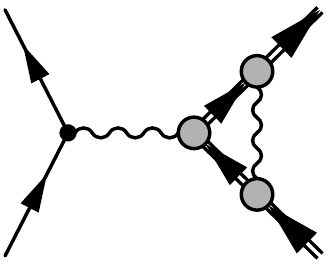}
  \includegraphics{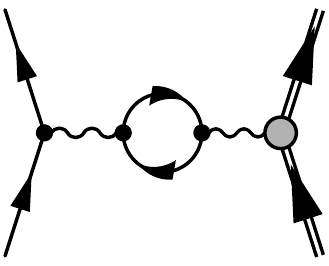}\\
  \includegraphics{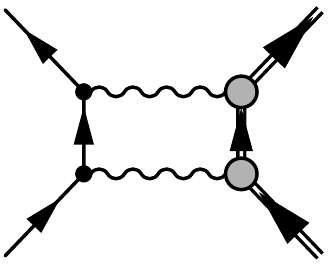}
  \includegraphics{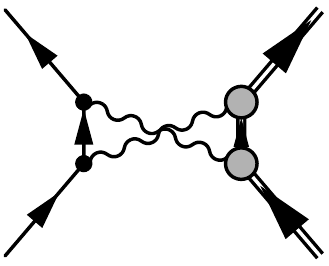}
\caption[The $\mathcal{O}(e^3)$ elastic diagrams]
{\label{fig:el_diagrams} There are several $\mathcal{O}(e^3)$ elastic diagrams, many of which have divergences.}
\end{figure}

Instead, we must model the shape of the ``radiative tail'' so that we can recover the 
one-photon exchange cross section as a function of how we define our acceptance for
elastic events. To do this, we should calculate the bremsstrahlung diagrams, shown in figure
\ref{fig:brems}, and integrate them over some window of electron momentum acceptance. 
This approach would work, except for the fact that this integral diverges. The divergence is 
cancelled by other divergences in the $\mathcal{O}(e^3)$ corrections to elastic scattering,
shown in figure \ref{fig:el_diagrams}. 
To account for soft bremsstrahlung, one must account for these other diagrams as well. 
After doing this, one arrives at a correction factor for recovering the one-photon exchange cross section:
\begin{equation}
\frac{d \sigma}{d \Omega}_{1\gamma} (\theta) = \frac{d \sigma}{d \Omega}_\text{meas.} (\theta,\Delta p) 
/ (1 + \delta(\theta,\Delta p)).
\end{equation}

In this chapter, I will first discuss the previous radiative corrections procedures for elastic
electron-proton scattering. I will then discuss the specific radiative correction needs
for OLYMPUS, including the details of the OLYMPUS radiative event generator that I, along with 
Rebecca Russell and Jan Bernauer, developed for OLYMPUS simulations. I will conclude with some 
discussion about the differences in the various correction models that the generator can simulate.

\section{Radiative Corrections to Elastic $ep$ Scattering}

\subsection{Diagrams}

\label{sec:ep_rad_corr}

The matrix element for elastic $ep$ scattering, $\mathcal{M}_ep$, can be expanded into a Feynman series in powers
of the fine structure constant $\alpha$:
\begin{equation}
\mathcal{M}_{ep} = \alpha \mathcal{M}_1 + \alpha^2 \mathcal{M}_2 + \alpha^3 \mathcal{M}_3 + \ldots \quad .
\end{equation}
The cross section, which depends on the square of the matrix element, can be similarly expanded:
\begin{equation}
\sigma_{ep} \propto |\mathcal{M}_{ep}|^2 = \alpha^2 |\mathcal{M}_1|^2 + 2\alpha^3\text{Re} \left [ \mathcal{M}_1 \mathcal{M}_2 \right] + \ldots \quad .
\end{equation}
The measured cross section will also have a contribution from bremsstrahlung, which goes as $\alpha^3$ at leading order. 
The bremsstrahlung diagrams are inelastic, so the contribution to the elastic cross section will be limited to the region
in which the photon energy is small and the kinematics are ``nearly elastic:''
\begin{equation}
\sigma_\text{meas.} \propto \alpha^2 |\mathcal{M}_1|^2 + 2\alpha^3 \left\{ 
\text{Re} \left [ \mathcal{M}_1 \mathcal{M}_2 \right] + \int_{\text{small~} E_\gamma} d\vec{k}_\gamma |\mathcal{M}_\text{brems.}|^2 \right\} \ldots \quad.
\end{equation}

Let's consider which specific diagrams contribute to the different terms. In the leading term, $|\mathcal{M}_1|^2$
is the square one-photon exchange term. At the order of $\alpha^3$, there are two groups of terms that
contribute. There are the $\mathcal{M}_1$ and $\mathcal{M}_2$ interference term,
Re$[\mathcal{M}_1 \mathcal{M}_2]$, and the bremsstrahlung term. The bremsstrahlung can be broken up
into groups based on the dependence on the lepton charge $z$:
\begin{itemize}
\item Lepton bremsstrahlung ($z^4$): $\left| \parbox[h][2cm][c]{2cm}{\includegraphics[width=2cm]{brems_ei.pdf}} + 
  \parbox[h][2cm][c]{2cm}{\includegraphics[width=2cm]{brems_ef.pdf}} \right|^2$
\item Lepton-proton bremsstrahlung interference ($z^3$): \\$\left( \parbox[h][2cm][c]{2cm}{\includegraphics[width=2cm]{brems_ei.pdf}} + 
  \parbox[h][2cm][c]{2cm}{\includegraphics[width=2cm]{brems_ef.pdf}} \right)^\dagger
  \left( \parbox[h][2cm][c]{2cm}{\includegraphics[width=2cm]{brems_pi.pdf}} + 
  \parbox[h][2cm][c]{2cm}{\includegraphics[width=2cm]{brems_pf.pdf}} \right)$
\item Proton bremsstrahlung ($z^2$): $\left| \parbox[h][2cm][c]{2cm}{\includegraphics[width=2cm]{brems_pi.pdf}} + 
  \parbox[h][2cm][c]{2cm}{\includegraphics[width=2cm]{brems_pf.pdf}} \right|^2$
\end{itemize}
All three of these groups diverge when integrated down to zero photon energy. These divergences
cancel against similar divergences in Re$[\mathcal{M}_1 \mathcal{M}_2]$. The lepton bremsstrahlung
divergence, proportional to $z^4$, cancels against a divergence in:
\[
2\text{Re}\left[ \parbox[h][2cm][c]{2cm}{\includegraphics[width=2cm]{ope.pdf}}^\dagger
  \parbox[h][2cm][c]{2cm}{\includegraphics[width=2cm]{ver_cor_e.pdf}} \right],
\]
the lepton vertex correction. The proton bremsstrahlung divergence, proportional to $z^2$, 
cancels against a divergence in:
\[
2\text{Re}\left[ \parbox[h][2cm][c]{2cm}{\includegraphics[width=2cm]{ope.pdf}}^\dagger
  \parbox[h][2cm][c]{2cm}{\includegraphics[width=2cm]{ver_cor_p.pdf}} \right],
\]
the proton vertex correction. The divergence in lepton-proton bremsstrahlung
interference cancels against a divergence in:
\[
2\text{Re}\left[ \parbox[h][2cm][c]{2cm}{\includegraphics[width=2cm]{ope.pdf}}^\dagger
  \left( \parbox[h][2cm][c]{2cm}{\includegraphics[width=2cm]{tpe_box.pdf}}
  +\parbox[h][2cm][c]{2cm}{\includegraphics[width=2cm]{tpe_cross.pdf}} \right)\right],
\]
the interference between one- and two-photon exchange. The only other term that contributes 
at order $\alpha^3$ is vacuum polarization:
\[
2\text{Re}\left[ \parbox[h][2cm][c]{2cm}{\includegraphics[width=2cm]{ope.pdf}}^\dagger
  \parbox[h][2cm][c]{2cm}{\includegraphics[width=2cm]{vac_pol.pdf}} \right],
\]
which does not contain a problematic divergence.

\subsection{Soft vs. Hard Two-Photon Exchange}

Two-photon exchange must be included in radiative corrections in order to cancel 
the divergence in lepton-proton bremsstrahlung interference. However, as mentioned
in chapter \ref{chap:background}, it cannot be calculated in a model-independent
way. However, it can be calculated exactly in the limit in which one of the 
photons carries no momentum, i.e., is soft. It is precisely this soft two-photon
exchange which has the divergence necessary to cancel that of bremstrahlung 
interference. The standard radiative corrections schemes (which will be outlined
in the following section) have proceeded by making an arbitrary distinction
between soft two-photon exchange and hard two-photon exchange. The soft two-photon
exchange is calculated and included. The hard two-photon exchange is neglected. 

The distinction between hard and soft two-photon exchange is completely arbitrary.
Not all radiative corrections schemes employ the same distinction. A measurement
must report which radiative correction scheme (and thus which soft definition) was
used in order for the result to be interpretable. 

Previous cross section measurements---the ones that see a form factor discrepancy---have
all used this radiative correction schemes employing this hard-soft distinction approach.
The form factor discrepancy exists despite accounting for soft two-photon exchange.
OLYMPUS aims to measure the contribution from \emph{hard} two-photon exchange, which
is the remaining piece that has been typically neglected.

\subsection{Previous Work on $ep$ Radiative Corrections}

Schemes for radiative corrections for electron scattering have been
devised as early as the 1940s \cite{Schwinger:1949ra}, and have generally
proceeded in the same fashion as my example from the introduction. The
goal is to recover the one-photon exchange cross section because form factors
are defined in the one-photon exchange frame work. All of the earliest
radiative correction schemes assume an inclusive experiment, in which only the
electron is detected, and that energy/momentum of the electron is measured
by the detector (and in the limit $m_e \rightarrow 0$, the energy and momentum are
the same). These schemes assume that the experimenter will define an energy window 
of width $\Delta E$, such that electrons with energies between the elastic energy
$E_{el.} \equiv E m_p / (m_p + 2E\sin^2\frac{\theta}{2})$ and $E_{el.} - \Delta E$ 
will count towards the cross section. These schemes prescribe a correction $\delta$
such that:
\begin{equation}
d\sigma_{meas.} = d\sigma_{1\gamma} \times (1+\delta(\theta,\Delta E)) .
\end{equation}
The experimenter chooses an appropriate $\Delta E$ given the resolution
limits of the detector, corrects the data, and recovers the one-photon
exchange cross section.

\begin{figure}[htpb]
\centering
\includegraphics{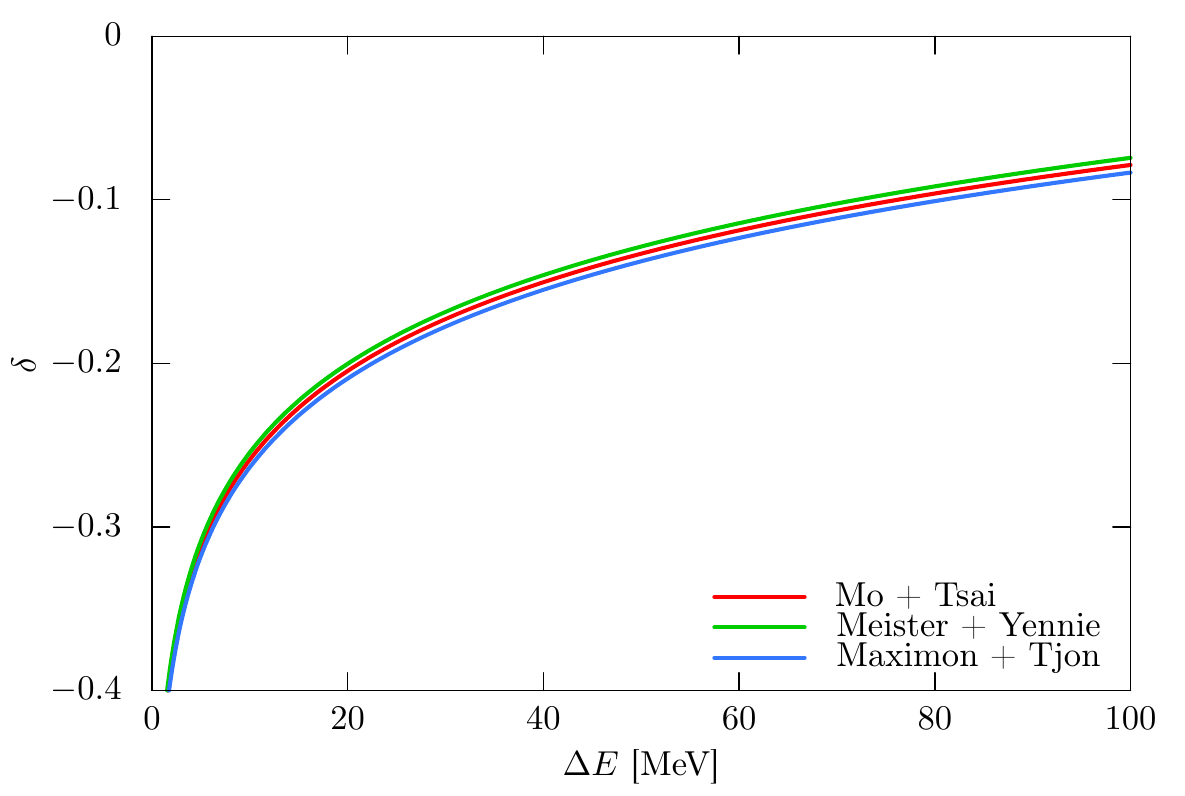}
\caption[The radiative correction $\delta$ from three standard prescriptions]
{\label{fig:rc_std} Standard radiative corrections are shown for 2~GeV electron beam,
and a scattering angle of $60^\circ$. The different corrections make slightly different
assumptions but make the same qualitative predictions: $\delta$ is negative, small, and monotonically
increasing as function of $\Delta E$. }
\end{figure}

The landmark radiative corrections procedure is that of Mo and Tsai \cite{Mo:1968cg} (1969),
which built on the earlier work of Tsai \cite{Tsai:1961zz} (1961). Meister and Yennie also 
produced a work based on Tsai, but with some approximations to make the 
the formula easier to calculate with computers of that era \cite{Meister:1963}. Though
Mo and Tsai's work is considered standard, there have been more recent
attempts to update it. In 2000, Maximon and Tjon produced a calculation \cite{Maximon:2000hm} that 
used fewer approximations, employed a different definition of soft two-photon exchange, 
and included a dipole model of the proton's form factors in order to add proton 
structure information to the proton vertex correction. A comparison of these
three corrections is shown in figure \ref{fig:rc_std}.

In 2001, Ent~et~al.\ published a radiative corrections procedure \cite{Ent:2001hm} 
used in the NE18 experiment that reworks Mo and Tsai's equations to produce a $\delta$ 
that depends on the missing energy $W$, instead of electron energy loss $\Delta E$, 
because $W$ is a more relevant quantity in coincidence experiments. The paper goes on to 
explain methods for applying radiative corrections via simulation, presaging
modern approaches. 

\subsection{Exponentiation}

We can ask: ``What does the function $\delta(\theta, \Delta E)$ look like qualitatively?'' 
There are several properties we can presume before we even calculate $\delta$. First, 
since $\delta$ represents a higher-order correction to the one-photon exchange cross
section, and since $\alpha$ is small, we expect $\delta$ also to be small compared to 1.
If it isn't, then that indicates a failure of the perturbation expansion. We also know
that the derivative $d \delta / d \Delta E$ must be positive; if we make our elastic
selection more permissive (i.e., make $\Delta E$ larger), than the measured cross
section must increase. Figure \ref{fig:rc_std} shows the behavior of three of the 
standard radiative correction prescriptions. The qualitative behavior is exactly like
we'd expect.  The corrections are negative, implying that the measured cross section will 
be lower than what we would expect in the one-photon exchange approximation. The correction 
grows with $\Delta E$. And for values of $\Delta E$ on the order of tens of MeV, $\delta$ is 
small compared to 1. 

\begin{figure}[htpb]
\centering
\includegraphics{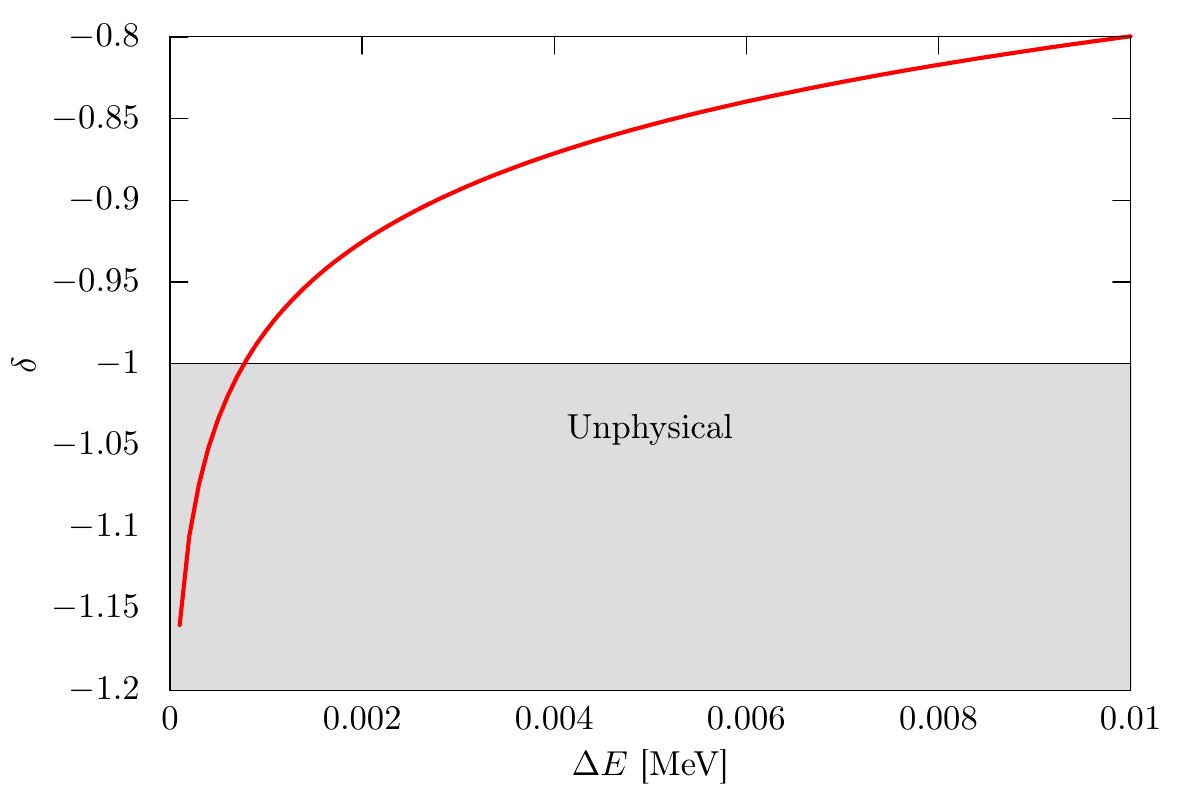}
\caption[Mo and Tsai correction at small $\Delta E_l$]
{\label{fig:mt_example} The Mo and Tsai correction is shown for a 2~GeV electron beam
and a scattering angle of $60^\circ$. At extremely low values of $\Delta E$, the correction becomes
unphysical because $(1+\delta)$ would be negative, implying a negative cross section.}
\end{figure}

However, figure \ref{fig:mt_example} reveals that $\delta$ diverges to negative infinity in the limit where 
$\Delta E$ is taken to 0. The first sign of trouble is when $\delta$ becomes less than -1. The multiplicative
correction factor $(1+\delta )$ is now negative, implying a negative cross section, which obviously is unphysical.
In this case, our perturbative expansion is breaking down. For very small cut-off values, higher order terms start
to matter a lot. The probability of the emission of many infinitesimally soft photon is infinite. 

Yennie, Frautschi and Suura demonstrated that higher order terms could be taken into account properly by
replacing the factor $(1+\delta)$ with $\exp(\delta)$ \cite{Yennie:1961ad}. The cut-off $\Delta E$ is 
reinterpreted as the sum of the energy of all of the emitted soft photons. This technique is called
exponentiation. Yennie et al.\ proved that exponentiation is correct in the soft limit, i.e.\ $\Delta E$ is small
and none of the soft photons make an appreciable change in the charged particle kinematics. At larger 
$\Delta E$, $\delta$ becomes small so that $\exp(\delta)$ becomes quite similar to $(1+\delta)$. 
At large values of $\Delta E$, the assumptions needed for exponentiation break down. 

Exponentiation is a tool that plays key role in the OLYMPUS radiative generator, described in the following
sections.

\section{OLYMPUS Radiative Corrections}

\label{sec:olympus_gen}

\subsection{Simulating Radiative Corrections with Monte Carlo}

Previous radiative corrections approaches are best suited to single-arm measurements with high momentum
resolution and well-defined acceptance apertures. OLYMPUS does not have any of these qualities. OLYMPUS
is a coincidence experiment: proton detection is built in to the trigger. The OLYMPUS spectrometer 
has a large continuous angular acceptance, but poor momentum resolution. The acceptance is complicated
since the OLYMPUS target is not-point like. 

An alternative approach is to use Monte Carlo simulation. In this approach, elastic $ep$ and bremmstrahlung 
events are randomly generated---that is, the momentum vectors of the outgoing particles are assigned---according 
to radiatively corrected cross sections. These events form the input of the OLYMPUS simulation software, which 
simulates particle trajectories through the magnetic field and through the matter of the detectors and simulates 
the signals that would be produced in the detectors. Rather than making a correction to a measurement, in this 
approach the measurement is compared with that from the simulated data set. 

This approach requires a great deal of software infrastructure. However it brings two advantages. The first
is that all of the complicated convolutions between radiative corrections, detector acceptance, and reconstruction
efficiency are handled automatically. The second is flexibility in event selection. To use previous radiative
corrections schemes, one's elastic event selection must be limited to picking a value for $\Delta E$ of the lepton
(or $W$ using Ent et al.). But by using a simulation, one can select elastic events with any cuts one wants and
apply them to both the measured and simulated data sets. This flexibility is necessary for a coincidence experiment
like OLYMPUS, since proton information plays a vital role in segregating elastic events from background.

The first step in this chain is to produce randomly generated events according to a cross section that has been
radiatively corrected. Software for this purpose is called a ``radiative event generator.'' This task isn't completely 
straightforward; there are several design choices and approximations that must be made. The rest of this section will 
cover the specifics of the OLYMPUS radiative event generator. 

Using Monte Carlo to simulate radiative corrections is not new with OLYMPUS. In fact, in high-energy particle 
physics this approach is now universal, and there are many widely distributed radiative generators for 
high energy processes. For the OLYMPUS experiment, we wrote an entirely new generator as a way to guarantee that the
priorities and features matched our needs. In parallel to our work, the Novosibirsk two-photon experiment
also wrote a brand new elastic $ep$ event generator, called ESEPP, described in a recent work \cite{Gramolin:2014pva}.
I will not cover ESEPP in detail in this thesis, but I will note that it was used to test the OLYMPUS generator during its 
development. 
 
\subsection{OLYMPUS Radiative Event Generator}

A new radiative event generator was written specifically for OLYMPUS. Though this required a great deal 
of work, which wouldn't have been necessary had we used an alternative suitable generator, such as ESEPP,
it did give us full understanding and control of what was going on ``under the hood.'' Additionally we
were able to build in features that were well-suited to the OLYMPUS analysis.

The most important of these features was the ability to have multiple weights. The OLYMPUS generator is
a weighted generator: each event carries a scalar weight describing its contribution to the Monte Carlo
integral. As a brief example, let's consider a generator which produces samples of a random variable $x$
aiming to reproduce a cross section $d\sigma / dx$. The generator might sample $x$ according to a probability
distribution $P(x)$, in which case each event will carry a weight of $d\sigma / dx \times (1/P(x))$. For 
the OLYMPUS analysis, we wanted to simulate several different radiative correction models, and ordinarily
this would require simulating the data set multiple times. The generation step is quite fast relative to
the simulation of how tracks propagate through the detector, and much faster compared to the track reconstruction.
Therefore, we designed the OLYMPUS generator so that each event carried a list of weights, one for each 
radiative corrections model we wanted to test. In the language of our earlier example, the list would have
weights: $\{ d\sigma_1 / dx \times (1/P(x)), d\sigma_2 / dx \times (1/P(x)), d\sigma_3 / dx \times (1/P(x)) \ldots \}$.
At the end-stage of the simulation, all of the events could be reweighted to get the results for each model
without having to re-propagate or re-track.

The radiative corrections choice with the biggest potential to affect the OLYMPUS result is whether or 
not to use exponentiation. For that reason, we have two big classes of weights. In what we call ``Method 1''
we extrapolate a soft exponentiated correction from the elastic peak out to large values of bremsstrahlung 
photon energy. In what we call ``Method 2'', we define a small arbitrary region of kinematic space around 
the elastic peak. Inside that region, we treat the event as fully elastic and use a non-exponentiated soft 
correction. Outside that region, we use the tree-level bremsstrahlung cross section. You can see that both
methods are more complicated than simply using exponentiation or not. However, in short hand, it will be
convenient to refer to Method 1 as the exponentiating method, and Method 2 as the non-exponentiating method.

I should note that Method 1 borrows heavily from the radiative corrections procedure developed by Jan Bernauer
for the A1 collabaration at Mainz, a description of which can be found in chapter 5 of his thesis \cite{bernauer:thesis}.
Method 1 is not a pure reproduction of the A1 generator in all areas, but follows some aspects exactly, namely
the prescription for the radiative cross section and the sampling of the photon direction. 

In this section, I will first describe our derivation of the radiative cross section used by Method 1. Next, 
I will describe our choice of probability distribution for sampling the kinematic variables, since this is 
important for guaranteeing good numerical behavior for Method 1. Then I will explain the cross section used 
for Method 2.

\subsubsection{Method 1: Exponentiation}

The quandary with exponentiation is that, while it provides a prescription for accounting 
for radiation to all orders, it only does so in the soft limit; that is, when the amount of
energy radiated away by bremsstrahlung photons is small. When a large amount of energy
has been radiated away, it makes more sense to think of the cross section in terms of that 
for hard bremsstrahlung. The prescription of method 1 attempts to interpolate between these 
two regions in a continuous way. 

To begin the derivation, let's start by describing what an ideal solution would look like. 
In OLYMPUS, the scattered lepton and recoiling proton are detected in coincidence. Therefore, the ideal 
cross section would be of the form:
\[
\frac{d^6 \sigma}{d\vec{p_l} d\vec{p_p}},
\]
where there is dependence on both the lepton momentum vector and the proton momentum vector. And ideally,
this cross section would account for radiative corrections to all orders. We do not have a prescription
for such an ideal cross section, so we will have to make some assumptions in order to approximate it.

The first practical assumption we can make is that the ideal cross section is dominated deep in the tail
by hard single-photon bremsstrahlung. Single photon bremstrahlung has a 3-body final state (lepton, proton,
and photon), and therefore occupies a 5-dimensional kinematic space. Our first assumption is that we can reduce
our six-fold differential cross section to a five-fold differential cross section; that is, by fixing five
kinematic variables, we've uniquely specified the kinematics of the final state. We can choose to fix any 
five independent variables. For this generator, our choice of five will be two for the lepton solid angle,
$\Omega_l$: ($\cos\theta_l,\phi_l$), two for the photon solid angle, $\Omega_\gamma$: ($\cos\theta_\gamma,\phi_\gamma$), 
and the energy loss of the lepton $\Delta E_l = E_l^\text{el.} - E_l$. We will specify a cross section of the form:
\[
\frac{d^5 \sigma}{d\Omega_l d\Omega_\gamma d\Delta E}.
\]
We can choose to write this differential cross section in the following way:
\begin{align}
\frac{d^5 \sigma}{d\Omega_l d\Omega_\gamma d\Delta E} &= \frac{\partial^3}{\partial \Omega_\gamma \partial \Delta E_l}
\left[ \frac{d\sigma}{d\Omega_l}_{1\gamma} e^{f(\Omega_\gamma, \Delta E_l | \Omega_l)} \right] \\
&= \frac{d\sigma}{d\Omega_l}_{1\gamma}  e^{f}
\left[ \frac{ \partial^3 f(\Omega_\gamma, \Delta E_l | \Omega_l) }{\partial \Omega_\gamma \partial \Delta E_l} \right],
\label{eq:rc_ass1}
\end{align}
where the one-photon elastic cross section has been factored out, and the dependence on $\Omega_\gamma$ and $\Delta E_l$ is
contained in an undetermined function $f(\Omega_\gamma, \Delta E_l | \Omega_l)$. The point of writing the cross
section in this form is to make an explicit connection to how a standard radiative correction is applied (with
exponentiation). The function $f$ takes the place of $\delta$. The dependence on $\Delta E_l$ in a standard correction
comes from integrating over the bremsstrahlung cross section with the elastic cross section factored out:
\begin{equation}
\delta(\Delta E_l | \Omega_l) = \left( \frac{1}{\frac{d\sigma}{d\Omega_l}_{1\gamma}}\right)
\int_{0}^{\Delta E_l} d\Delta E_l' \int_{4\pi}d\Omega_\gamma 
\frac{d^5 \sigma}{d\Omega_l d\Omega_\gamma d\Delta E_l'}_\text{brems.},
\end{equation}
so that by differentiating with respect to energy, we get:
\begin{equation}
\frac{\partial \delta}{\partial \Delta E_l} = \left( \frac{1}{\frac{d\sigma}{d\Omega_l}_{1\gamma}}\right)
\int_{4\pi}d\Omega_\gamma \frac{d^5 \sigma}{d\Omega_l d\Omega_\gamma d\Delta E_l}_\text{brems.}.
\end{equation}
A standard $\delta$ has no dependence on $\Omega_\gamma$, since this is integrated out. But this brings us to 
our second assumption, that we can approximate the derivatives of our underdetermined function $f$ with the
tree-level bremsstrahlung cross section:
\begin{equation}
\left[ \frac{ \partial^3 f(\Omega_\gamma, \Delta E_l | \Omega_l) }{\partial \Omega_\gamma \partial \Delta E_l} \right]
\longrightarrow 
\left( \frac{1}{\frac{d\sigma}{d\Omega_l}_{1\gamma}}\right) \frac{d^5 \sigma}{d\Omega_l d\Omega_\gamma d\Delta E_l}_\text{brems.}.
\end{equation}
Putting this back into equation \ref{eq:rc_ass1}, we get:
\begin{equation}
\frac{d^5 \sigma}{d\Omega_l d\Omega_\gamma d\Delta E} = e^{f} \frac{d^5 \sigma}{d\Omega_l d\Omega_\gamma d\Delta E_l}_\text{brems.}.
\end{equation}
Our third and final assumption is that we can replace $f$ in $e^f$ with a standard correction $\delta$ of our choosing. That brings
us to the expression for the Method 1 cross section:
\begin{equation}
\frac{d^5 \sigma}{d\Omega_l d\Omega_\gamma d\Delta E} = e^{\delta} \frac{d^5 \sigma}{d\Omega_l d\Omega_\gamma d\Delta E_l}_\text{brems.}.
\end{equation}

In Method 1, there are a few things to consider. First, this approach works only in the near-elastic limit.
Exponentiation is only valid in this limit, and in addition, standard corrections all use the 
``Soft-Photon Approximation'', which assumes that so little energy is carried away by bremsstrahlung photons 
that the elastic scattering matrix element can be analytically factored out of the bremsstrahlung matrix element. 
It is a mistake to trust that this cross section will be accurate deep in the radiative tail. 
Second, this method aims to account for radiation to all orders, but uses a single-photon bremsstrahlung cross
section to predict the dependence on $\Omega_\gamma$. This assumption is probably pretty good because
the bremsstrahlung cross section blows up for very soft emitted photons. If multiple photons are emitted summing
to a total energy of $E_\gamma$, then the dominant mode will be one photon carrying nearly $E_\gamma$, with 
all other photons being soft. Third, for near-elastic kinematics, $\delta$ is negative, but increases with 
increasing $\Delta E_l$. $e^\delta$ is a suppression factor for the bremsstrahlung cross section and implies that
higher-order radiative effects slightly reduce what we'd expect for the bremsstrahlung cross section.
For very large values of $\Delta E_l$, i.e., far from the elastic peak, $\delta$ will cross over and become 
positive, and $e^\delta$ is an enhancement factor. As mentioned earlier, the region deep in the tail should
not be trusted. In the same way, an enhancement of bremsstrahlung in the deep tail is not necessarily believable.
If our result depends on events in these kinematics, then Method 1 is probably not a good prescription for us.

\subsubsection{Sampling Distribution}

Now that we have a prescription for our radiatively-corrected cross section, all that a radiative generator
needs to do is to produce random samples of the variables $(\cos\theta_l, \phi_l, \cos\theta_\gamma,\phi_\gamma,\Delta E_l)$
and then to weight each event $i$ with weight $w_i$:
\begin{equation}
  w_i = \frac{ e^{\delta} }{P(\cos\theta_l, \phi_l, \cos\theta_\gamma,\phi_\gamma,\Delta E_l)} 
\frac{d^5 \sigma}{d\Omega_l d\Omega_\gamma d\Delta E_l}_\text{brems.},\
\end{equation}
where $P$ is the probability distribution that is being sampled. In an ideal world of infinite computing power,
the choice of $P$ does not affect the accuracy of the generator. However, from a practical standpoint, the 
choice of $P$ can have a large impact on the performance of the generator. There are three issues to keep in
mind.
\begin{enumerate}
\item \textbf{Weight uniformity}: The simulated data will eventually be binned according to some set of criteria,
  and the weights of the events in the bin will be summed. The statistical uncertainty in the bin contents for 
  a given Monte Carlo sample is proportional to the variance of the weights of the events in the bin. If the choice
  of $P$ produces a large variance in weights, then more simulated data will be needed to attain equivalent precision. 
\item \textbf{Ease of sampling}: If sampling from $P$ is computationally difficult, then the generator will be slower
  and the simulation will have less precision for an equivalent run time.
\item \textbf{Numerical stability}: If the weights or components needed for calculating the weights get very large or 
  very small, care must be taken to avoid numerical instability. A careful choice of $P$ may help reduce instability.
\end{enumerate}
To make the weights completely uniform, all one needs to do is sample from a distribution that is proportional
to the underlying cross section. However, this is often computationally difficult. The underlying cross section
is also highly non-uniform. The cross section varies strongly with $\cos\theta_l$, has an integrable divergence
as $\Delta E_l \longrightarrow 0$, and has a non-trivial dependence on $\Omega_\gamma$. In short, there are many
opportunities for numerical instability. Keeping these concerns in mind, let's look at the OLYMPUS radiative 
generator's choice of $P$. 

First, the cross section is fully azimuthally symmetrical, so $\phi_l$ can be sampled isotropically. We choose also
to sample $\cos\theta_l$ isotropically since we are interested in results as a function of $\theta_l$. Even though
weights will vary significantly over the range of $\theta_l$, we will usually be binning by $\theta_l$ (or by some
monotonically related variable, like $Q^2$), so we will not be introducing a large variance into any individual bin.
Once we have chosen $\cos\theta_l$ and $\phi_l$, then we can work on sampling $\Delta E_l$. 

For $\Delta E_l$, we want a sampling distribution that matches the cross sections dependence on $\Delta E_l$, given
the constraints of numerical stability and ease of computation. Fortunately, we know that the $\Delta E_l$ dependence
is approximately $\exp{\delta}\frac{\partial \delta}{\partial \Delta E_l}$, so we can use this as our sampling
distribution. In just about every standard correction, $\delta$ takes the form
\begin{equation}
f + \frac{t}{2}\log\left[ u \left(\frac{\Delta E_l}{E^\text{el.}}\right)^2\right]
\end{equation}
for some constants $f$,$t$,$u$ that depend on the lepton angle. Our probability distribution will be:
\begin{align}
P(\Delta E_l | \Omega_l) \sim& e^\delta \frac{\partial}{\partial \Delta E} \left\{ 
f + \frac{t}{2}\log\left[ u \left(\frac{\Delta E_l}{E^\text{el.}}\right)^2\right] \right\} \\
\sim & e^f \left[ u^{t/2} \left( \frac{\Delta E_l}{E^\text{el.}}\right)^t \right] \frac{t}{u\Delta E_l}\\
\sim & \left(\Delta E_l \right)^{t-1},
\end{align}
that is, a power law. To determine the normalization, let's integrate over the range $[0,\Delta E_\text{cut}]$:
\begin{align}
1 =& N \int_0^{\Delta E_\text{cut}} \left(\Delta E_l\right)^{t-1} \\
=& \frac{N\left(\Delta E_\text{cut}\right)^t}{t}
\end{align}
\begin{align}
N =& \frac{t }{\Delta E_\text{cut}^t}\\
P(\Delta E_l | \Omega_l) =& \frac{t}{\Delta E_l} \left(\frac{ \Delta E_l}{\Delta E_\text{cut}}\right)^t
\end{align}

Let's take a look at what the event weights will be. Of course, we haven't yet specified $P(\Omega_\gamma|\Omega_l \Delta E_l)$,
but we can start to get an idea. In appendix \ref{chap:brems}, the tree-level bremsstrahlung cross section
is worked out. Specifically, we'll need the result in equation \ref{eq:brems_tree_level}. When we combine
all of the elements, we find that the weight for an event with kinematics ($\Omega_l$, $\Omega_\gamma$, $\Delta E_l$)
is:
\begin{align}
w &= e^{\delta} \times \frac{d^5 \sigma}{d\Omega_l d\Omega_\gamma d\Delta E_l}_\text{brems.}\times 
\frac{1}{P(\Omega_l) P(\Delta E_l | \Omega_l)P(\Omega_\gamma|\Omega_l \Delta E_l)} \\
&= e^f u^{t/2} \left( \frac{\Delta E_l}{E_3^\text{el.}}\right)^t \frac{ \alpha^3 \left\langle|\mathcal{M}'|^2\right\rangle }{ 64 \pi^2 E_1 m_p}
  \frac{E_3}{\left|E_4 + E_3 + k\cos\theta_{l\gamma} - E_1\cos\theta_l \right|} \frac{\tilde{\mathcal{J}}}{\Delta E_l}
\frac{\frac{\Delta E_l}{t}\left(\frac{ \Delta E_l}{\Delta E_\text{cut}}\right)^{-t}}{P(\Omega_l) P(\Omega_\gamma|\Omega_l \Delta E_l)} \\
&= \frac{e^f u^{t/2}}{t} 
\left( \frac{\Delta E_\text{cut}}{E_3^\text{el.}}\right)^t \frac{ \alpha^3 \left\langle|\mathcal{M}'|^2\right\rangle }{ 64 \pi^2 E_1 m_p}
  \frac{E_3}{\left|E_4 + E_3 + k\cos\theta_{l\gamma} - E_1\cos\theta_l \right|}
\frac{\tilde{\mathcal{J}}}{P(\Omega_l) P(\Omega_\gamma|\Omega_l \Delta E_l)} . \label{eq:weight} 
\end{align}
The problematic divergence of the type $(\Delta E_l)^{-1}$ from the cross section has been cancelled by 
the factor of $\Delta E_l$ in the sampling distribution. Our weight will be numerically stable, even for
small values of $\Delta E_l$.

Fortunately, it is very easy to produce random numbers in a power law distribution. 
To produce random numbers in the distibution $P(\Delta E_l | \Omega_l)$, we'll need to invert the cumulative
distribution $C(\Delta E_l | \Omega_l) = (\Delta E_l / \Delta E_\text{cut})^t$. 
Given a random number $r$, uniformly sampled on $[0,1]$, then:
\begin{equation}
\Delta E_l = \Delta E_\text{cut} r^{1/t}.
\end{equation}

\begin{figure}[htpb]
\centering
\includegraphics{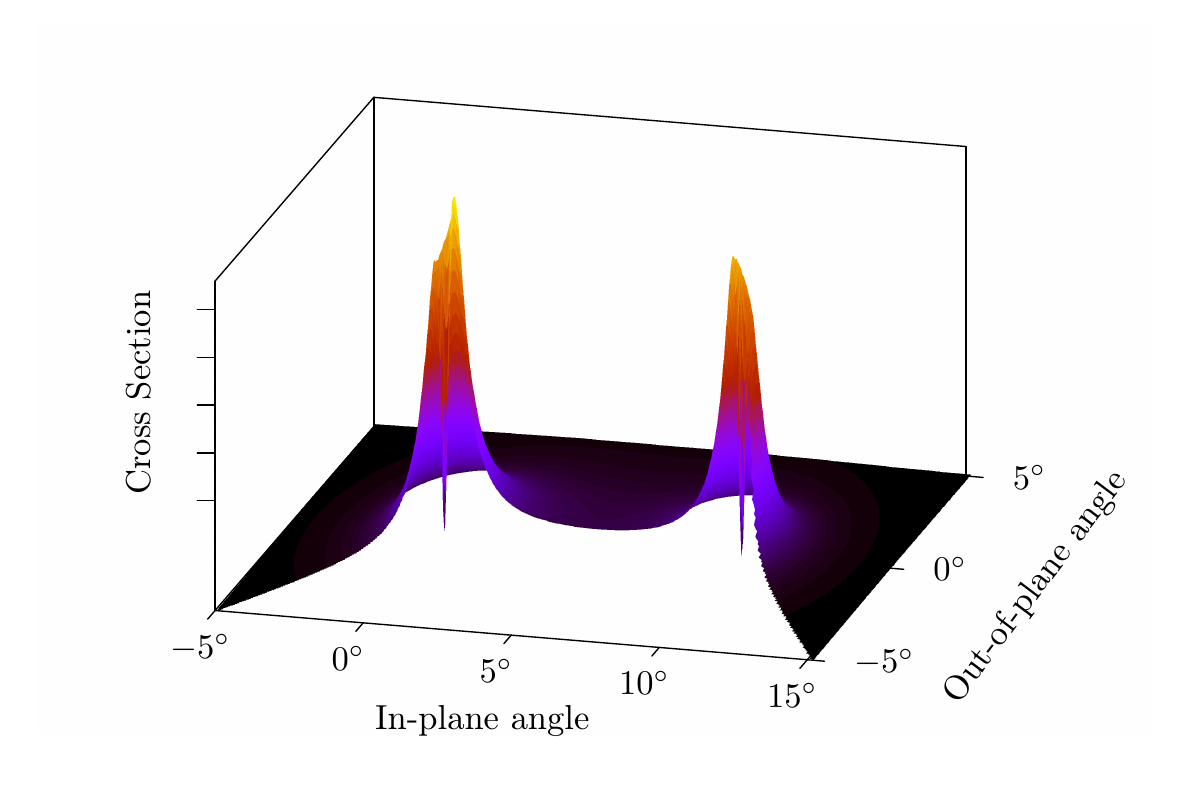}
\caption[Photon angle distribution in bremsstrahlung]{\label{fig:photon_peaks}
The bremsstrahlung cross section for fixed $\theta_l=10^\circ$, $\Delta E_l = 100$~MeV is plotted
as a function of the photon angles. The cross section is strongly peaked in the direction of the
incoming electron ($0^\circ,0^\circ$) and the outgoing electron ($10^\circ,0^\circ$). The generator
will have the best performance if the photon angle sampling distribution matches those peaks.}
\end{figure}

The final piece of the sampling distribution is the photon direction distribution, $P(\Omega_\gamma | \Omega_l, \Delta E_l)$.
The photon direction distribution in bremsstrahlung has two very large peaks: one in the same direction as the 
beam momentum, the other in the direction of the scattered lepton's momentum\footnote{It should be noted that 
when one does not neglect the electron mass, the cross section plummets 
when the lepton and photon momentum vectors are exactly colinear. The peaks in the distribution could be 
described as volcano-like: rising steeply, but suddenly dropping at the exact center.
These drops are barely visible in figure \ref{fig:photon_peaks}.}.
These can be seen in figure \ref{fig:photon_peaks}. In experiments at higher energies than OLYMPUS, bremsstrahlung 
can also be peaked in the direction of the recoiling proton, though the high mass of the proton broadens this peak.

The highly-peaked nature of the bremsstrahlung distribution can pose problems for our radiative generator.
If we sample the photon direction isotropically, the large change in cross section will produce a large
variance in weights, hurting the generator performance. Since in every analysis we do, we integrate out
the photon direction, we want to sample from a distribution that matches the cross section as closely
as possible. However, sampling from the cross section directly is difficult because we can't invert, and
because it is fairly computationally difficult to calculate it. It's feasible to calculate it a few times
per event to produce a weight; it is not feasible to calculate it thousands of times per event.

\begin{figure}[htpb]
\centering
\includegraphics{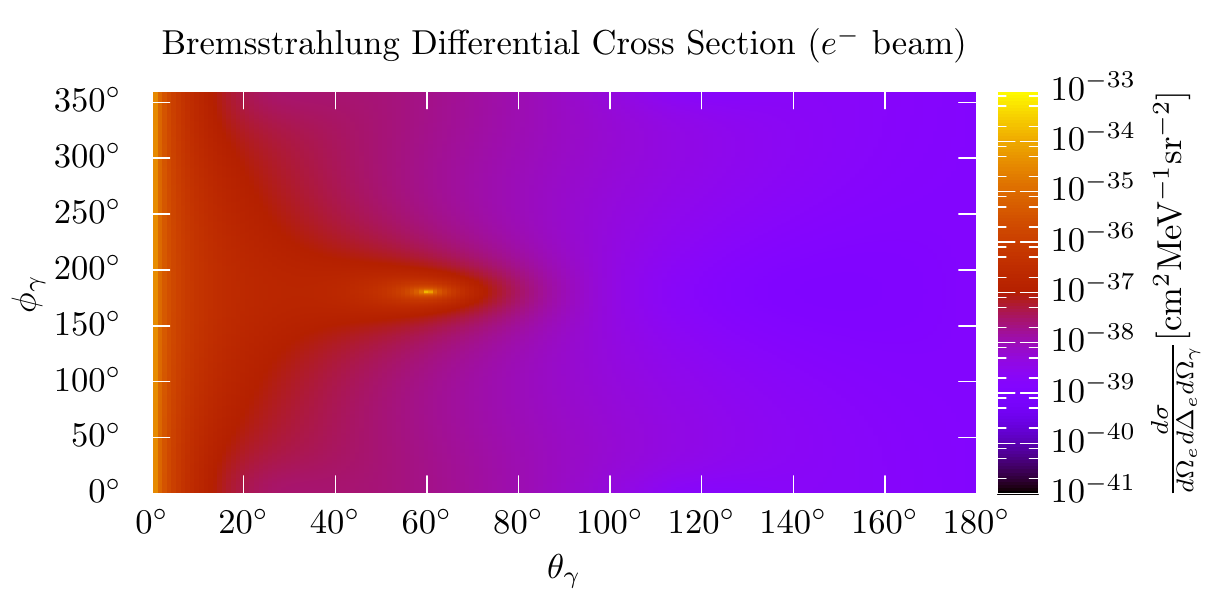}\\
\includegraphics{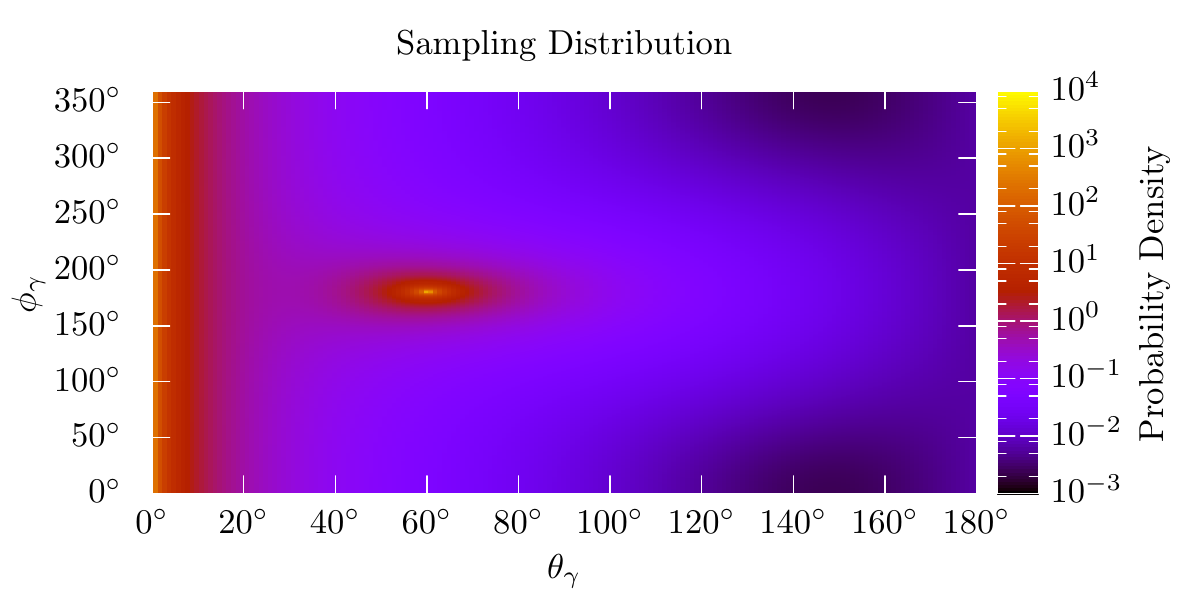}\\
\includegraphics{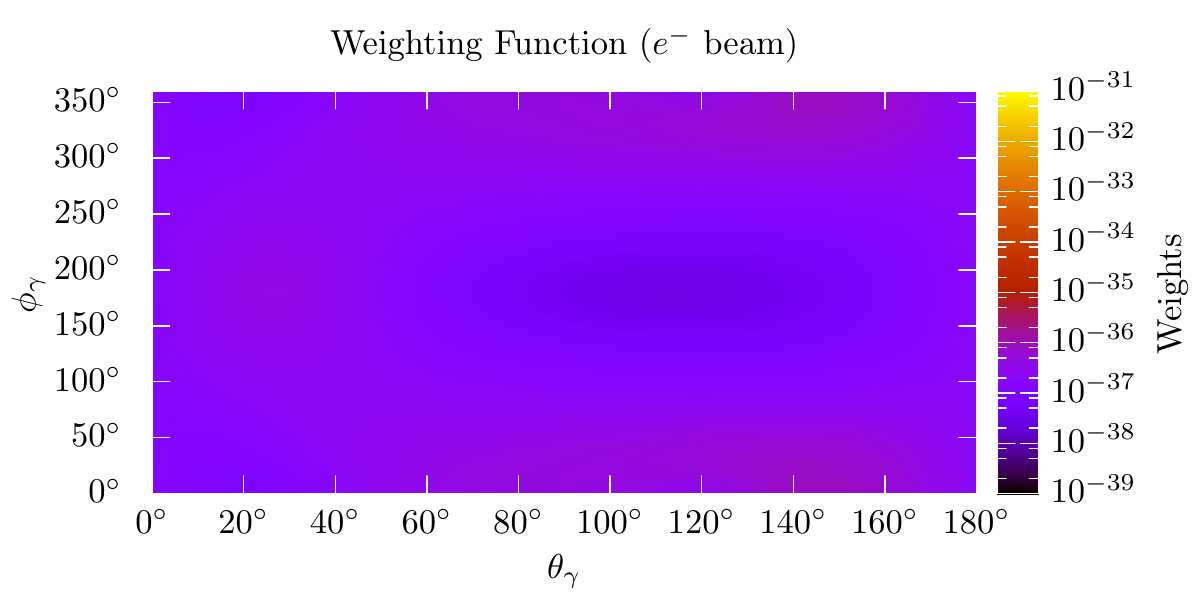}
\caption[Photon angle sampling to reduce weight variance]
{\label{fig:brem_photon_weights} The bremsstrahlung cross section (top plot) varies over several 
orders of magnitude with respect to the photon direction. We sample the photon direction from the distribution
shown in the middle plot. The resulting weights (bottom plot) are much more uniform.}
\end{figure}

Our solution (illustrated in figure \ref{fig:brem_photon_weights}), is to model two peaks in the 
angular distribution with the function:
\begin{align}
P(\cos\theta) &= \frac{1}{2\frac{E_l}{p_l}\log\left[ \frac{E_l + p_l}{E_l - p_l} \right] - 4}
\times \frac{1 - \cos^2\theta}{\left( \frac{E_l}{p_l} - \cos\theta \right)^2 } \\
&= \frac{1}{N} \times \frac{1 - \cos^2\theta}{\left( \frac{E_l}{p_l} - \cos\theta \right)^2 },
\end{align}
where $\theta$ is the angle between the photon and lepton, $E_l$ is the energy of the lepton,
and $p_l$ is the norm of the three-momentum of the lepton ($E_l$ is very nearly equal to $p_l$). 
Sampling from this distribution requires us to invert the cumulative distribution:
\begin{equation}
C(\cos\theta) = \frac{1}{N}\int_{-1}^{1} d\cos\theta \frac{1 - \cos^2\theta}{\left( \frac{E_l}{p_l} - \cos\theta \right)^2 }.
\end{equation}
Let's substitute $x= \frac{E_l}{p_l} - \cos\theta$:
\begin{align}
C(\cos\theta) &= \frac{1}{N}\int_{\frac{E_l}{p_l}-1}^{\frac{E_l}{p_l}-\cos\theta} \frac{dx}{x^2}
\left[ \left(1 -\frac{E_l}{p_l}^2\right) + 2\frac{E_l}{p_l}x - x^2 \right] \\
&= \frac{1}{N}\left[  \left(1-\frac{E_l}{p_l}^2\right) \left\{ \frac{1}{\frac{E_l}{p_l} - 1} - \frac{1}{\frac{E_l}{p_l} - \cos\theta}\right\}
  + 2\frac{E_l}{p_l}\log\left\{ \frac{\frac{E_l}{p_l}-\cos\theta}{\frac{E_l}{p_l} - 1} \right\}  + \cos\theta - 1 \right]\\
&= \frac{1}{N}\left[ \cos\theta  + 2\frac{E_l}{p_l}\log\left\{ \frac{\frac{E_l}{p_l}-\cos\theta}{\frac{E_l}{p_l} - 1} \right\}
  - 2 - \frac{E_l}{p_l} - \frac{  \left(1-\frac{E_l}{p_l}^2\right) }{ \frac{E_l}{p_l} - \cos\theta } \right].
\end{align}
This can be inverted numerically using bisection. 

Our approach on each event is to pick either the incoming lepton or the outgoing lepton with 
equal probability. Then we sample $\cos\theta$ as described above, while we sample $\phi$ 
isotropically. We then must transform these angles from the coordinate system relative to
lepton to the global coordinate system. With five kinematic variables now fixed, the kinematics
of the entire event are determined. We can specify the four-vectors for all of the incoming and
outgoing particles and can calculate all of the components needed to form the weight in equation
\ref{eq:weight}.

\subsubsection{Method 2: Non-Exponentiation}

An alternative approach to Method 1 is to avoid using exponentiation, and to simply aim
to produce a cross section that is accurate at next-to-leading order. The advantage is
that the problems with extrapolating an exponentiated correction far from the elastic
peak are avoided, but the cost is that radiation is not treated to all orders. The
non-exponentiated approach was added to the OLYMPUS radiative generator as an alternate 
weight, and is known as Method 2. This was largely the work of my classmate Rebecca Russell, 
who deserves, if not all of the credit, at least all minus $\epsilon$.

Method 2 weights are calculated for each event, along with method 1 weights. That means
that Method 2 has to make do with the same sampling distribution as described in the
previous section. This posed some minor complications, for which effective solutions 
have been found.

Method 2 works by segregating events into those it considers elastic, and those it
considers inelastic. The segregation criterion is the energy emitted by the bremsstrahlung
photon, $k$. If the photon energy is less than some arbitrary boundary $k_b$, then
the event is considered elastic, and given weight:
\[
w_\text{el.}=\frac{d\sigma}{d\Omega_l}_{1\gamma} \times (1 + \delta(k_b)) \times \frac{1}{P(\Omega_l,\Omega_\gamma,\Delta E_l)}.
\]
If $k > k_b$, then the event is considered inelastic, and given weight:
\[
w_\text{inel.} = \frac{d\sigma}{d\Omega_l d\Omega_\gamma d\Delta E_l}_\text{brems.} \times \frac{1}{P(\Omega_l,\Omega_\gamma,\Delta E_l)}.
\]
Elastic events are essentially weighted according to a standard correction, and 
inelastic events are weighted according to the tree-level bremsstrahlung cross section.
Radiative effects at the next-to-leading level are reproduced. The choice of boundary $k_b$
should not have an effect on the result. It must be chosen to be much smaller than the
detector resolution, but larger than the point at which $\delta$ becomes negative (see
figure \ref{fig:mt_example}). Typically, we chose $k_b = 1$~MeV.

This method runs into trouble because the so-called elastic events have a large variance
in their weights. To make Method 1 work, we sampled the photon angle from a sharply peaked
distribution. The cross section for these events, $(1+\delta) d\sigma/d\Omega$ has no
dependence on the photon direction. Dividing by the sampling distribution produces an
enormous variance.

To make Method 2 work, individual elastic weights were replaced by average elastic
weights:
\[
w_\text{el.} \longrightarrow \langle w_\text{el.} \rangle = 
\frac{d\sigma}{d\Omega_l}_{1\gamma} \int_{k<k_b} P(\Omega_\gamma,\Delta E_l|\Omega_l) d\Omega_\gamma d\Delta E
\frac{(1 + \delta(k_b))}{P(\Omega_\gamma,\Delta E_l|\Omega_l)}.
\]
These average elastic weights were precomputed for a wide range of lepton angles. When
computing a weight, the generator would interpolate between precomputations at the nearest
scattering angles. This significantly reduced the variance in the weights.

Another minor complication comes from the choice of elasticity criterion. Most standard
corrections define $\delta$ as a function of $\Delta E_l$, the energy lost by the 
lepton relative to elastic energy. In a coincidence measurement, the photon energy $k$,
which is equivalent to the missing energy $W$, makes more sense as an elasticity criterion. 
To make Method 2 work, a correction $\delta(k)$ was needed. Ent et al.\ describe such a
formulation \cite{Ent:2001hm}, but neglect the electron mass. Gramolin et al., in their
description of the ESEPP generator, describe a prescription that, while not analytic, 
does preserve the electron mass \cite{Gramolin:2014pva}. The OLYMPUS generator follows the 
ESEPP prescription.

We also chose a third alternate weight using $\Delta E_l$ as the elasticity criterion. 
We refer to this as ``Method 3'', though this method is philosophically identical to 
Method 2, and should probably have been named ``Method 2a.''  The use of $\Delta E_l$ 
as the elasticity criterion greatly simplifies the algebra, but can produce somewhat 
undesirable effects. For instance, it is possible for events with a large photon
energy to be considered elastic, simply because the proton has lost energy, not the lepton.
Still, this method costs almost nothing, so we chose to add it to our ensemble of weights.
The average weight in the elastic region is trivial:
\begin{align}
\langle w_\text{el.} \rangle &= \frac{d\sigma}{d\Omega_l}_{1\gamma} \int_{4\pi} d\Omega_\gamma \int_0^{\Delta E_b} d\Delta E_l
P(\Omega_\gamma,\Delta E_l|\Omega_l) \frac{(1 + \delta(\Delta E_b))}{P(\Omega_\gamma,\Delta E_l|\Omega_l)} \\
&= 4\pi \Delta E_b (1 + \delta(\Delta E_b)).
\end{align}

\section{Discussion}

\begin{figure}[htpb]
\centering
\includegraphics{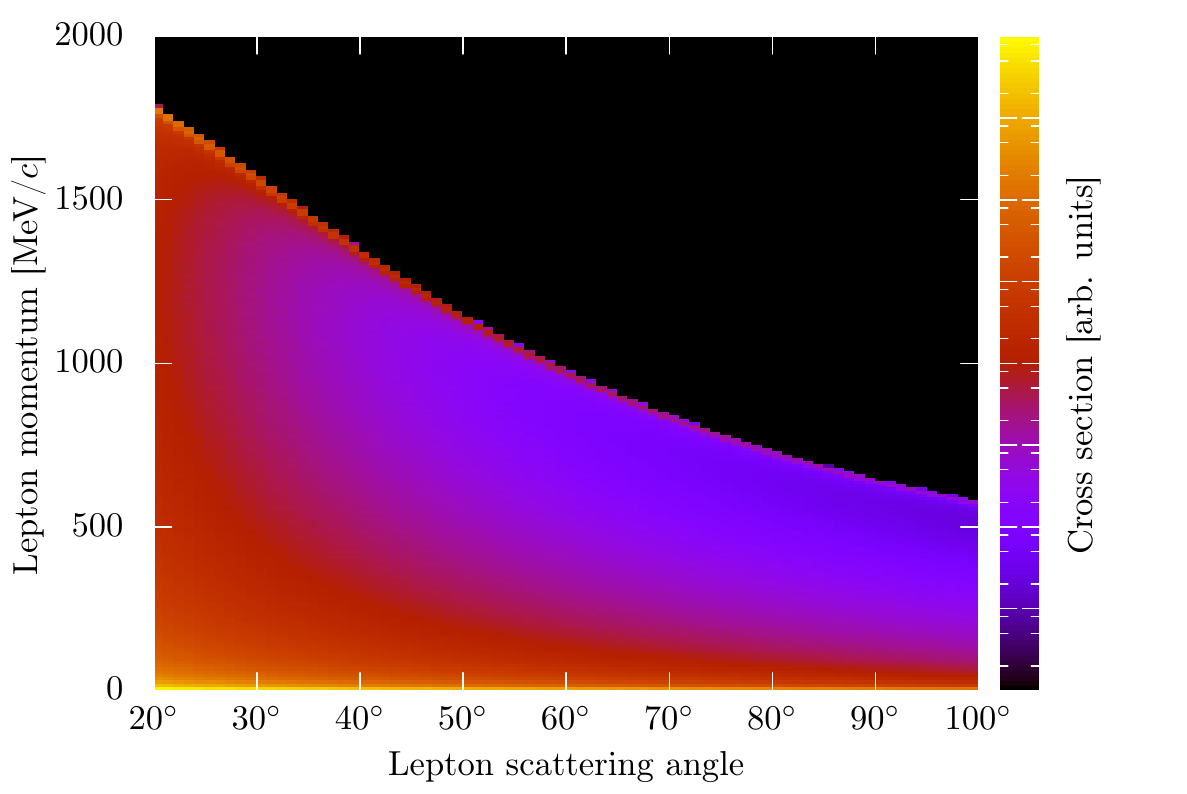}\\
\includegraphics{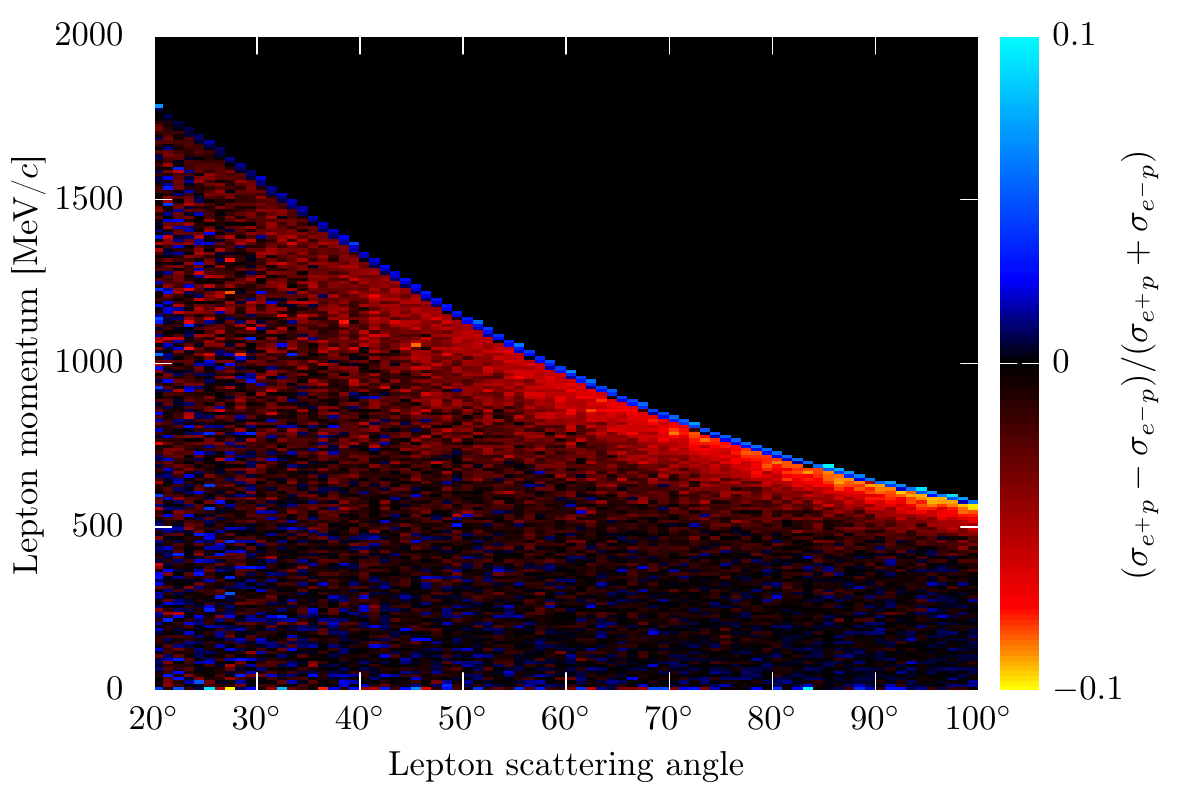}
\caption[Generator output]{\label{fig:gen_output} Events from the generator are binned by the lepton angle
and the lepton momentum. The top plot shows the cross section produced with an electron beam.
The bottom plot shows the ratio of positron and electron cross sections. The kinematic
dependence of the ratio is non-trivial.}
\end{figure}

Plots of the generator output are shown in figure \ref{fig:gen_output}, histogrammed by lepton
angle and momentum. Of course, the generator cross section is five-dimensional, but for the purposes
of visualization, I've had to choose two variables. In these plots, the elastic ridge is the top of the 
colored region, above which the kinematics are forbidden. The top plot shows the cross section with an 
electron beam. There is considerable cross section for very low photon momenta at all photon angles. 
This is caused by initial state radiation, which will be discussed in subsection \ref{ssec:isr}. 
The bottom plot shows the ratio of the positron-proton cross section to the electron-proton cross section.
The ratio varies significantly over the kinematic range shown, and does so in a non-trivial way. Close to the
elastic ridge, positrons are favored; further in the tail, electrons are favored. If the generator
has any errors in its implementation, it would be very easy to influence the ratio in a spurious way.

We took a number of steps to validate the radiative generator, and I will discuss some of these in this
section. The most complicated part of the generator, the bremsstrahlung matrix element, was validated
against that of the ESEPP generator. The generator output was also tested against the radiative tails
from standard radiative corrections.

\subsection{Matrix Element Comparison with ESEPP}

The bremsstrahlung matrix element was a very large source of potential bugs. As a cross check,
we compared our matrix element against that of the ESEPP generator. Our two different implementations
of the matrix element should give the exact same results (up to the choice of definition for the off-shell 
proton current, a very minor effect). Our initial comparisons revealed some typing mistakes on our part,
and after fixing these, our matrix elements agreed to a very high level. Our lepton terms, which have no
off-shell proton vertices, agreed to the level of the numerical precision of our computer, and the proton 
term and interference term agreed to better than one part in $10^5$. After this, we were satisfied that 
our matrix element implementation was correct.\footnote{Caveat emptor: unfortunately this doesn't completely
guarantee the accuracy of the transcription of the matrix element (equation \ref{eq:brems_tree_level}).} 

\subsection{Comparisons with Maximon and Tjon}

After we had confidence in our implementation of bremsstrahlung, we then turned to the task of verifying
the generator as a whole. One straightforward test is to use the generator to integrate over the photon 
direction and compare the result (which depends on $\theta_l$, and $\Delta E_l$) with the predictions
of a standard correction. By default, our generator uses the $\delta$ from Maximon and Tjon, and that
particular choice was natural for comparison. 

\begin{figure}[htpb]
\centering
\includegraphics{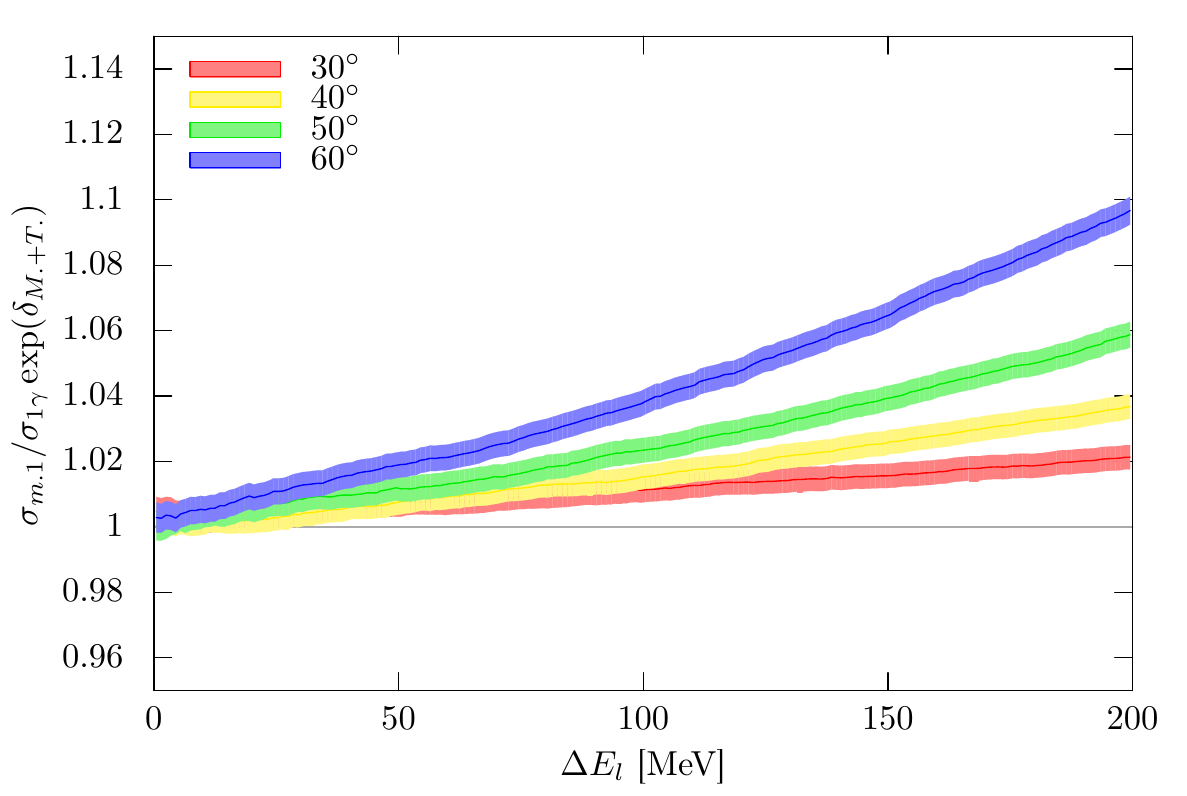}\\
\includegraphics{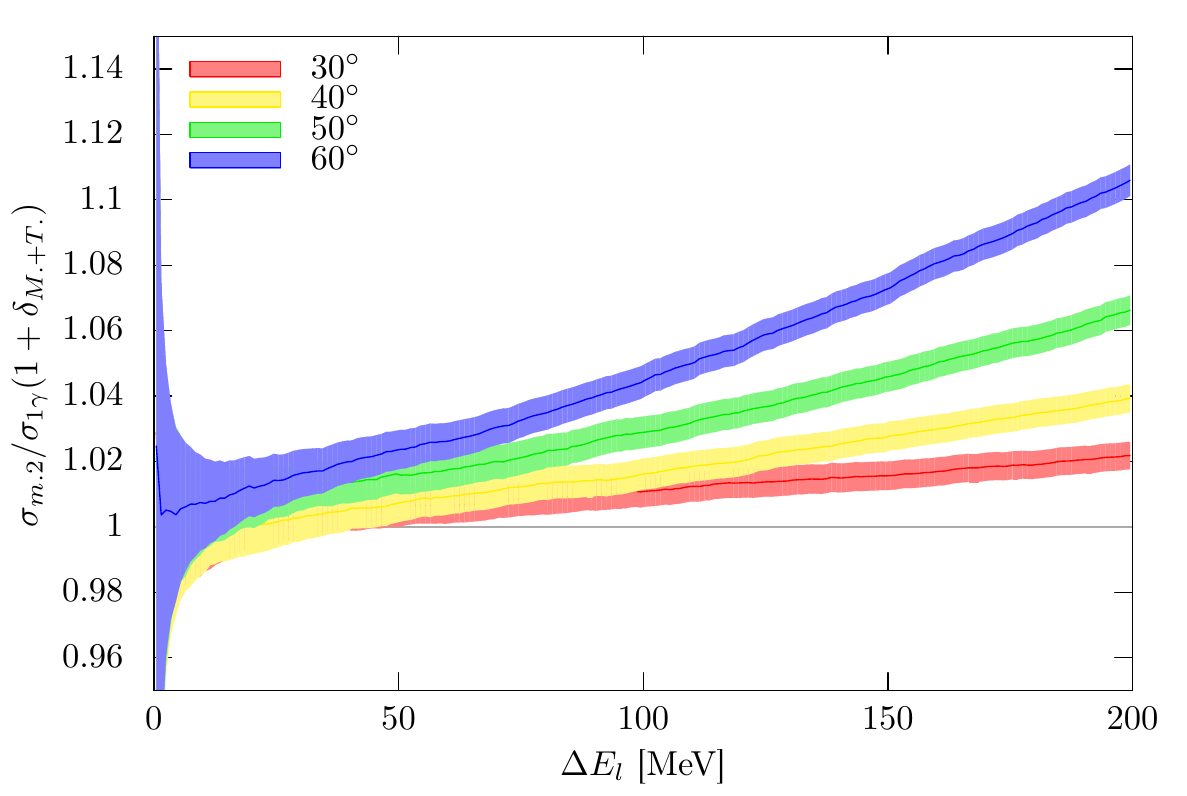}
\caption[Generator output divided by Maximon and Tjon]
{Our generator converges to the Maximon and Tjon prediction as $\Delta E_l \rightarrow 0$, for both
Method 1 (top) and Method 2 (bottom), just as we expect. \label{fig:gen_vs_mt_ratio}}
\end{figure}

Figure \ref{fig:gen_vs_mt_ratio} shows the results from the generator, divided by the prediction of
Maximom and Tjon, plotted as a function of $\Delta E_l$. The $y$-axis is the generator cross section,
integrated over photon direction and over $\Delta E_l$ up to the point marked on the $x$ axis. This is
then divided by that same cross section as predicted by Maximon and Tjon. 
In the case of Method 1 (top plot) we are comparing to an exponentiated $\delta$, and in the
case of Method 2 (bottom plot) we are comparing to a non-exponentiated $\delta$. The results confirm
that the generator is behaving as expected. Both Method 1 and Method 2 converge to Maximon and Tjon
when $\Delta E_l$ is small, for multiple angles. We expect convergence at low $\Delta E_l$ because
Maximon and Tjon (and all standard corrections) calculate bremsstrahlung in the soft limit. When the 
energy carried away by the bremsstrahlung photon becomes large (and consequently $\Delta E_l$ grows),
Maximon and Tjon become less accurate (compared with the full tree-level calculation in our generator.)

It is noticeable that, for very small values of the $\Delta E_l$ cut-off, the uncertainty from Method 2
increases substantially. This is caused by the sampling distribution. The uncertainty is driven by how
many samples fall within Method 2's elastic region.

\begin{figure}[htpb]
\centering
\includegraphics{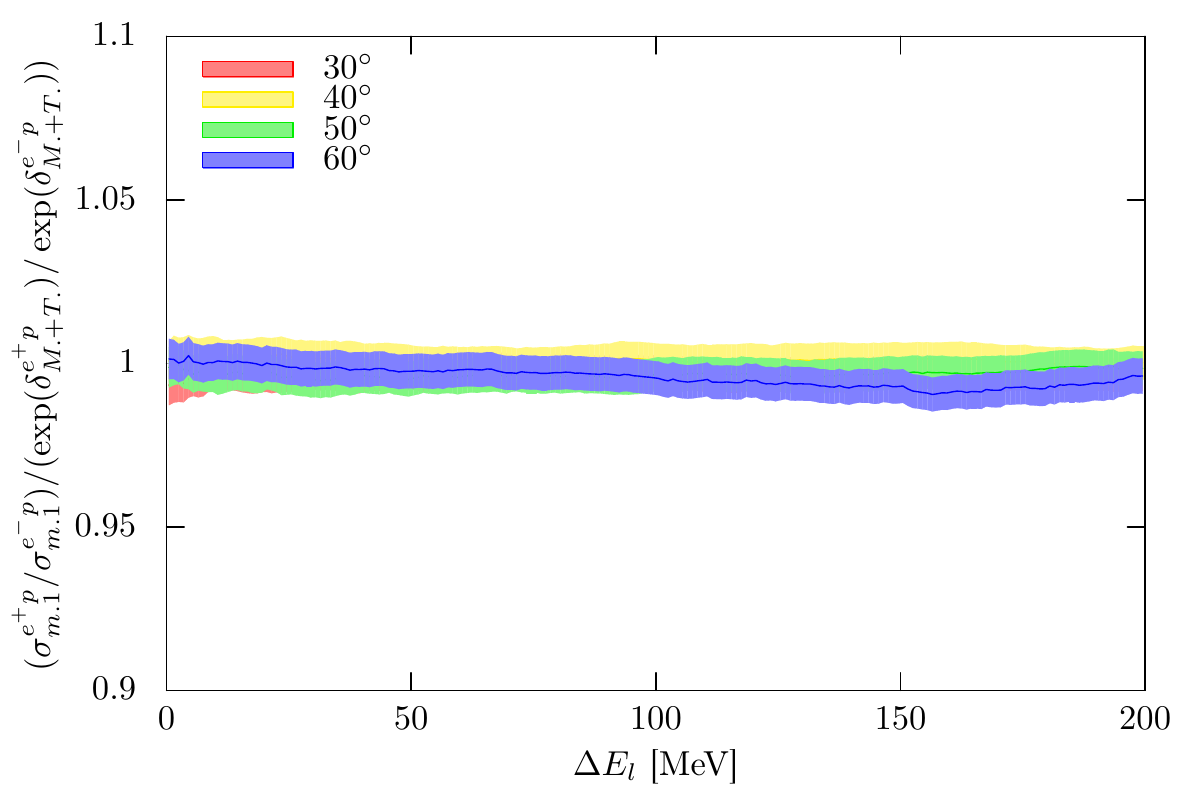}\\
\includegraphics{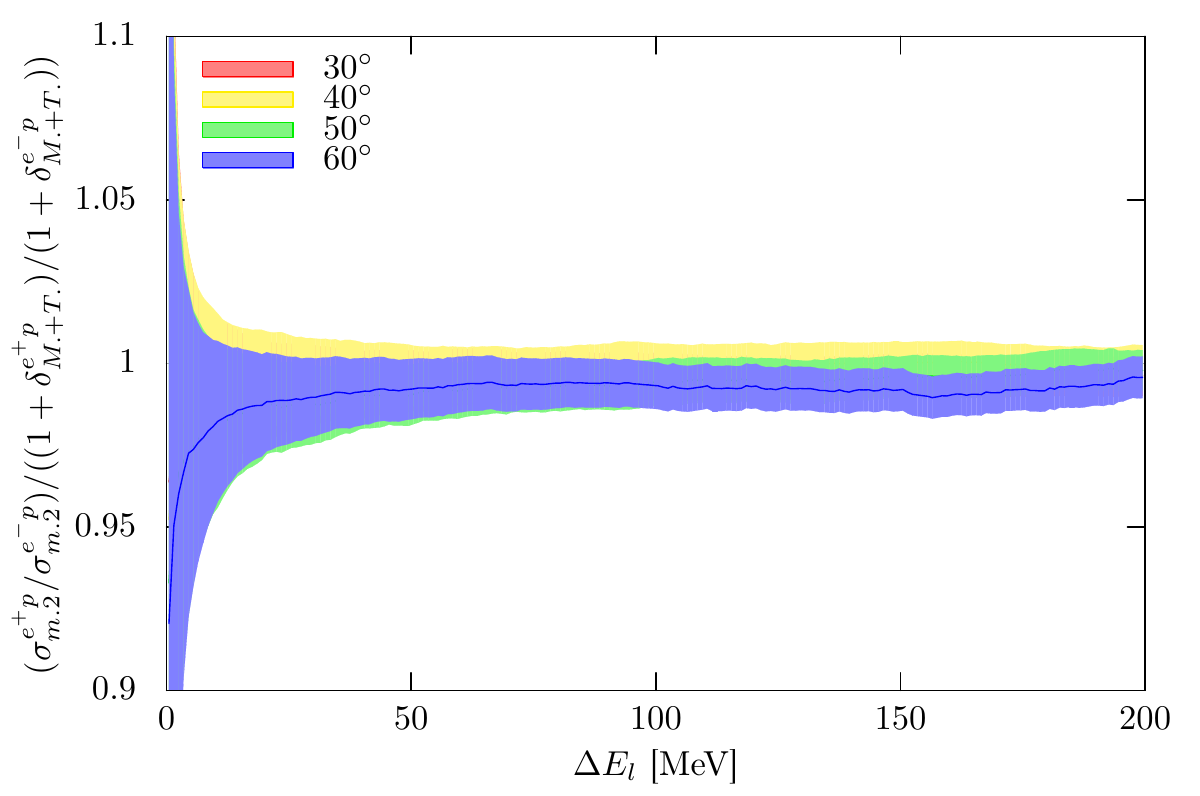}
\caption[Species-ratio predicted by the generator and Maximon and Tjon]{ \label{fig:gen_vs_mt_spec_ratio}
The cross section ratio predicted by the generator is almost identical to that predicted by Maximon
and Tjon, and it's largely insensitive to the $\Delta E_l$ cut-off.
}
\end{figure}

The goal of OLYMPUS is to compare electrons and positrons, so we must also compare the behavior of the
generator with respect to both lepton species. Figure \ref{fig:gen_vs_mt_spec_ratio} shows the cross
sectio ratio predicted by the generator divided by that predicted by Maximon and Tjon. As expected,
both methods converge to Maximon and Tjon for small values of the $\Delta E_l$ cut-off (although
the uncertainty of Method 2 is large for very small cut-off values). What is also reassuring is that
the agreement is largely independent of $\Delta E_l$. The magnitude of the radiative correction introduced
into the positron-electron ratio by our generator will not deviate from Maximon and Tjon regardless
of the effective $\Delta E_l$ cut-off used in our analysis.

\begin{figure}[htpb]
\centering
\includegraphics{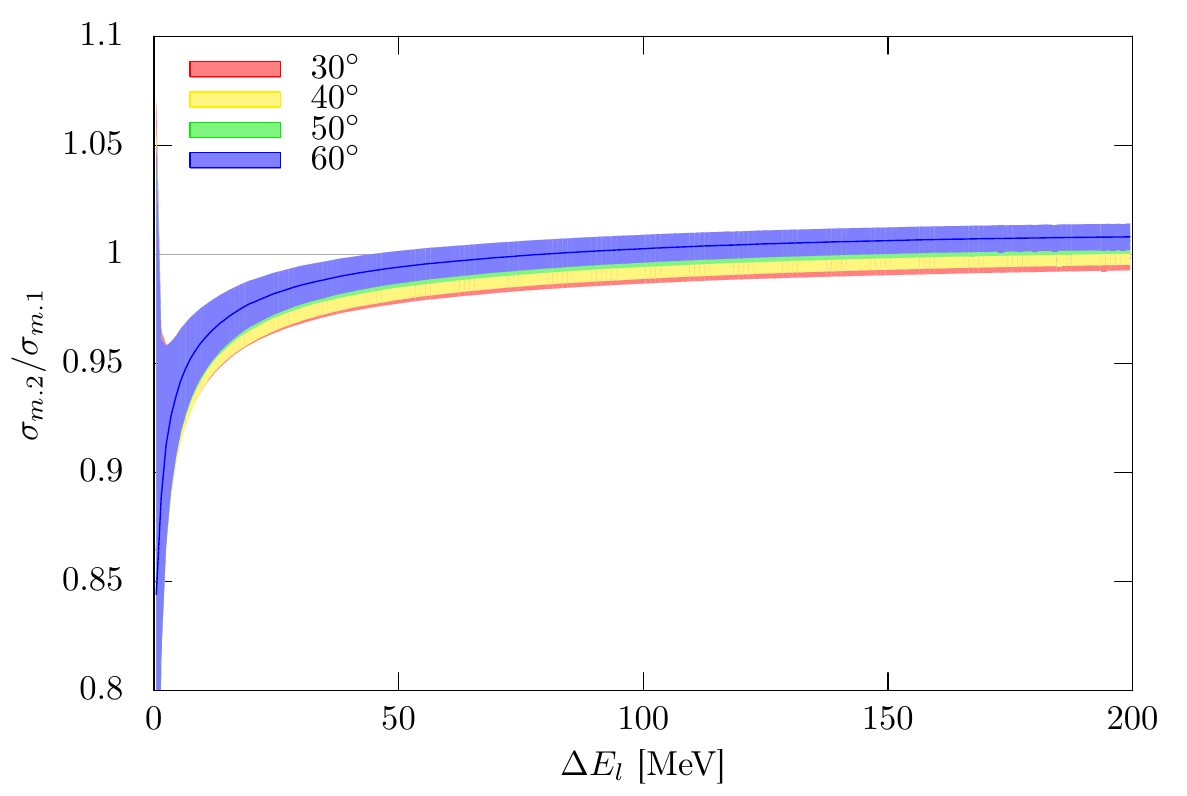}
\caption[Ratio of Method 1 and Method 2 corrections]{\label{fig:gen_m1_m2_ratio}
Method 2 and Method 1 are very similar for large cut-off values. The differences between the two 
methods are only large at very small cut-off values. The effective cut-off for OLYMPUS will probably
be on the order of 100--200~MeV.
}
\end{figure}

We can also ask how Method 1 and Method 2 compare with each other as a function of angle and cut-off. 
The ratio of the two corrections is shown in figure \ref{fig:gen_m1_m2_ratio}. At large values of the
$\Delta E_l$ cut-off, Method 1 and Method 2 converge. This makes sense. As $\Delta E_l$ grows, $\delta$
gets closer to 0, and in that limit, there is little difference between $\exp(\delta)$ and $(1+\delta)$.
There are some persistent differences at the percent-level. These differences will inform our systematic
uncertainty from radiative corrections.

\subsection{Enhancement Due to Initial State Radiation}

\label{ssec:isr}

\begin{figure}[htpb]
\centering
\includegraphics{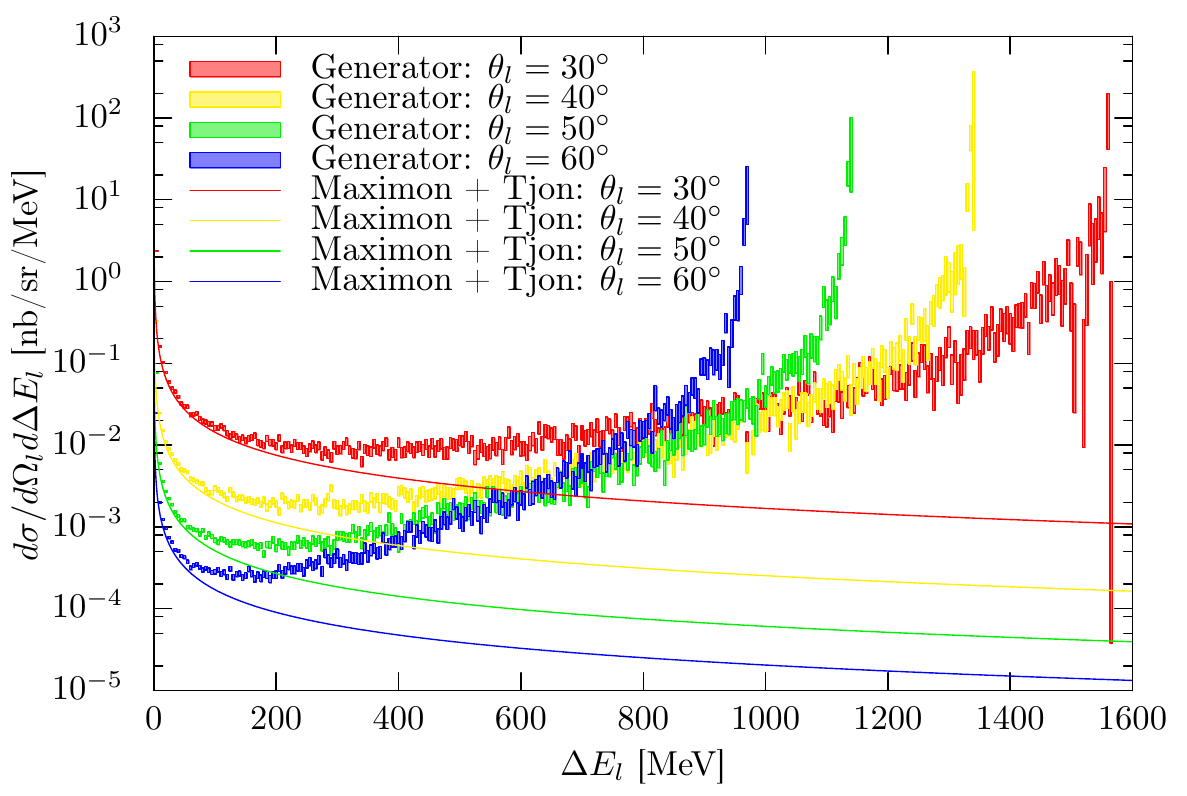}
\caption[Radiative tails predicted by Maximon and Tjon and the generator]
{\label{fig:gen_vs_mt} The cross section \emph{increases} at large values of $\Delta E_l$ because
of initial state radiation. Bremsstrahlung from the initial lepton leg effectively reduces the incoming
beam energy, enhancing the cross section.}
\end{figure}

There is interesting behavior in figure \ref{fig:gen_vs_mt_ratio}: the generator cross section rises
relative to Maximon and Tjon with increasing $\Delta E_l$. This implies that the generator cross section
has a heavier radiative tail than Maximon and Tjon predict. Figure \ref{fig:gen_vs_mt} shows the radiative
tails predicted by Method 1 (bins), as well as those of Maximon and Tjon (lines). I could, in principle,
show Method 2 as well, but the differences are imperceptible on a log scale. The radiative tails are
fatter in the generator than in Maximon and Tjon.

What is even more surprising is that the cross section reaches a minimum and then starts rising. It continues
to rise all the way to the maximum value\footnote{
The scattered lepton can lose all of its kinetic energy but cannot lose its rest mass.}:
$\Delta E_l = E_3^\text{el.}-m_e$. The origin of this phenomenon is initial state radiation (ISR). If the 
bremsstrahlung photon is emitted in the incoming lepton direction, the dominant diagram is the one where the 
photon is emitted from the initial lepton leg. The photon carries energy away from the lepton, effectively 
reducing the beam energy. The elastic scattering cross section is larger at smaller beam energies. Our generator,
which uses a full tree-level bremsstrahlung matrix element, captures this phenomenon. Maximon and Tjon, which 
calculates bremsstrahlung in the soft-limit, cannot. 

While this was a very satisfying discovery to make, it exposes a potential problem in our sampling distribution.
To make our weights uniform, we sampled $\Delta E_l$ according to Maximon and Tjon (the lines in figure
\ref{fig:gen_vs_mt}), aiming to reproduce the distribution predicted by generator's cross section (bins in figure
\ref{fig:gen_vs_mt}). Since the two distributions diverge, the weights will become increasingly large as $\Delta E_l$
increases. If we were ever to do a study in which we integrated over the full range of $\Delta E_l$, there would
be an enormous variance in the weights. Now certainly an elastic event selection will reject events with large 
$\Delta E_l$, and so any use of the simulation for such studies will not be hampered by weight variance. However,
we tried to devise a solution in case any simulations deep in the tail were needed. 

Our solution is to add a second sampling distribution for $\Delta E_l$ that rises with $\Delta E_l$, in order to 
capture the ISR enhancement. We propose modifying the sampling distribution:
\begin{equation}
P(\Delta E_l | \Omega_l) \longrightarrow \chi \frac{t}{\Delta E_l} \left(\frac{ \Delta E_l}{\Delta E_\text{cut}}\right)^t +
(1 - \chi)P_\text{ISR}(\Delta E_l | \Omega_l),
\end{equation}
where the variable $\chi$ represents the fraction of events drawn in the ``soft'' distribution, while
$(1-\chi)$ represents the fraction drawn in the ``hard'' ISR distribution. What functional form should
we then use for $P_\text{ISR}(\Delta E_l | \Omega_l)$? The one-photon-exchange cross section scales 
roughly with $E^{-2}$ at low beam energies. Therefore, a reasonable choice would be:
\begin{equation}
P_\text{ISR}(\Delta E_l | \Omega_l) = \frac{E(E - \Delta E_\text{cut})}{\Delta E_\text{cut} (E - \Delta E_l )^2}.
\end{equation}

The generator weight will need to be reworked with this new distribution. Whereas before we needed to weight
with 
\[
\frac{1}{P(\Delta E_l | \Omega_l)} = \frac{1}{\frac{t}{\Delta E_l} \left(\frac{\Delta E_l}{\Delta E_\text{cut}}\right)^t },
\]
we will now need to weight by
\begin{align}
\frac{1}{P(\Delta E_l | \Omega_l)} =& \frac{1}{\frac{\chi t}{ \Delta E_l} \left(\frac{\Delta E_l}{\Delta E_\text{cut}}\right)^t  
+ (1-\chi) \frac{1}{\frac{t}{\Delta E_l} \left(\frac{\Delta E_l}{\Delta E_\text{cut}}\right)^t }} \\
=& \frac{ \Delta E_l \left( \frac{\Delta E_\text{cut}}{\Delta E_l} \right)^t }
{\chi t + (1-\chi) \frac{ E (E - \Delta E_\text{cut})}{(E - \Delta E_l)^2} \left( \frac{\Delta E_l}{\Delta E_\text{cut}}\right)^{1-t} }.
\end{align}
Just as before, the numerator is cancelled by factors in the cross section. The second term in the denominator has 
a bare $\Delta E_l^{1-t}$, but since $0<t<1$, this term becomes very small when $\Delta E_l$ goes to 0, and thus insignificant
compared to the $\chi t$ term. So this modification preserves the numerical stability of the weights. 

A useful property with this modification is that $\chi$ is an adjustable parameter for the user. Setting $\chi=1$ restores
the unmodified probability distribution, suitable for simulating elastic event selection. Setting $\chi < 1$ can increase
the number of large $\Delta E_l$ events that the generator produces, smoothing out the weights deep in the tail. Described
in another way: $\chi$ tunes the statistical precision of the elastic peak versus the ISR peak in a given simulation. Setting 
$\chi$ close to one enhances the precision of the elastic peak, while setting $\chi$ close to zero gives priority to the ISR
peak. $\chi=0.5$ is a middle ground.

\section{Running the Generator}

In this section, I'll describe the settings used for our simulation of the main analysis. First,
the sampling distribution was controlled by two tunable parameters:
\begin{enumerate}
\item $\chi=0.5$
\item $\Delta E_\text{cut} = E_3^\text{el.}-m_e$, the maximum value at each angle. 
\end{enumerate}
We balanced the statistics between the elastic peak and the ISR peak, in case there any ISR events contaminated our
elastic event selection. $\Delta E_\text{cut}$ was set to the maximum value at each angle; we simulated the entire
radiative tail.

Second, each event contained a vector of weights, one for each different cross section model.
Broadly these can be categorized as using Method 1 and Method 2; within those categories, there
were several choices we could make.
\begin{enumerate}
\item Form Factor Models
  \begin{itemize}
  \item Point-like proton
  \item Standard dipole
  \item Global fit by Kelly et al. \cite{Kelly:2004hm}
  \item Global fit by Bernauer et al., including $\pm$ fit uncertainties \cite{Bernauer:2013tpr}
  \end{itemize}
\item Standard Correction Model
  \begin{itemize}
  \item Maximon and Tjon \cite{Maximon:2000hm}
  \item Mo and Tsai \cite{Mo:1968cg}
  \item Meister and Yennie \cite{Meister:1963}
  \end{itemize}
\item Vacuum Polarization Model
  \begin{itemize}
  \item Leptonic correction only
  \item Leptonic correction + hadronic correction extracted from data \cite{vacpol:web,vacpol:pres}
  \end{itemize}
\item Full vs.\ soft bremsstrahlung
\end{enumerate}
Each combination of all of the different options leads to an individual weight. In analysis,
the effect of each choice can be studied be reweighting the simulated histograms. This can
be done without resimulating individual events.

\chapter{Determining Luminosity from SyMB Multi-Interaction Events}

\label{chap:lumi}

\section{Introduction}

The OLYMPUS data set is technically two data sets: one with an electron
beam, the other with a positron beam. In order to make comparisons between
the two sets, they both must be normalized by their respective integrated 
luminosities. OLYMPUS was designed with multiple luminosity monitoring systems 
with this in mind.

To measure an asymmetry between electron and positron cross sections, strictly
speaking, only the relative integrated luminosity between the data sets is needed. 
It is not crucial that the OLYMPUS luminosity monitors be extremely accurate
in their absolute determination of luminosity. It is, however, crucial that the
systematic differences in the monitor performance between electron and positron
running modes be as small as possible. Uncertainty in the relative luminosity 
determination between beam species (which I will frequently refer to as ``species-relative luminosity''
in this chapter) leads directly to uncertainty in the cross section asymmetry
that OLYMPUS measures. 

OLYMPUS collected data with three independent monitoring systems; the slow
control system, the $12^\circ$ tracking telescopes, and the symmetric
M\o ller/Bhabha calorimeters (SyMBs). The slow control system, while 
simple and less accurate, was extremely useful because it could be used online 
to immediately give a rough estimate of the luminosity of the data being collected.
The $12^\circ$ telescopes recorded the rate of forward elastic $ep$ scattering
from which integrated luminosity could be extracted. This method had an
intrinsic limitation due to contributions from hard two-photon exchange. The SyMBs avoided
this problem by looking at purely QED processes: M\o ller and Bhabha scattering.
Originally, this detector system was designed to monitor the scattering rate
at the symmetric angle by detecting the two final-state leptons in coincidence. After multiple
years of analysis, it was found that this method had a systematic uncertainty
on the order of 3\% for the species-relative luminosity, and thus was not useful for 
OLYMPUS. However, an alternate analysis method developed by the author---one which 
makes use of multi-interaction events---was found to have an accuracy better than 0.3\%, small enough to be
used effectively by OLYMPUS to achieve its physics goals. This alternate analysis, as well as its results, are 
the subject of this chapter.

\subsection{Slow Control System}

The OLYMPUS slow control system handled the sending of instructions to the
myriad of high voltage supplies, vacuum pumps, and gas flow valves, as well
as recording the data from a variety of sensors linked to these pieces of
equipment. The word ``slow'' refers to the fact that this sending and recording
took place on timescales much slower than the trigger rate. For example, 
temperature sensors only needed to record variations over time scales of seconds,
not microseconds. The data recorded by the slow control system could be used to reconstruct
luminosity. The system recorded the beam current in the storage ring, as well
as the flow rate of gas into the target and the temperature of the target cell.
From this information, the thickness of the target (in units of atoms per area)
could be reconstructed, and the luminosity determined.

The slow control system was useful because it was simple. No high-level
analysis was needed, and an estimate of luminosity could be determined 
almost immediately. However, this method lacked the accuracy necessary to
make a percent-level asymmetry measurement. In measuring absolute luminosity,
the system was estimated to be accurate to the level of 20\%, limited
mostly by knowledge of the target cell's gas conductance. 
Of course, the important figure of merit was the species-relative luminsosity,
for which the slow control system was estimated to be accurate to the 
level of about 3\%. The limiting factor was the knowledge of the temperature
profile of the gas in the target. Temperature was estimated using a sensor
attached to the outside of the target cell, which might have recorded a slightly
different temperature than that of the gas inside cell. The gas might also have
had a temperature gradient along the length of the cell. Any errors in the 
gas temperature might have been slightly different for the two beam species.
The background environment and beam quality were certainly different between
the two species, and this could have altered the equilibrium temperature of 
the target, either directly, or indirectly through, for example, wake field
heating.

\subsection{$12^\circ$ Tracking Telescopes}

\label{ssec:12deg}

The $12^\circ$ tracking telescopes monitored luminosity by measuring the rate
of elastic $ep$ scattering at forward angles. In events of interest, the scattered
lepton would pass through the telescope at an approximately $12^\circ$ scattering
angle, while the recoiling proton would be detected in a time-of-flight scintillator
bar on the opposite sector. Most (on the order of 90\%) of the protons also passed 
through the drift chamber acceptance, though this was not a requirement for the $12^\circ$ trigger. 
Each $12^\circ$ telescope had two scintillator planes, three multi-wire proportional 
chambers (MWPCs), and three gas electron multiplier (GEM) detectors. Each MWPC and
GEM detector could determine a two-dimensional tracking point. The GEMs had much
better position resolution, of approximately 100~$\mu$m. Ideally, 
a passing lepton would produce six two-dimensional tracking points. The magnetic field in the region
permitted a momentum analysis of the lepton tracks. For events in which the proton was tracked
in the drift chambers, the momentum of the lepton could be correlated with the momentum
of the proton for a clean selection of elastic events. In events without a proton
track, elastic events could still be selected using the azimuthal angle of the proton
(as determined in the time-of-flight scintillator), as well as the lepton's angle-momentum
correlation.

A very thorough analysis of the $12^\circ$ telescope data was performed by 
Brian Henderson and the interested reader is encouraged to learn the details 
from his thesis \cite{henderson:thesis}. I will summarize his main findings here. 
First, he found that the GEM detectors had correlated time-dependent inefficiencies;
that is, in some events all three GEMs failed to register detections for a particle
that the MWPCs and scintillators saw unequivocally. This behavior seemed to happen
semi-randomly and varied over time. For the sake of making a stable luminosity 
determination, the GEM information had to be ignored. This, unfortuntaely, sacrificed 
the resolution advantage of the GEMs, but was necessary for the sake of the analysis\footnote{
The GEM resolutions still were extremeley useful in making high-resolution maps
of the MWPC and scintillator efficiencies.}. Second, he found, through simulation,
that the rates in the $12^\circ$ telescopes were very sensitive to the spatial 
distribution of gas in the target. Furthermore, due to the magnetic field, the 
telescopes had acceptance for different parts of the target for different beam
species. In order to make a species-relative luminosity determination, it was important
to model the density distribution of the target as accurately as possible. This 
prompted Brian to perform a detailed microscopic conductance calculation, which
is described in his thesis. By incorporating the results of this calculation into
the OLYMPUS simulation, Brian achieved much better agreement between simulation and
data in several reconstructed distributions. Third, after improving the target distribution
in simulation, Brian found that the $12^\circ$ luminosity extraction gave very similar
results to that of the slow control system. Since the $12^\circ$ system had a smaller
systematic uncertainty than the slow control system, this last result was essentially
a validation that the slow control system was accurate. 

The $12^\circ$ luminosity results should be considered cautiously because the
the species-relative luminosity includes the effects of hard two-photon exchange.
The $12^\circ$ telescopes make a measurement at $\epsilon=0.98$, and the species asymmetry is constrained
to be zero at $\epsilon=1$, so two-photon exchange cannot be a large effect for the telescopes. Still, it is
unknown. Any results normalized to the luminosity determined by the $12^\circ$ telescopes
are relative to the amount of hard two-photon exchange at $\epsilon=0.98$ and $Q^2=0.175$~GeV$^2/c^2$. 
If one had an independent measure of luminosity, the $12^\circ$ telescopes could be used to
make lepton sign asymmetry measurement at that kinematic point, effectively increasing
the kinematic reach of OLYMPUS to higher $\epsilon$. 

\subsection{Symmetic M\o ller/Bhabha Calorimeters}

The symmetric M\o ller/Bhabha calorimeters (SyMBs) were designed to provide a
luminosity monitor that avoided the problem of high-$\epsilon$ two-photon exchange
all together. The SyMBs recorded the rate of elastic scattering of beam leptons
by atomic electrons in the target---M\o ller scattering when running with an electron
beam and Bhabha scattering when running with a positron beam. Only quantum electrodynamics
(QED) is needed to calculate these cross sections, so no hadronic two-photon exchange
effects were involved in making a luminosity determination. The SyMB calorimeters were 
positioned in order to detect both final state leptons in coincidence for scattering 
at the symmetric angle, approximately $1.29^\circ$ for a 2~GeV beam energy. 

An extensive analysis along these lines was performed by my colleague Colton O'Connor.\footnote{I should add,
that this analysis depended on a separate radiative M\o ller/Bhabha radiative generator, designed
and implemented by Charles Epstein \cite{Epstein:2016lpm}.}
In this thesis, I will refer to that work as the main SyMB analysis method. As with the 
$12^\circ$ telescopes, I will not try to repeat every detail of Colton's work, and instead
I encourage the interested reader to consult his thesis \cite{oconnor:thesis}. His principal
finding can be summarized simply: extracting an accurate luminosity from the rate
of symmetric M\o ller/Bhabha coincidences was not possible. First, the method is 
susceptible to large systematic uncertainties, the most significant of which stems
from uncertainties in the position of the beam. Colton estimated the systematic
uncertanties at the time of this writing to be nearly 3\% on the species-relative
luminosity. Second, the species relative luminosity from the main SyMB analysis differed 
by several percent relative to the slow control and $12^\circ$ systems, which agree with each 
other to within 1\%. The slow control and $12^\circ$ results already presented a two-against-one 
scenario, but the main SyMB analysis also lost its remaining credibility because the magnitude
of the discrepancy with other systems changed significantly over the OLYMPUS run. For early
parts of the fall run, the species-relative discrepancy was only approximately 3\%, but this
grew to 5\% by the end of the fall run. While it might be possible to 
entertain scenarios in which the slow control and $12^\circ$ systems were both inaccurate by
the same amount, it would be a stretch to argue that both changed their level of inaccuracy
at the same point in time by the same amount. It was decided that the main SyMB analysis was
untenable and not worth pursuing further. There are a number of possible sources of error
which can contribute to the discrepancy in its results, but none of them can be determined
conclusively, and they cannot be corrected for.

The subject of this chapter is an alternate method for extracting luminosity from the SyMB
data. This method uses multi-interaction events (MIE): multiple beam leptons from the 
same bunch having interactions in the target. This method was not in any way part of
the original SyMB design. It was pursued in the hope that it would shed light on 
the source of problems in the main SyMB analysis. But after pursuing it further, 
we discovered that the MIE method is better suited than the main analysis for making
a species-relative luminosity determination. We estimate the systematic error to be 0.27\%.
This method would be quite effective for luminosity monitoring in future lepton sign asymmetry 
measurements.

\section{Overview of the SyMB Calorimeters}

\subsection{SyMB Hardware}

\begin{figure}[htpb]
\centering
\includegraphics[width=14cm]{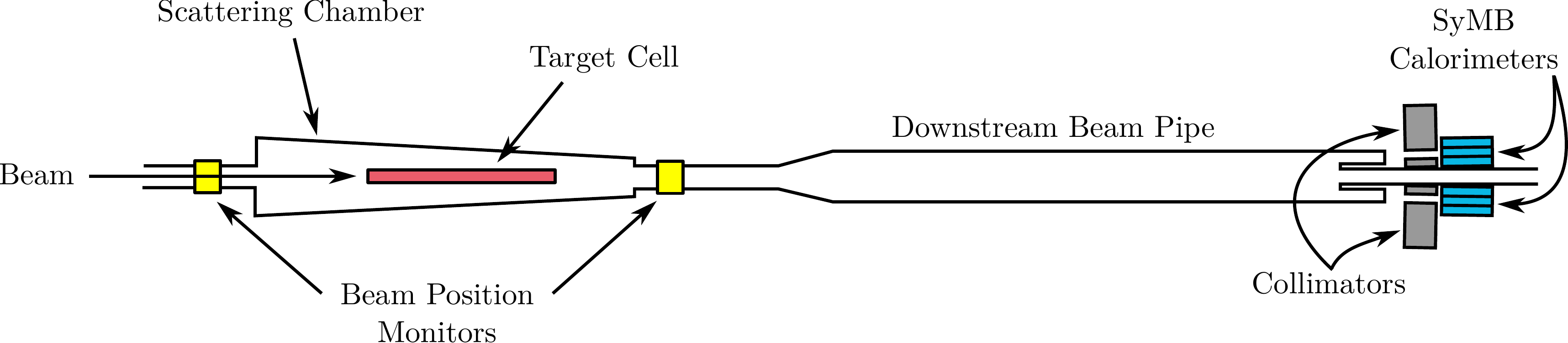}
\caption[Beamline schematic]{\label{fig:bl_schematic} The SyMBs are shown relative to the OLYMPUS beam line, roughly
to scale, from above. The important pieces of hardware for this report are highlighted in color: the target in red,
the beam position monitors (BPMs) in yellow, and the calorimeters in blue.}
\end{figure}

The symmetric M\o ller/Bhabha calorimeters sat approximately 3~m downstream from the
scattering chamber, a few centimeters from the beamline, one to the left, the other
to the right. A schematic that shows the position of the SyMBs relative to the OLYMPUS
beamline is shown in figure \ref{fig:bl_schematic}. Each calorimeter was composed of nine
lead fluoride crystals, whose Cherenkov light was detected by photomultiplier tubes (PMTs).
The calorimeters were shielded by collimators: thick lead blocks with a cylindrical
aperture that allowed particles from the target to enter the calorimeters. The downstream
beam pipe passed between the calorimeters.

\begin{figure}[htpb]
\centering
\includegraphics[width=12cm]{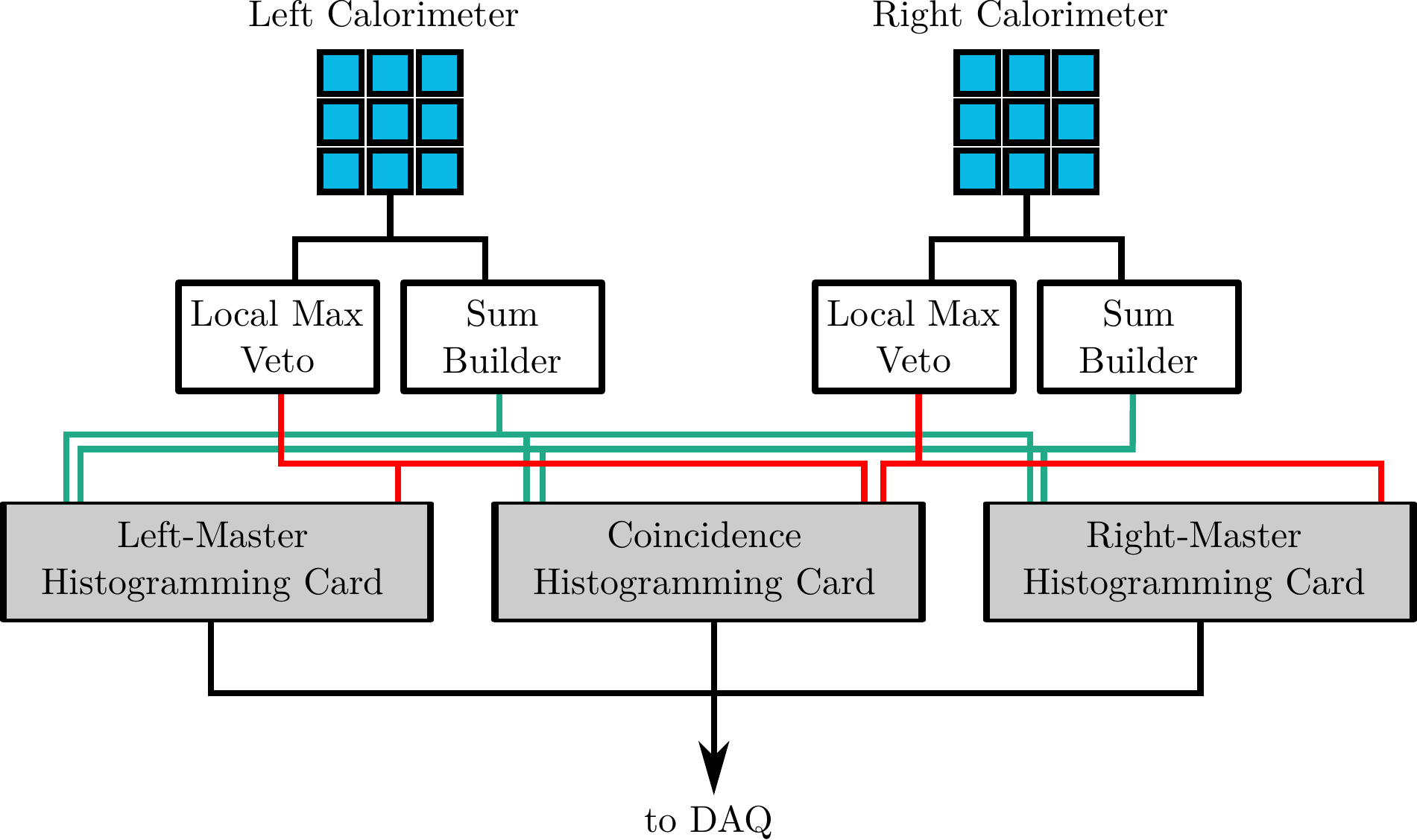}
\caption[SyMB readout]{\label{fig:symb_readout} The signals from the calorimeters were fed to
three fast histogramming cards, which took as input the sum of energy from
all nine crystals of each calorimeter. The cards could be vetoed by the
Local maximum veto. Both vetoes were applied to the Coincidence Histogram.}
\end{figure}

A schematic of the SyMB readout is shown in figure \ref{fig:symb_readout}. For each
calorimeter, the signals from each of the nine PMTs were summed in a sum-builder,
and also sent to a logic unit called the local maximum veto. A veto signal was produced
if any of the eight peripheral crystals had a larger signal than the central
crystal. The purpose of this veto was to eliminate background which did not originate
from the target. Particles from the target would pass through the collimator
aperture and deposit the majority of their energy in the central crystal.

The sums and vetos were sent to three fast histogramming cards, called the left
master, right master, and coincidence histogram cards. The left master required that the trigger
conditions be met by the left calorimeter. The right master required that
the trigger conditions be met by the right calorimeter. The coincidence card
required a trigger from both calorimeters. Each card stored a two-dimensional
histogram, in which one dimension represented the energy sum from the left
calorimeter, and the other dimension represented the energy sum from the right
calorimeter.

The calorimeters were subject to high rates, on the order of tens of kilohertz,
but were designed to operate without any dead-time. The amount of time needed
to digitize and record the energy deposited from an electromagnetic shower was
shorter than the DORIS bunch spacing (approximately 100~ns). That meant that 
the SyMBs could easily differentiate events in adjacent bunches. The bunch 
width in DORIS was less than 2~ns, so multiple interactions within the same 
bunch crossing of the target could not be distinguished at all. In such multi-interaction events, 
the calorimeters recorded a sum of the energy deposited from all of the interactions. 
Given these sort of time scales, the SyMB essentially measured the total energy
deposited from each bunch crossing. Rather than measuring the number of events
in a given amount of time, the SyMBs measured the number of bunch crossings
in which an event of interest took place. 

\subsection{SyMB Data}

\begin{figure}[htpb]
\centering
\includegraphics{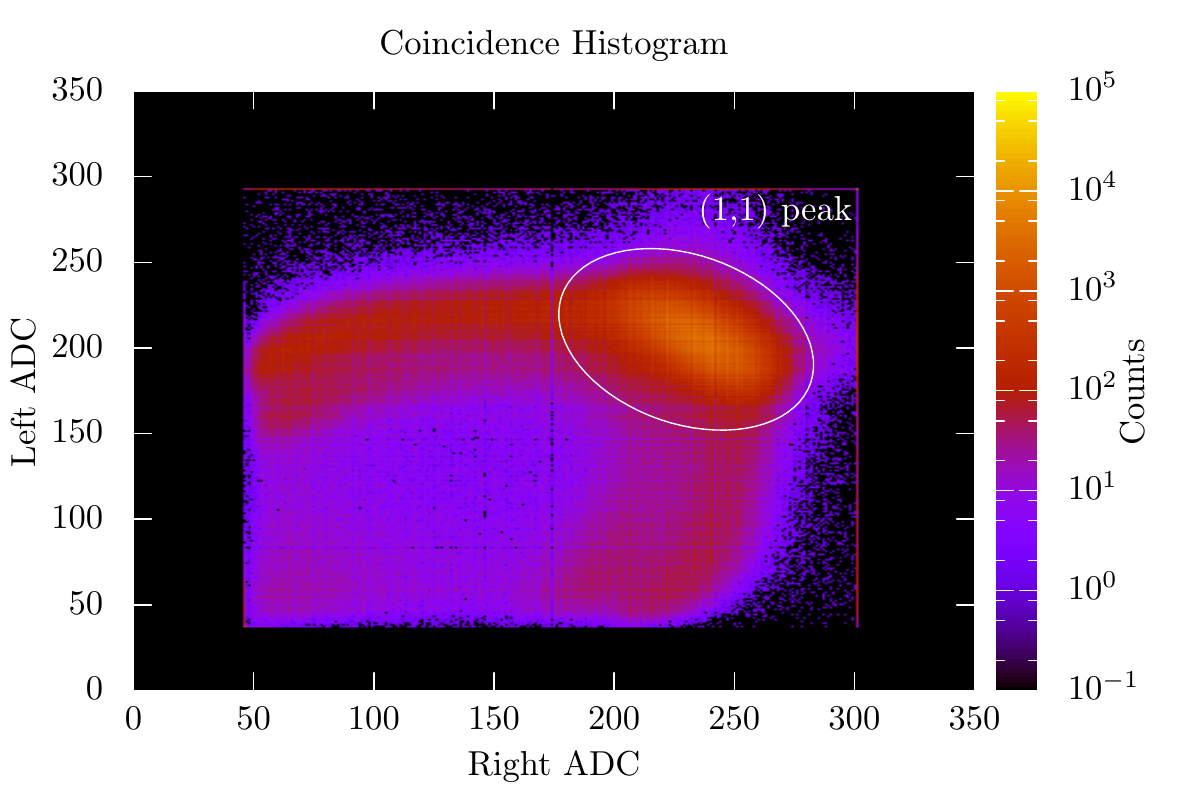}
\caption[SyMB coincidence histogram]{\label{fig:symb_co_hist}
The coincidence histogram was filled when the threshold and local-maximum conditions were satisfied by
both calorimeters.}
\end{figure}

\begin{figure}[htpb]
\centering
\includegraphics{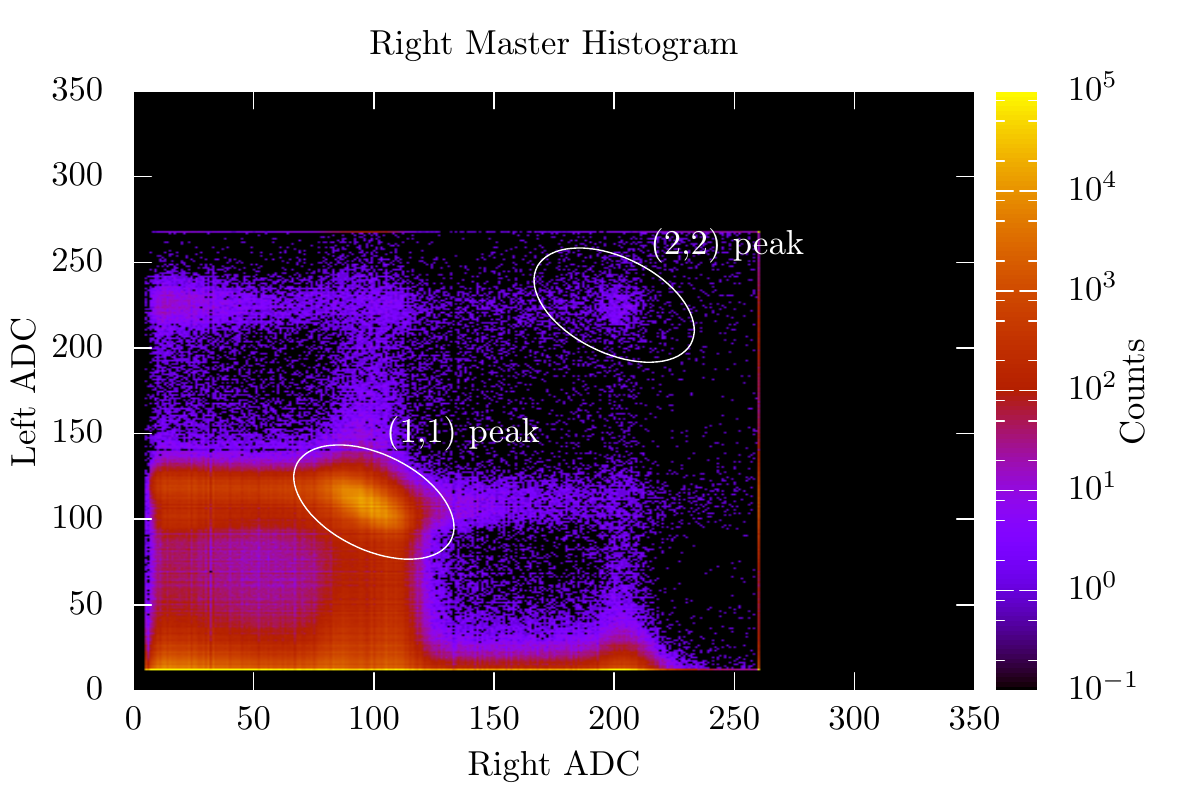}
\caption[SyMB right-master histogram]{\label{fig:symb_rm_hist}
The right-master histogram was filled when the threshold and local-maximum conditions were satisfied by
the right calorimeter.}
\end{figure}

\begin{figure}[htpb]
\centering
\includegraphics{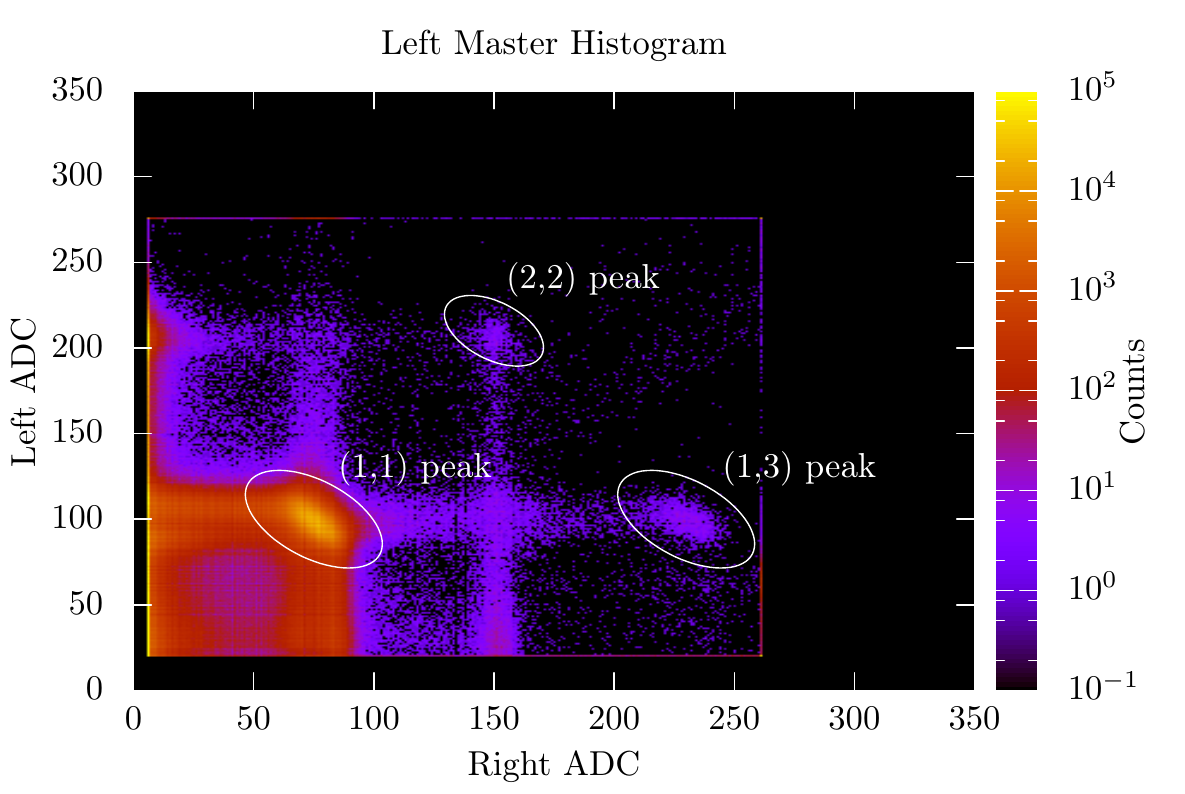}
\caption[SyMB left-master histogram]{\label{fig:symb_lm_hist}
The left-master histogram was filled when the threshold and local-maximum conditions were satisfied by
the left calorimeter.}
\end{figure}

The SyMB data came in the form of three two-dimensional histograms, one from each
histogramming card. The high rates in the SyMBs would have swamped the OLYMPUS data 
acquisition system (DAQ), so rather than reading out every event, the DAQ would  
periodically (every 30,000 main triggers, approximately once per minute of wall-time)
read out and reset the SyMB histograms. This meant that event-by-event information was 
not recorded. Instead, the SyMB histograms show the aggregation of the events over the
period between readouts. The three histograms are shown in figures \ref{fig:symb_co_hist}, 
\ref{fig:symb_rm_hist}, and \ref{fig:symb_lm_hist}, for a typical run. Each count
in the histograms corresponds to one bunch crossing of the target.

To good approximation, there were only two sorts of events which deposited energy in the SyMB
calorimeters. The first was symmetric lepton-lepton scattering---M\o ller scattering ($e^-e^- \rightarrow e^-e^-$)
during electron running and Bhabha scattering ($e^+e^- \rightarrow e^+e^-$) during positron
running. In symmetric M\o ller/Bhabha scattering, approximately 1~GeV of energy was deposited in the left
calorimeter and 1~GeV of energy was deposited in the right calorimeter. The second type of event was elastic lepton-proton
scattering, in which the scattered lepton deposited 2~GeV of energy in one of the calorimeters.
Neglecting cases where leptons lost energy from scattering in the lead collimator before entering the calorimeter, the
calorimeters saw integer numbers of GeV at each bunch crossing. Since multiple interactions could
occur within one bunch crossing, this number could sometimes exceed the beam energy (2~GeV). In this thesis, I will
refer to the various signal peaks using a two-dimensional coordinate system where the first coordinate
refers to the number of GeV deposited in the left calorimeter, and the second coordinate refers to the number of GeV deposited
in the right calorimeter. For example, if 1~GeV were deposited in the left calorimeter, and 3~GeV were deposited
in the right, this event would fall in the (1,3) signal peak. The probability of any events occuring in a
single bunch was low, so the likeliest energy deposition in any given bunch was (0,0).

The three histograms had different dynamics ranges, so not all of the histograms had all of the 
same signal peaks. Furthermore, not all of the peaks had count-rates close to what was expected 
from simulation. The cause of these discrepant count-rates was not conclusively determined, although
we formulated a hypothesis that is consistent with the observed discrepancies \cite{comparator:saturation}. 
I will summarize here the different peaks that are present in the three histograms, the problems
with each, and the explain the specific combination of count-rates which can produce a useful
luminosity extraction.

The histograms did have underflow and overflow bins for events in which the energy was outside of the
histogram's dynamic range. In principle, the information in these bins could have been used to extract
luminosity information. For example, the (2,0) peak in the left master histogram could have been used
to extract the rate of elastic $ep$ scattering. In practice, the rates of the underflow and overflow
bins were anomalous---in some cases too high and in other cases too low compared to simulation---and
we came to conclusion that these bins did not function properly. In the interest of brevity, I will
not cover the underflow and overflow signal peaks in this section.

\subsubsection{Coincidence Histogram}

The coincidence histogram, shown in figure \ref{fig:symb_co_hist}, had the smallest dynamic range of the three
histograms, in both the left and right axes. Only the (1,1) signal peak was visible below the overflow.
In the main SyMB analysis, luminosity was extracted by comparing the number of counts in the (1,1) signal
peak to simulation. In doing so, we found that the ratio $\mathcal{L}_{e^+p}/\mathcal{L}_{e^-p}$ is 5\%
smaller than that of the slow control system and the $12^\circ$ telescopes during the latter part
of the fall run. (The discrepancy varies from 2\% to 5\% depending on which part of the fall run is considered.)
This discrepancy, as  well as the large systematic uncertainties from the method itself, prevent us
from trusting the main SyMB analysis.

\subsubsection{Right Master Histogram}

In the right master histogram, shown in figure \ref{fig:symb_rm_hist}, the dynamic range was slightly wider so
that both the (1,1) and (2,2) signal peaks were visible. The (1,1) peak had a similar rate to the (1,1) peak in 
the coincidence histogram, i.e., it had the same species-dependent discrepancy.

The (2,2) peak was populated by two different classes of multi-interaction events. In the first class,
there are two symmetric M\o ller/Bhabha interactions in the same bunch crossing. Two leptons hit the left calorimeter,
and two hit the right calorimeter. In the second class, there are two $ep$ elastic events in which the
leptons hit opposite calorimeters. By using the simulated rates of $ee$ and $ep$ scattering, we estimated how
how many (2,2) counts we should expect. We found that the right master histogram has only 50--60\% of the
expected rate, and therefore cannot be used for a trustworthy luminosity measurement.

The right master histogram had a lattice-like structure caused by coincidence events in which
one particle lost some energy (say, in the collimator) before entering the calorimeter. There were
cross-like structures at (1,2) and (2,1), where lattice lines intersected. Events falling at
these positions were multi-interaction events in which one event was a M\o ller/Bhabha interaction with only one
lepton entering a calorimeter. The SyMB apertures had a large degree of overlap; that is, if one
lepton passed through the left aperture, the chances were good that the other lepton would make it
through the right aperture. The consequence was that the (1,2) and (2,1) peaks were not very clearly
defined and were not readily useable for analysis.

\subsubsection{Left Master Histogram}

The left master histogram, shown in figure \ref{fig:symb_lm_hist}, included, in addition to the (1,1) and (2,2) peaks,
a third signal peak at (1,3). Events in this peak were produced by a symmetric M\o ller/Bhabha interaction
occuring in the same bunch as an $ep$ scattering in which the lepton entered the right calorimeter. The
M\o ller/Bhabha interaction put 1~GeV in both calorimeters, while the $ep$ lepton deposited an
additional 2~GeV in the right calorimeter.

The (1,1) rate matched what is observed in both the right master and coincidence histograms; it had the same
species-dependent discrepancy when compared with slow control and the $12^\circ$-system. The (2,2) rate was
slightly lower than that of the right master histogram. The peak had about 40\% of the counts one would expect
given the simulated $ep$ and $ee$ rates.

The rate in the (1,3) peak was very close to what would be predicted from the M\o ller/Bhabha and $ep$ cross
sections. In a given run period, the number of (1,3) events should, to lowest order, follow the relation:
\begin{equation}
N_{(1,3)} = \frac{ \mathcal{L}^2 \sigma_{(1,1)} \sigma_{ep\rightarrow R}}{N_b} = \frac{\mathcal{L} N_{(1,1)} \sigma_{ep\rightarrow R} }{N_b}
\end{equation}
where $N_b$ is the number of bunch crossings and $\mathcal{L}$ is the integrated luminosity during the run
period. For the sake of simplicity, we can estimate $\mathcal{L}$ using the slow control system. If one uses
simulation to estimate $\sigma_{(1,1)}$ and $\sigma_{ep\rightarrow R}$, then the ratio of (1,3) data rate to
prediction shows a slight beam-species dependence, which is strikingly similar to the beam-species
dependence of the M\o ller/Bhabha luminosity extraction. If, instead, one estimates $\sigma_{(1,1)}$
directly from the (1,1) peak in data, then the ratio of data to prediction in the (1,3) peak has
practically no apparent beam-species dependence.

It seemed that whatever was causing the beam-species dependence in the main M\o ller/Bhabha luminosity
extraction, caused the same beam-species dependence in the (1,3) rate. That meant that the
(1,3) and (1,1) peaks together could be used to extract a reasonable estimate of luminosity.

\section{Multi-Interaction Event Analysis}

To extract a luminosity from the Left Master (1,1) and (1,3) peaks, we need to make a few assumptions
about the cause of the problems in the main SyMB analysis. To summarize the main problem, the rate of M\o ller
events in data is about 5\% too low (relative to expectation from simulation and slow control) whereas
the rate of Bhabha events in data is not. That is:
\begin{equation}
\frac{ N^\text{data}_{e^-e^-}}{\sigma^\text{sim.}_{e^-e^-} \times \mathcal{L}^\text{s.c.}_{e^-e^-}}
\times
\frac{\sigma^\text{sim.}_{e^+e^-} \times \mathcal{L}^\text{s.c.}_{e^+e^-}}{ N^\text{data}_{e^+e^-}}
\approx 0.95.
\end{equation}

We assume that there are two possible ways for this discrepancy to occur. One possibility is that
the M\o ller/Bhabha simulation may be inaccurate in a way that over-predicts the
M\o ller cross section, i.e., $\sigma^\text{sim.}_{e^-e^-}$ is too large. The other possibility
is that the calorimeters or their DAQ system may have some sort of inefficiency during electron-beam
running, i.e., $N^\text{data}_{e^-e^-}$ is too low. We would like that our luminosity
extraction be robust to both of these causes. We should avoid using $\sigma^\text{sim.}_{e^\pm e^-}$, and
any use of $N^\text{data}_{e^\pm e^-}$ should be made relative to $N^\text{data}_{(1,3)}$ so that
any effective inefficiency will cancel.

\subsection{Derivation}

In a bunch crossing, there is a large number of beam particles, but the probability of any
one beam particle having an $ee$ or $ep$ interaction is small. Therefore, we can use the
Poisson distribution to describe the number of interactions in a bunch crossing. Specifically,
the probability of having $N$ interactions in a bunch with integrated luminosity $\mathcal{L}_j$ is
given by:
\begin{equation}
P(N) = \frac{e^{-\sigma \mathcal{L}_j} (\sigma \mathcal{L}_j)^N}{N!}
\end{equation}
where $\sigma$ is the cross section for an interaction.

Let us classify three types of interactions which can put energy into the SyMBs. M\o ller/Bhabha interactions,
denoted by $ee$, put 1~GeV in each calorimeter. An $ep$ interaction where the lepton deposits 2~GeV of energy
in the left calorimeter will be denoted by $ep\rightarrow L$. And $ep$ interactions where the lepton deposits
2~GeV of energy in the right calorimeter will be denoted by $ep\rightarrow R$.

The probability for a bunch crossing to result in energy deposition of the form (1,1) is given by:
\begin{align}
P(1,1) =& P_{ee}(1) \times P_{ep\rightarrow L}(0) \times P_{ep\rightarrow R}(0) \\
= & e^{-\mathcal{L}_j (\sigma_{ee} + \sigma_{ep\rightarrow L} + \sigma_{ep\rightarrow R})} \times \sigma_{ee} \mathcal{L}_j\\
= & e^{-\mathcal{L}_j \sigma_\text{tot.} } \times \sigma_{ee} \mathcal{L}_j.
\end{align}
In the same way, we can define the probability for energy deposition of the form (1,3):
\begin{align}
P(1,3) =& P_{ee}(1) \times P_{ep\rightarrow L}(0) \times P_{ep\rightarrow R}(1) \\
= & e^{-\mathcal{L}_j (\sigma_{ee} + \sigma_{ep\rightarrow L} + \sigma_{ep\rightarrow R})} \times \sigma_{ee} \sigma_{ep\rightarrow R} \mathcal{L}_j^2 \\
= & e^{-\mathcal{L}_j \sigma_\text{tot.} } \times \sigma_{ee} \sigma_{ep\rightarrow R} \mathcal{L}_j^2.
\end{align}

To find the number of counts in a given peak from a given run period, we must sum these probabilities
over the number of bunches in the run. If the total number of bunches in a run period is $N_b$, and the
total integrated luminosity for the run period $\mathcal{L}$ is given by $\mathcal{L} = \sum_j^{N_b} \mathcal{L}_j$,
then we find, in the large $N_b$ limit, that:
\begin{align}
N_{(1,1)} =& \sum\limits_{j=0}^{N_b} P(1,1) =  \sum\limits_{j=0}^{N_b} \left[
  e^{-\mathcal{L}_j\sigma_\text{tot.} } \times \sigma_{ee} \mathcal{L}_j \right] \\
N_{(1,3)} =& \sum\limits_{j=0}^{N_b} P(1,3) =  \sum\limits_{j=0}^{N_b} \left[
  e^{-\mathcal{L}_j \sigma_\text{tot.}} \times
  \sigma_{ee} \sigma_{ep\rightarrow R} \mathcal{L}_j^2 \right].
\end{align}

For the cross sections and bunch luminosities we are dealing with, $\mathcal{L}_j \sigma_\text{tot.} \ll 1$.
It makes sense to recast the exponentials as series in powers of $\mathcal{L}_j \sigma_\text{tot.}$, which
can be truncated. Now, dividing the two rates, and dropping any terms beyond first order in
$\mathcal{L}_j \sigma_\text{tot.}$, we get:
\begin{align}
\frac{N_{(1,3)}}{N_{(1,1)}} &= \frac{ \sigma_{ep\rightarrow R} \sum\limits_{j=0}^{N_b} \mathcal{L}_j^2
\left[ 1 - \mathcal{L}_j\sigma_\text{tot.} \right]}
{\sum\limits_{j=0}^{N_b} \mathcal{L}_j \left[ 1 - \mathcal{L}_j\sigma_\text{tot.} \right]} \\
&= \frac{ \sigma_{ep\rightarrow R} N_b \left[ \langle \mathcal{L}_j^2 \rangle - \sigma_\text{tot.} \langle \mathcal{L}_j^3 \rangle \right] }
{\left[ \mathcal{L} - N_b \sigma_\text{tot.} \langle \mathcal{L}_j^2 \rangle \right]}\\
&= \frac{ \sigma_{ep\rightarrow R} N_b }{\mathcal{L}}
\left[ \langle \mathcal{L}_j^2 \rangle - \sigma_\text{tot.} \langle \mathcal{L}_j^3 \rangle \middle]
\middle[ 1 + \frac{N_b}{\mathcal{L}} \sigma_\text{tot.} \langle \mathcal{L}_j^2 \rangle \right].
\end{align}

From here, we can use the relation that the variance in luminosity per bunch, $v_b$, is equal to
$\langle \mathcal{L}_j^2 \rangle - \langle \mathcal{L}_j \rangle ^2$, or equivalently
$\langle \mathcal{L}_j^2 \rangle - \mathcal{L}^2/N_b^2$. We'll proceed with some rearranging, and
drop the $\sigma_\text{tot.}^2$ term:
\begin{align}
\frac{N_{(1,3)} }{N_b N_{(1,1)} \sigma_{ep\rightarrow R}} &=
\frac{1}{\mathcal{L}} \left[ \left(v_b + \frac{\mathcal{L}^2}{N_b^2}\right) - \sigma_\text{tot.} \langle \mathcal{L}_j^3 \rangle \middle]
\middle[ 1 + \frac{N_b}{\mathcal{L}} \sigma_\text{tot.} \left(v_b + \frac{\mathcal{L}^2}{N_b^2}\right) \right] \\
&= \frac{\mathcal{L}}{N_b^2} + \frac{v_b}{\mathcal{L}} + \frac{N_b \sigma_\text{tot.}}{\mathcal{L}^2}
  \left\{ \left(v_b + \frac{\mathcal{L}^2}{N_b^2}\right)^2 - \frac{\langle \mathcal{L}_j^3 \rangle \mathcal{L}}{N_b } \right\}.
\end{align}

We can rearrange this to get an expression for luminosity:
\begin{equation}
\mathcal{L} = \frac{N_{(1,3)}N_b }{N_{(1,1)} \sigma_{ep\rightarrow R} }
- \frac{v_b N_b^2}{\mathcal{L}} - N_b \sigma_\text{tot.}
  \left\{ \left(\frac{v_b N_b}{\mathcal{L}} + \frac{\mathcal{L}}{N_b}\right)^2 - \frac{\langle \mathcal{L}_j^3 \rangle N_b}{\mathcal{L}}\right\}.
\label{eq:mie_result}
\end{equation}
\sloppy
Equation \ref{eq:mie_result} shows that luminosity can be estimated from a main
term, $N_{(1,3)} N_b / N_{(1,1)} \sigma_{ep\rightarrow R}$, with some corrections.
In this derivation, I only kept terms to first order in $\mathcal{L} \sigma_\text{tot.}$,
but, if necessary, corrections can be calculated to arbitrary order.

Let us consider the meaning of these corrections. The first correction term, $v_b N_b^2 / \mathcal{L}^2$,
describes the effect of luminosity variance between bunches. High-luminosity bunches are
much more likely to have multi-interaction events than low-luminosity bunches. If there is variance
between the bunches, this will have a small effect on the multi-interaction event rate. The second correction
term, which has a leading factor of $\sigma_\text{tot.}$, accounts for the fact that there
may be three interactions in a bunch crossing. Some small fraction of would-be (1,3) events
will fall outside of the (1,3) peak because of additional energy deposited by a third
interaction in the same bunch.

\begin{figure}[htpb]
\centering
\includegraphics{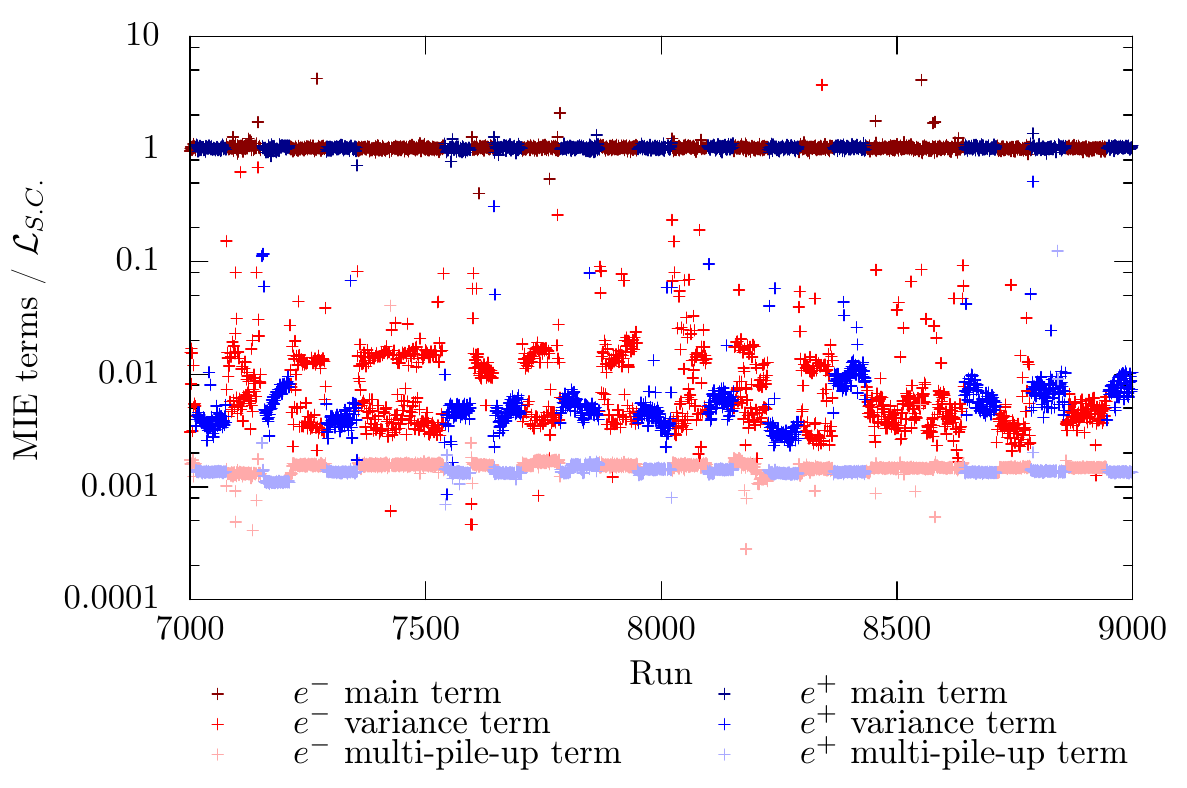}
\caption[Relative sizes of the MIE correction terms]{\label{fig:mie_terms}
Bunch luminosity variance is a percent-level correction. Higher order multi-interactions matter at the
per-mil level. This justifies neglecting subsequent terms in the $\mathcal{L} \sigma_\text{tot.}$
expansion.}
\end{figure}

Figure \ref{fig:mie_terms} shows the scale for these correction terms relative to the luminosity estimate
from slow control $\mathcal{L}_{S.C.}$. The variance term is highly species-dependent, and corrects
the luminosity extraction by a few percent. This term is absolutely necessary to achieve
percent-level accuracy desired for OLYMPUS. The correction for higher order multi-interaction events is smaller,
on the order of one to two tenths of a percent. It stands to reason that subsequent higher-order terms will
be even smaller and will not be significant for a percent-level luminosity estimate. This justifies
the decision to truncate our power expansion at first order.

The result in equation \ref{eq:mie_result} shows the relationship between fundamental quantities like
cross sections, count rates, and luminosities. To apply this relation to a useable luminosity
extraction, we must decide how to estimate these fundamental quantities, consistent with the
assumptions laid out at the beginning of this section. Let us first look at the quantites 
in the main term. $N_{(1,3)}$ and $N_{(1,1)}$ naturally come
from data; we must integrate the left master histogram around the signal peaks. $N_b$ is known
to good accuracy from the slow control live-time and the DORIS average bunch spacing (96.8~ns).
In principle, $\sigma_{ep \rightarrow R}$ could be estimated from data, but since this signal shows
up in the unreliable histogram underflow, we cannot trust it. Instead, we can estimate this using
simulation.

I want to point out that equation \ref{eq:mie_result} isn't the result of solving for $\mathcal{L}$.
In fact, $\mathcal{L}$ appears on both sides of the equation. However, $\mathcal{L}$ has been
eliminated from the main term. In the correction terms, we can estimate $\mathcal{L}$ from the
slow control system. Any slight deviations from the real luminosity will have a negligible effect
on already small corrections.

The bunch luminosity variance, $v_b$, as well as $\langle \mathcal{L}_j^3 \rangle$ require some
work to estimate. Both depend on the variation of current within the 10 bunches of DORIS, as well
as the variation in time of those currents. To estimate these, I use a readout from the DORIS
computer which shows the individual bunch currents as a function of time. I then use the OLYMPUS
slow control system to estimate the target density so that I can reconstruct the individual
bunch luminosities.

The last remaining quantity is $\sigma_\text{tot.}$, which has the components $\sigma_{ep \rightarrow L}$,
$\sigma_{ep \rightarrow R}$, and $\sigma_{ee}$. The $ep$ cross sections must come from
simulation as discussed earlier. $\sigma_{ee}$ can be estimated using either simulation
or from the (1,1) peak in data. Because the higher-order multi-interaction term is so small, the difference
between using simulation or data to estimate $\sigma_{ee}$ is negligible. Putting this all
together, our estimate of the luminosity becomes:

\begin{equation}
  \mathcal{L} = \frac{N_{(1,3)}^\text{data}N_b }{N_{(1,1)}^\text{data} \sigma_{ep\rightarrow R}^\text{sim.} }
  - \frac{v_b N_b^2}{\mathcal{L}_{S.C.}} - N_b \sigma_\text{tot.}^\text{sim.}
  \left\{ \left(\frac{v_b N_b}{\mathcal{L}_{S.C.}} + \frac{\mathcal{L}_{S.C.}}{N_b}\right)^2 -
  \frac{\langle \mathcal{L}_j^3 \rangle N_b}{\mathcal{L}_{S.C.}}\right\}.
\end{equation}

\subsection{Analysis Procedure}

\label{ssec:symb_procedure}

Extracting a count rate from a signal peak in a histogram requires integrating over the peak between
some bounds. The placement of the bounds obviously affects the result of the integral, and this is
problematic. The best solution to this problem is to simulate every histogram as a reference, and to
integrate the data histogram and simulation histogram over the same bounds. That avenue isn't available
to us because we want to avoid using the M\o ller/Bhabha simulation.

At the same time, the lack of a reference simulation is freeing. We will have to place arbitrary bounds,
so we can do so without too much angst. At the end, we must test that our choice of bounds has a minimal
effect on our extraction of the species-relative luminosity, even if that choice has a large effect on the
various integrals. For this reason, in my analysis I have chosen to make simple box cuts, centered around
the various signal peaks.

\begin{figure}[htpb]
\centering
\includegraphics{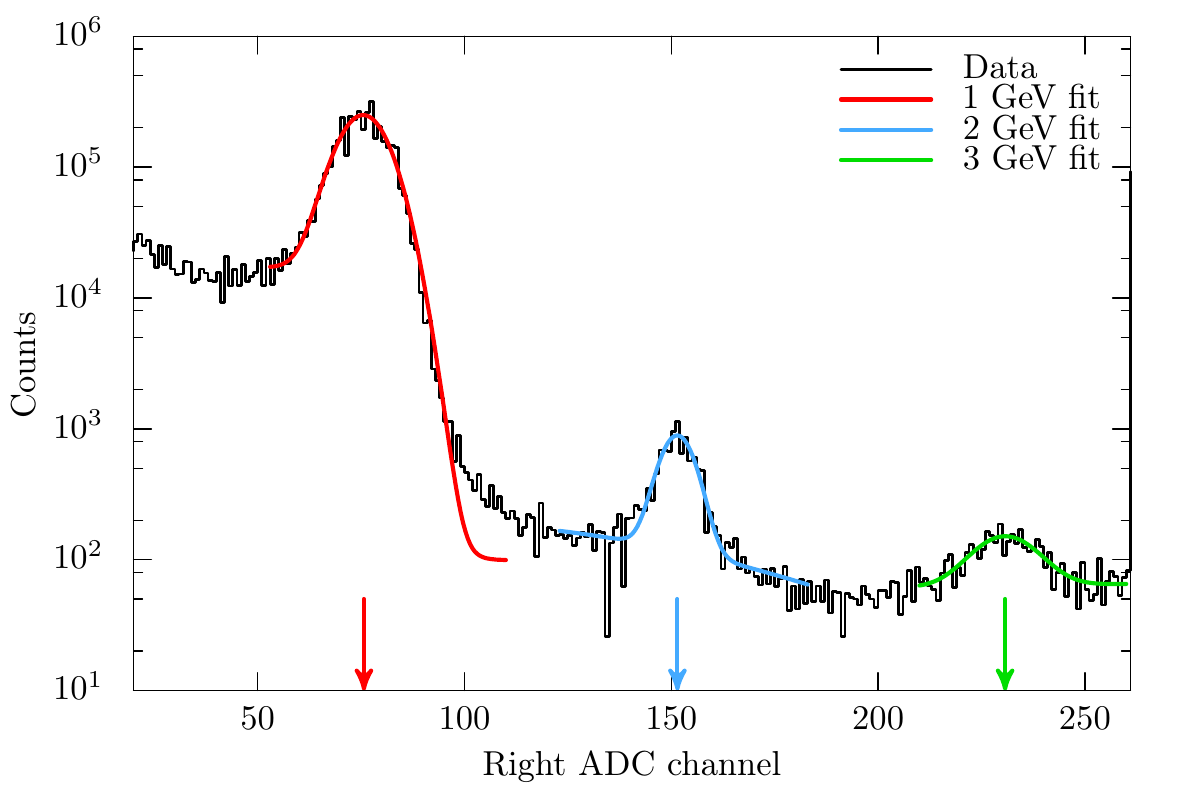}
\caption[Fits establishing the SyMB signal peak positions]
{ \label{fig:symb_centroid_fits} I fit the one-dimensional ADC spectrum to find the peak positions.}
\end{figure}

To ascertain the centroids of the signal peaks, I fit the 1-dimensional ADC spectra with a
signal-plus-background model:
\begin{equation}
  f(x|A,\mu,\sigma,B,\tau,C) = A e^{ -\frac{(x-\mu)^2}{2\sigma^2}} + \frac{B}{ 1 + e^{\frac{x-\mu}{\tau}}} + C.
\end{equation}
I take the parameter $\mu$ to be the peak position. I do this for the 1 and 2 GeV peaks in the left ADC
spectrum and the 1, 2, and 3 GeV signal peaks in the right ADC spectrum, as shown in figure \ref{fig:symb_centroid_fits}.

\begin{figure}[htpb]
\centering
\includegraphics{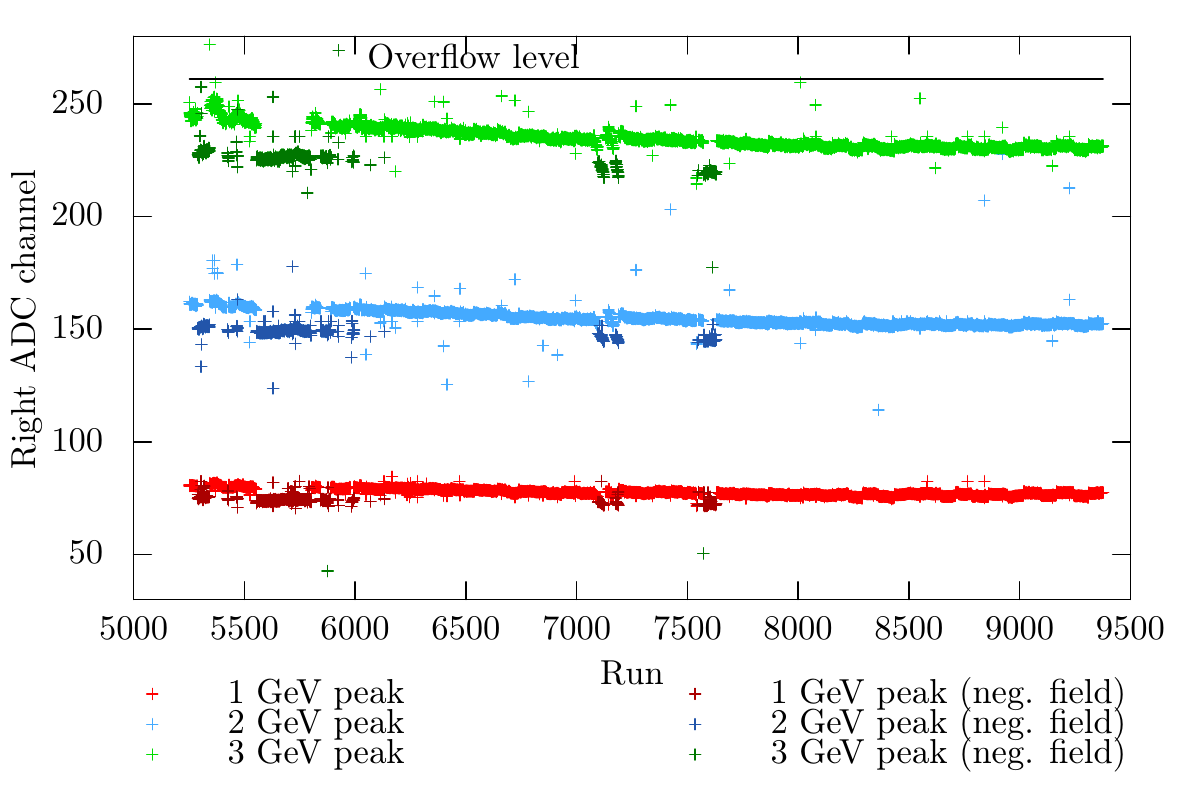}
\caption[Peak position variation during data taking]
{ \label{fig:symb_peak_pos} The signal peak positions changed slightly over the fall run.
The 3 GeV peak in the right ADC is very close to the over-flow level, especially early on. If
the box cut size is too large, the box will run off the end of the histogram, throwing off the 
luminosity determination. The solution is to use tighter boxes for the earlier runs. 
Additionally, one can see that when using the negative toroid polarity, the performance of the 
SyMB's PMTs was altered enough to noticeably shift the peak positions downward.}
\end{figure}

Figure \ref{fig:symb_peak_pos} shows how the peak positions for the right ADC (in the left master
histogram) evolve over the course of the fall run. The peak positions shift over time to
lower channel numbers. In the early part of the run, the 3~GeV peak is quite close to the
overflow boundary. To avoid clipping, any studies involving these runs need quite tight
boxes.

Since all three peak positions change, it makes sense to set the box width based on a fixed
energy scale, rather than a fixed number of ADC channels. The ADC spectra must then be calibrated
to develop a conversion between units of energy and ADC channels. For the results I show in this note,
I used a linear calibration for the left master's left ADC (using the 1 and 2~GeV peak positions),
and a quadratic calibration for the right ADC (using the 1, 2, and 3~GeV peak positions). I integrated
over a fixed window with a half-width of 286~MeV for both the left and right ADCs. The cross sections
$\sigma_{ep\rightarrow L}$ and $\sigma_{ep\rightarrow R}$ were found by integrating the simulated
histograms over bins [200,250] and [185,230] respectively.

\section{Systematic Uncertainties}

The accuracy of the MIE luminosity determination is limited by a handful of systematic effects.
Slight inaccuracy in the beam position, the detector geometry, the magnetic field, or of the 
energy of the beam in simulation can produce slight inaccuracy in the calculation of $\sigma_{ep}$. 
The box cut procedure can affect the results of $N_{(1,3)}$ and $N_{(1,1)}$. The specific radiative 
corrections procedure used in simulation can change the definition of $\sigma_{ep}$. In this section, 
I will go through each of these and estimate their effect on the MIE determination of the species 
relative luminosity.

The numbers I will present in this section are the percent uncertainties in the determination of
the species relative luminosity, which I define as $\mathcal{L}_{e^+p} / \mathcal{L}_{e^-p}$. The
uncertainty in this quantity, which I'll call $\delta_\mathcal{R}$, produces a uniform baseline
uncertainty for all of the asymmetry measurement points, in the form: $\delta_A \approx \delta_R / 2$.
That is, a two percent uncertainty in the relative luminosity produces a one percent uncertainty
in the asymmetry in every bin. 

Whereas, in the main SyMB analysis the systematic uncertainties proved to be so large to make the
analysis unusable, the MIE result is much more robust with respect to systematic uncertainties.
I estimate the systematic uncertainty
on the species-relative luminosity to be 0.27\%. A breakdown of the various systematics is shown in
table \ref{table:mie_sys}.

\begin{table}
\centering
\begin{tabular}{ | l | l | }
\hline
Systematic & Value (\%) \\
\hline
\hline
Beam Position & 0.21 \\
Geometry & 0.13 \\
Box Sizes & 0.10 \\
Magnetic Field & 0.05 \\
Radiative Corrections & 0.03 \\
Beam Energy & 0.01 \\
\hline
\hline
Total & 0.27 \\
\hline
\end{tabular}
\caption[MIE systematic uncertainties]{\label{table:mie_sys} I estimate the systematic uncertainy of the multi-interaction event species-dependence to be
0.27\%, making this system viable for the OLYMPUS analysis.}
\end{table}

\subsection{Beam Position}

\begin{figure}[htpb]
\centering
\includegraphics{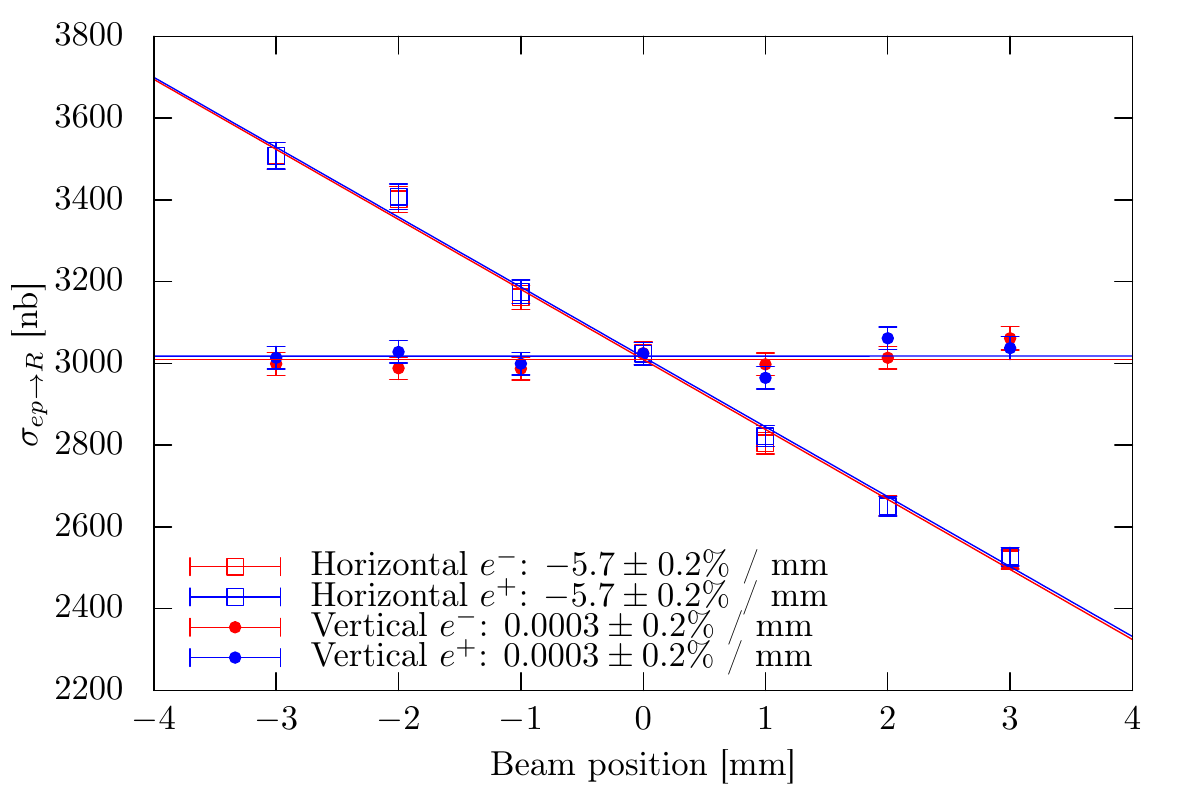}
\caption[The effect of beam position on $\sigma_{ep\rightarrow R}$]
{\label{fig:symb_bpm_pos} The cross section $\sigma_{ep\rightarrow R}$ is unaffected by changes in the beam's
vertical position, but changes in the horizontal position have a 5.7\%/mm effect, which is independent of beam species.}
\end{figure}

\begin{figure}[htpb]
\centering
\includegraphics{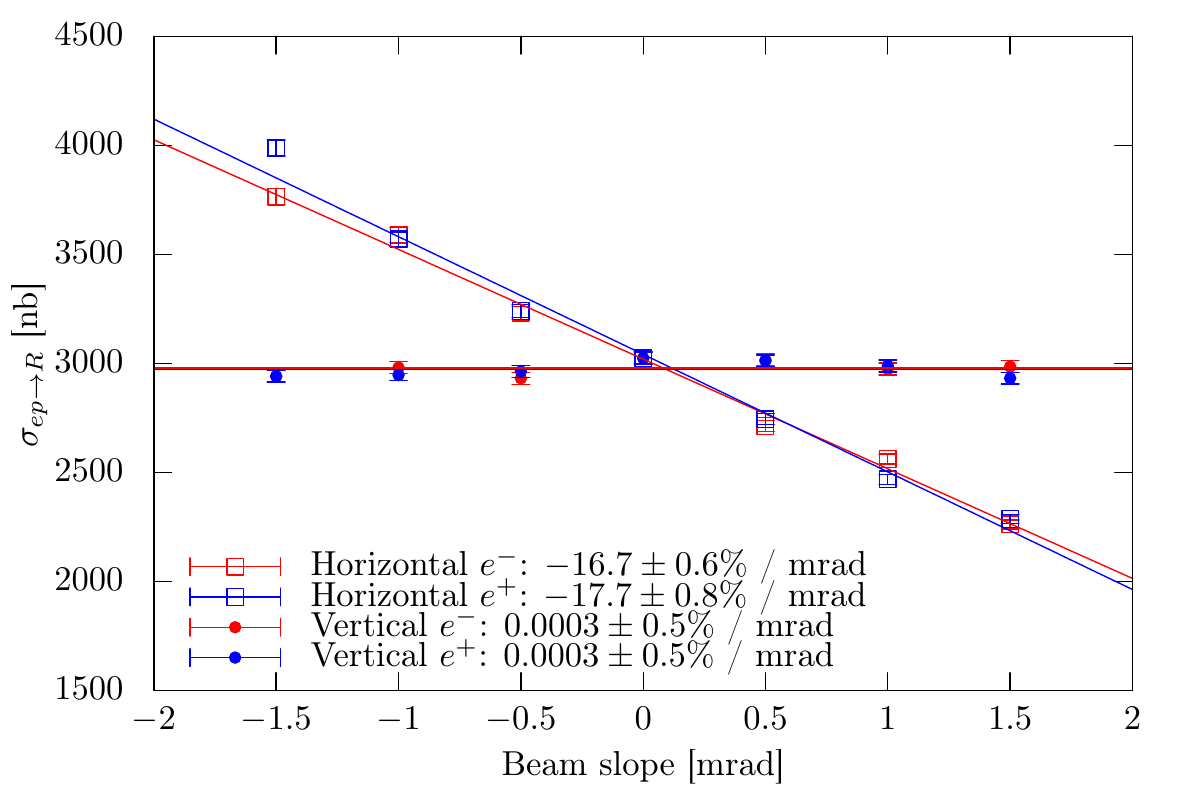}
\caption[The effect of beam slope on $\sigma_{ep\rightarrow R}$]
{\label{fig:symb_bpm_slo} The cross section $\sigma_{ep\rightarrow R}$ is unaffected by changes in the beam's
vertical slope, but changes in the horizontal slope have a 17\%/mrad effect, which is nearly independent of beam species.}
\end{figure}

If there were an inaccuracy in our knowledge of the beam position, this would lead us to calculate an erroneous value
of $\sigma_{ep\rightarrow R}^\text{sim.}$. We would be simulating the beam in one position, when in reality it occupied
another. We can estimate the effect by altering the beam position in simulation and looking at the change $\sigma_{ep\rightarrow R}^\text{sim.}$.
Figures \ref{fig:symb_bpm_pos} and \ref{fig:symb_bpm_slo} show the results of simulations with varying beam position offsets and slopes.
Generally, the cross section $\sigma_{ep\rightarrow R}^\text{sim.}$ is unaffected by changes in the beam's vertical parameters,
but has a large dependence on the beam's horizontal parameters. This dependence essentially doesn't change with beam species. However
there are ways that a species-dependent systematic might manifest itself.

\begin{figure}[htpb]
\centering
\includegraphics{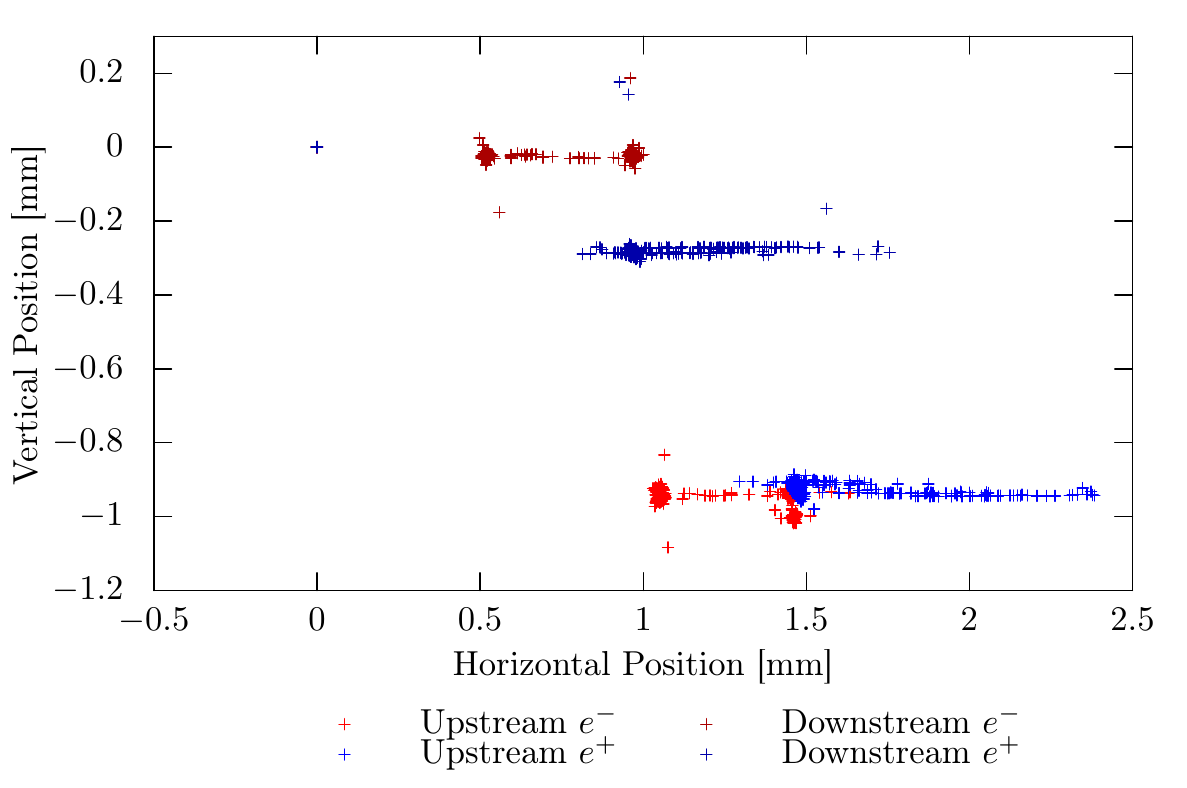}
\caption[Beam position by species]
{\label{fig:symb_beam_by_spec} The electron beam tended to sit about 400~$\mu$m to the left of the positron beam at both BPMs.
  While upstream the two beams had roughly the same vertical position, this grew to about 300~$\mu$m by the downstream BPM.}
\end{figure}

Let us consider two ways in which a species-dependent systematic might occur:
\begin{enumerate}
\item The survey of the BPM positions inaccurately captures the true BPM axes, such that the true axes are rotated
from the supposed axes. Since the two species have slightly different vertical beam positions (shown in figure \ref{fig:symb_beam_by_spec}),
this error would introduce an additional species-dependent horizontal offset for which we do not account.
\item The BPMs themselves have some species-dependent inaccuracy.
\end{enumerate}

In the following section, I will present our estimates for the accuracy of the BPMs. Following that I will calculate
the magnitude of the two sources of species-dependent systematic effects.

\subsubsection{Accuracy of the BPMs}

\label{sssec:bpm}

The accuracy of the beam position in simulation is determined by two factors:
\begin{itemize}
\item The accuracy of the survey of the beam position monitors (Were the BPMs physically located where we believed them to be?)
\item The accuracy of the BPM readout calibration (Are we interpreting the BPM data correctly?)
\end{itemize}
Let’s address these in turn.

The accuracy of the BPM survey has a species-independent effect on the simulation. We can estimate the
survey accuracy by looking at the residuals of the fits to survey data. The residuals indicate that the
BPMs are constrained in position to 200~$\mu$m. However, the orientation of the BPM axes are not as well
constrained. During the survey, a target ball was moved along the four cap faces of each monitor to
establish the planes of the caps in order to fix the orientation of the monitor axes. For one of the BPMs,
these survey points are only coplanar on the level of 500~$\mu$m accuracy. Therefore, we estimate the relative
orientation of the axes of the two monitors is good to within a rotation of 500~$\mu$m over a 55~mm radius,
or approximately $0.52^\circ$.

The accuracy of the BPM readout is much better than 200~$\mu$m, so the species-independent part of this effect
can be safely neglected in comparison to the BPM survey. The crucial number is thus the species-dependent
readout accuracy. For the majority of later runs, the Libera readout was available. The Libera readout produces
a position measurement that is independent of the current direction, and therefore species independent to a high
degree. Looking at the non-linearity of the fits to the calibration data, and by attributing all of that to a
possible species-dependent effect, we estimate that the BPMs have a species-dependent accuracy of 20~$\mu$m.
We believe that the majority of any species-dependent inaccuracy (caused by the effect of the beam profile,
or caused by the choice of functional form used to fit the calibration data) will be correlated between the
BPM modules. Uncorrelated species-dependent inaccuracy could also be problematic, as it can introduce changes
in measured beam slope bewteen the two species. Therefore, we conservatively estimate a limit of 10~$\mu$m for
any uncorrelated, species-dependent inaccuracies of the BPM calibration.

\subsubsection{BPM Axis Rotation}

In subsubsection \ref{sssec:bpm}, I estimated the accuracy of the relative orientation of the BPM axes to be 0.52$^\circ$.
As seen in figure \ref{fig:symb_beam_by_spec}, the two species have a slightly different vertical position and slope. At the
center of the target, the two beams are vertically offset by about 140~$\mu$m. The systematic uncertainty from
an introduced horizontal offset is:
\[
\delta_\text{rot. pos.} = 140 \mu\text{m} \times \sin\left( 0.52^\circ \right) \times \frac{ 5.7\% }{\text{mm}} = 0.007\%.
\]

A rotation of the BPM axis could also introduce an artificial horizontal slope. The measured difference in vertical slopes
between the two species is on the order of 300~$\mu$m over the 1.464~m between the two BPMs. The systematic uncertainty is:
\[
\delta_\text{rot. slo.} = \frac{300 \mu\text{m}}{1.464~\text{m}} \times \sin\left( 0.52^\circ \right) \times \frac{ 17.7\% }{\text{mrad}} = 0.03\%.
\]

\subsubsection{Species-Dependent BPM Inaccuracy}

In subsubsection \ref{sssec:bpm}, I estimated that the BPMs have a 20~$\mu$m species-dependent uncertainty
that is fully correlated between the two monitors, and a 10~$\mu$m species-dependent uncertainty that is uncorrelated
between the two monitors. The uncertainty from an introduced position offset would be:
\[
\delta_\text{corr. pos.} = 20 \mu\text{m} \times 5.7\%/\text{mm} = 0.11\%
\]
from the correlated uncertainty, and:
\[
\delta_\text{pos.} = 10 \mu\text{m} \times \frac{\sqrt{z_1^2 + z_2^2}}{(z_1+z_2)^2} \times 5.7\%/\text{mm} = 0.04\%
\]
from the uncorrelated uncertainty, where $z_1=-0.801$~m and $z_2=0.663$~m are the positions of the two beam position monitors.
The uncertainty from an introduced slope (produced only by the BPM's uncorrelated uncertainty) would be:
\[
\delta_\text{slo.} = \frac{10 \mu\text{m}\sqrt{2}}{1.464\text{m}} \times 17.7\%/\text{mm} = 0.17\%.
\]

\subsubsection{Totals}

We can combine these uncertainties to get a total uncertainty from the BPMs.
\[
\delta_\text{BPM} = \sqrt{ \delta_\text{rot. pos.}^2 + \delta_\text{rot. slo.}^2 + \delta_\text{corr. pos.}^2 + \delta_\text{pos.}^2 +
\delta_\text{slo.}^2 } = 0.21\%
\]

\subsubsection{Remark on the Neumann Readout}

Between runs 7357 and 7358, the Libera BPM readout was connected. Prior to run 7358, only the Neumann readout was available.
However, only the Libera readout was used for the BPM calibration. Generally, we trust the Libera readout more than the Neumann
since its calibration is beam species independent. Up until now, I have limited my discussion to analysis of runs in which the
Libera readout was available. At some point, it may become valuable to extract luminosities from the earlier Neumann runs. I have
studied the differences between the two systems over the runs that they overlap and I believe the change in readout contributes negligibly
to the accuracy of the analysis.

To simulate runs prior to 7358, a few intermediate steps are needed to estimate the beam position. By looking at the data from later runs
when both readouts are present, we fit a map between the Libera and Neumann data. This map is species dependent. When simulating
and early run, the map is first applied to the Neumann data to produce fake Libera data. These data are then analyzed using the
BPM survey fits.

I tried to study the effect of not having Libera readout by looking at the later runs. I looked at the difference in the beam position
reconstructed from Libera data to the beam position reconstructed from the Neumann data (through the intermediate steps).
The differences in position are on the order 1~$\mu$m, and the differences in slope are on the order of 1~$\mu$rad. Even if these
differences were maximally species dependent, they would be insignificant compared to the those from $\delta_\text{BPM}$. Therefore
I think there is no need to worry about using the early data.

\subsection{Geometry}

\begin{figure}[htpb]
\centering
\includegraphics{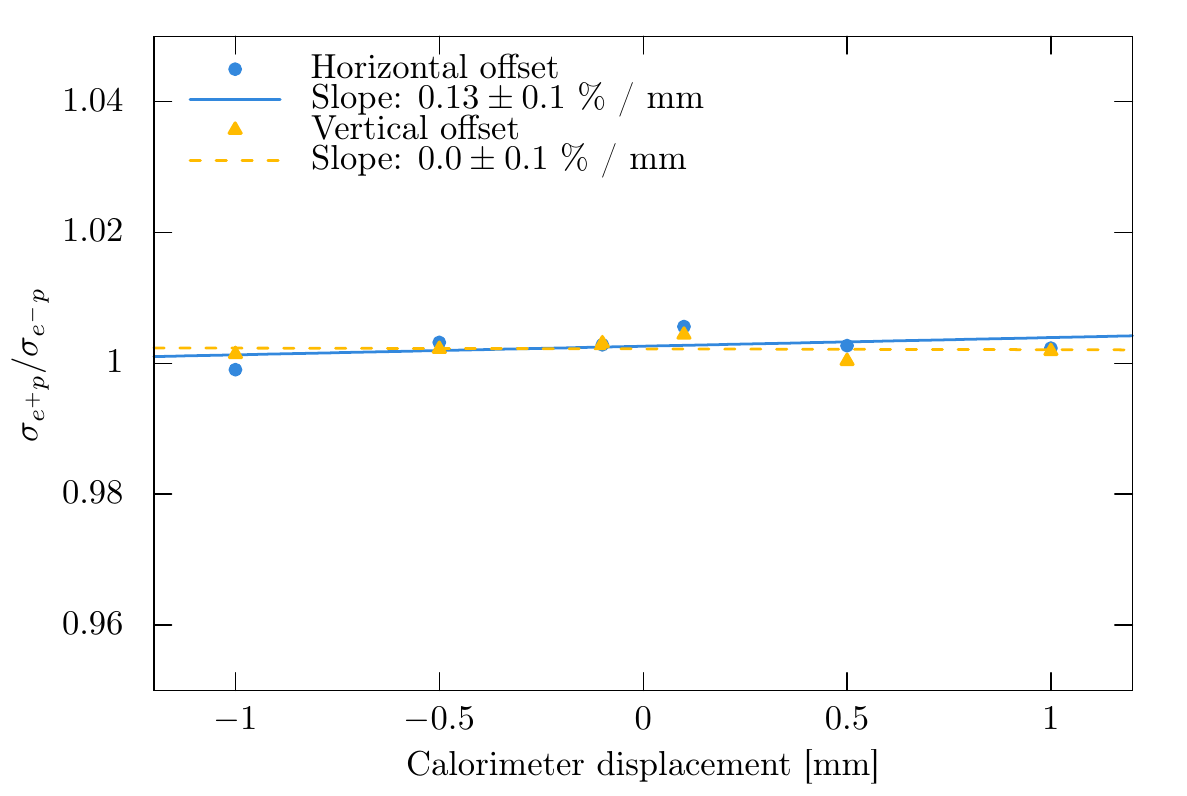}
\caption[The effect of displacing the SyMBs]
{\label{fig:symb_disp} Moving the right SyMB 1~mm in $x$ produces a 0.13\% change in the species ratio. Moving the
right SyMB in $y$ produces no significant effect.}
\end{figure}

\begin{figure}[htpb]
\centering
\includegraphics{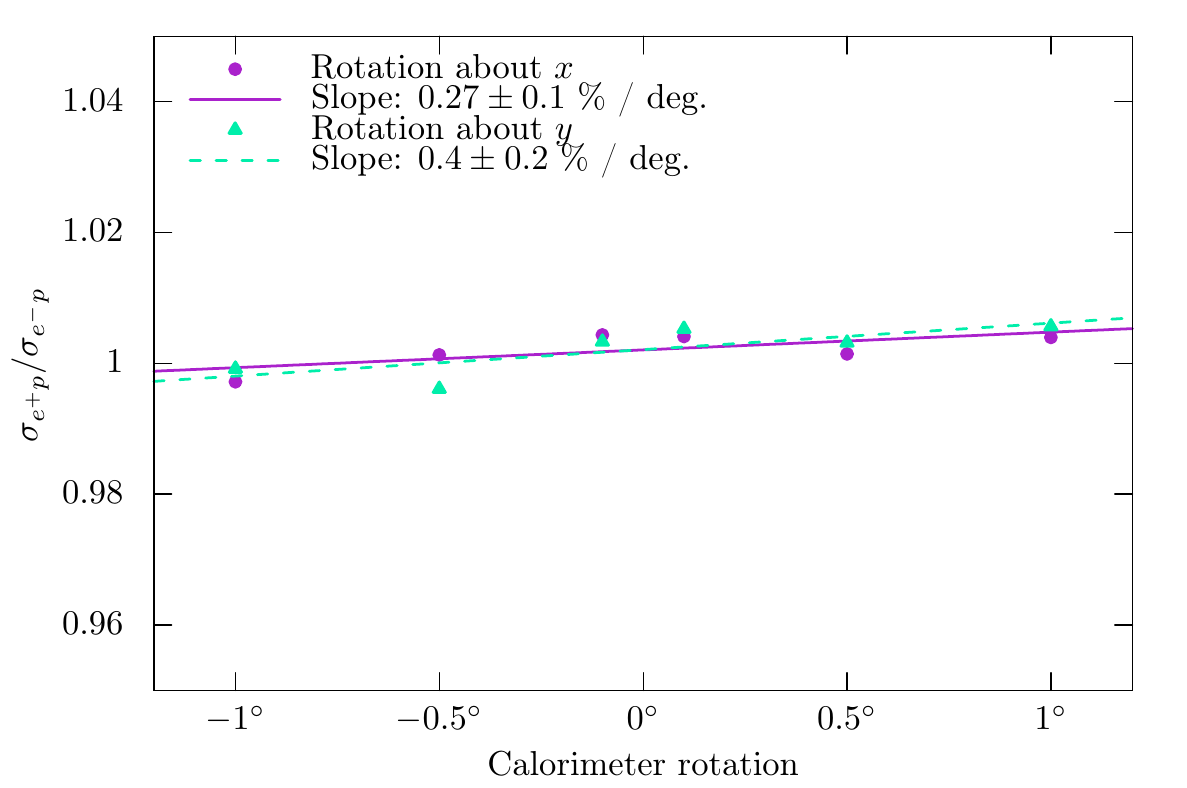}
\caption[The effect of rotating the SyMBs]
{\label{fig:symb_rot} Rotating the right SyMB 1$^\circ$ about the $x$-axis produces a 0.27\% change in the species ratio.
Rotating the right SyMB 1$^\circ$ about the $y$-axis produces a 0.4\% change in the species ratio.}
\end{figure}

If the SyMB detectors sit in a slightly different place in simulation compared to reality, this might distort our calculation of
$\sigma_{ep\rightarrow R}^\text{sim.}$. To estimate this effect, I ran simulations with the SyMBs in slightly different positions
to see how this might affect the cross section, specifically looking at species dependence. The results are shown in figures
\ref{fig:symb_disp} and \ref{fig:symb_rot}.

We estimate that the survey of the SyMBs was accurate to within 500~$\mu$m in position and to within 0.2$^\circ$
in rotation. Therefore, we estimate the systematic errors to be:
\[
\delta_{x\text{~pos.}} = 500 \mu\text{m} \times \frac{0.13\%}{\text{mm}} = 0.07\%,
\]
\[
\delta_{x\text{~rot.}} = 0.2^\circ \times \frac{0.27\%}{\text{deg.}} = 0.05\%,
\]
\[
\delta_{y\text{~rot.}} = 0.2^\circ \times \frac{0.40\%}{\text{deg.}} = 0.08\%.
\]
The systematic uncertainty from the $y$ position is harder to estimate because the result
of the simulation shows zero effect, within the precision of the study. In this case, we
can conservatively assume that the size of the effect is at most equal to the fit uncertainty:
\[
\delta_{y\text{~pos.}} = 500 \mu\text{m} \times \frac{0.10\%}{\text{mm}} = 0.05\%.
\]
Combining these systematics, we get:
\[
\delta_\text{geom.} = \sqrt{\delta_{x\text{~pos.}}^2 + \delta_{y\text{~pos.}}^2 + \delta_{x\text{~rot.}}^2 + \delta_{y\text{~rot.}}^2}
= 0.13\%.
\]

\subsection{Box-Cut Procedure}

\begin{figure}[htbp]
\centering
\includegraphics{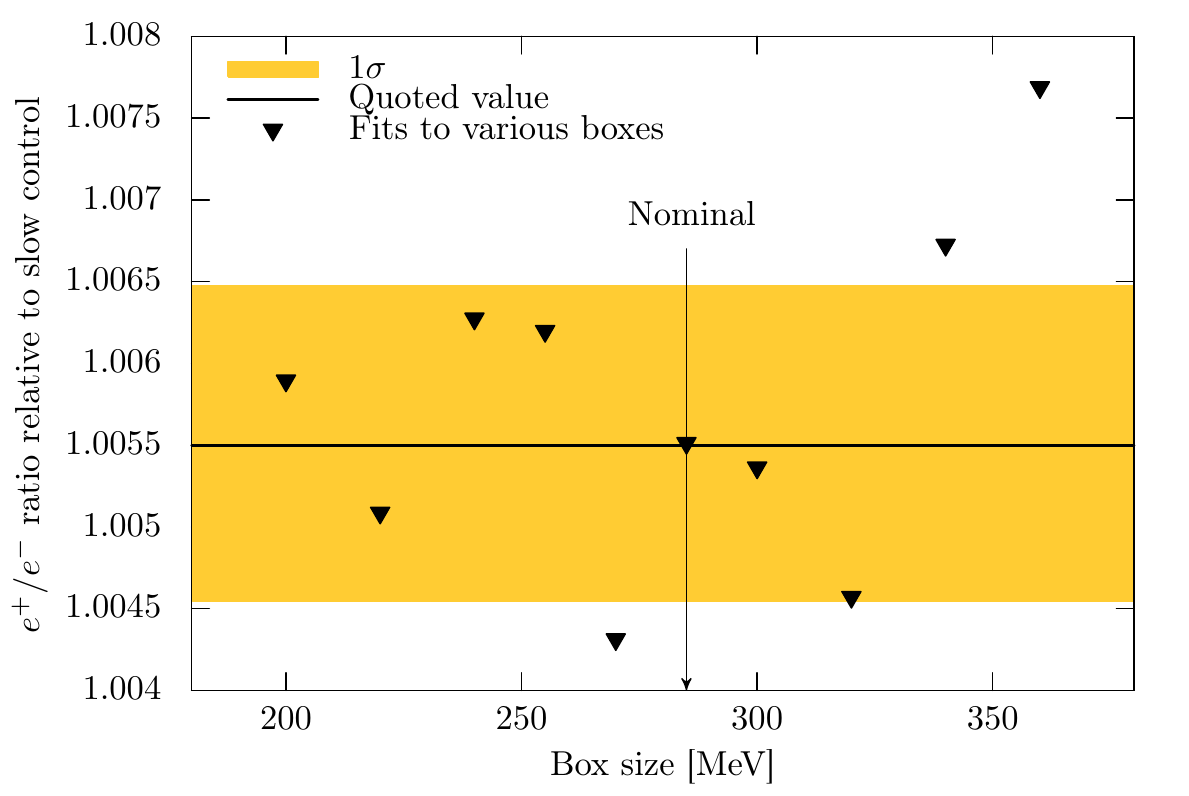}
\caption[The effect of changing the box-cut size]
{\label{fig:symb_box} Changing the box cut size does not have a significant impact on the luminosity species ratio.}
\end{figure}

As mentioned in section \ref{ssec:symb_procedure}, care must be taken that the choice of cuts
does not influence the result. To study this, I produced a species ratio for many
different box cut sizes to look for any trends. The results are shown in figure
\ref{fig:symb_box}. Fortunately, even large changes in the box cut size have a minimal impact on the luminosity species ratio.
To estimate a systematic uncertainty, I calculated the standard deviation of the ratios for all of the different box
sizes I tested: 0.10\%.

\subsection{Magnetic Field}

The magnetic field can have systematic offsets stemming from a variety of sources. The field may have a different
value than what was measured. The position of the measurement point may be different than what was recorded. The
interpolation procedure may introduce some residual error. All of these can contribute to distorting our calculation
of $\sigma_{ep\rightarrow R}^\text{sim.}$.

\begin{figure}[htbp]
\centering
\includegraphics{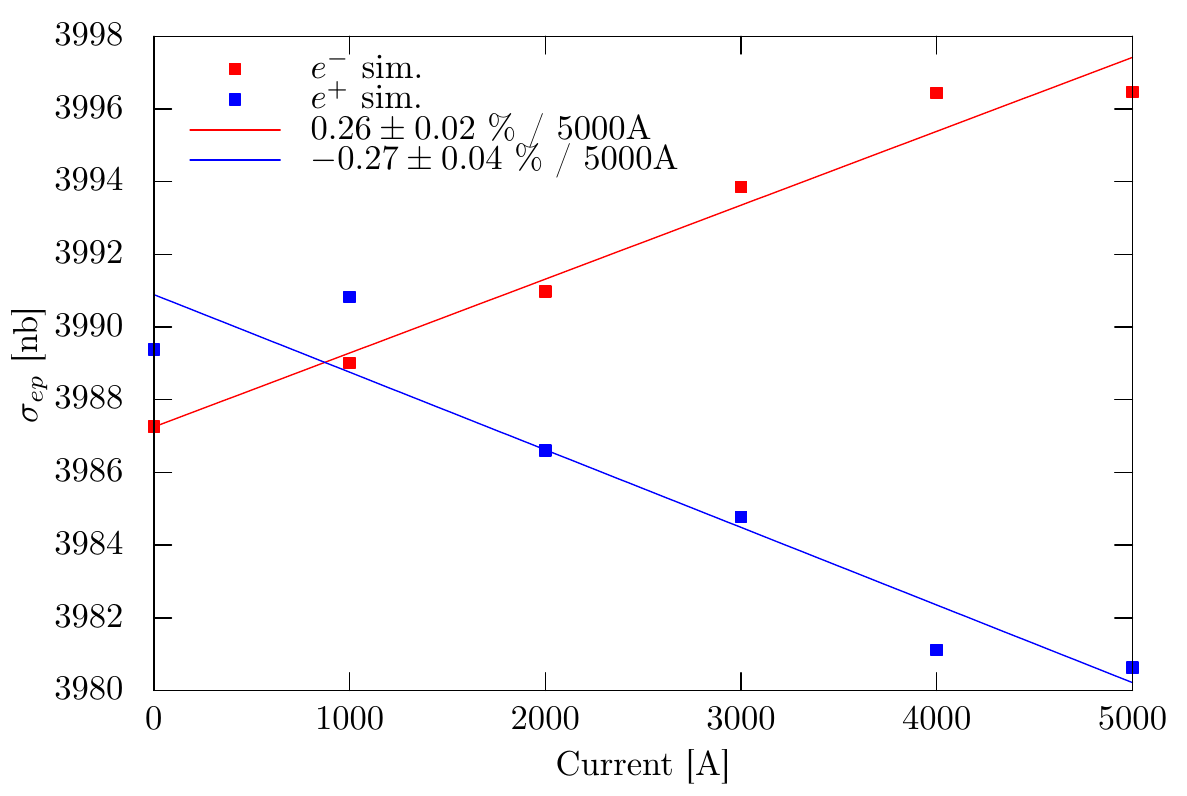}
\caption[The effect of the magnetic field on $\sigma_{ep\rightarrow R}$]
{\label{fig:symb_magnet} The magnetic field has the effect of raising the $e^-p$ cross section in the right
SyMB while decreasing the $e^+p$ cross section. The cross sections are slightly different at zero current because
of radiative effects.}
\end{figure}

As a first step, I calculated $\sigma_{ep\rightarrow R}^\text{sim.}$ for both beam species over a range of magnetic
currents. This can give us a sense of the full scale of the effect of the magnetic field. The results, shown in
figure \ref{fig:symb_magnet}, indicate that the field raises the $e^-p$ cross section and decreases $e^+p$ cross section,
changing the species ratio by 0.53\%. The magnitude of this effect does not indicate any sort of systematic error; if the
field map is correct then this effect is accounted for in simulation. Rather, we need to judge if there could be any
error in our understanding of this effect.

Ideally, we could use a data-driven method to test the slopes in figure \ref{fig:symb_magnet} and assign a systematic uncertainty.
I looked at several possibilities, but unfortunately none were successful. One way to judge the accuracy of the
magnetic field would be to look at the size of residuals between the field map and the magnetic field survey data.
However, the magnetic field map in the beamline region (the only region relevant for the SyMBs) is estimated by
directly interpolating the field survey data points. Consequently we have no residuals to use. Another possibility
is to look at the runs with different current settings to see if a slope can be extrapolated. Looking at the (0,2)
peak counts did not yield anything useful because of problems with the histogram underflow bins. I also looked at
the ratio of (1,3) to (1,1) peaks with the assumption that the slow control luminosity is accurate, but there are
not enough runs with differing magnet currents and the statistics are poor.

Without a data driven approach to testing the slopes in figure \ref{fig:symb_magnet} I will resort to arguments of scale.
Conservatively, I have a high degree of confidence that our magnetic field is accurate to within 10\%. Since the
full effect of the magnetic field is to introduce a species difference of 0.5\%, I estimate a systematic for the
magnetic field to be 0.05\%.

\subsection{Radiative Corrections}

The radiative $ep$ generator is used in the calculation of $\sigma_{ep\rightarrow R}^\text{sim.}$.
There are several choices that can be made in calculating radiative corrections in which the options
aren't necessarily better or more accurate, but can still affect the results. Let's consider a
few of these:
\begin{itemize}
\item \textbf{Form Factor Model} -- The form factors are used as an input to calculate proton
  vertex functions for all of the radiative diagrams, including bremsstrahlung. Ignoring hard
  two-photon exchange, which is not included in the OLYMPUS generator, the choice of form factor
  can influence the species ratio through the bremsstrahlung interference term.
\item \textbf{Hard vs. Soft Definition} -- Soft two-photon exchange must be included since its
  divergence is needed to cancel the divergence in lepton-proton bremsstrahlung interference.
  How one artificially defines the soft region of the two-photon contribution will affect the
  species ratio. Specifically, Maximon and Tjon define the soft region in a slightly different
  way than Mo and Tsai.
\item \textbf{Exponentiation vs. Non-exponentiation} -- Naively, radiative corrections involve
the application of a factor $(1+\delta)$ to the one-photon cross section. The correction $\delta$,
which is negative, accounts for next-to-leading order effects, and depends on the elastic cut-off.
In the soft limit, one can account for radiative effects to all orders by substituting
$(1 + \delta) \longrightarrow \exp(\delta)$. Certainly exponentiation is more accurate for very
small cut-off values. But it is not clear which method better reflects reality for large cut-offs.
\end{itemize}

Of these three choices, the only one that makes any detectable difference at the kinematics relevant
to the SyMB is the choice of whether or not to exponentiate. Using the multiple weights in the OLYMPUS
radiative $ep$ generator, I was able to reweight my simulation runs to see the effect of using
exponentiation or not. The value of $\sigma_{ep\rightarrow R}^\text{sim.}$ is 0.05\% higher when exponentiating.

To estimate a systematic uncertainty, I assumed that the difference between exponentiating represented the full
width of our uncertainty about the true radiative correction, and that the quoted uncertainty should
be half of that difference. I chose to round up to 0.03\%.

\subsection{Beam Energy}

If the electron and positron beams have slightly different energies, then this will introduce
a species difference to all of the cross sections involved. The change in M\o ller/Bhabha cross
sections will cancel in the $N_{(1,3)}$ to $N_{(1,1)}$ ratio, but we do need to take into account
the effect of the $ep$ cross section changing by species.

We estimate that the two beams were within 100~keV of energy of each other. The difference in
the Rosenbluth cross sections at $1.27^\circ$, 100~keV apart in energy, is 0.01\%, which I will
quote as a systematic uncertainty.

\section{Results}

\subsection{Time Dependence}

\begin{figure}[htb]
\centering
\includegraphics{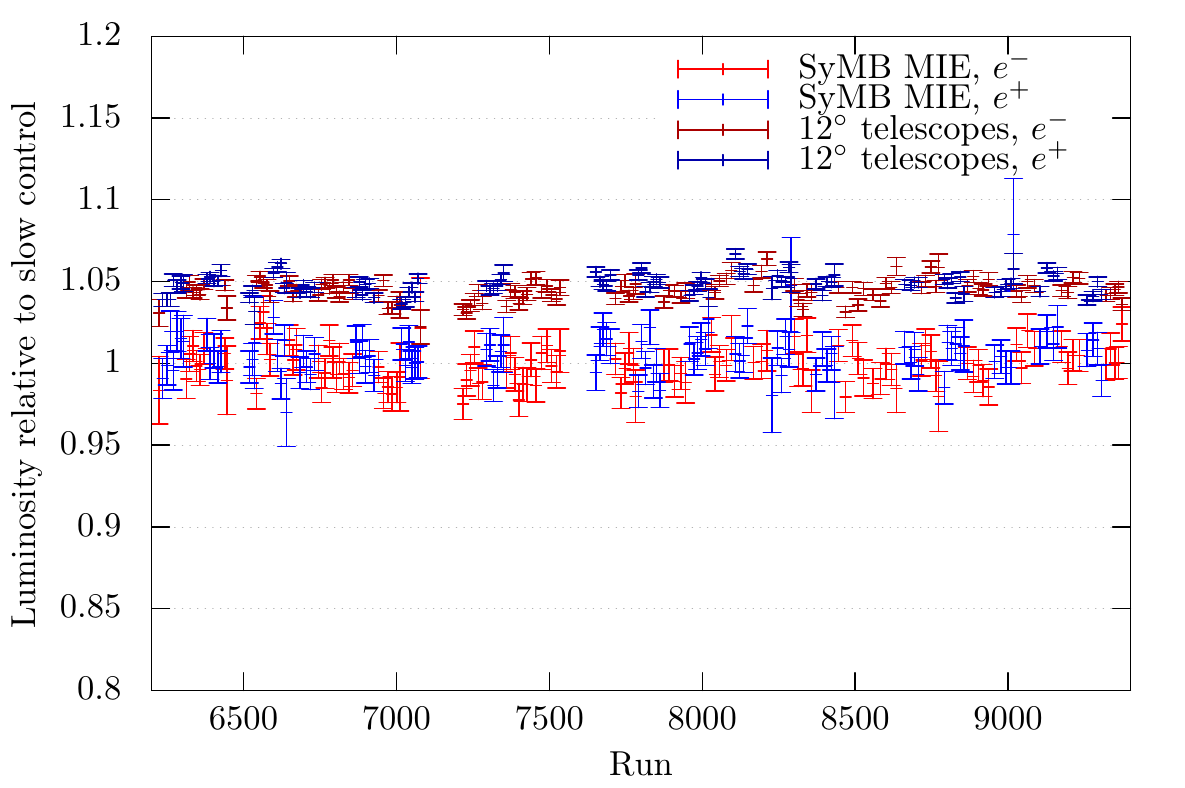}
\caption[MIE results]{\label{fig:mie_results} The MIE and $12^\circ$ luminosity determinations have 
many of the same time-dependent variations when compared to the slow control system. That
suggests that the slow control determination varied in time, probably due to temperature
fluctuations in the target.}
\end{figure}

Figure \ref{fig:mie_results} shows both the MIE luminosity determination and the $12^\circ$ tracking
telescope luminosity determination relative to the luminosity determined by the slow control system, 
plotted over the set of runs used in the OLYMPUS analysis. The absolute luminosity is not determined by 
the MIE analysis (loosening or widening the box cuts changes the absolute luminosity determination) so the 
exact placement of the MIE points on the $y$ axis is arbitrary. The errorbars are statistical only, and 
one can see that the MIE analysis has a worse statistical precision than the $12^\circ$ system. The precision
of the MIE analysis is largely limited by the statistics on $N_{(1,3)}$. Multi-interaction events happen 
more rarely than single-interaction events, so it comes as no surprise that the MIE method sacrifices precision.
In this figure, each data point corresponds to ten runs aggregated together, to reduce the statistical errors
per point.

The key result in figure \ref{fig:mie_results} is that the MIE analysis and the $12^\circ$ systems see most
of the same time-dependent variations when compared to the slow control system. This fact implicates the 
slow control system as being the source of most of the time-variation. This is probably caused by slight
fluctuations in the target gas temperature, which are not entirely captured by the target cell temperature
sensor. That the MIE analysis and $12^\circ$ analysis agree on many of the small scale time variations
is a strong validation of both analyses.

\subsection{Correction Factor to Slow Control}

\begin{figure}[htpb]
\centering
\includegraphics{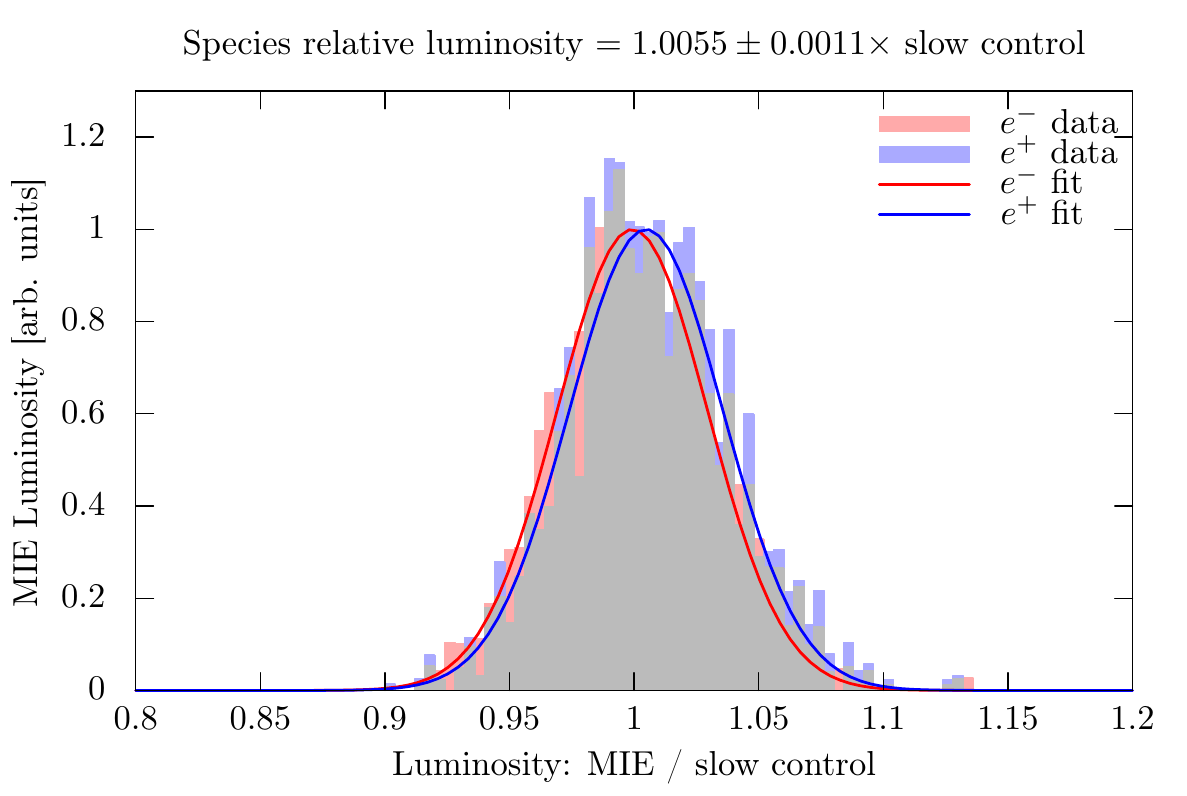}
\caption[Projection of MIE luminosity by species]
{\label{fig:symb_projections} The MIE analysis suggests that the slow control luminosity for positrons
is 0.55\% too low relative to the slow control luminosity for electrons.}
\end{figure}

Since there are no major time-dependent trends in the MIE and slow control luminosity extractions,
it is appropriate to use a luminosity correction factor to translate from the slow control luminosity
(used by the OLYMPUS simulation) and the MIE luminosity, which we believe is more accurate. To determine
this correction factor, the data in the top plot of figure \ref{fig:mie_results} were projected onto the
$y$ axis and histogrammed so that the distributions by species could be studied. The projections and 
gaussian fits are shown in figure \ref{fig:symb_projections}. The distributions by species have slightly
different means, indicating that the slow control luminosity has a slight species-dependent inaccuracy. 
The fits suggest that the slow control measures positron running luminosity to be 0.55\% too low relative
to electron running luminosity. 

There is a statistical uncertainty on this correction factor of 0.11\%. Despite the low statistics on 
$N_{(1,3)}$, when aggregating over many runs, the statistical uncertainty is still well below the
systematic uncertainty of 0.27\%. Adding these uncertainties in quadrature produces a total error
on the species-relative luminosity of 0.29\%. This meant that the OLYMPUS asymmetry measurements will
have a global uncertainty of 0.15\% (rounding up) due to luminosity. 

\subsection{Luminosity Normalization Point}

\begin{figure}[htbp]
\centering
\includegraphics{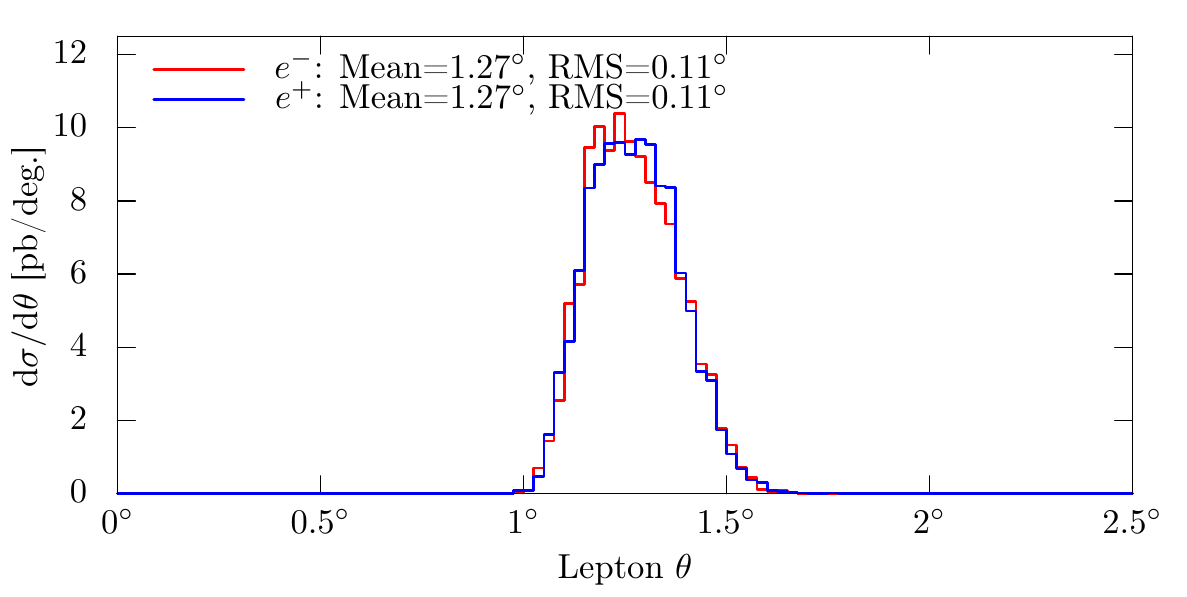}\\
\includegraphics{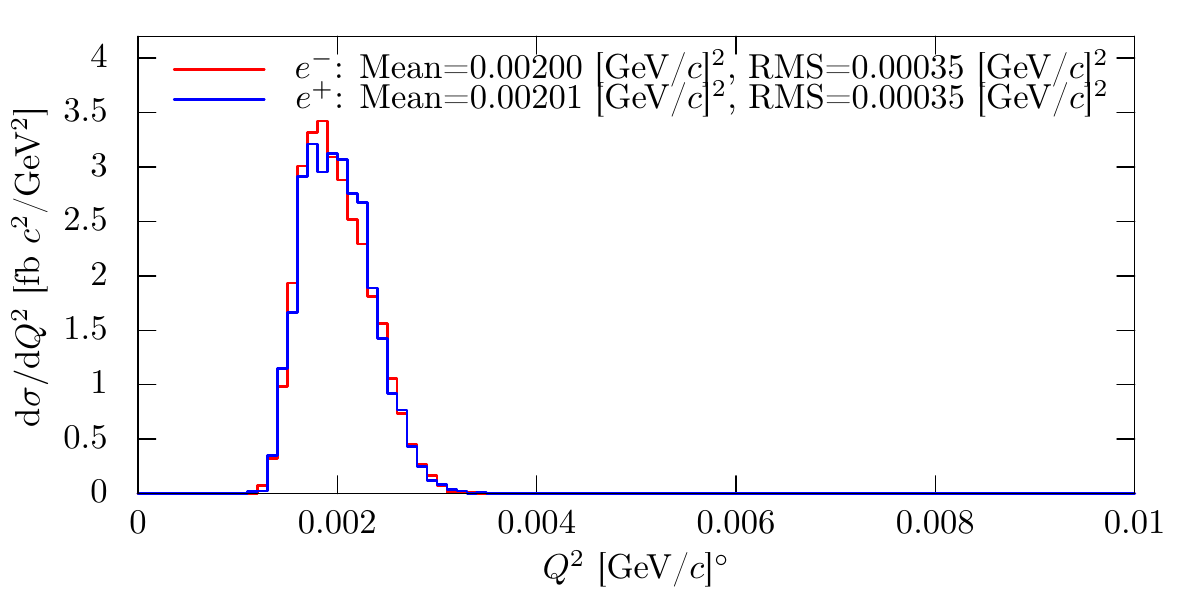}\\
\includegraphics{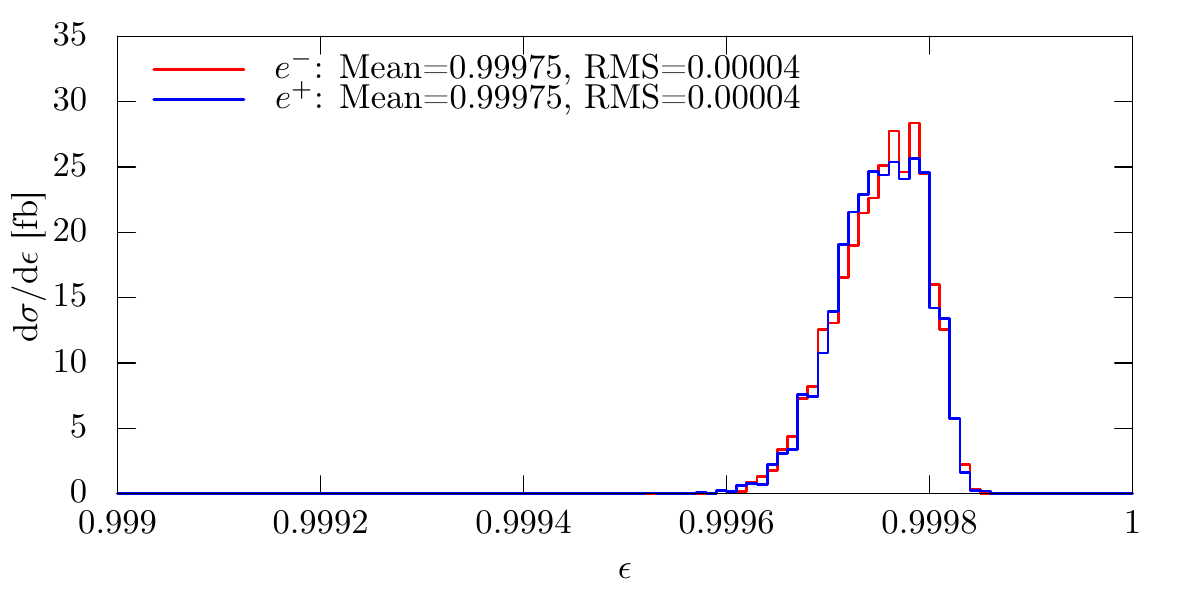}
\caption[MIE luminosity normalization point]
{\label{fig:mie_lnp} Any physics result normalized to this analysis will take as an LNP:
$\theta=1.27^\circ$, equivalent to $Q^2 = 0.002$~(GeV/$c$)$^2$, and $\epsilon=0.99975$. }
\end{figure}

The multi-interaction event analysis depends on the elastic $ep$ cross section at the forward
scattering angle of the right SyMB calorimeter (specifically, the collimator aperture). There is
some contribution at that angle, however small, from hard two-photon exchange.
Therefore, any result normalized to the MIE analysis will be relative to the lepton sign asymmetry
at this angle. This angle is what is called a ``Luminosity Normalization Point'' (LNP).

To determine the precise location of the LNP, I looked at the generator output for simulated
$ep$ events which entered the right calorimeter. The results are shown in figure \ref{fig:mie_lnp}.
The LNP falls at a scattering angle of $1.27^\circ$, corresponding to a $Q^2$ of 0.002~(GeV/$c$)$^2$,
and to value of $\epsilon$ equal to 0.99975. These values are very similar for both beam species.

\section{Discussion}

It was surprising to find that the MIE analysis, which was explored as an afterthought, had 
the smallest systematics of any of the OLYMPUS luminosity monitors. However, after comparing 
systematics with the main SyMB analysis, it became apparent that the MIE analysis had several
inherent advantages. I will conclude this chapter by discussing these advantages, and discussing
what improvements could be made if a luminosity monitor were designed around the MIE method.

\subsection{Inherent Advantages of MIE}

The single biggest advantage of the MIE method is that its luminosity extraction comes from
the ratio of two count-rates, rather than a single countrate. Any pernicious systematics
which affect electron running differently from positron running, must also affect both 
count rates differently. If the detectors have any inefficiency, or if the data acquisition
system occasionally fails to add an event to the histogram, these effects will cancel in the
ratio of count-rates. Clearly, some effect of this kind troubled the main SyMB analysis and
produced a discrepancy in the species-relative luminosity. The MIE analysis was immune.
There are examples of systematics which do affect the two count-rates differently: beam
position and survey geometry come to mind. But these systematics largely affect the two
beam species in the same way and the system is only barely sensitive to them when extracting
a species relative luminosity.

Another advantage, which was in a way a happy accident, is that the two signal peaks were
treated identically by the trigger. The (1,3) peak visible in data was only visible in the
left master histogram (shown previously in figure \ref{fig:symb_lm_hist}). Only signals in the
left calorimeter were considered in the trigger, and the (1,1) and (1,3) peaks have identical
energy deposition on the left. One hypothesis for the cause of problems in the main analysis
and of the deficits in the (2,2) peaks is inappropriate vetos \cite{comparator:saturation}; the MIE analysis is
completely immune to such problems.

\begin{figure}[htpb]
\centering
\includegraphics{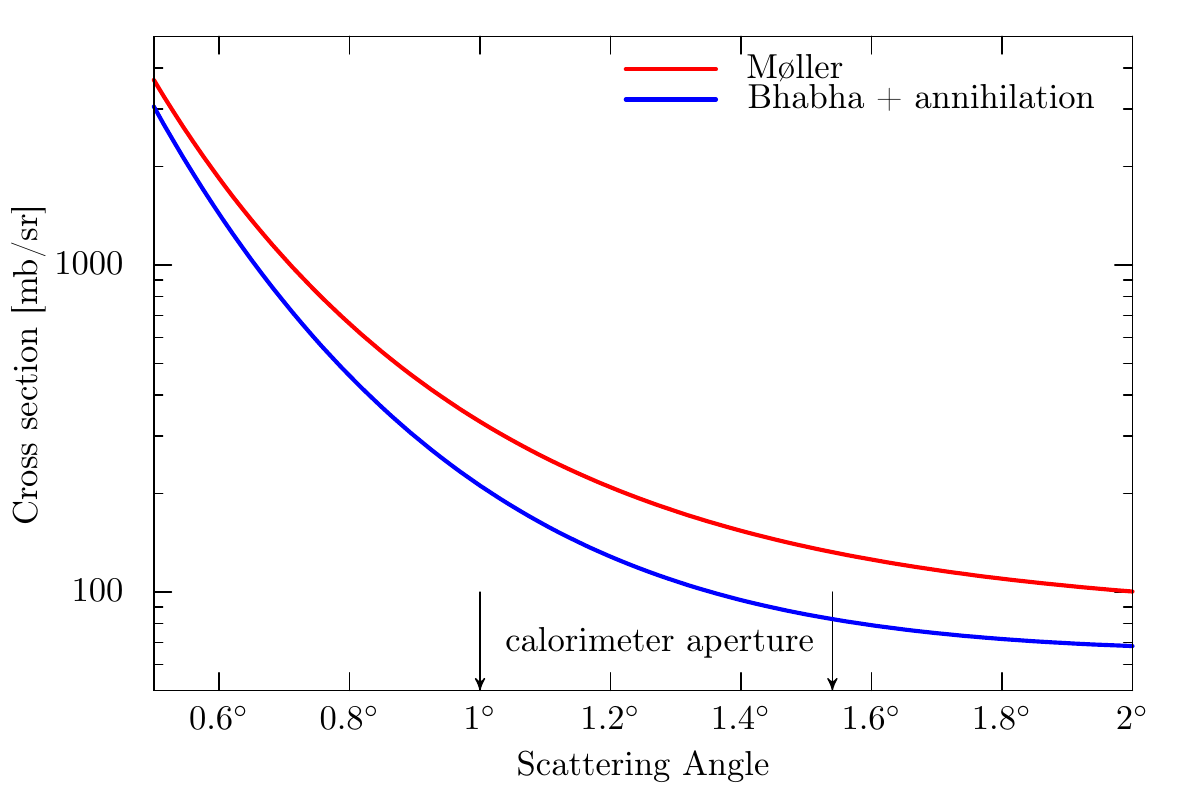}
\caption[Comparison of the M\o ller and Bhabha cross sections]
{\label{fig:mb_compare} The cross section for events in the SyMB calorimeters
changes by almost a factor of two between electron running (red) and positron running
(blue). This is not ideal when making a relative luminosity measurement.}
\end{figure}

A third advantage is that the actual measured quantity does not change greatly in scale
between electron running and positron running. In the MIE analysis, the ratio of countrates
is practically the same between species because the $e^-p$ cross section is equal to the
$e^+p$ cross section, absent radiative corrections. In the main analysis, the M\o ller and Bhabha 
cross sections differ by nearly a factor of two, as can be seen in figure \ref{fig:mb_compare}.
Making a relative luminosity measurement with a given accuracy requires determining that difference 
with the same accuracy. This is easier to do if the difference is small. Making an accurate
determination of the large cross section difference in the main analysis ended up not being possible.

\subsection{Improvements for Future Implementations}

OLYMPUS has been an excellent demonstration that the method is sound, and that future lepton sign 
asymmetry experiments would do well to take advantage of this techinque for luminosity monitoring. 
However, the MIE analysis was still only an afterthought. It could have been even more accurate had it
been the focus of the SyMB design. The obvious changes to the SyMB hardware that could have improved
the MIE analysis all involve the dynamic range of the histograms. If the histogram ranges had all been
slightly wider, the (1,3) peak could have been visible in other histograms. This would have allowed
cross checks of the effect of the trigger and veto conditiions. Furthermore, the left/right symmetry
of the detector system could have been exploited had the (3,1) peak been visible. Extracting luminosity
from two different MIE analyses could have provided constraints on beam position and geometry systematics.

\chapter{Track Reconstruction}

\label{chap:recon}

\section{Overview}

Track reconstruction is the task of determining particle trajectories, ``tracks'' in shorthand,
from the unprocessed OLYMPUS data. The task can be divided into two
component problems: first, determining how many tracks are in an event and second, estimating, 
for each track, the particle's vertex and initial momentum vector. I will refer to the first 
component problem as ``tracking finding'' and the second as ``track fitting'', and describe our
solutions to each in the following sections.

The drift chambers were the main tracking detectors in the OLYMPUS spectrometer. The drift chamber
data for each event came in the form of TDC times associated with the different sense wires in the
chambers. These times were translated into definite positions along a trajectory, first with some
initial processing (described in section \ref{sec:tdc_proc}) and then using a time-to-distance (TTD)
function (described in section \ref{sec:ttd}). Some of the information from the ToFs was also used
in track reconstruction. The specific ToF bar hit by a particle gave some position information about
its scattering angle, and the time difference between TDC signals from the top and bottom PMTs was
used to locate the vertical position of the track. The track reconstruction algorithms combined the
ToF and drift chamber data in order to find tracks and fit their initial conditions.

\section{Processing the Drift Chamber Data}
\label{sec:tdc_proc}

\begin{figure}[htpb]
\centering
\includegraphics{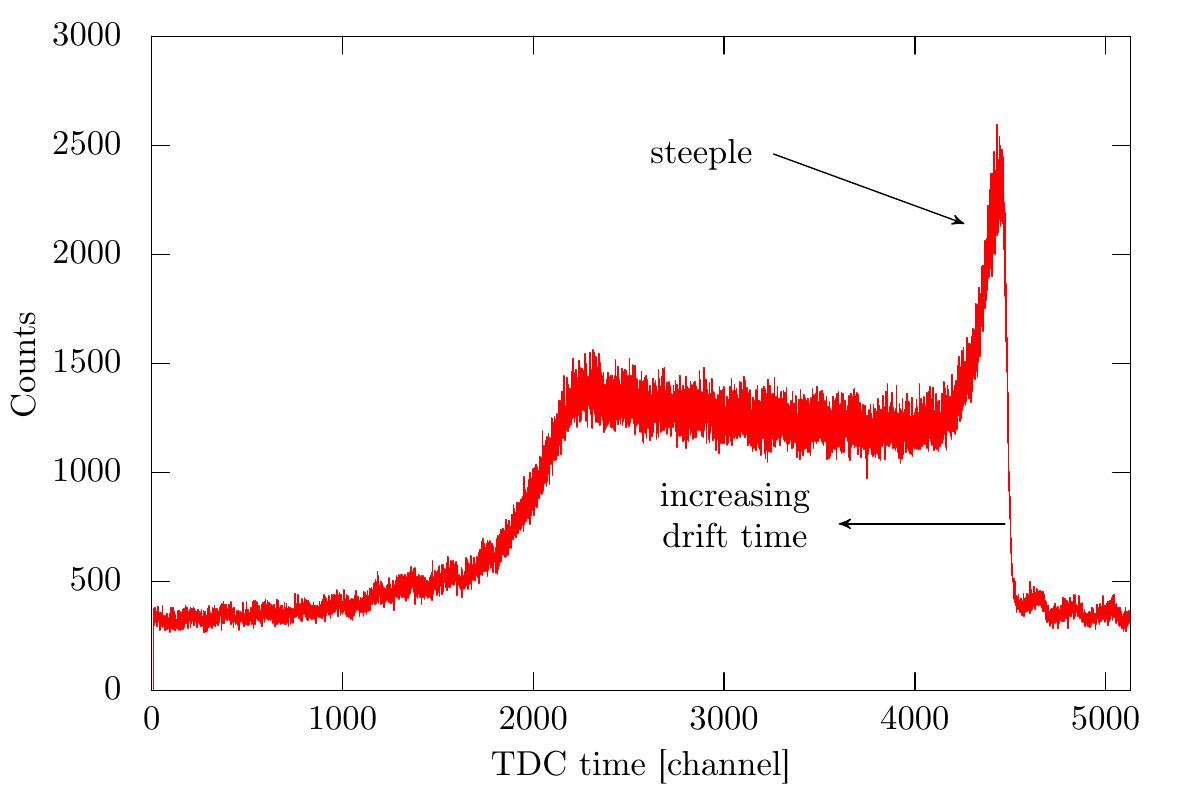}
\caption[Example drift chamber TDC distribution]
{\label{fig:tdc_church_raw} The TDC distribution from 100 runs of data is shown for sense wire 317. 
The distribution has the characteristic ``church'' shape.}
\end{figure}

The drift chamber data came in the form of a list of wires that had signals, and then a list of
TDC times corresponding to each wire. Some initial processing was needed to convert those TDC times
into true drift times. In this section, I will give a description of the initial processing steps. 

The distribution of TDC times for a single wire over a run had a characteristic ``church'' shape, shown
in figure \ref{fig:tdc_church_raw}.
There was a relatively constant noise floor, a ``steeple'' at large TDC times, and a ``roof'' extending
to lower TDC times. Because the TDCs were operated in common stop mode, smaller TDC times correspond
to longer drift times. The steeple has times produced by tracks that passed very close to the wire.
the other end of the church distribution is populated by tracks that passed far from the wire. 

The TDC distributions of most wires had, in addition to the church, a pattern of oscillating noise 
that was in-phase with and had the same frequency as the bunch clock. Our best guess as to the origin
of this noise was background originating from adjacent bunches. This noise was not problematic but 
needed to be accounted for when calibrating the $t_0$ times for each wire, as will be discussed in the
following sections.

\subsection{Correcting Trigger Jitter}

\begin{figure}[htpb]
\centering
\includegraphics{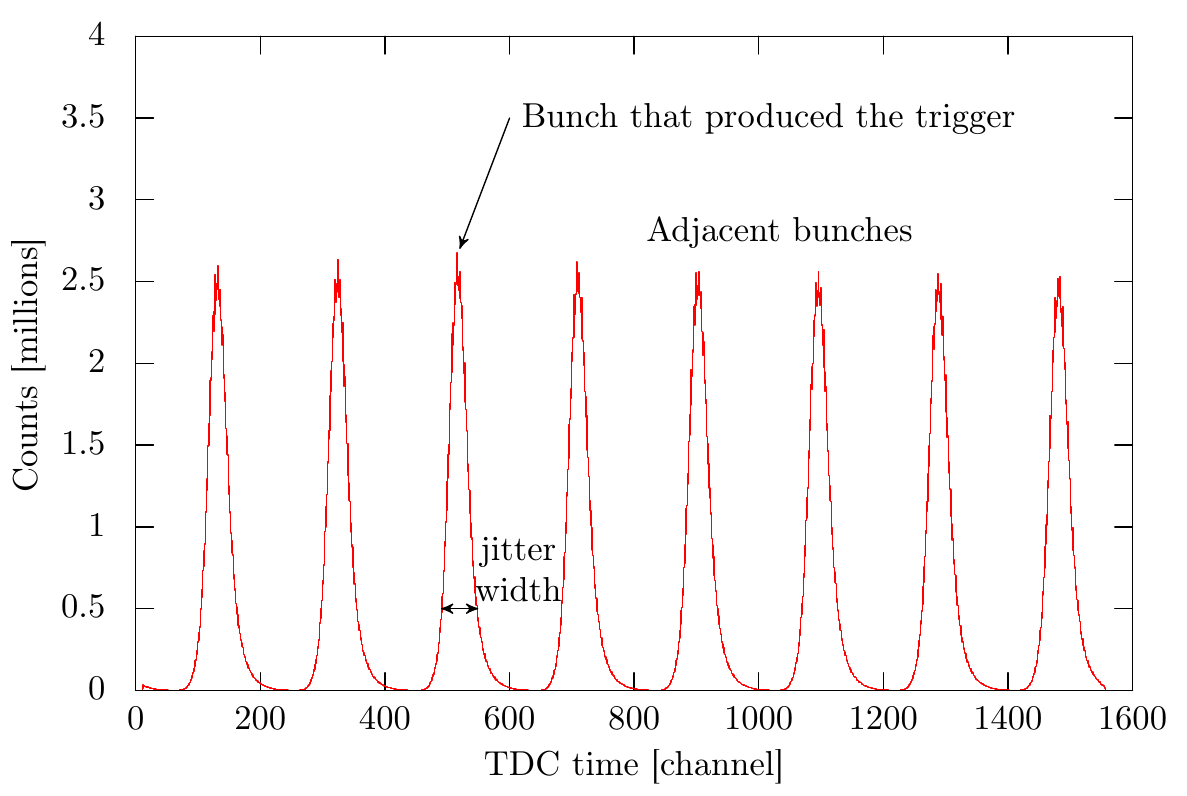}
\caption[Bunch clock TDC distribution]{\label{fig:tdc_bunch_clock} The trigger had some time jitter, as can be seen in the bunch clock 
TDC spectrum. By subtracting the TDC time for bunch clock pulse corresponding to the bunch that produced
the trigger, the common trigger jitter can be removed.}
\end{figure}

\begin{figure}[htpb]
\centering
\includegraphics{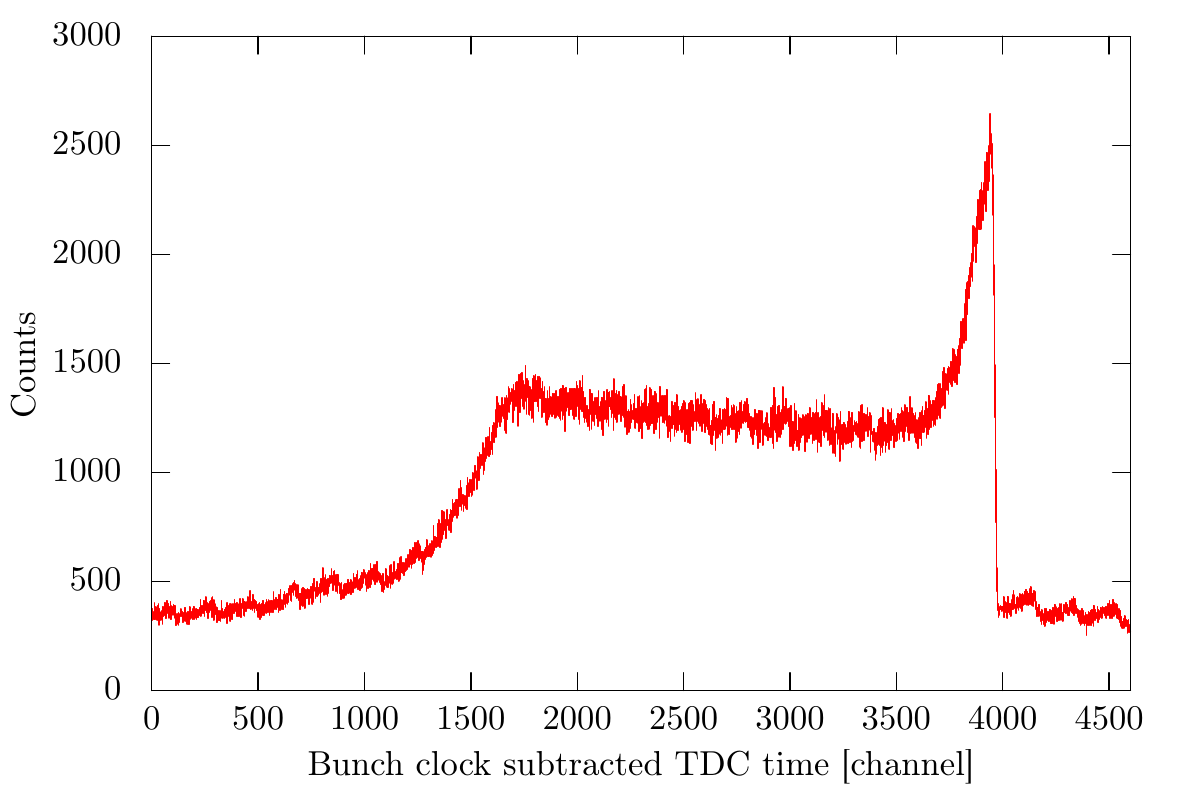}
\caption[Jitter-corrected drift chamber TDC distribution]
{\label{fig:tdc_church_sub} The jitter-corrected TDC distribution is shown for sense wire 317.}
\end{figure}

The first processing step was to correct for the time jitter introduced by the trigger. 
The start signal for the TDC was a current pulse on the sense wire. The stop signal was provided by
a delayed trigger signal. The trigger signal had some time jitter, which worsened the drift
time resolution. Fortunately, this time jitter could be corrected by using the bunch clock, a simple
clock pulse that marked when an accelerator bunch reached the center of the target. We connected the 
bunch clock as the start signal of an open TDC channel with the same common stop. This produced a distribution
like the one shown in figure \ref{fig:tdc_bunch_clock}. The jitter was removed from the TDC times by subtracting 
the TDC time of the clock pulse for the bunch that produced the trigger. The jitter-corrected TDC distribution
for wire 317 is shown in figure \ref{fig:tdc_church_sub}, and can be compared with the raw TDC distribution
of figure \ref{fig:tdc_church_raw}.

\subsection{$t_0$ Calibration}

\begin{figure}[htpb]
\centering
\includegraphics{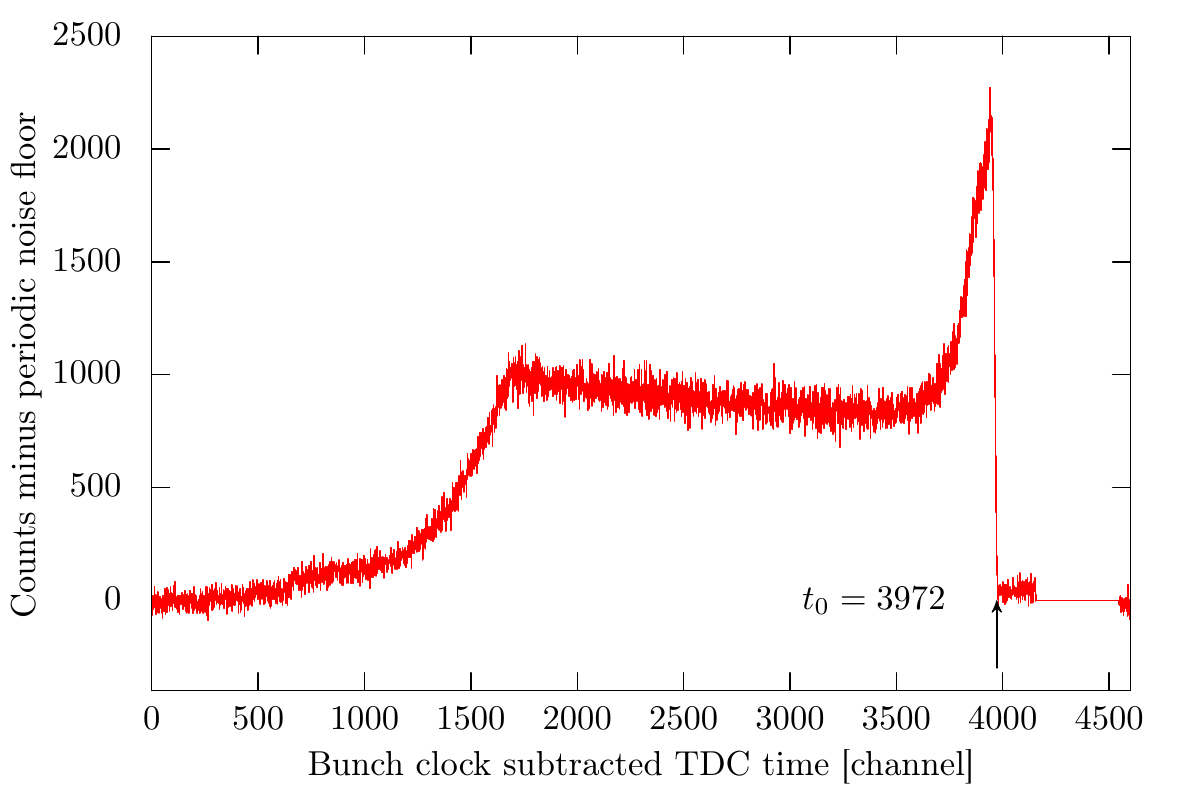}
\caption[Noise-subtracted drift chamber TDC distribution]
{\label{fig:tdc_church_clean} The clean TDC distribution (shown for sense wire 317) was produced 
by subtracting the periodic noise from between channels 4156 and 4544. This allowed a more stable $t_0$ fit. }
\end{figure}

After subtracting the bunch clock, the next step was to find the time of zero drift, or $t_0$, for
each wire. This time corresponded to tracks that passed directly next to the wire and had zero 
effective drift distance and zero effective drift time. $t_0$ was the baseline TDC time for calculating
drift times. 

We estimated $t_0$ to be the TDC time at which the steeple met the noise floor in the TDC distribution.
This point was obscured by the oscillating 10~MHz noise. Since the noise had a well-known frequency and
phase, we could subtract it from the TDC distribution, create a ``clean'' distribution, shown in figure
\ref{fig:tdc_church_clean}. We then found the point on the steeple corresponding to 80\% and 20\% of the 
steeple height, and extrapolated to the noise floor to find $t_0$.

\section{Track Finding Using Pattern Matching}

\begin{figure}[htpb]
\centering
\includegraphics[width=\textwidth]{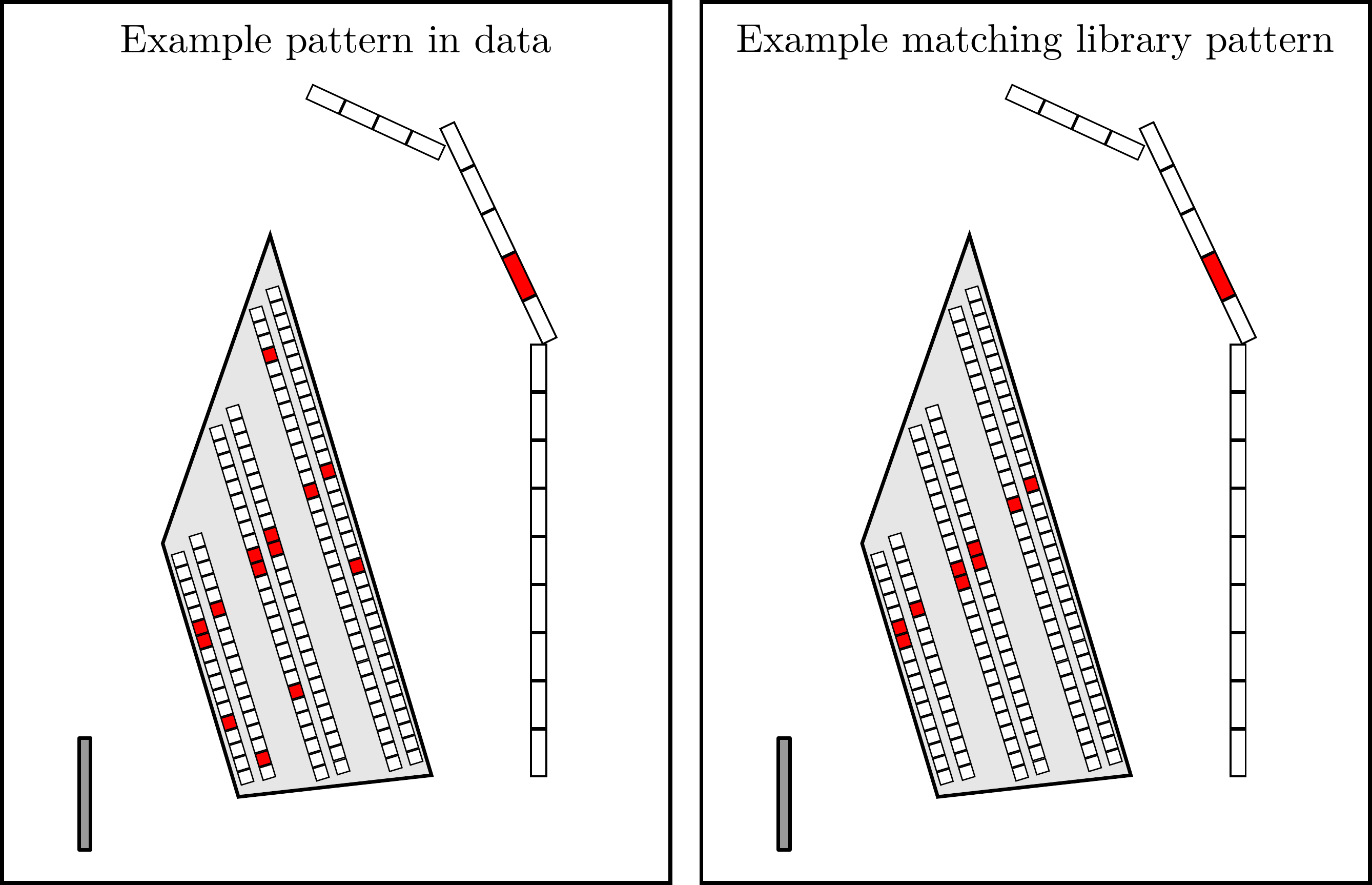}
\caption[Pattern examples]{\label{fig:pattern_example} This example data pattern (left panel)
has one track as well as several noise hits. The library pattern it matches (right panel) only
contains the cells (and ToF bars) hit by the track. }
\end{figure}

We chose to approach track finding in a different way than the method pursued at BLAST. At OLYMPUS,
the beam energy was higher, the accelerator had a lower duty factor, and the beam halo produced by
the target was greater. As a consequence, the background environment was worse, especially in the
inner chambers of both sectors. We needed some way to reject noise hits that did not come from tracks
emerging from the target, and to do so quickly, without trying every combination of hits in all
chambers. Our track finding approach was based on the fast pattern matching algorithm of Dell'Orso 
and Ristori \cite{Dell'Orso:1990cj}. In their paper, they modeled a particle physics detector 
as having segments which either made a detection or did not. In our implementation, shown in the 
example in figure \ref{fig:pattern_example}, each segment
was a single drift cell. If at least one wire produced a valid TDC time, then the cell was considered
hit. By using cells as the smallest discrete element, we avoided mixing the time-to-distance problem
with the pattern matching problem. In addition to drift cells, we also added the ToFs into the pattern 
matching scheme; if both PMTs in a bar produced a signal, then the bar was considered hit. A single 
sector of the spectrometer had 159 drift cells and 18 ToF bars. Consequently, our patterns were 177 bits long. 

The pattern matching algorithm required a library of patterns corresponding to valid tracks. We generated
a library using simulated tracks (see section \ref{ssec:prop} for more detail) with a wide range of initial
vertices, angles, and momenta. Each library pattern had at least one hit cell in each drift chamber super-layer,
and at least one hit ToF bar. Initially, we planned to match separate pattern libraries for the different 
particle species ($e^+$, $e^-$, $p$, $\pi^+$, etc.), but the libraries overlapped so extensively that we
combined them into a single pattern library. The libraries for the two sectors, which were very nearly 
identical, were also combined. The result was a library of approximately 270,000 patterns.

\begin{figure}[htpb]
\centering
\includegraphics[width=\textwidth]{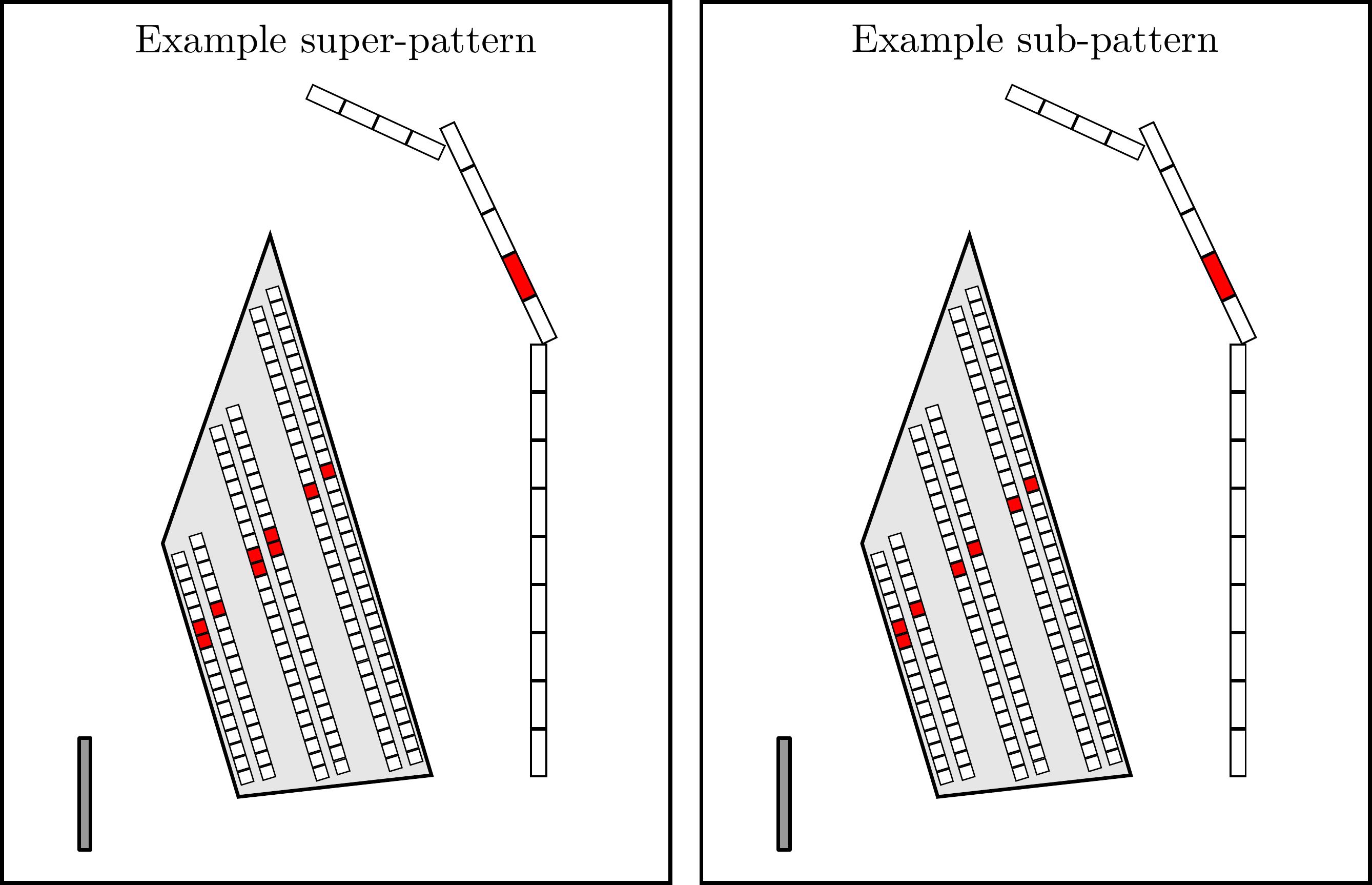}
\caption[Super- and sub-patterns]{\label{fig:subpattern_example} The sub-pattern (right panel) contains
a subset of the hit cells in the super-pattern (left panel). Only the super-pattern need be considered
by the track fitter.}
\end{figure}

When reconstructing a single event, after the initial data processing and identifying which wires and ToF bars
had valid TDC times, pattern matching was performed on each sector. The data pattern was compared against the 
simulated pattern library, and a list of library matches was produced. In order for a library pattern to be
a match, all of its hit cells needed to be hit in data. Stated another way, the bitwise-AND of the library
pattern and the data pattern needed to be equal to the library pattern. Each matching library pattern was 
considered to be a track candidate, with the exception of matches that were
``sub-patterns'' of other matches. Sub-patterns contained a subset of the hits of another matching pattern,
and there was thus no need to produce an additional track candidate. An example of a sub-pattern is shown
in figure \ref{fig:subpattern_example}.

\subsection{Modifications to Handle Inefficient Regions}

This pattern matching scheme needed to be modified in order to cope with drift chamber cells that were 
inactive, or slightly inefficient. There were two drift cells that were completely inactive. Though these
cells never produced any signals, for the purposes of pattern matching these cells were always considered
to have been hit. Tracks that passed through the dead cells were still successfully matched. 

Several regions of the drift chambers were found to be slightly inefficient. This presented a potential
problem in pattern matching because the library patterns might expect a hit in an inefficient region that
might not be present in the data. To cope with this, we modified the criterion for a successful match. 
Instead of requiring that the bitwise-AND of the data pattern and library pattern equal the library
pattern, we required that the bitwise-AND of the data pattern and library pattern have at most one
cell missing from the library pattern.

\section{Track Fitting with the Elastic Arms Algorithm}

Each track candidate---the set of TDC times from the cells (and ToFs) matching a library pattern---was fit
using an algorithm based on the Elastic Arms Algorithm (EAA) of Ohlsson, Peterson, and Yuille 
\cite{Ohlsson:1991eh, Ohlsson:1992kg}. EAA represents an approach different from what was used 
successfully at BLAST and was chosen to cope with the challenges specific to OLYMPUS, namely
worse position resolution due to higher drift velocity in the drift chambers, and an increase
in noise. At BLAST tracks were reconstructed using a microscopic approach. Hits were linked within
a super-layer based on linearity and combined with the adjacent super-layer if the resulting track
``segment'' pointed back toward the target. In this way, smaller units were aggregated until the 
track was found. At OLYMPUS, left-right ambiguities (first mentioned in section \ref{ssec:drift_chambers})
in a cell were not always resolvable given the limited position resolution. Our algorithm used 
a macroscopic approach that considered all of the hit positions simultaneously in order to find 
the likeliest initial conditions for the track.

The problem of track fitting was to estimate, for a given candidate, the likeliest initial 
conditions. The space of initial conditions was four dimensional; neglecting any transverse
motion of the beam, a track was determined by its vertex position $z$, its scattering angle
$\theta$, its azimuthal angle $\phi$, and its momentum $p$. The fit of these four parameters
was guided by the vector of hit positions $\vec{x} = \{x_0, x_1, \ldots x_i \ldots\}$ from the
various wires and ToF bars in the candidate pattern. A simple fit to find the values of $\{z,\theta,\phi,p\}$
that best match the hit positions $\vec{x}$ could not work because of the problem of left-right
ambiguities. A single TDC time in the drift chambers could correspond to two different hit positions,
one to the left of the wire and one to the right. Furthermore, some track candidates had some 
noise hits that did not truly belong to the track, but were part of the matched pattern. 
EAA was equipped to handle noise rejection. An extension of the original EAA algorithm was 
specifically designed to work with ambiguities in hit positions \cite{Lindstrom:1995ue}.

\begin{figure}[htpb]
\centering
\includegraphics{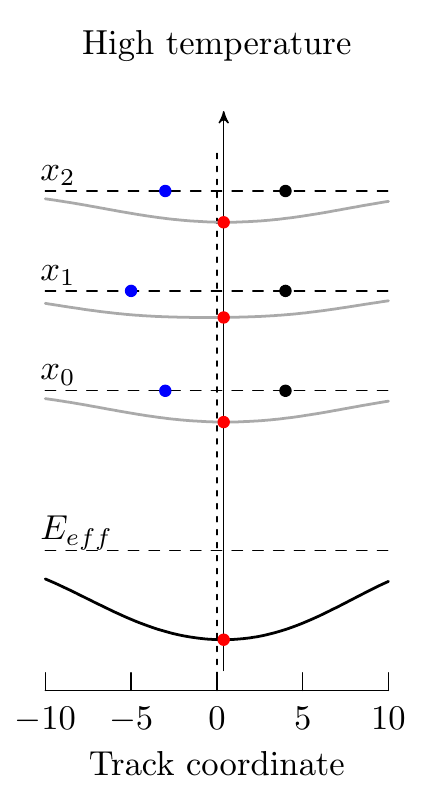}
\includegraphics{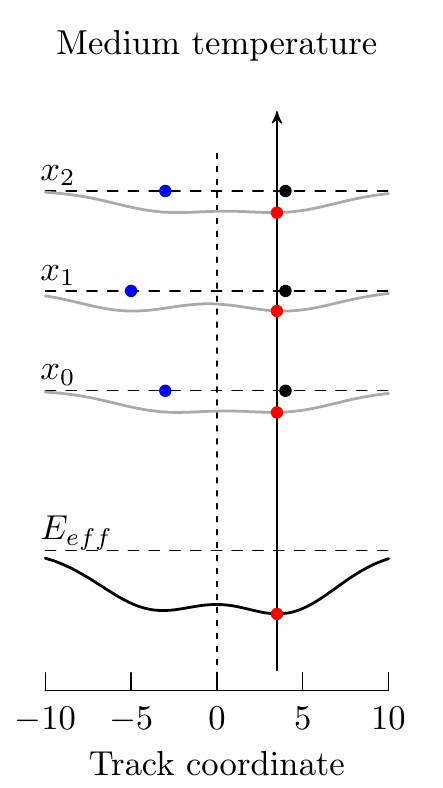}
\includegraphics{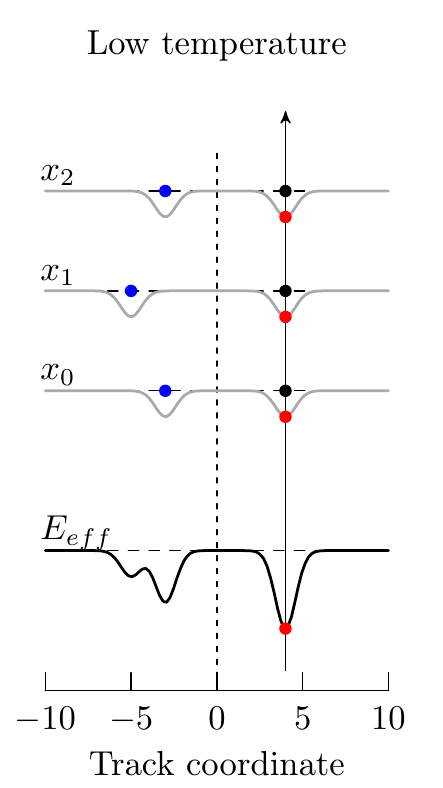}
\caption[Demonstration of Elastic Arms Algorithm]{\label{fig:eaa} The Elastic Arms Algorithm is demonstrated
with a simplified example. The space of track initial conditions is one dimensional, represented by the $x$ axis. 
There are three detector planes; hit positions on these planes are denoted $x_0$, $x_1$, and $x_2$. The detector
planes have a left-right ambiguity; the true hit positions are shown as black circles and the false hit positions
are shown as blue circles. The sense wires in this example have some staggering, causing the false hit positions
to no longer appear in a straight line. The algorithm finds the track coordinate that minimizes an effective potential
energy, $E_{eff}$, shown at the bottom of each panel. At high temperature (left panel) there is a single global minimum, 
which suggests a track that is halfway between the false and true hits. At medium temperature (middle panel) the
minimum splits in two, and the true hits are slightly preferred. At low temperature, the most likely track points
through the true hits, and the false hits are rejected as noise. As a general trend, as the temperature decreases,
there become more local minima, but the minima become narrower. }
\end{figure}

The EAA algorithm, demonstrated with a simplified example in figure \ref{fig:eaa}, 
is a type of annealing filter. In analogy to the physical process of 
annealing a crystal, the algorithm has a ``temperature'' parameter, $\beta$, which varies 
over the course of a fit. Early on, $\beta$ is small, implying a high temperature; the 
algorithm is more tolerant of hit positions that do not quite match those of the best-fit track.
As the algorithm progresses, $\beta$ is increased, reducing the temperature; the algorithm
becomes more stringent and rejects hits that are too far from the best-fit track position.
In addition to $\beta$, there is also a parameter $\lambda$, which determines the cost for
rejecting a hit as noise. Early on, the penalty is high, encouraging the algorithm to be
more tolerant and giving every hit a chance to influence the fit. Gradually this penalty is
reduced, allowing the algorithm to reject the hits that do not lie along the track.\footnote{ The 
algorithm does not specify exactly how $\beta$ and $\lambda$ should evolve during a fit.
My colleague, Rebecca Russell, performed the studies necessary to find a suitable $\beta$
and $\lambda$ evolution for OLYMPUS track fitting \cite{russell:thesis}. } The idea behind
the algorithm is that the likeliest initial conditions are specified by the low-temperature 
global minimum of the fit. However, at low temperatures there are many local minima. At
very high temperature, there is only one minimum. By finding the high-temperature global
minimum, and by staying in the global minimum as temperature is reduced, the correct
low-temperature global minimum can be found. 

The word ``elastic'' in ``Elastic Arms Algorithm'' does not refer in any way to elastic
scattering. Rather, an ``elastic arm'' is the name given to a deformable track template, essentially a map from 
initial parameters $\{z,\theta,\phi,p\}$ to track positions $\vec{x}$. The template should
be ``elastic'' in the sense that $x_i$ is a continuous and differentiable function over the 
domain of the initial condition space $\{z,\theta,\phi,p\}$. In an experiment without a magnetic field, particle tracks are 
straight, and it easy to specify track positions as a function of initial conditions.
At OLYMPUS, the magnetic field was non-uniform, we needed a more sophisticated approach.
We considered using a deterministic numerical simulation of the trajectories, but this
proved to be too computationally intensive and severely limited the speed of the track
fitting. Instead, we pre-computed trajectories on a grid over the four-dimensional space
of initial conditions, and used cubic spline functions to interpolate between grid points. 
The result was a set of continuous and differentiable functions $x_0(z,\theta,\phi,p)$, 
$x_1(z,\theta,\phi,p)$, $x_2(z,\theta,\phi,p) \dots$ that allowed us to forward propagate
a guess of initial conditions to estimates of the hit positions on each detector layer. 
We named this procedure of determining track positions with pre-computed functions ``FastTrack.'' 
The FastTrack functions were suitable elastic arms for EAA. 

Since the energy loss of electrons and protons in passing through matter is slightly different, we developed
a different set of FastTrack functions for each particle type. We found that we could combine
the electron and positron functions by remapping the initial conditions. If we replaced $p$
with $1/p$, then an electron with infinite momentum and a positron with infinite momentum
both mapped to $1/p=0$. By allowing $1/p$ to have positive values for positrons and negative
values for electrons, the fit could start with a lepton of ambiguous charge and fit the 
lepton charge to the data. We used this procedure as well for proton fits, acknowledging that any 
track that had anti-proton-like curvature was most likely that of an electron.

\subsection{Escaping False Minima Using Jump-Scans}

We could study the performance of the track fitter by attempting to reconstruct simulated data. 
We found that even if the drift chambers had perfect time-resolution, some fraction of events were
mis-reconstructed. In this case, by mis-reconstructed, I mean that the algorithm resolved at least
one left-right ambiguity incorrectly, and fit the track to at least one false position, skewing 
the estimated initial conditions. Rebecca Russell is certainly the expert on matters relating
to track reconstruction improvement \cite{russell:thesis}, but I did contribute one idea which
was implemented for the final track reconstruction of the OLYMPUS data. I named this component
of the fitting algorithm ``Jump-scan''. 

In a typical mis-reconstructed event, the fit becomes trapped in a local-minimum. The global minimum
is still the correct solution; however, the high-temperature global minimum is not connected to the
low-temperature global minimum. To try to prevent this, the algorithm pauses at an intermediate 
temperature, then tries jumping out of its current minimum and scanning the neighborhood to find 
a better minimum. After these attempts, EAA proceeds using the best minimum found. 

\begin{figure}[htpb]
\centering
\includegraphics{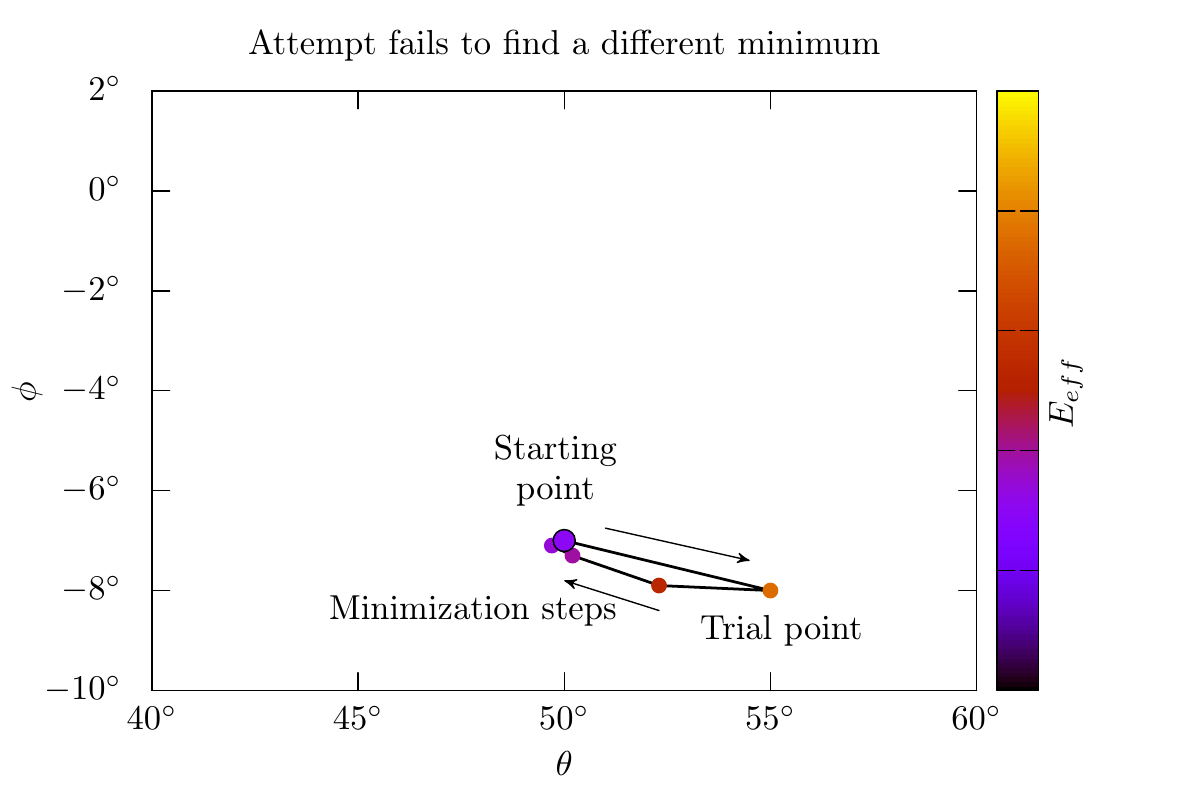}\\
\includegraphics{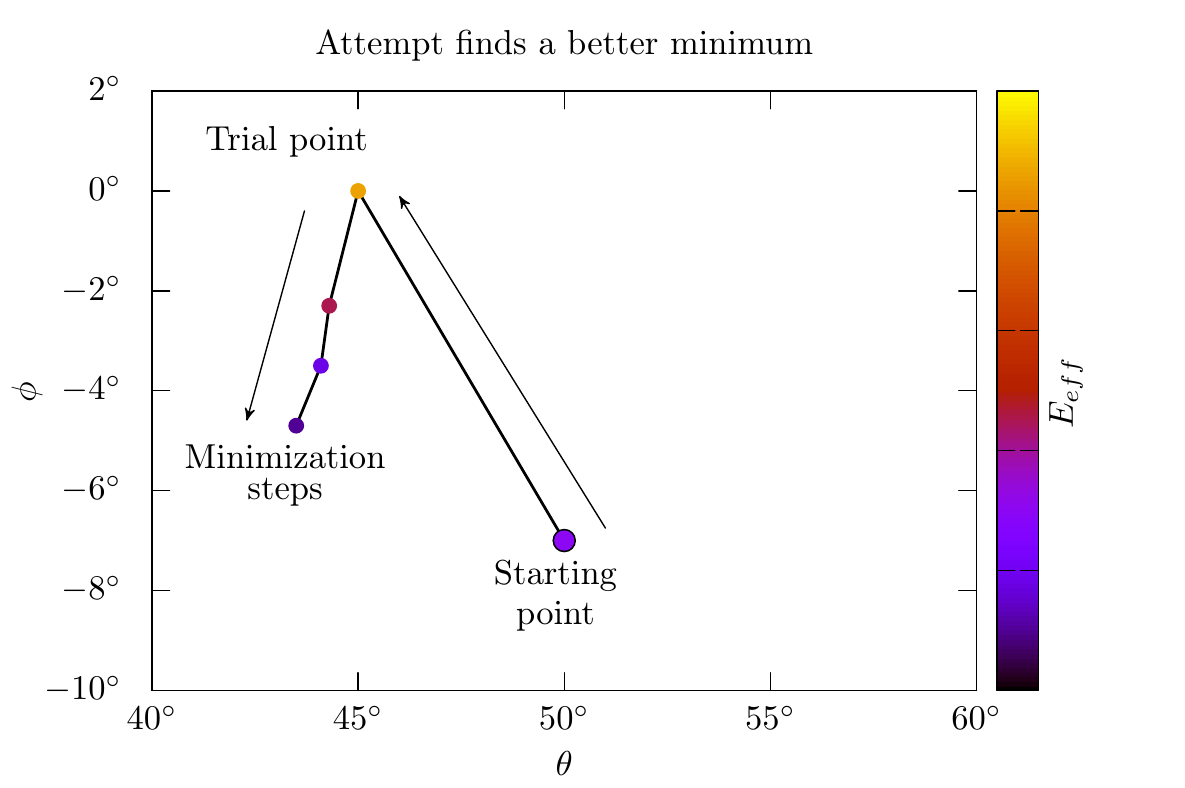}
\caption[Demostration of Jump-scan]{\label{fig:jumpscan} This figure shows two trials (out of 48) 
during Jump-scan, with each step being represented as a circle on the $\theta,\phi$ plane. The trial 
in the top plot fails to find a better minimum, returning to the minimum of the starting point. 
The trial in the bottom plot finds a better minimum, and EAA will continue from this minimum at the
conclusion of Jump-scan. }
\end{figure}

Jump-scan works in the following way. At the chosen intermediate temperature $\beta_j$, the algorithm
will have a best guess for the initial conditions, $\{z_j, \theta_j,\phi_j, p_j\}$. The algorithm will
try minimizing from 48 nearby points to see if a better minimum can be found. The algorithm will jump
from $\theta_j$ to $\theta_j -10^\circ$, $\theta_k-5^\circ$, $\theta_j+5^\circ$, and $\theta_j+10^\circ$,
since mis-reconstructed tracks often have reconstructed scattering angles that are similar to their 
true scattering angles. Reconstructed $\phi$ and $p$ are often wildly off, so rather than jumping to nearby
values, we scan over $\phi=-8^\circ, 0^\circ, 8^\circ$, and $1/p = -1/0.083$~GeV$^{-1}$, $-1/0.250$~GeV$^{-1}$, $1/0.250$~GeV$^{-1}$, 
$1/0.083$~GeV$^{-1}$. Reconstructed values of $\theta$ and $z$ are heavily correlated, so we choose only
to vary $\theta$, and to keep $z=z_j$. This reduced the search space to three dimensions instead of four. 
At each of the 48 trial points, we minimize to the nearest local minimum. We then compare the minima, 
move to the best one, and continue the EAA procedure. An example of how this might proceed is shown
in figure \ref{fig:jumpscan}.

Jump-scan was an inelegant fix, but it was pragmatic. We found that it improved the perfect reconstruction rate of 
simulated tracks from about 90\% to over 99\%. The cost, however, was that it dramatically slowed down
the track reconstruction rate. Jump-scan had to be applied to every track in order to be effective. There was
no way to know (without trying Jump-scan) if the reconstruction had gone awry and a better minimum could be found. 
We never succeeded in finding a better fix, balancing the goals of fast reconstruction and a low mis-reconstruction
rate. 

\section{Time-to-Distance in the Drift Chambers}
\label{sec:ttd}

\subsection{Overview}

In order for track fitting to be successful, it was crucial to have an accurate a set of 
time-to-distance (TTD) functions that could translate drift times into drift distances. 
These functions are approximately linear over a large range of times; a slightly longer
drift time implies a slightly longer drift distance. However, close to the sense wires
and far from the sense wires, drift time and drift distance cease having a linear relationship.
There are several other complications. TTD functions have a dependence on the incident angle, 
$\alpha$, of the track relative to the wire plane as well as a weak dependence on the azimuthal angle $\phi$
(the functions can vary slightly in the direction parallel to the sense wires). 
The magnetic field introduces what is called a Lorentz angle to the drift direction; the drift 
electrons travel at an oblique angle to the electric field, and this angle depends on the magnetic
field strength. Since the magnetic field of the spectrometer is non-uniform, every wire has a 
different time-to-distance relationship. Even the left and right sides of a wire have different
TTD functions. What we want to specify is therefore a function $d$ for each side $s$ of every
wire $w$:
\begin{equation}
  d_{w,s} (t,\alpha, \phi).
\end{equation}

There are two properties of the TTD functions $d_{w,s} (t,\alpha, \phi)$ that can be easily 
surmised:
\begin{enumerate}
\item $d_{w,s} (0,\alpha, \phi) = 0$, i.e., zero drift time implies zero drift distance, regardless of $\alpha$ and $\phi$, 
\item $d_{w,s} (t,\alpha, \phi)$ should monotonically increase with $t$, i.e., longer drift time should always imply longer
drift distance. 
\end{enumerate}

\begin{figure}[t]
\centering
\includegraphics{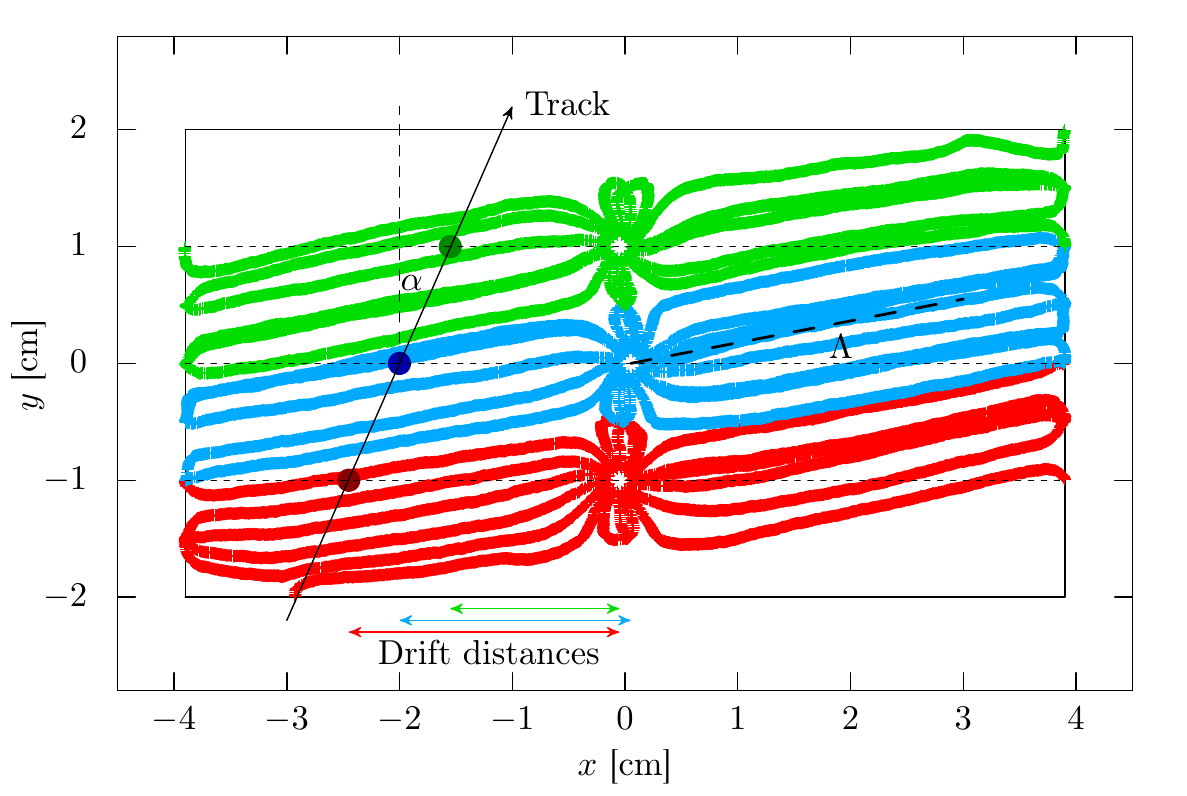}
\caption[Drift lines within a cell]{\label{fig:drift_lines} This figure shows the drift lines in a
drift chamber cell, as calculated by a numerical simulation. The drift distance in the TTD function is 
reckoned from the point at which the track crosses the sense wire plane, shown in circles for an
example track. The TTD function depends on $\alpha$, the incidence angle of the track. The magnetic
field caused the electrons to drift at an oblique angle, $\Lambda$, with respect to the electric field,
which points along $x$ in this coordinate system. Reversing the magnetic field direction would also 
reverse the direction of the Lorentz angle.}
\end{figure}

For the reader to become familiar with the TTD in the context of OLYMPUS, I show a simulation
of the lines of drift in a typical drift chamber cell in figure \ref{fig:drift_lines}. As in figures
\ref{fig:cell_cartoon} and \ref{fig:drift_cell_potential}, I show a cross section of the drift cell,
so that the wires point in and out of the page. The electric field points parallel to the $x$ direction.
However the drift lines are rotated by the Lorentz angle, which I'll denote as $\Lambda$, relative to
the $x$ direction. An example track is shown with an angle of incidence, $\alpha$ relative to the 
wire planes. We define the drift distance, $d$, as the distance between the sense wire and the intersection
of the track with the wire plane. The drift time, $t$, is the time between the passing of the track and the 
arrival of the first ionization electrons at the sense wire. 

In the following sections I will describe three approaches we tried in order to ascertain accurate TTD functions.
The first two approaches were unsuccessful, but the third succeeded. In the first approach, we attempted to calculate 
the TTD functions using an open-source software package called Garfield$++$ \cite{Garfieldpp}. In the second
approach, we attempted to ascertain TTD functions from data, and to describe them using general spline functions.
In the third, successful approach, we developed a simple parameterization for the shape of a TTD function for the 
OLYMPUS drift chambers, and fit the parameters of the model using data. 

\subsection{TTD from Garfield}

Garfield$++$ is an open source software package developed by CERN to allow customized simulations of tracking detectors.
It is written in C$++$ and builds on a previous implementation written in Fortran.
There is a wide range of simulations that Garfield$++$ can perform, both with two and three dimensional geometries. We 
attempted to calculate TTD functions using the following approach. First, we specified the gas mixture and used Garfield$++$'s 
implementation of the software Magboltz \cite{Magboltz} to calculate the response of ionization electrons to electric and 
magnetic fields in that mixture. Next, we specified the two-dimensional geometry of a row of drift cells, the potentials 
on all of the wires, and the local magnetic field. Then, we used Garfield$++$ to simulate the two-dimensional drift of 
ionization electrons from every part of the cell to the sense wires. Lastly, we synthesized the 
drift time information from the simulations to create TTD functions. We performed this procedure for every drift cell,
taking care to specify the local magnetic field of each cell. 

This approach did not produce good results. First, we did not precisely know the gas mixture, since the ethanol
concentration varied over time (see figure \ref{fig:ethanol_church_length}). We ran simulations with a range of
ethanol concentrations, and tried to match different parts of the dataset with different concentrations, but this
never yielded a convincing set of TTD functions. Our calculated TTD functions also suffered because we had to assume a nominal 
description of the geometry. The slight differences in wire positions and voltages between reality and the simulation
contributed to the TTD functions' inaccuracy. In addition to these challenges, our choice of a two-dimensional simulation
prevented us from calculating any of the TTD dependence on $\phi$. We could have attempted a three-dimensional 
simulation, but without having confidence in our approach, we moved on to a different strategy: extracting TTD functions
from data. 

\subsection{Spline Fits to Data}

\begin{figure}[htpb]
\centering
\includegraphics{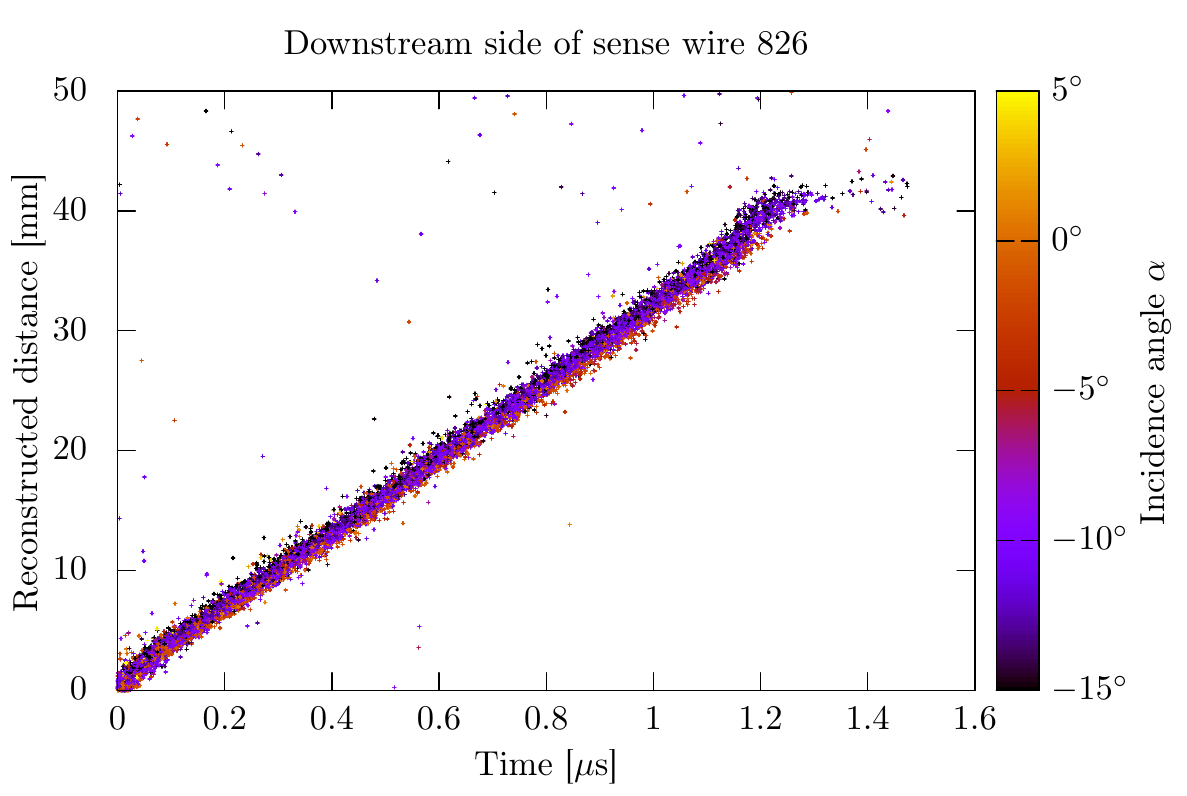}
\caption[Time-to-distance relationship indicated from data]{\label{fig:ttd_data} By making a guess at the 
TTD functions, we could run the track reconstruction and compare the resulting track positions with the
times from data to improve our guess. In this plot, color indicates the incidence angle, $\alpha$, of the
reconstructed track.}
\end{figure}

The strategy of our second approach was to iteratively fit TTD functions to data. We started with a set of guess TTD
functions so that we could run the track reconstruction over data and produce a set of reconstructed tracks. The 
track positions could be compared with the drift times in the data to update our TTD functions. An example of
such a comparison, made for one side of one sense wire, is shown in figure \ref{fig:ttd_data}. This procedure could
be iteratively performed until the TTD functions converged to a solution.

For this approach to work, we needed to choose a functional form for our TTD functions. We chose cubic spline functions,
an extremely general choice that put no constraints on the shape that the TTD function could take. Our hope was that
the spline functions could easily accommodate slight deviations produced by differences in the wire positions or changes
in the magnetic field along the length of the wire. 

\begin{figure}[p]
\centering
\includegraphics{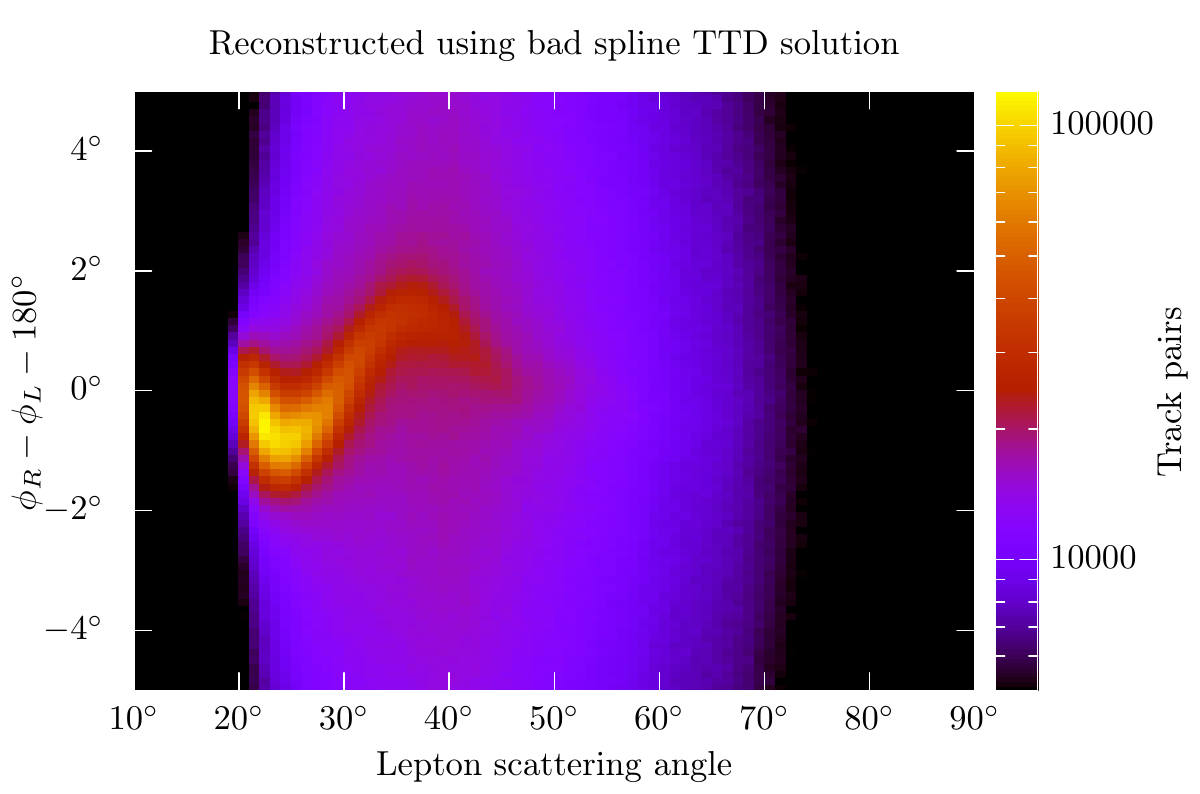}\\
\includegraphics{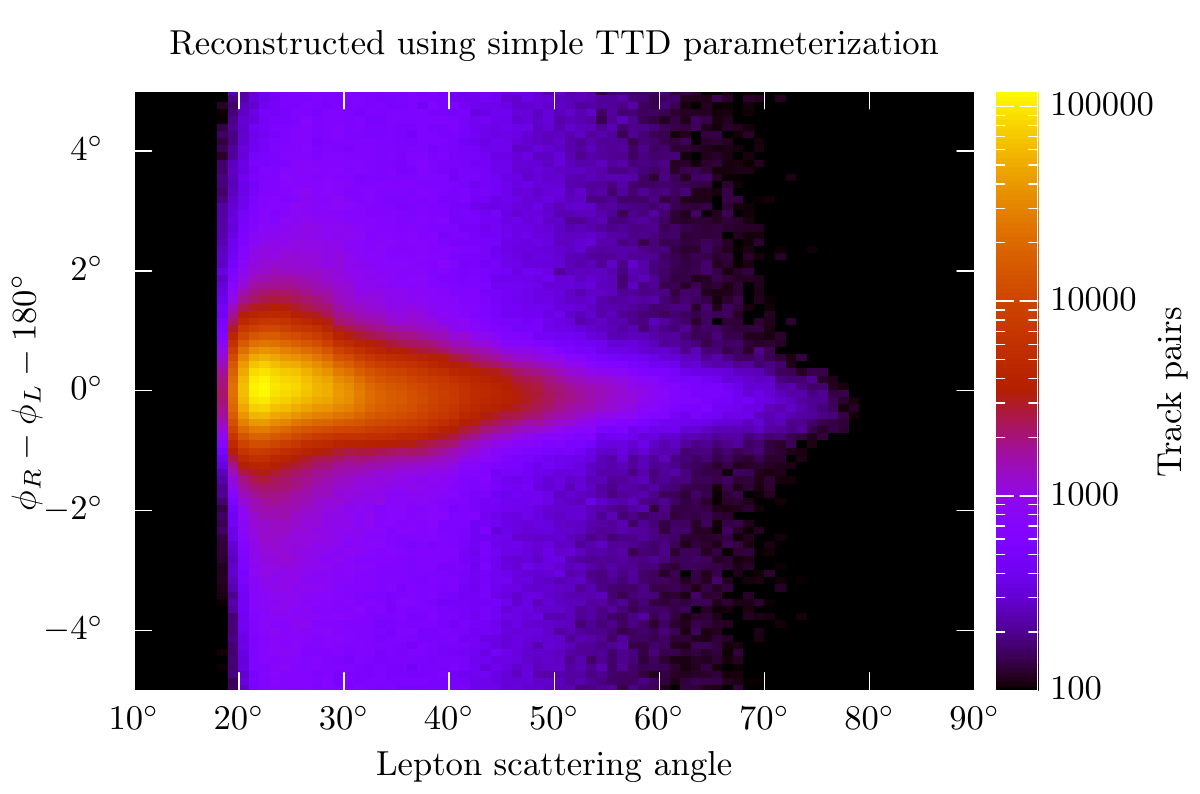}
\caption[Oscillating coplanarity distribution produced by bad TTD functions]
{\label{fig:coplanarity_wiggles} The top plot shows the coplanarity distribution of 
track pairs for one particular spline solution. The TTD function introduced a bizarre
oscillation into the distribution as a function of lepton-scattering angle. For comparison,
the bottom plot shows the coplanarity distribution produced when using the TTD function described in 
section \ref{ssec:ttd_axel}.}
\end{figure}

Using this approach, the TTD functions always converged to a solution after a few iterations. However, the quality of these 
solutions was suspect. With the functions having so much freedom, it was easy for them to converge to a solution that was
not a global optimum. As an example, in one such solution the mean of the coplanarity distribution of lepton and proton
track pairs oscillated by several degrees as a function of the lepton scattering angle, shown in the top plot of figure
\ref{fig:coplanarity_wiggles}. The origin of this oscillation was determined to be a set of faulty ToF timing calibrations 
that biased the track reconstruction. When the calibrations were corrected, and TTD fitting iterations were resumed, the
oscillations persisted. This indicated that stable non-optimal solutions were possible for the TTD parameters, which meant that 
quality of the initial TTD guess was critically important. We would have had more confidence in the fitting procedure if the
many different initial guesses all converged to the same solution. Instead, we began to worry about how to gauge the quality
of various TTD guesses. 

\begin{figure}[htpb]
\centering
\includegraphics{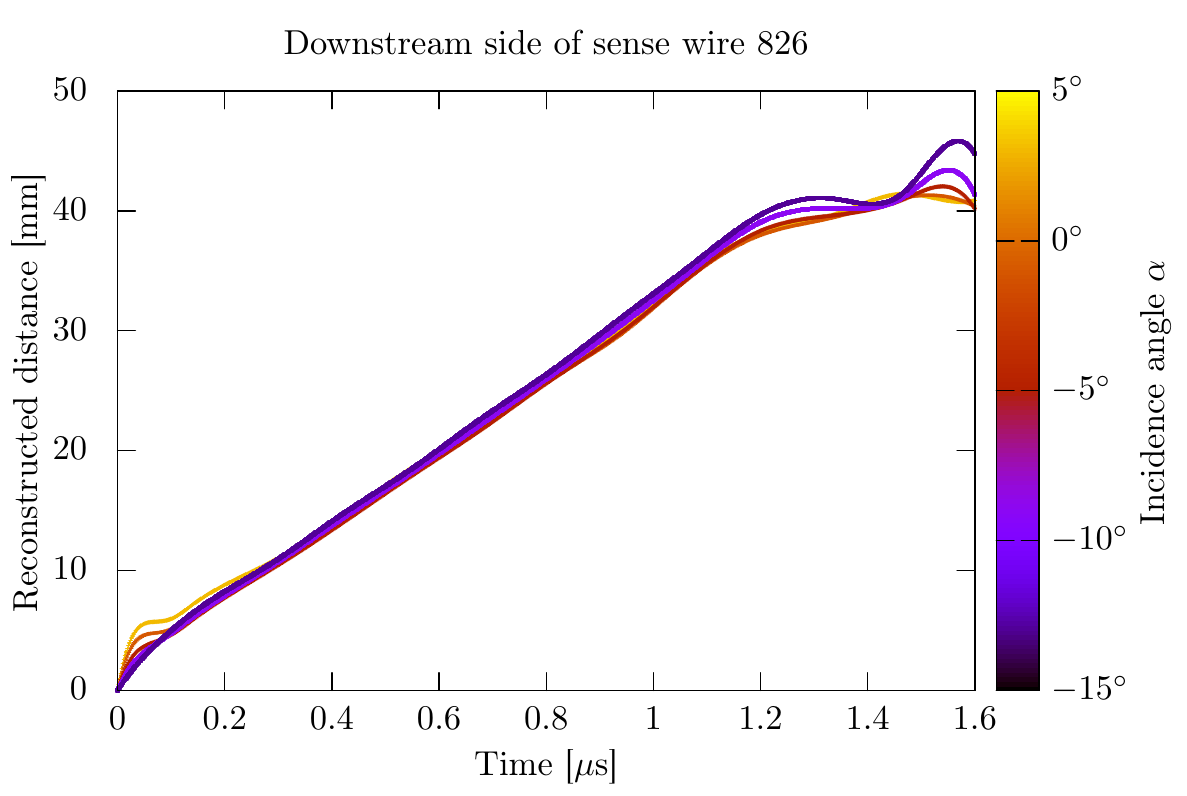}
\caption[Time-to-distance relationship in a spline fit]{\label{fig:ttd_spline} The spline fits could over-fit
the data in regions where the data was sparse. For this particular wire, the function does not monotonically
increase, which is problematic.}
\end{figure}

Another reason to be suspicious of the spline functions was that they did not necessarily monotonically increase with 
increasing time. The spline functions would frequently have regions, especially with large drift times, in which longer times
could imply shorter distances. This was caused by the lack of residual data with large drift times. In regions with
dense data, the splines were constrained to be well-behaved. In regions where the data was sparse, the splines had the
freedom to overfit the data and exhibit wild behavior. This can be seen in figure \ref{fig:ttd_spline}, which shows the
spline fit to the data of figure \ref{fig:ttd_data}.  If we had found an effective way to regularize the splines, then 
perhaps this approach would have yielded accurate TTD functions, but instead we chose to pursue a third approach, which was 
ultimately successful.

\subsection{Simple Parameterization Fit to Data}

\label{ssec:ttd_axel}

In our most successful approach for determining time-to-distance functions, we first made a simple parameterization
for the shape of a function, and then fit the parameters for each side of each wire to reconstructed data. We tried
to choose a shape that matched all of the qualitative features that we observed in reconstruction data, but that lacked the 
freedom to overfit sparse regions of data. To guide our parameterization, we started with a simple two-dimensional geometric model,
and then allowed the parameters of that model to vary quadratically in $\phi$. A derivation of the parameterization is 
given in appendix \ref{app:axelttd}. 

The simple parameterization is defined piece-wise into three distinct regions. For times smaller than 0.2~$\mu$s, the function
was a cubic polynomial in $t$. For intermediate times, the function was linear in $t$. For large times, the function 
had a constant value, $d_\text{max}$. The parameterization had 10 free parameters, with physical interpretations:
\begin{align}
v(\phi) &= v_0 + v_1\phi + v_2\phi^2 & \text{ representing the drift velocity} \\
w(\phi) &= w_0 + w_1\phi + w_2\phi^2 & \text{ representing the width of the jet of drift lines} \\
\Lambda(\phi) &= \Lambda_0 + \Lambda_1\phi + \Lambda_2\phi^2 & \text{ representing the Lorentz angle} \\
r & & \text{representing the radius of the non-linear region}.
\end{align}

\begin{figure}[htpb]
\centering
\includegraphics{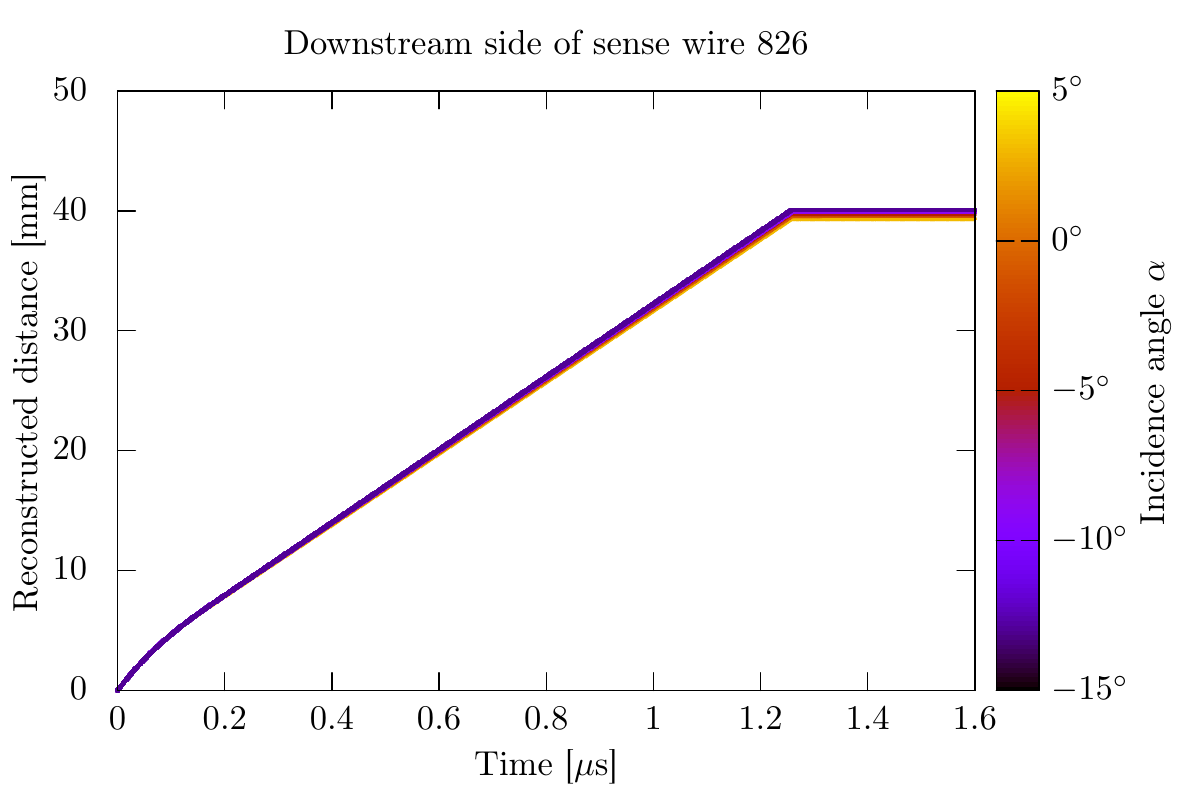}
\caption[Simple TTD parameterization]{\label{fig:ttd_axel} The TTD functions of the simple parameterization were 
well-behaved even after fitting the free parameters to data.}
\end{figure}

The parameters were iteratively fit to reconstructed tracks. We started by fitting the parameters to results of
a Garfield$++$ simulation. Then using that initial guess, we reconstructed a set of tracks, and then updated the TTD
parameters. Each side of each sense wire was allotted its own set of parameters. Furthermore, we 
divided the OLYMPUS data into segments in time based on the approximate ethanol concentration in the drift chambers
and fit the TTD parameters of each segment separately. We found that the TTD parameters converged to a stable solution
within about five iterations. An example TTD function is shown in figure \ref{fig:ttd_axel}. 

The simple parameterization had desirable properties. The parameters could vary to accommodate slight differences
between drift cells, but the TTD functions remained constrained and well-behaved. The kinematic distributions of
reconstructed tracks were sensible; the coplanarity distribution peaked at $0^\circ$ as expected (shown in the bottom
plot of figure \ref{fig:coplanarity_wiggles}) and no longer had bizarre oscillations. The parameterization yielded 
hit position resolutions on the order of 0.5--1.0~mm, depending on the wire. 

The simple parameterization yielded the most effective set TTD functions that we produced. These functions were used
for reconstructing tracks for the analysis discussed in chapter \ref{chap:analysis} and the results shown in chapter 
\ref{chap:results}.

\chapter{Analysis}

\label{chap:analysis}

\section{Analysis Strategy}

\begin{figure}[htpb]
\centering
\includegraphics[width=\textwidth]{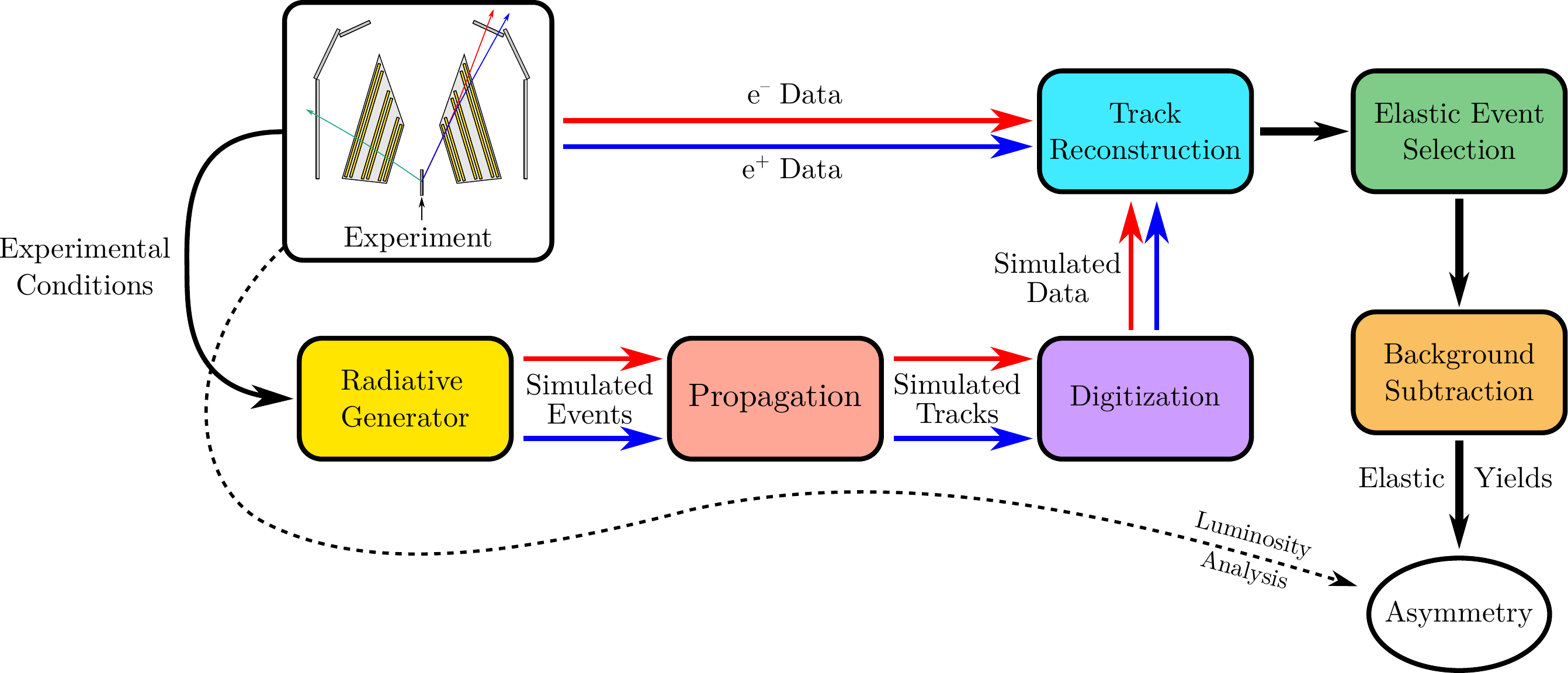}
\caption[Schematic of the analysis chain]
{\label{fig:analysis_schematic} The track reconstruction, elastic event selection, and background subtraction
  were performed on both data and simulated data using the same software. The final asymmetry I will present is that 
  of data minus that of simulation.}
\end{figure}

In this chapter, I want to lay out the steps that take the raw data from the experiment and convert them to 
a lepton sign asymmetry. Some of these steps (for example, the radiative corrections) were described in previous
chapters. The goal of this chapter is to both cover the remaining steps, and to orient the reader as to how
the entire analysis chain fits together. 

In chapter \ref{chap:rc}, I discussed how the OLYMPUS radiative corrections were necessarily convolved with
detector attributes: acceptance, efficiency, resolution. Our solution to this problem is to simulate these
convolutions numerically using Monte Carlo, and then to report an asymmetry that is adjusted for the asymmetry
we expect to find through simulation. For the sake of clarity, let's look at some examples in detail, starting with
the generator. The events generated in simulation have a slight asymmetry (from soft TPE and from bremsstrahlung 
interference), which varies with $Q^2$. This asymmetry is convolved with, among other things, inefficiency over an area
of the detector---say an inactive cell in the drift chambers---since the magnetic field bends electrons and positrons at the same 
$Q^2$ to different parts of the detector. The simulated asymmetry is not caused by hard two-photon exchange, and so
must be subtracted from the measured asymmetry.

The general road map for the analysis is shown in the diagram in figure \ref{fig:analysis_schematic}. 
In this map the chain moves from the experiment itself, top left, to the result, the asymmetry, at the
bottom right. The experimental conditions (beam position, target density, detector settings, etc.) from
the experiment serve as input settings to the radiative generator, described in section \ref{sec:olympus_gen}.
The generator produces simulated events: lists of particles, their vertex, and their momentum vectors.
At the next stage, the ``propagation'', the trajectories of these particles are simulated, taking into account
the particles' response to magnetic fields and the physical matter of the spectrometer. The simulated trajectories,
also called ``tracks,'' pass next to the ``digitization'', which simulates the detector signals that would be produced 
in order to create simulated data in the exact same format as the experimental data. The data are fed to the 
track reconstruction (the subject of chapter \ref{chap:recon}), which tries to invert the processes of digitization and 
propagation, attempting to find the likeliest of initial
conditions for a particle, given the set of detector signals in an event. The reconstructed events
are fed next to a piece of software that performs an elastic event selection. This software acts like a filter,
accepting events whose reconstructed initial conditions fit those of an elastic collision and rejecting those
that look like background. This filter is not perfect, and some residual background must be estimated 
and subtracted. The result is yields of elastic events, binned as a function of $Q^2$ (or $\epsilon$ or $\theta_l$, or
whatever kinematic variable one chooses.) The experimental and simulated yields are combined with the 
results from the luminosity analysis to form an asymmetry, according to:
\begin{equation}
  \label{eq:asym_def}
  A_{2\gamma} = \frac{ \frac{N^\text{data}_{e^+p}}{\mathcal{L}_{e^+p}} - \frac{N^\text{data}_{e^-p}}{\mathcal{L}_{e^-p}} }
  { \frac{N^\text{data}_{e^+p}}{\mathcal{L}_{e^+p}} + \frac{N^\text{data}_{e^-p}}{\mathcal{L}_{e^-p}} }
   - \frac{ \sigma^\text{sim.}_{e^+p} - \sigma^\text{sim.}_{e^-p} }{\sigma^\text{sim.}_{e^+p} - \sigma^\text{sim.}_{e^-p} }.
\end{equation}

In the following sections, I will explain, with more detail, how each of these components in the analysis 
chain works. I will spend particular detail on the elastic event selection, which is solely my own work. 
A number of my colleagues in the collaboration have designed independent elastic event selection software,
with the hope that each independent procedure can help cross check the others. It is anticipated that some
combination of these procedures will be used to produce the results that will soon be submitted for publication.
I see this chapter as an opportunity to document my design and explain my approach. I will conclude this chapter
by mentioning two cross checks of the analysis that can be performed without biasing the result.

\section{Simulation Chain}

In this section, I will describe that chain of software that we used to produce our simulated data set. 
Even for the simulated data, the first stage in the chain was the experiment. The running conditions
of the experiment changed over time while we collected data; the beam position varied slightly, the 
flow rate of gas into the target was varied, and the settings of the different detectors were occasionally 
adjusted. To account for this variation, we chose to vary the running conditions of
the simulation in exactly the same way as they varied during the experiment. The experimental data are 
naturally broken up into segments, called
``runs'', and we matched the simulation run-by-run to the experiment. For data run 8652, we also produced
a simulated run 8652. If the beam moved slightly midway through run 8652, we adjusted the simulated beam 
position midway through the simulated run 8652. The conditions of each simulated run matched the conditions
of each data run.

That does not mean that the events in simulated run 8652 corresponded to any of events in data run 8652. 
The specific kinematics of each simulated event were generated randomly, and only the running conditions 
matched the data run. Furthermore, we did not want our simulation to hamper the statistical precision
of our result, so the number of elastic events generated for simulation far out-numbered the elastic 
events in our experimental data set. 

The first piece of software in the simulation chain was the radiative generator, which was covered extensively
in chapter \ref{chap:rc}. The output of the generator was, for each event, a vertex position, the momentum
vectors for the lepton and proton (as well as the radiated photon, which, for all analyses except that 
of the SyMB, could be ignored), and a list of weights. These events then passed down the chain to the 
propagation and digitization stages, which will be described in the following sections.

\subsection{Propagation}

\label{ssec:prop}

The next step after generating events is to simulate how those events would propagate through the material
and magnetic field of the spectrometer. This propagation step takes a particle's initial vertex and momentum
vector and simulates its trajectory. The equations of motion are non-trivial and must be solved numerically.
The magnetic field produces curvature in the trajectories, which depends on a particle's momentum. As 
particles pass through matter, they lose kinetic energy and, with some probability, can experience secondary 
scattering. The trajectory is therefore non-deterministic. The simulation should also exhibit this same 
non-determinism to fully reproduce the characteristics of the experimental data.

The engine of the propagation is an open-source software library, GEANT4 \cite{Agostinelli:2002hh}, which
can numerically and non-deterministically simulate the passage of particles through matter and electromagnetic
fields. As inputs to GEANT4, we provide the geometric map of the detector (discussed in section \ref{sec:survey}),
and the description of the magnetic field (from the interpolation scheme of section \ref{ssec:magnet_spline_interp}),
as well as the list of particles to simulate for each event. We ask, as output from GEANT4, the positions and momentum
vectors at important points along a particle's trajectory (e.g.\ the wire planes of the drift chambers) as well as  other
quantities of interest (e.g.\ the kinetic energy loss in the ToF scintillators). 

\subsection{Digitization}

The digitization step converts simulated trajectories from the propagation step and calculates the resulting
digital signals that would be produced by our detector electronics. The digitization simulates the TDC and ADC
values that our detectors would produce from the simulated track. Just like the generation and propagation
steps, this procedure also needs to be non-deterministic.

Each detector system has its own digitization procedure, and, rather than discussing any single digitization
scheme in detail here, I will present some illustrative examples of what may be included in such a scheme. 
For the drift chambers, the task is to take trajectory positions and to determine the corresponding TDC values
for drift times to the relevant wires. This step requires inverting the time-to-distance function and applying
the correct smearing to the resulting time. For the SyMBs, the digitization requires estimating the ADC signals
from the light yield by Cherenkov radiation in the lead fluoride crystals (see Colton O'Connor's thesis for more
detail \cite{oconnor:thesis}). For the ToFs, the digitization step includes the simulation of both TDC and ADC
signals. The TDC signals must reflect the time-of-flight of the particle as well as the arrival time of scintillation
light at the PMTs. The ADC signals must reflect the amount of scintillation light produced and account for attenuation
of that light as it travels to the PMTs (see the theses of Lauren Ice and Rebecca Russell for more detail
\cite{ice:thesis, russell:thesis}).

All detectors have a probability for being inefficient, i.e., failing to record a signal from a particle that 
passed through them. This inefficiency is generally random, but can often be correlated over a region of a detector. For
example, in many drift chamber cells the three sense wires either all recorded hits from a passing track or
all failed to record a hit. Spatial maps of detector inefficiencies, which accounted for correlations, were an 
important ingredient to the digitization. These maps were estimated from the experimental data, either via
alternate or less stringent triggers, and/or by using the track reconstruction to localize particle trajectories
at various positions in the detector, which either recorded a signal or did not (for more detail, see Brian
Henderson's thesis \cite{henderson:thesis}).

\section{Elastic Event Selection}

\label{sec:elastic_event_selection}

The software that performs elastic event selection acts like a filter, allowing elastic events, the signal, to pass while
rejecting background. Generally, it is more important to preserve signal than it is reduce background, simply because if 
some signal is being discarded, it is difficult to prove that the same fraction is being discarded for both electron and 
positron data, and in both the experimental and simulated data sets. 

Still, removing as much background as possible (without cutting into signal) is important. The final step 
in the analysis is background subtraction, and an analysis is hardly convincing if the majority of the elastic yield
is made up of background contamination. The relationships between the kinematic variables are well-known for the elastic
signal (particles have momenta that are strictly correlated with their angles), but the background can come
from many different sources, and can have kinematic correlations that are harder to discern. It is better
to remove as much background as possible at the elastic event selection phase, than to have to subtract
large amounts of an unknown background away.

The most conceptually simple way to segregate signal and background is by using some selection criterion to make
a cut, illustrated by the cartoon in figure \ref{fig:rc_example}. For example, in elastic scattering the sum of 
the momenta of the lepton and proton must equal the momentum of the beam lepton ($\approx 2$~GeV/$c$ in the beam 
direction). I choose to set a cut at $1.4$~GeV/$c$: 
if the reconstructed momenta in the beam direction sum to less than 1.4~GeV/$c$, then I label that event a 
background event and discard it from my elastic sample. In this section, I will describe the various cuts
I choose to make to reduce background.

One pitfall to be avoided, when designing a cutting procedure, is that the cross section changes dramatically
over the full range of the OLYMPUS acceptance. If looking at a distribution in some variable,
say the sum of reconstructed momenta in the beam direction, the distribution will be dominated by low-$Q^2$ events.
The rarer high $Q^2$ events will be completely obscured. A cut at the ideal position for low-$Q^2$ events might
remove some high-$Q^2$ signal. For that reason, the choice of cut placement must take into account the 
change in a variable's distribution over the entire range of $Q^2$. I remind the reader that, in elastic scattering
at fixed beam energy, there is only one free kinematic variable. Therefore, it is equivalent to consider evolution
over $Q^2$, or over the proton polar angle, or over any one kinematic variable one chooses. The specific choice of variable is not important,
but it is crucial that evolution over some kinematic variable be considered.

Another pitfall to be avoided is the fact that distributions can be slightly different between $e^-$ and $e^+$ running,
between experimental and simulated data, and between the two sectors of the spectrometer. The way that I choose to 
address this is by making what I call ``fitted cuts''. For example, the distribution for the energy of the beam as 
reconstructed from scattering angles,
\[
E_1(\theta_l,\theta_p) \equiv m_p  \left( \cot \frac{\theta_l}{2} \cot \theta_p  - 1 \right),
\]
changes depending not only on which slice in $Q^2$ one considers, but also on the beam species, on the sector that
the lepton traverses, and on whether one is considering data or simulation.  The shape of the distribution is approximately
gaussian and so by fitting this distribution with gaussian function for many slices of $Q^2$, for all of these 
different conditions, a cut can be made at some number of standard deviations from the mean for each distribution.
The advantage is that slight differences in distributions by sector, species, etc.\ are accounted for. The downside
is that the cut position is dependent on the success of a fit, which may be influenced by things like the
background contribution, the choice of functional form, or the statistics present in a $Q^2$ slice. In some
areas of the event selection I believe that this approach is worth the costs, while in some other areas I make
simple axis-parallel cuts that are independent of beam species, etc. 

In the following subsections, I will go into detail about the procedure I designed. 

\subsection{Finding the Best Pair of Tracks in an Event}

\begin{figure}[htpb]
\centering
\includegraphics[width=\textwidth]{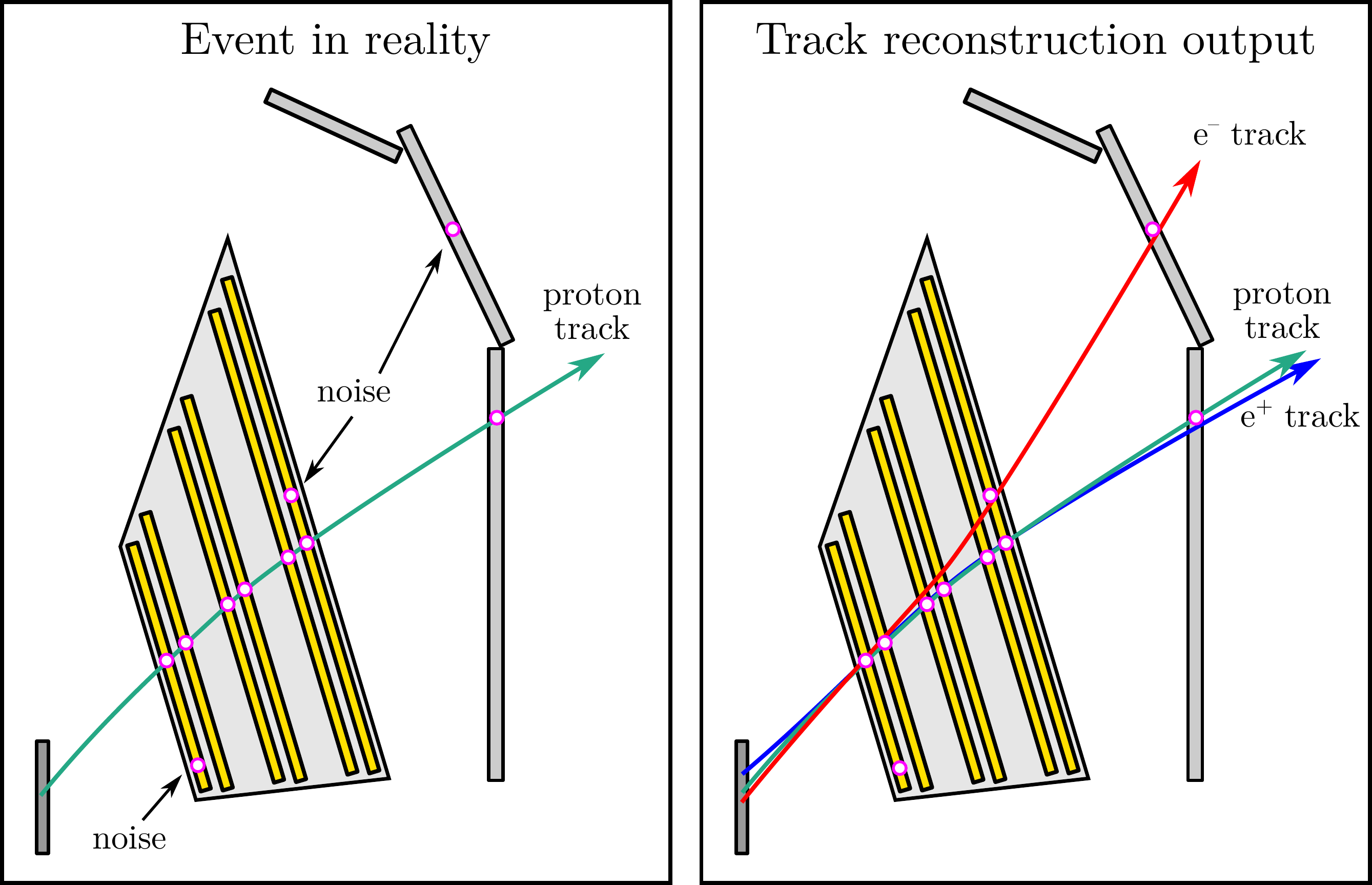}
\caption[Illustration of typical output from the track reconstruction]
{\label{fig:tracks_vs_truth} This cartoon attempts to illustrate the output of the track
reconstruction. The left panel shows an event in which a proton traverses the right sector of the
spectrometer. The proton induces signals in the detector all along the track. There are also noise
signals. The track reconstruction has to interpret those signals and find the likeliest trajectory
that could have produced them. In an event such as this, our track reconstruction might output 
three tracks, as shown in the right panel. The reconstruction identifies the correct hits for the
true track, and fits them with both a lepton template (resulting in a positron track) and a proton
template. The proton track is a better fit to the detector signals than the positron track. Two of the
noise hits also confuse the track reconstruction into thinking that a second particle has passed through
the right sector. The reconstruction attempts to fit these signals with a lepton template (resulting in an 
electron track), and proton template. In this particular example, the reconstruction fails to find a
proton solution. The event selection software must cope with the fact that multiple tracks may all
correspond to the same original particle.}
\end{figure}

The first stage of the elastic event selection is to find the best pair of tracks in an event. As a consequence
of the track reconstruction procedure, multiple tracks might be found, all corresponding to one true trajectory.
Figure \ref{fig:tracks_vs_truth} shows a cartoon which attempts to illustrate the behavior of the track
reconstruction. Since the reconstruction attempts
to fit every track candidate as if it were a lepton and as if it were a proton, the tracks typically come in pairs. Sometimes
the combination of signals in data, or some spurious noise close to a trajectory, fool the reconstruction into thinking 
two particles passed through a sector when only one did. The first stage of the elastic event selection is to
sort out the tracks found by the reconstruction and find the best-fit lepton-proton pair from the jumble. This is
done by looping over every possible combination of left-sector lepton and right-sector proton, as well as every
possible combination of right-sector lepton and left-sector proton, in order to find the best pair.

\subsubsection{Loose Cuts}

\begin{figure}[htpb]
\centering
\includegraphics{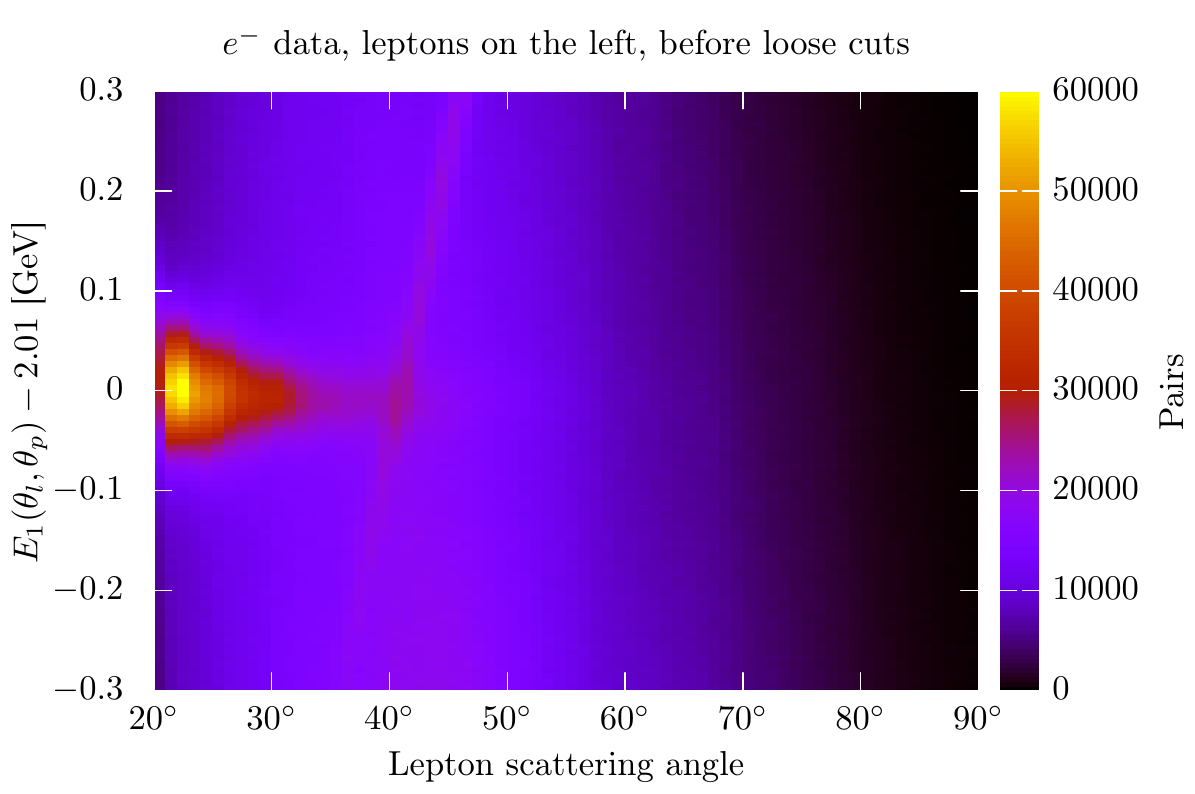}\\
\includegraphics{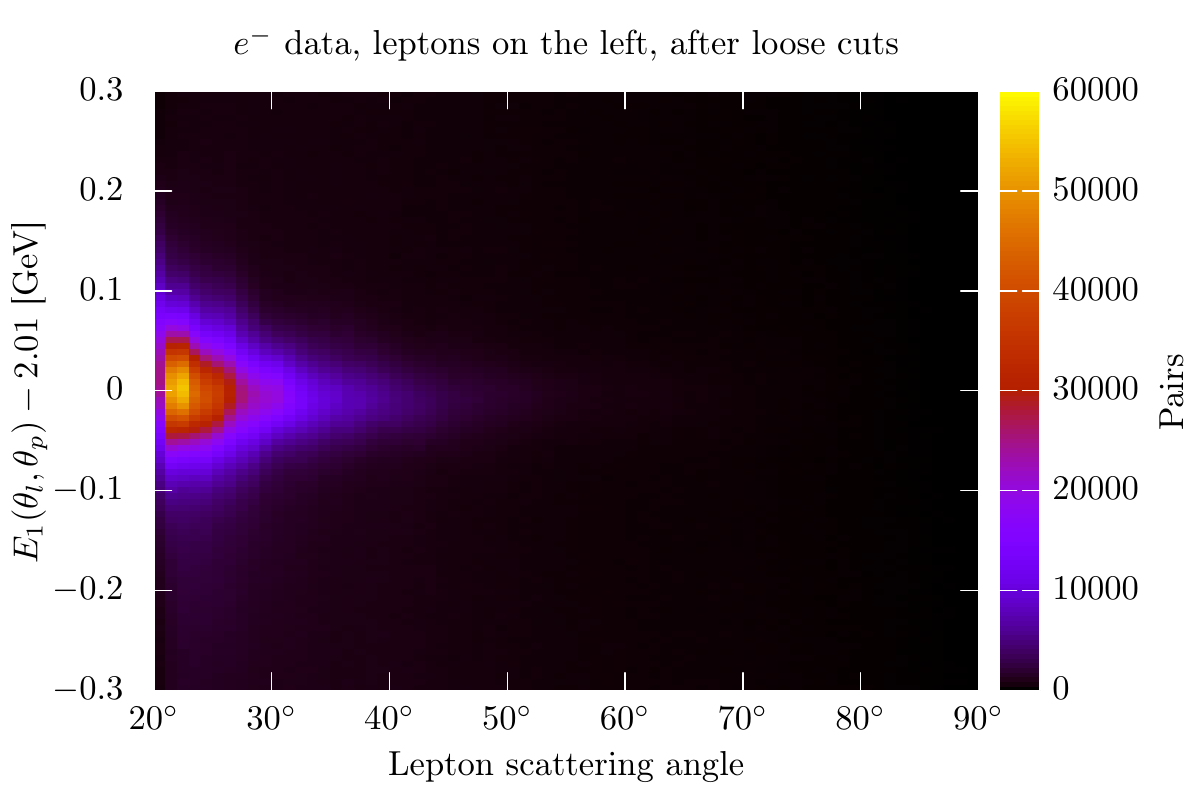}
\caption[The effect of making loose elastic cuts]{\label{fig:loose_cuts} Making a set of loose cuts in seven variable dramatically reduces
the non-elastic background so that elastic distributions can be more easily fit. The top plot shows
all electron-proton pairs, while the bottom plot shows pairs surviving the loose cuts.}
\end{figure}

The first step I take at this stage is to make a set of loose cuts in a set of constrained track
variables, the effect of which can be seen in figure \ref{fig:loose_cuts}. 
The purpose of this is to reduce the combinatoric background with no loss of signal so that
the distributions of elastic events become prominent over background. The space of constrained variables is seven dimensional;
making cuts on seven independent variables is sufficient to span the space. For clarity, let us do some dimensional counting.
The track reconstruction estimates the momentum vector, and vertex position of a track, for a total of four
quantities per track. The time-of-flight scintillator additionally provides the flight time for a total of five per track.
An elastic scattering event has two tracks for a total of ten quantities. However, elastic scattering can 
occur at any value of polar and azimuthal angles, and the vertex position may be anywhere along the target.
Of the ten quantities that the reconstruction provides, three of them are free, leaving seven available for 
making selection cuts. I choose to make loose cuts in the following seven variables (in units where $c=1$):
\begin{enumerate}
\item Vertex correlation: $|z_L - z_R| < 100$~mm
\item Azimuthal correlation: $|\phi_R - \phi_L - 180^\circ| < 6^\circ$
\item Beam energy from angles: $|E_1(\theta_l, \theta_p) - 2.01~\text{GeV}| < 0.3$~GeV
\item Lepton mass-squared from ToF timing: $| |\vec{p}_l|^2 (1/\beta_l^2 - 1) - m_e^2 | < 1$~GeV$^2$
\item Proton mass-squared from ToF timing: $| |\vec{p}_p|^2 (1/\beta_p^2 - 1) - m_p^2 | < 1.5$~GeV$^2$
\item Lepton inverse momentum\footnote{I choose to cut in the inverse of momentum rather than momentum, 
  since sagitta, which is proportional to inverse momentum, is the quantity measured by the drift chambers.}: 
  $| 1/p_l - 1/p_l(\theta_l)| < 1$~GeV$^{-1}$
\item Proton inverse momentum: $| 1/p_p - 1/p_p(\theta_p)| < 2$~GeV$^{-1}$.
\end{enumerate}
The placement of these cuts was designed to be at least five standard deviations from the means
over the full range of lepton scattering angles and in most cases the placement is closer to 7--10 
standard deviations. After these loose cuts are performed, the elastic events are prominent in the any distribution of kinematic 
variables, and this allows stable fitting of elastic peaks. 

\subsubsection{Particle-ID}

\begin{figure}[htpb]
  \centering
  \includegraphics{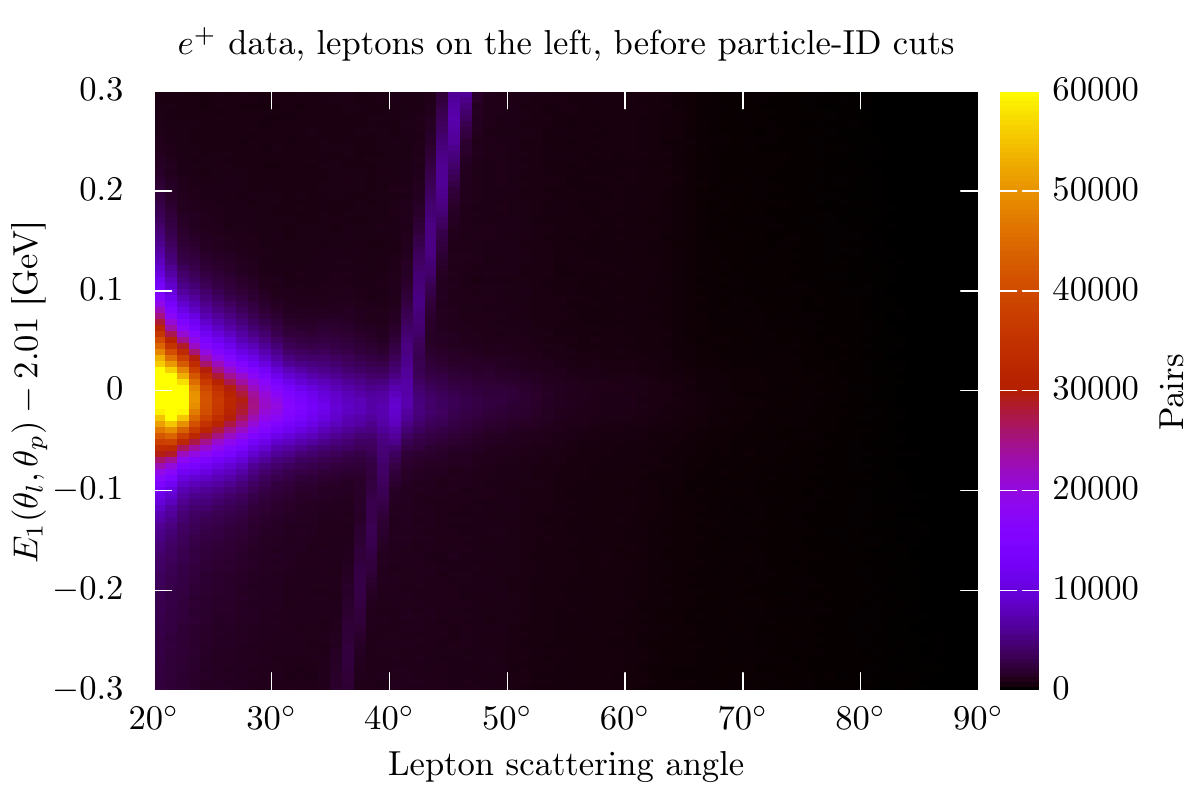}\\
  \includegraphics{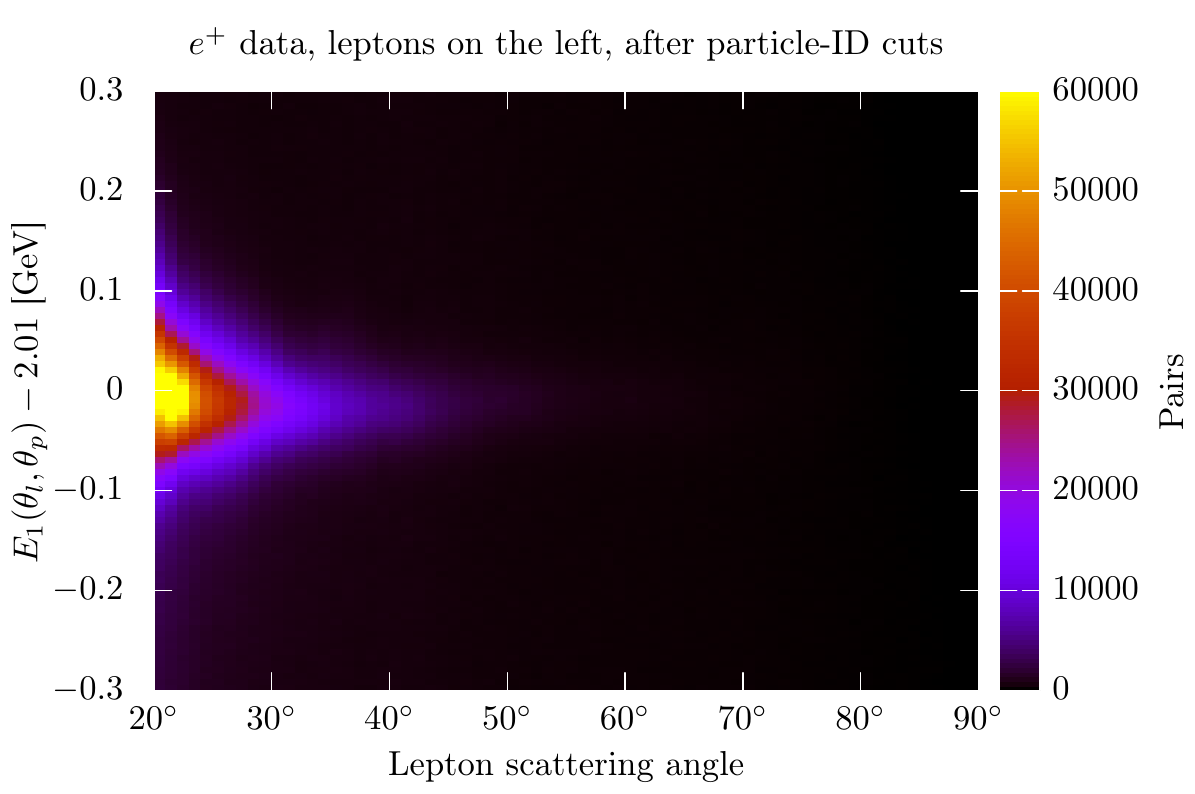}
  \caption[The effect of making PID cuts]{\label{fig:pid_cuts} The top plot shows data with positrons in the left sector, after loose cuts
    have been made. The stripe running through the distribution, at approximately $40^\circ$, is made up of combinatoric
    pairs in which the particle-ID assignments are backwards. This is a problem in positron running, since the positron
    and proton cannot be distinguished from curvature direction. In the bottom plot, which shows the distribution after
    particle-ID cuts have been applied, the stripe has been effectively removed. }
\end{figure}

The next step is to remove combinations in which the particle-ID assignments are reversed. The track reconstruction
does not attempt to make any particle-ID decisions; every track candidate is fit first as a lepton and again as
a proton, and both tracks are provided as output. Therefore, as I step combinatorially over all lepton-proton track
pairs, some will consist of a true lepton, tracked as a proton, matched to a true proton, tracked as a lepton. To 
exclude these combinations, I use timing information from the ToFs to reconstruct a particle's squared mass from 
its time-of-flight and momentum. 

\begin{figure}[htpb]
\centering
\includegraphics{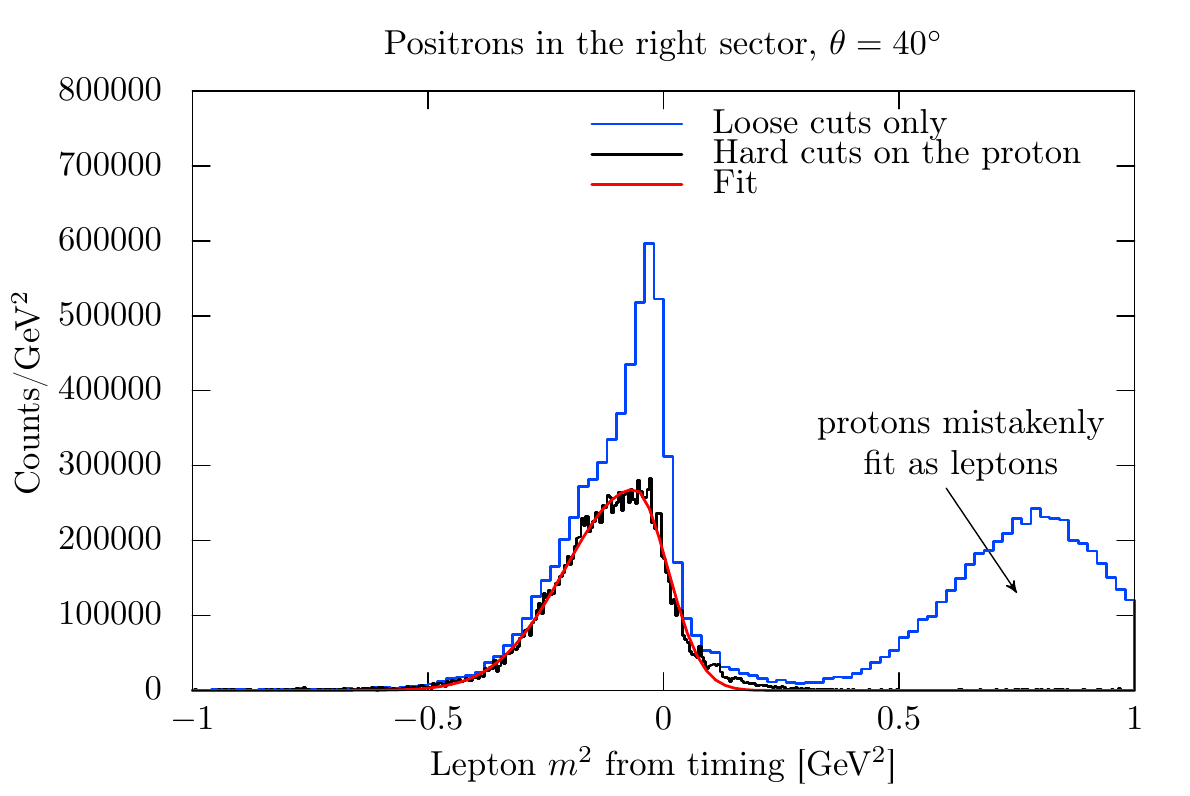}
\caption[The effect of hard proton arm cuts on the lepton $m^2$ distribution]
{\label{fig:hard_pro_pid_study} By making hard cuts on the proton in a pair, the lepton mass-squared
distribution cleans up dramatically. The peak caused by protons mis-identified as leptons disappears. Notice
also that at this angle, the lepton mass-squared reconstructs lower than its expected value of $m_e^2 \approx 0$.
This is just one example of distributions that are slightly offset from expectation due to slight bias in the
track reconstruction.}
\end{figure}

I choose to make a fitted cut, and furthermore, since the track reconstruction in one sector is independent
of the track reconstruction of the other, I know that the lepton mass-squared is reconstructed independently
from the proton mass-squared. To fit the distributions, I look at mass-squared as a function of angle in the
set of events where the other sector has been heavily restricted. For example, to fit the lepton mass-squared
in the left sector, I look at events in which the right sector has a proton which matches the following criteria:
\begin{itemize}
\item $| |\vec{p}_p|^2 (1/\beta_p^2 - 1) - m_p^2 | < 0.4$~GeV$^2$
\item $| 1/p_p - 1/p_p(\theta_p)| < 0.05$~GeV$^{-1}$.
\end{itemize}
This dramatically cleans up the lepton sample in order to make stable fits. To fit the proton distributions,
I require that the opposite sector lepton match:
\begin{itemize}
\item $| \vec{p}_l|^2 (1/\beta_l^2 - 1) - m_e^2 < 0.2$~GeV$^2$
\item $| \vec{p}_l|^2 (1/\beta_l^2 - 1) - m_e^2 > -0.3$~GeV$^2$
\item $| 1/p_l - 1/p_l(\theta_l)| < 0.03$~GeV$^{-1}$.
\end{itemize}
I fit the distributions in slices of $\theta$ with an asymmetric gaussian plus a constant background. 
An illustration of a fit for right sector positrons at $\theta_l=40^\circ$ is shown in figure \ref{fig:hard_pro_pid_study}.

I construct a score for each pair based on the how well the reconstructed masses match the means ($\mu$) and widths
($\sigma$) of the fitted distributions:
\begin{align}
s_l \equiv & \frac{\text{max}( m^2_l - \mu_l(\theta_l),0)}{\sigma_l(\theta_l)}\\
s_p \equiv & \frac{\text{min}( m^2_p - \mu_p(\theta_p),0)}{\sigma_p(\theta_l)}\\
s \equiv & \sqrt{ s_l^2 + s_p^2 }
\end{align}
and reject pairs for which $s > 5$.

\subsubsection{Selecting the Best Pair}

The last step at this stage is to pick the best pair of the remaining combinations. 
It is extremely rare for two independent elastic reactions to occur in the same bunch, and I 
can safely neglect this case. If two or more pairs survive the loose cuts and the particle-ID
selection, usually one of two things is going on. In the first case, the track reconstruction has found
two tracks of the correct particle type when only one particle passed through that sector of the spectrometer. 
This is not terribly worrisome because the track reconstruction usually finds nearly identical momentum 
vectors for the two tracks. In the second case, there was a simultaneous background track, which
should be removed. My approach is to select the best pair based on vertex correlation of the tracks
in the two sectors. A background track will have an uncorrelated vertex. In the case of two 
tracks originating from the same particle, I reason that the one with a vertex that better matches
the track on the opposite sector was probably a more successful reconstruction.
To select the best pair, I assign a vertex correlation score based on fits to the distributions
of vertex correlation, sliced by lepton scattering angle. 

\subsection{Reducing the Inelastic Contribution}

Other than the loose cuts, the selection of the best pair enforces no requirements about the 
elasticity of the event. The particle-ID cuts require that one track was really produced
by a lepton and that the other track was really produced by a proton, and the selection
based on vertex correlation enforces that the tracks emerged from the same vertex, but
the sample at this stage contains both elastic $ep$ events as well as inelastic events.
The second stage of the event selection is to reduce the inelastic background. There are 
still four kinematic criteria that have yet to be leveraged: the beam energy
reconstructed from angles, the momentum of the lepton, the momentum of the proton, and
the coplanarity (or azimuthal correlation) of the tracks. 

\begin{figure}[htpb]
\centering
\includegraphics{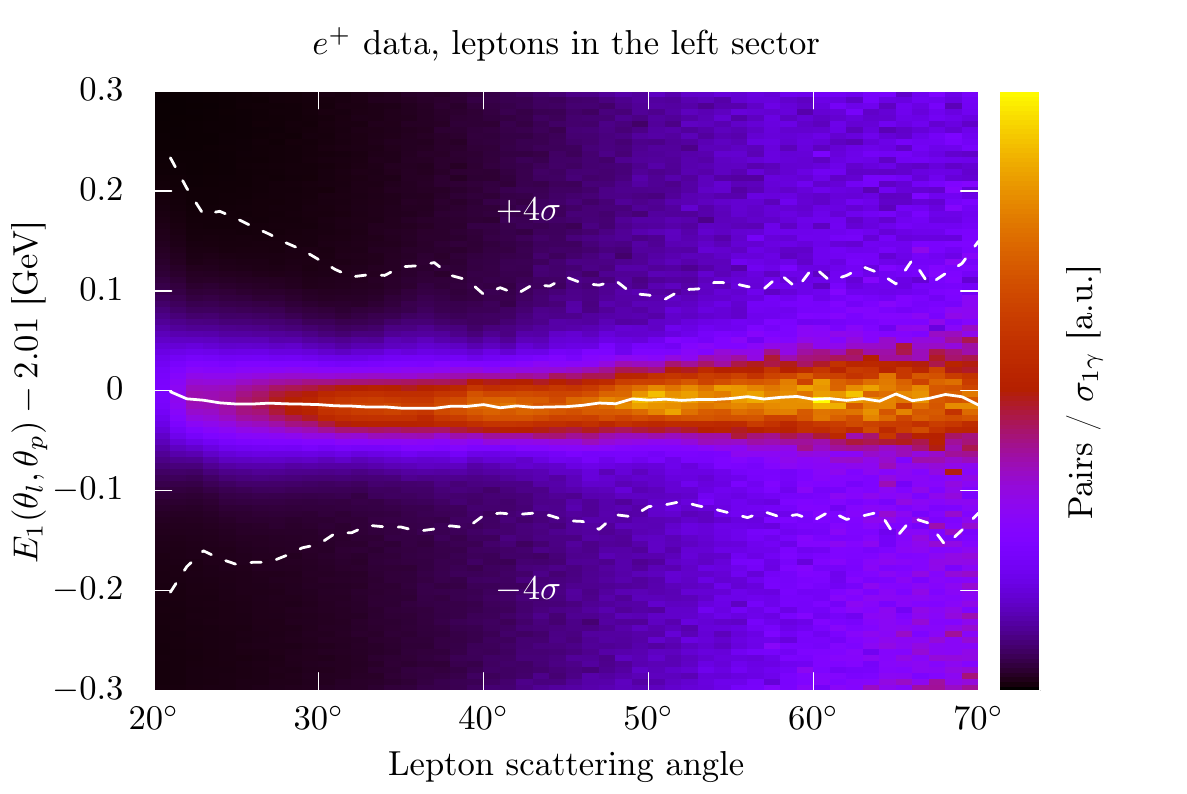}
\caption[Cut on energy as reconstructed from the lepton and proton angles]
{\label{fig:e_ang_cut} I make a fitted cut on the beam energy, as reconstructed from angles
of the lepton and proton. I've illustrated the cut here for positron data in the left sector.}
\end{figure}

The next cut I make is a fitted cut on beam energy reconstructed from angles, illustrated in 
figure \ref{fig:e_ang_cut}. I fit the reconstructed beam energy distribution in slices of 
lepton scattering angle with an asymmetric gaussian plus a constant background. I discard 
events in which:
\[
\left| \frac{ E_1(\theta_l,\theta_p) - \mu(\theta_l) }{\sigma(\theta_l)} \right| > 4.
\]

\begin{figure}[htpb]
\centering
\includegraphics{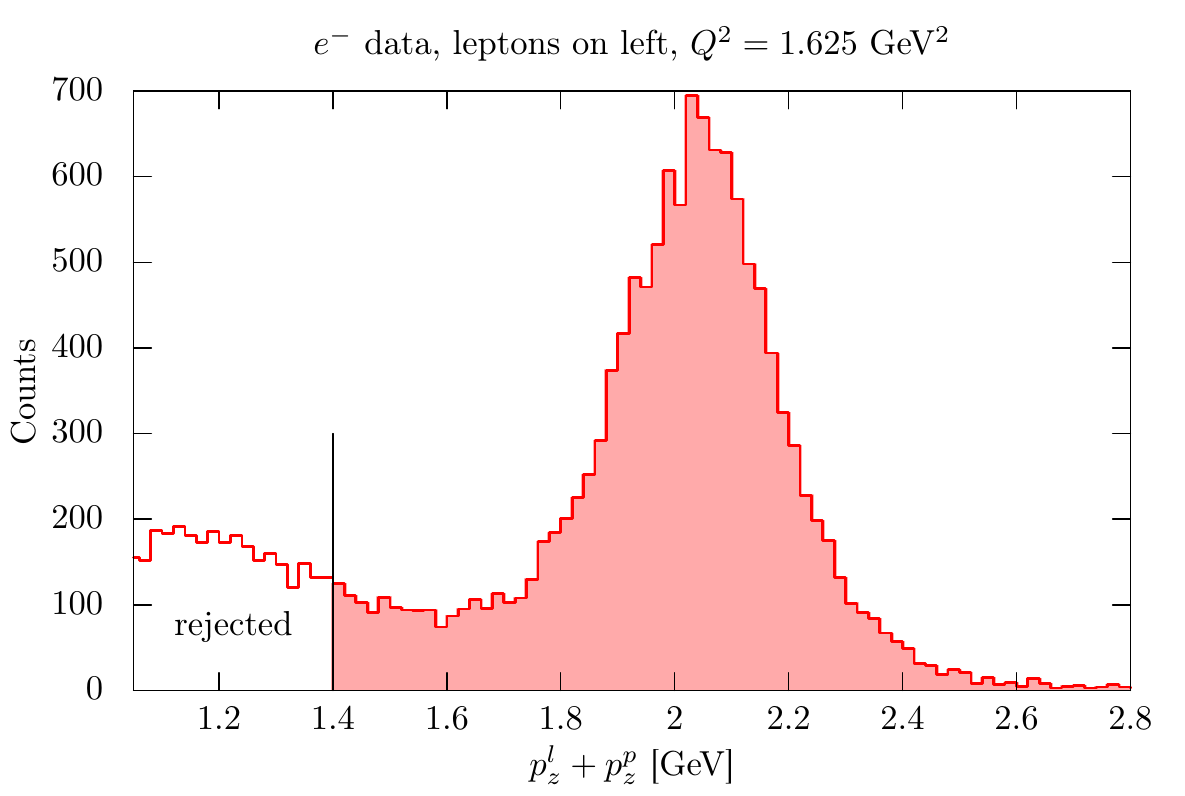}
\caption[Cut on the reconstructed momentum in the $z$ direction]
{\label{fig:pz_cut_slice} I find that an axis parallel cut that requires $\sum p_z > 1.4$~GeV
is a simple way to remove inelastic background. }
\end{figure}

\begin{figure}[htpb]
\centering
\includegraphics{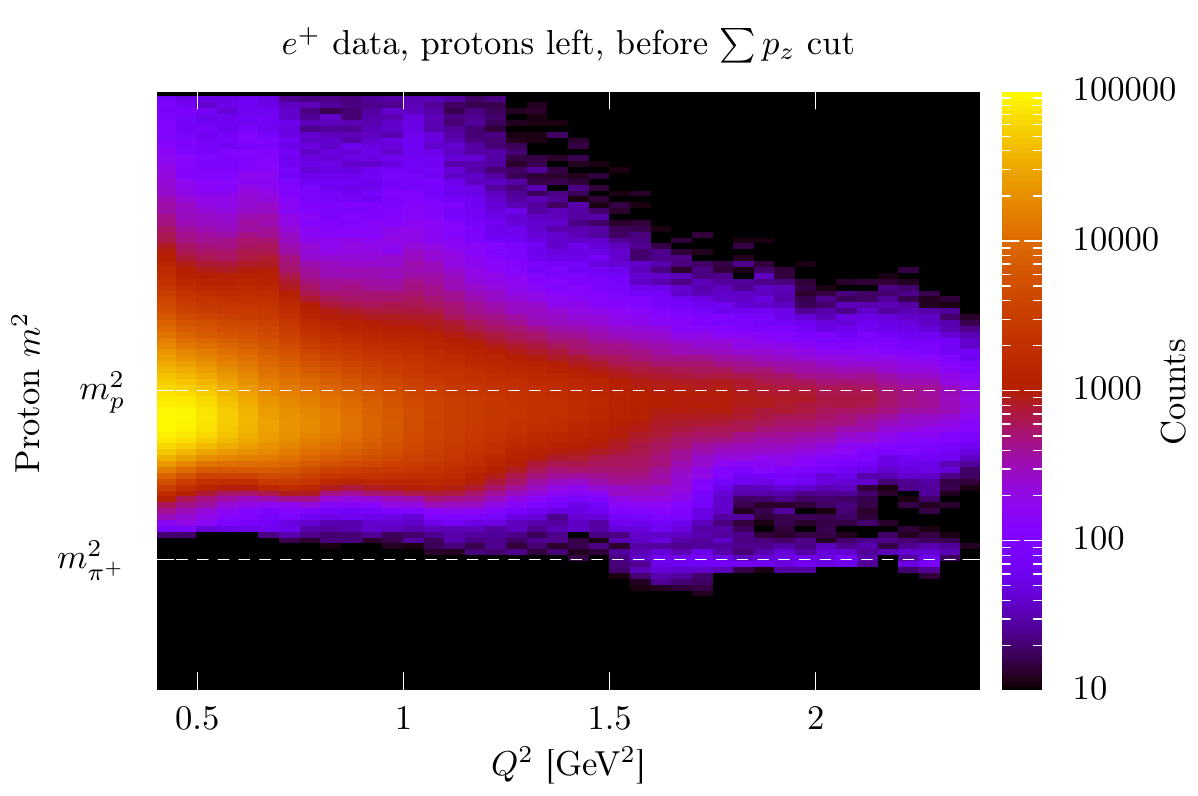}\\
\includegraphics{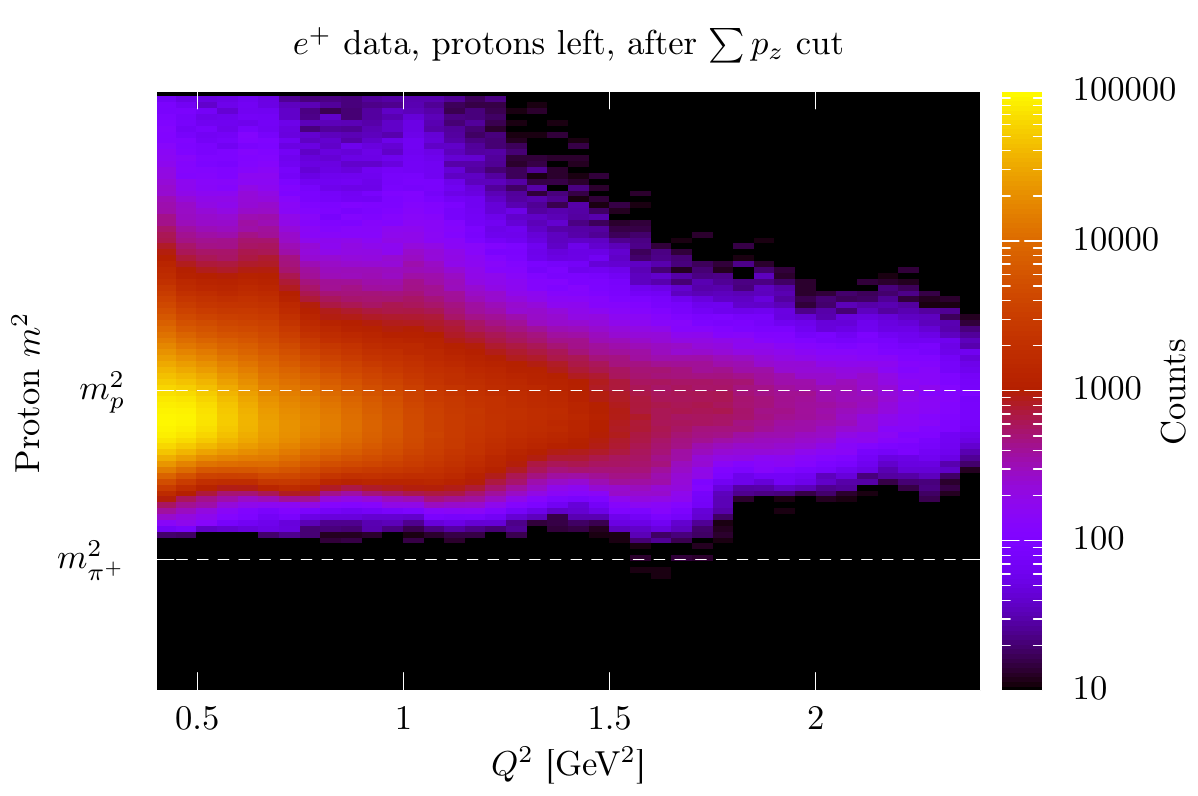}
\caption[The removal of $\pi^+$ background with a $\sum p_z$ cut]
{\label{fig:pz_cuts} Making a cut on $\sum p_z$ helps remove the background in which
a pion is reconstructed as a proton. Before the cut (top plot), there is some residual background
with a reconstructed mass close to $m_\pi^2$. After the cut (bottom plot), this background has
been eliminated.}
\end{figure}

Ideally, I would next use two more fitted cuts based on the momenta of the lepton and proton,
but I never found a stable way to fit the elastic peaks in the momentum distributions. In both
the lepton and proton momentum distributions, the elastic peaks are not well separated from
the inelastic background. However, I did find a combination that was effective at separating
background: a cut on the sum of momenta in the direction of the beam (the $z$ direction in the
OLYMPUS coordinate system). I make an axis-parallel cut:
\[
p_z^l + p_z^p > 1.4~\text{GeV},
\]
as shown in figure \ref{fig:pz_cut_slice}. I find that this cut is effective at removing the 
remaining $e^\pm p \rightarrow e^\pm \pi^+ n$ background that persists in the elastic sample.
This can be seen in the distribution of the proton mass-squared, shown in figure \ref{fig:pz_cuts}.

Rather than make a cut on coplanarity, I reserve that distribution in order to estimate the
remaining background in my elastic sample. 

\subsection{Fiducial Cuts}

\begin{figure}[htpb]
\centering
\includegraphics{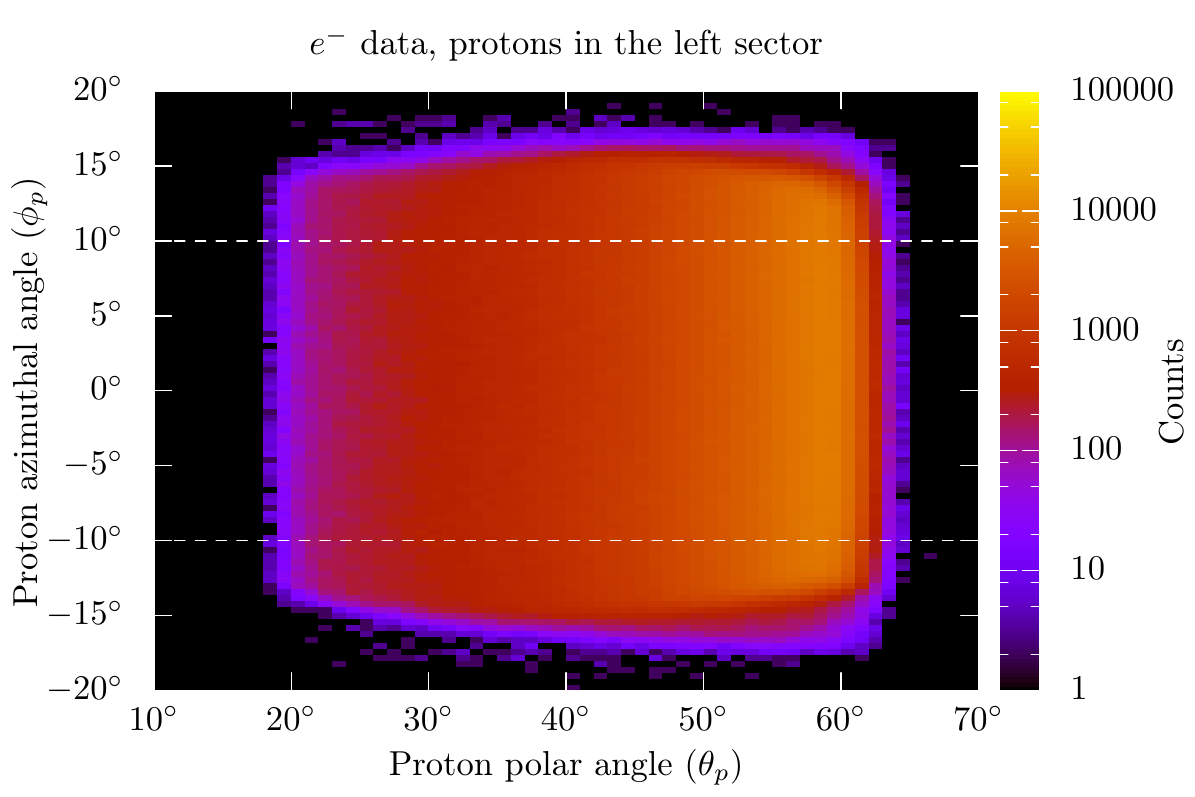}\\
\includegraphics{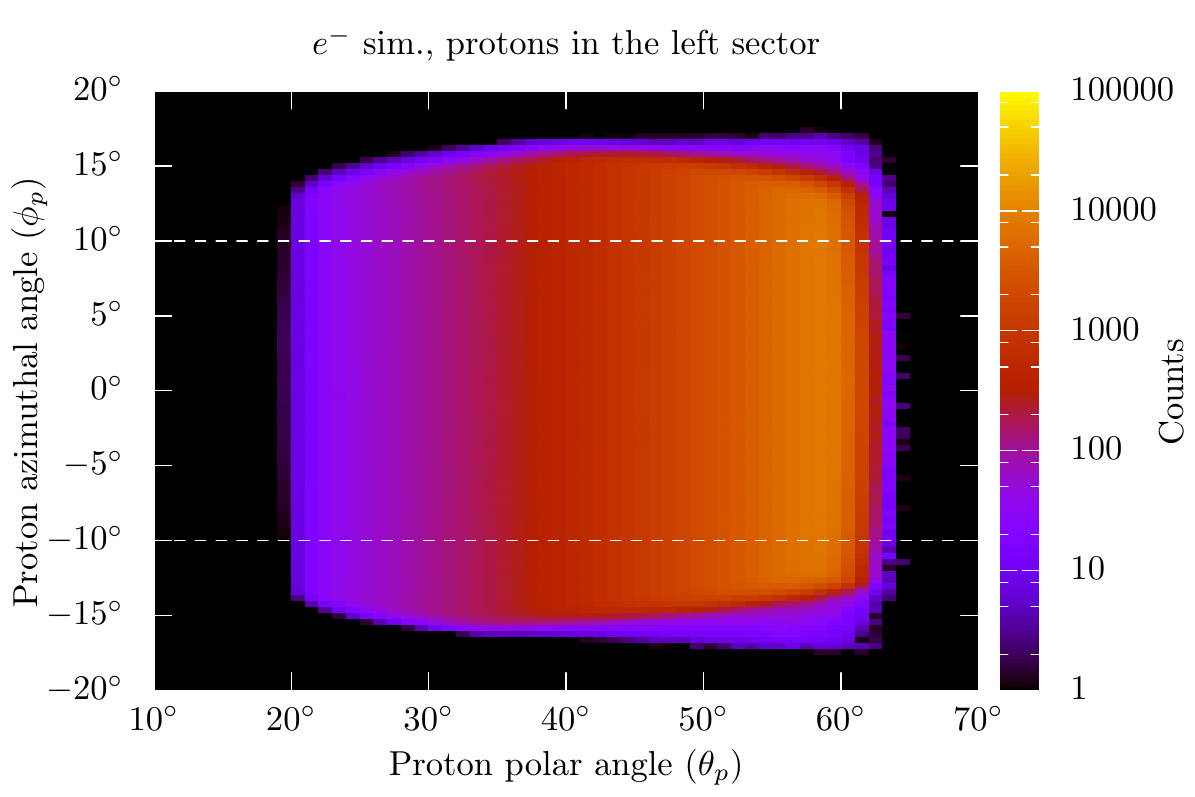}
\caption[Fiducial cuts]{\label{fig:fiducial_cuts} I make conservative fiducial cuts: the proton azimuthal
angle, $\phi_p$, was required to fall within $10^\circ$ of the horizontal plane, i.e.,
between the dashed lines. This prevents problems from the slight differences between the
experimental acceptance (shown in the top plot) and the simulated acceptance (bottom plot).}
\end{figure}

The last stage of the elastic event selection is to make fiducial cuts. The purpose of these cuts
is to prevent the effects of acceptance edges from introducing perturbations in the final yields.
I only make fiducial requirements on the proton, because the proton reconstruction should be
identical for both electron and positron running. I make two fiducial requirements on the elastic 
events. The first requirement is that the azimuth of the proton be within $\pm 10^\circ$ of the 
horizontal plane ($\phi_p=0^\circ$ on the left sector and $\phi_p=180^\circ$ on the right sector.
This is shown in figure \ref{fig:fiducial_cuts}. 
The second requirement is that proton vertex be within 350~mm of the center of the target. The
target cell extends to $\pm$300~mm, but I allow an extra 50~mm since the reconstruction of the
vertex has some non-zero resolution.

\section{Background Subtraction}

\label{sec:background_subtraction}

The final step in the analysis chain is to estimate and subtract the remaining background from
the elastic yields. For this step, I use the coplanarity distribution. Elastic events are coplanar,
while background events need not be. An advantage of using the coplanarity distribution is that
the left and right tails of the distribution are the same size. The elastic peak in coplanarity 
has no skewness. 

My approach to estimating the background under the elastic peak is to interpolate between the
side-bands of the distribution. The challenge then is to pick a suitable model for the shape
of the side-bands. If the background were made up of tracks with random azimuthal angles, then
we could expect the background to have a triangular coplanarity distribution (since a triangle 
is the convolution to two rectangular distributions). If the background still had significant pion 
contamination, we might expect the coplanarity distribution to be more peaked. What I find is
that the side-bands look very flat. This is surprising, and I cannot explain why this might
be true. But it certainly makes background estimation a lot easier.

\begin{figure}[htpb]
\centering
\includegraphics{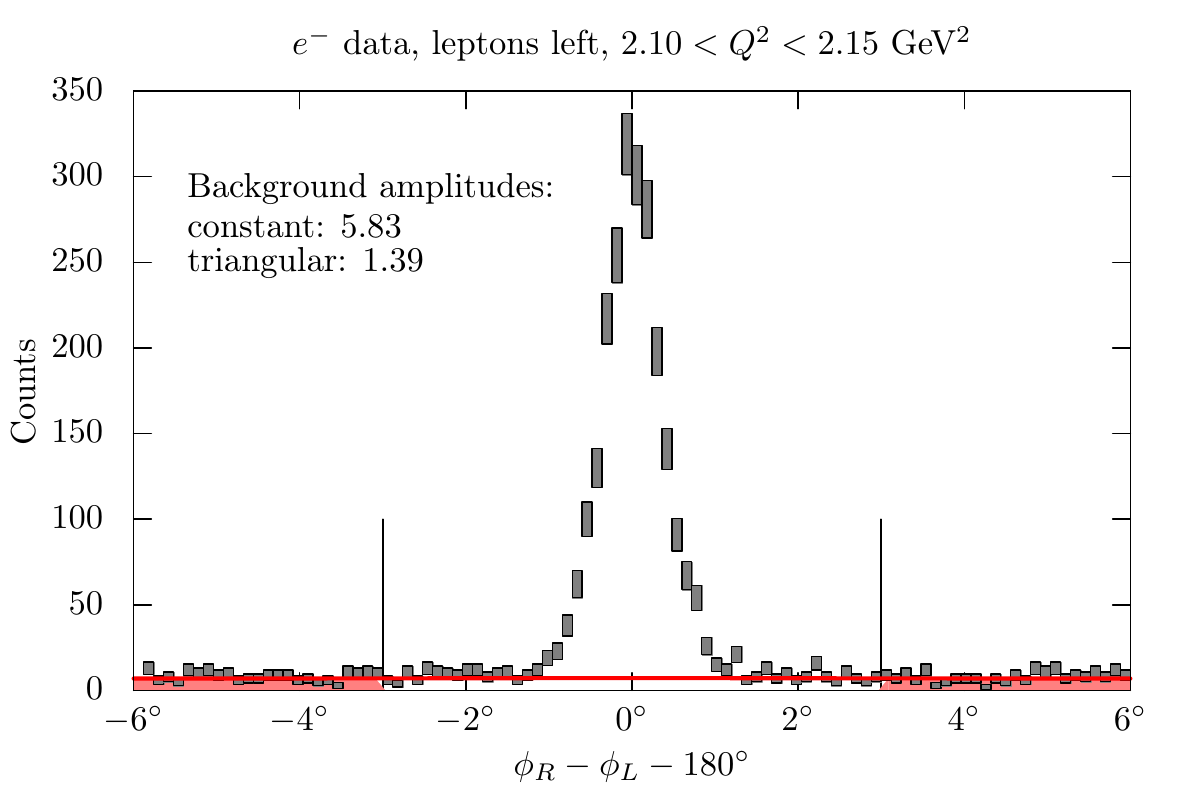}
\caption[Background subtraction]{\label{fig:bkg_sub_slice} The side-bands of the coplanarity distribution are
very flat. The fit indicates a small triangular component as well. }
\end{figure}

My background estimation procedure is as follows. First, I divide the yields into bins in the
kinematic variable of interest (typically $Q^2$, $\epsilon$, or $\theta_l$). Much like fitted
cuts, I fit the background separately for the two species and for the two sectors. For each bin,
I fit the elastic peak with a gaussian in order to determine the peak position. I
do this to guard against any potential inaccuracy in the track reconstruction. Once I find
the peak position, I then fit the side-bands (which I define as $-6^\circ$ to $-3^\circ$
and from $3^\circ$ to $6^\circ$ in coplanarity) with a constant-plus-triangle model. I fix
the peak of the triangle to be at the same position as the coplanarity peak. I then estimate
the background in the signal region (which I define as $-3^\circ$ to $3^\circ$) by interpolating
the constant-plus-triangle model between the side-bands. In practice, I find that the triangular component
in the side-band fits is small compared to the constant component, but I still include it
because there is no detriment; the fits are very stable. The results of one background fit
are shown in figure \ref{fig:bkg_sub_slice}.

Even though the simulation does not include background subtraction, I perform the same procedure
with simulated data anyway. Bremsstrahlung in the radiative generator will cause some elastic
events to fall in the side-band region. I want to subtract the same amount of these radiative
events from both data and simulation. 

\begin{figure}[htpb]
\centering
\includegraphics{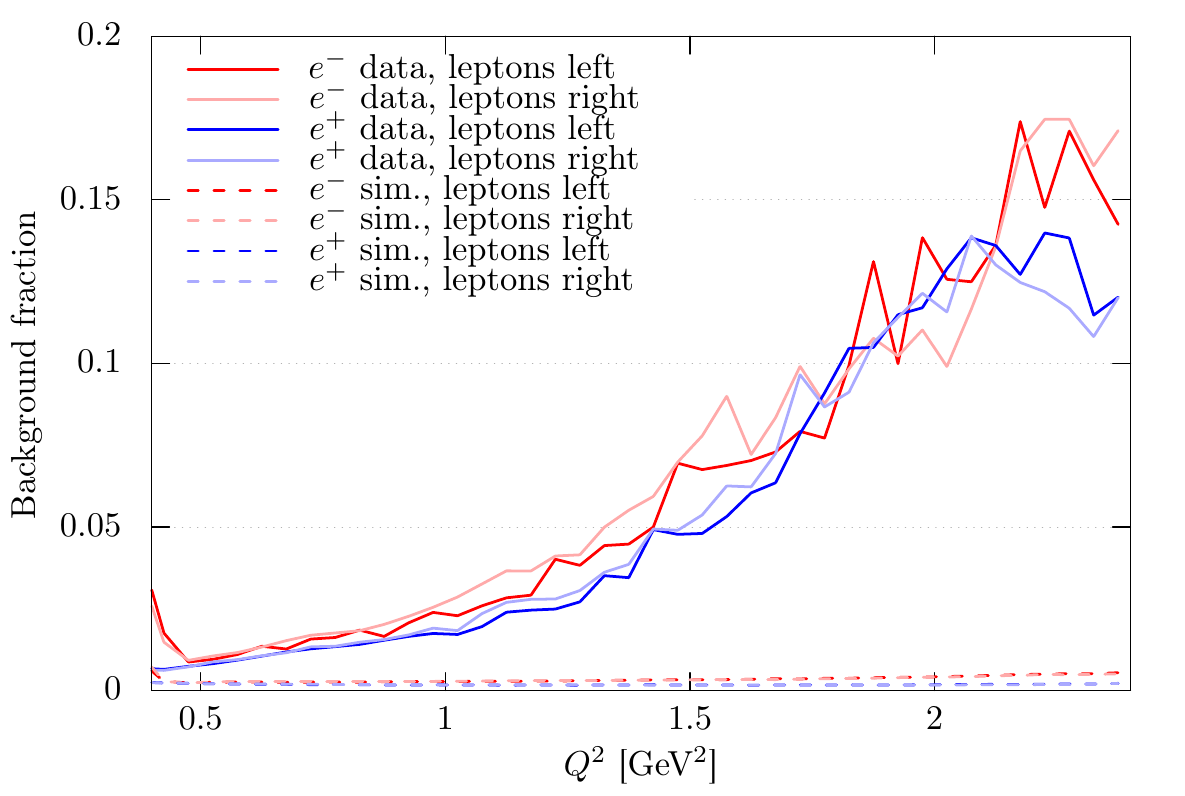}
\caption[Background fraction]{\label{fig:bkg_fraction} The background fraction is higher at larger $Q^2$, but never
exceeds 20\%. The background fraction for simulation never exceeds 1\%, which is reassuring because
the simulation has no background, other than radiative bremsstrahlung events.}
\end{figure}

The fractions of background events in both data and simulation are shown in figure \ref{fig:bkg_fraction}. 
The background represents a great fraction of the events at larger $Q^2$, but never exceeds 20\%. 
The background fraction has the same behavior for both species and both sectors. The background fraction
in simulation is much smaller, never exceeding 1\%.

\begin{figure}[htpb]
\centering
\includegraphics{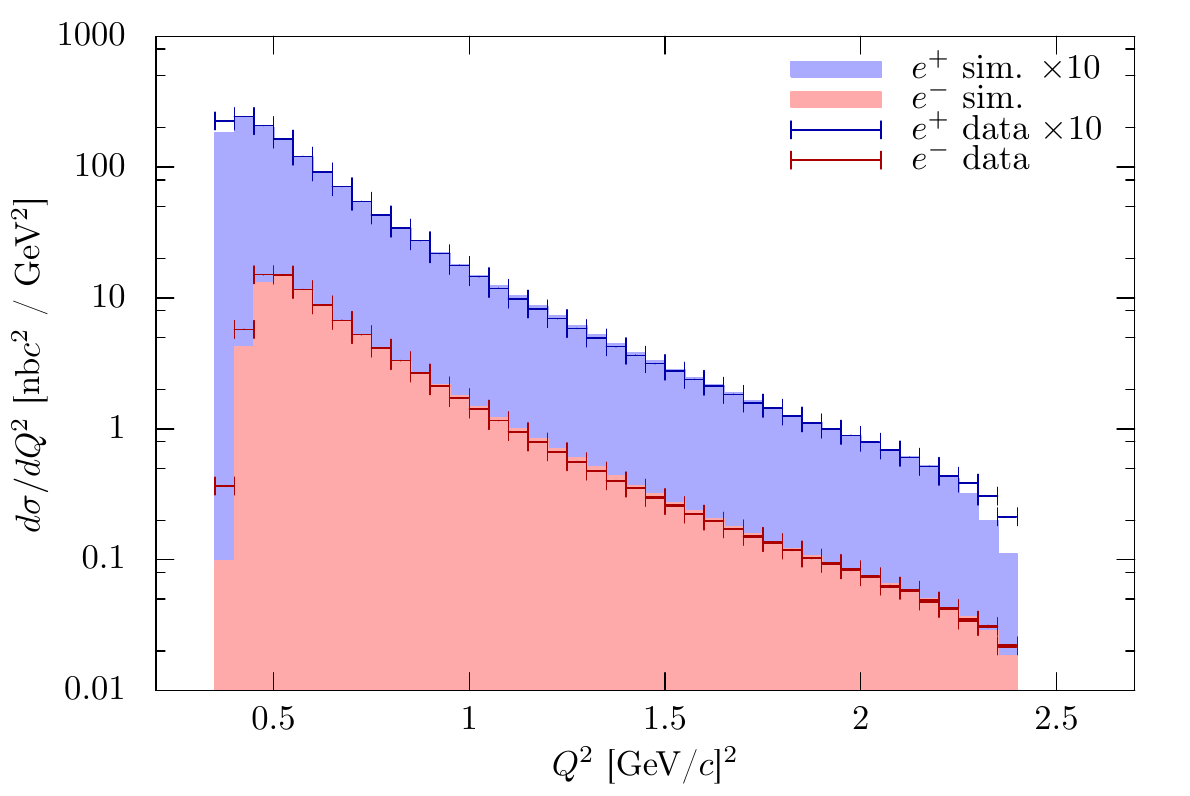}
\caption[Elastic yields]{\label{fig:elastic_yields} The yields of elastic events for data and simulation 
track each other closely over several orders of magnitude.}
\end{figure}

The result of background subtraction are final elastic yields (shown in figure \ref{fig:elastic_yields}),
which can be combined with luminosity information (according to equation \ref{eq:asym_def}) to form an asymmetry.

\section{Cross Checks}

\label{sec:analysis_checks}

It is important to be able to cross check the accuracy of the analysis and to do
so without biasing the result. There are numerous opportunities for mistakes to be made:
the simulation might fail to match the experiment in some crucial aspect, the track reconstruction
can fail to find tracks, the elastic event selection might accidentally throw away some portion of
the signal, or the background subtraction fits might fail. Over the years that this analysis chain
has been developed and tested, these errors have been identified (and corrected) by investigating the
intermediate results: the output from the generator, propagation, digitization, track reconstruction,
and event selection. However, once the intermediate results look consistent, it is still possible
for problems to be lurking in the final result. How can we build confidence in the asymmetry we
report? In this section I will present two ways to cross check the analysis without biasing the 
final result.

\subsection{Lepton-Averaged Data vs. Simulation}

Agreement between experimental and simulated yields is an obvious test of the analysis. However, the asymmetry we
want to measure, will manifest itself as a difference between the experimental data, which may have significant hard two-photon exchange,
and simulation, which does not include hard two-photon exchange. Tuning the simulation to match data, in essence, tunes the result. 
A way to avoid this is by averaging the electron and positron data sets. In the average the hard two-photon exchange will
cancel, and then tuning the simulation will not bias the result. Peaks or discontinuities in the ratio of
data to simulation indicate that there is still a problem in the analysis that must be solved.

This technique is ultimately limited by our knowledge of the form factors. The lepton-averaged yield amounts, 
essentially, to an absolute cross section measurement. The absolute cross section for $ep$
scattering is known only as well as the form factors are known, which, in the $Q^2$-range of OLYMPUS, is only
at the level of about 5--10\%. While this limits our ability to cross-check the analysis, it implies that OLYMPUS may be able
to make an absolute cross section measurement, as long as the systematic errors can be brought under control.

\subsection{Comparisons of Left and Right Sector Measurements}

The OLYMPUS spectrometer is nearly left/right symmetric. Therefore, one can naturally divide the data
between events in which the lepton was tracked in the left sector (and the proton in the right)
and events in which the lepton was tracked in the right sector (and the proton in the left). Any measurement
made for lepton-left events should match that for lepton-right events. A ratio of left and right
measurements should be consistent with unity, while also hiding the actual result of the measurement. 
Deviations from unity indicate areas with problems. By tuning the analysis to reduce left/right
discrepancies, one can improve the analysis without biasing any results.

\chapter{Results and Discussion}

\label{chap:results}

\section{Binning}

In this chapter, I present preliminary results from the OLYMPUS experiment. The data were analyzed according
to the procedure laid out in chapter \ref{chap:analysis}, specifically with elastic event selection (section \ref{sec:elastic_event_selection})
and background subtraction (section \ref{sec:background_subtraction}) methods that I designed. As I mentioned
in chapter \ref{chap:analysis}, multiple procedures for elastic event selection and background subtraction have
been developed in parallel so that each procedure can serve as cross check of the others. The results of this
chapter have not had the benefit of those cross checks, so are therefore preliminary. It is intended that the
results presented here will be superseded by final OLYMPUS results that will be submitted for publication soon.

The results I present have been binned, and before presenting numbers, I want to discuss my choice of binning.
The first choice to make is where to place fiducial limits in $Q^2$. The spectrometer only accepts a finite
$Q^2$ range of scattering angles, but the range is not clear cut because the OLYMPUS target is 60~cm long. 
At very large $Q^2$ and also at small $Q^2$, there can be acceptance edge effects which produce artificial
asymmetries. Rather than trusting the full range of $Q^2$ in which elastic events were collected, I choose
to limit the results to a narrower range of $Q^2$ where I have confidence that the magnitude of any acceptance
effects are small. 

\begin{figure}[htpb]
\centering
\includegraphics{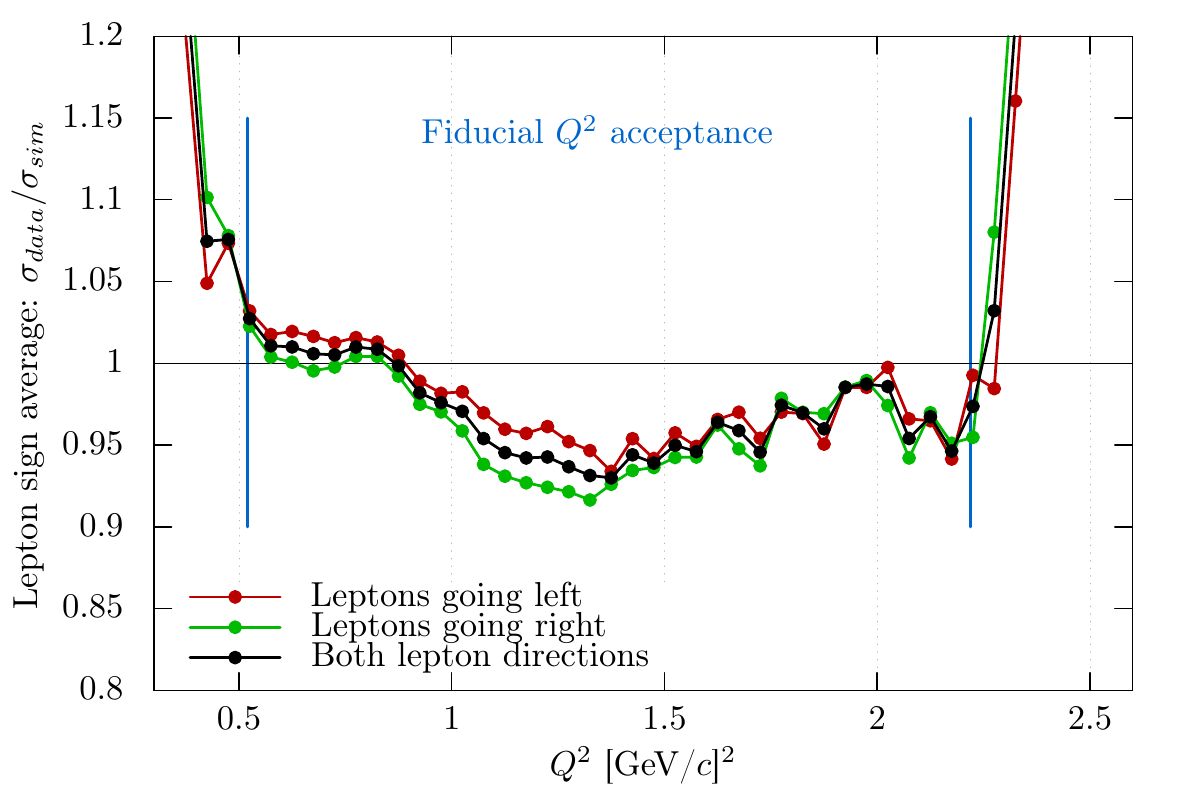}\\
\includegraphics{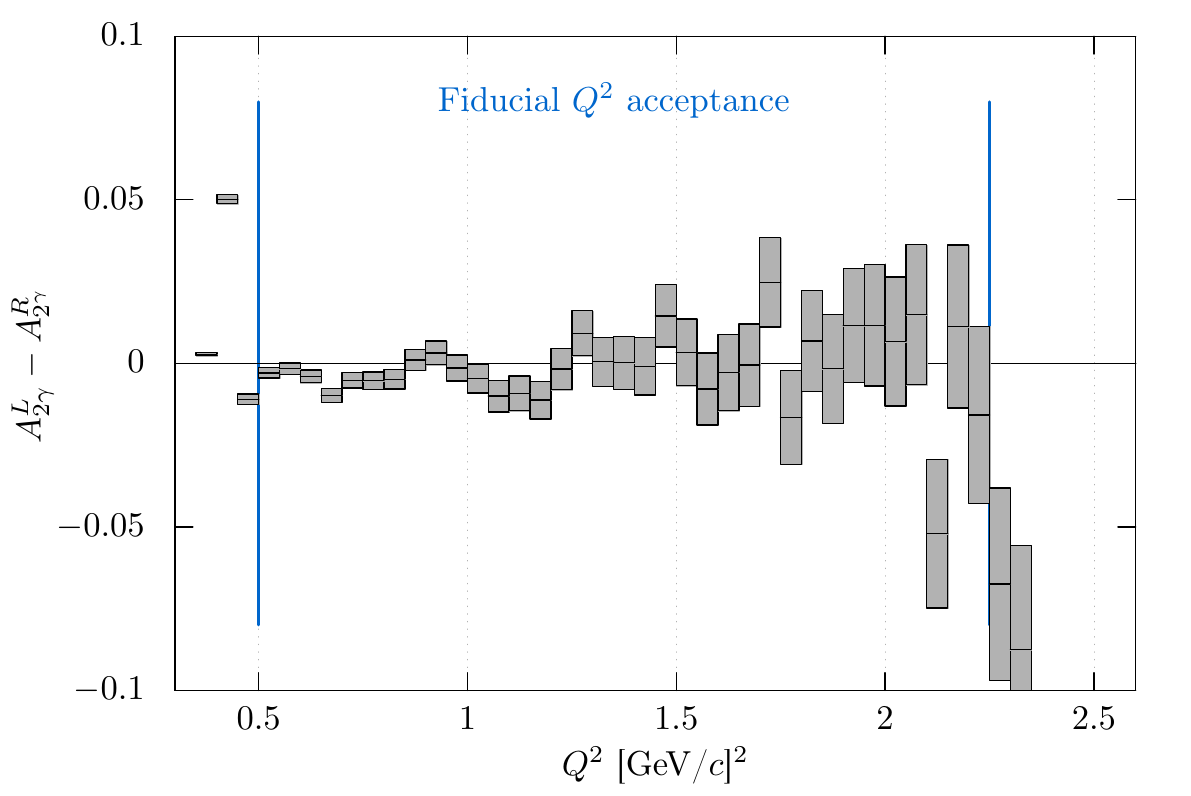}
\caption[Acceptance edge effects]{\label{fig:results_binning} I have chosen the fiducial bounds in $Q^2$ by looking for acceptance edge
effects in the lepton-averaged data divided by simulation (top plot) and in difference in asymmetry between
the lepton-left and lepton-right data sets (bottom plot). In the bottom plot, the error bands represent statistical
uncertainties only.}
\end{figure}

To gauge the placement of fiducial limits in $Q^2$, I use the cross checks discussed in section \ref{sec:analysis_checks},
and shown in figure \ref{fig:results_binning}. Using these cross checks, I can make sensible limits on the $Q^2$ range without
considering the result itself, thereby avoiding a potential source of bias in the results.
The first cross check, the lepton-averaged ratio of experimental and simulated cross sections, is shown in the top plot
 and separated by the lepton sector. For $Q^2$ below 0.5~GeV$^2$ and above 2.25~GeV$^2$, the ratio rises suddenly, 
indicating edge effects. Outside of these bounds, the experimental 
acceptance no longer matches the simulated acceptance and the results are not reliable. At intermediate $Q^2$, the
ratio dips by about 5\% below 1. This is not cause for alarm; it reflects uncertainty in the simulated $ep$ cross
section. The left lepton ratio mirrors the trend in the right lepton ratio, suggesting the underlying simulated form
factors are the cause of the dip. In contrast, for the sudden rise in the ratio at $Q^2=2.25$~GeV$^2$, the left ratio
diverges from the right ratio, signaling an acceptance edge effect.

The second cross check, the asymmetry for left leptons minus the asymmetry for right leptons, is shown in the bottom
plot of figure \ref{fig:results_binning}. In a perfect analysis, the difference between the two asymmetries is zero.
As in the lepton averaged ratio, sudden rises or drops indicate acceptance edge effects. From the results of the two
cross checks, I believe only results in the range 0.5~GeV$^2$~$< Q^2 <$~2.25~GeV$^2$ are reliable, and so I will only
present results in that range.

\begin{figure}[htpb]
\centering
\includegraphics{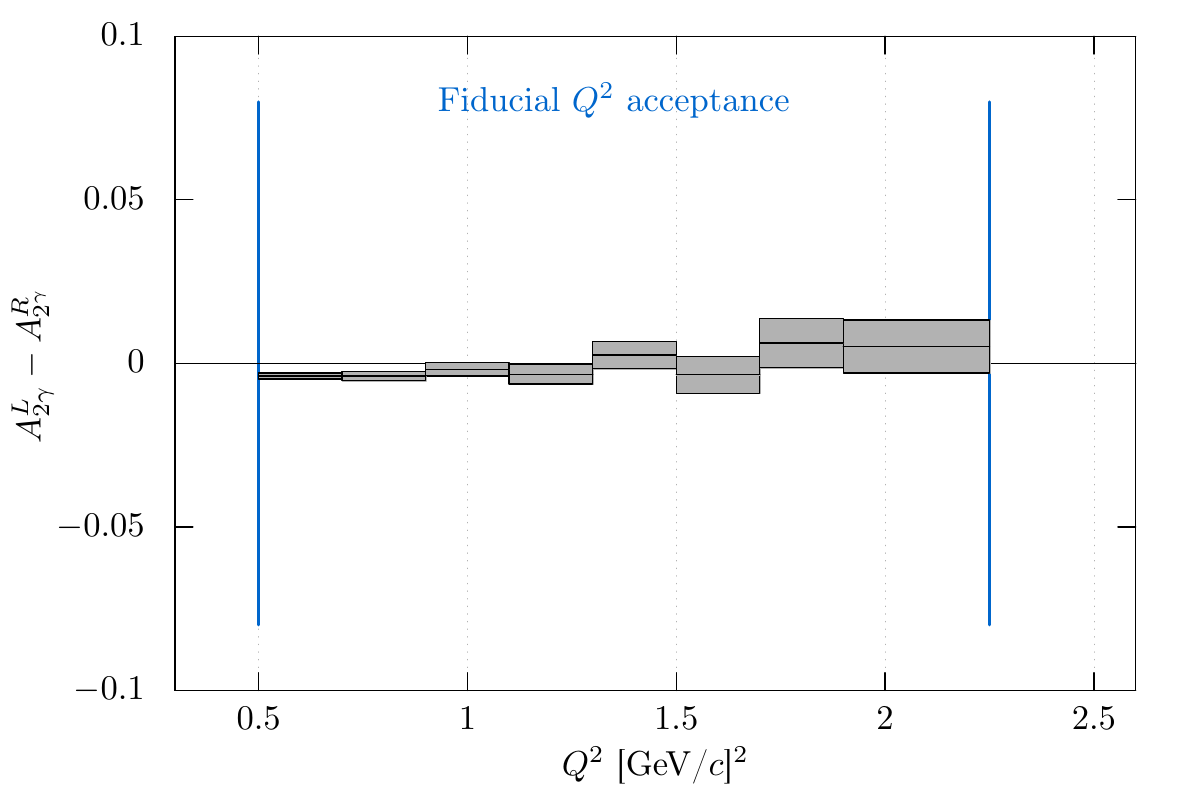}
\caption[Asymmetry difference between left and right spectrometer arms]
{\label{fig:results_asym_LR_final} The asymmetry difference between left leptons and right leptons
is presented in the binning scheme I use to present results. The fluctuations are much smaller.}
\end{figure}

For these cross checks, I have chosen to look at bins that are 0.05~GeV$^2$ wide. This is useful for looking at 
small-scale structures and for acceptance edges, but small bins have the potential for fluctuations, both from
statistics, but also from analysis. Slight efficiency dips from dead channels or fluctuations in side-band fits
during background subtraction can introduce deviations in small bins that would be washed out in larger bins. 
Larger bins insulate the result from small scale asymmetries that are not caused by two-photon exchange. This comes
at a cost; there is less information about the kinematic trends in the asymmetry if the data are aggregated into
fewer wider bins. As a compromise between these two positions, I have chosen to present the result in 8 bins, shown
as bins 1--8, in table \ref{table:binning}. Bins 1--7 are 0.2~GeV$^2$ wide, while the last, to accumulate more statistics, 
has a width of 0.35~GeV$^2$. The difference in asymmetry between left leptons and right leptons in this binning scheme 
is shown in figure \ref{fig:results_asym_LR_final}. 

\begin{table}
  \centering
  \begin{tabular}{| c c c c c c |}
    \hline
    Bin & $Q^2$ & min. $Q^2$ & max. $Q^2$ & max. $\epsilon$ & min. $\epsilon$ \\
    & (GeV/$c$)$^2$ & (GeV/$c$)$^2$ & (GeV/$c$)$^2$ & & \\
    \hline \hline
    0 & 0.165 & 0.095 & 0.235 & 0.9877 & 0.9675 \\
    \hline
    1 & 0.60 & 0.50 & 0.70 & 0.9221 & 0.8813 \\
    2 & 0.80 & 0.70 & 0.90 & 0.8813 & 0.8346 \\
    3 & 1.00 & 0.90 & 1.10 & 0.8346 & 0.7818 \\
    4 & 1.20 & 1.10 & 1.30 & 0.7818 & 0.7230 \\
    5 & 1.40 & 1.30 & 1.50 & 0.7230 & 0.6581 \\
    6 & 1.60 & 1.50 & 1.70 & 0.6581 & 0.5874 \\
    7 & 1.80 & 1.70 & 1.90 & 0.5874 & 0.5113 \\
    8 & 2.075 & 1.90 & 2.25 & 0.5113 & 0.3667 \\
    \hline
  \end{tabular}
  \caption[Binning scheme for the results]{\label{table:binning} This table shows the bin boundaries for the bins in which I will
    present the OLYMPUS results. The bins are nominally demarcated by $Q^2$, and the values of 
    epsilon are calculated assuming a 2.01~GeV beam energy. Bin 0 represents the data point from
    the $12^\circ$ tracking telescopes.}
\end{table}

In addition to the eight bins from my analysis of the main spectrometer data, I also include a bin
(Bin 0 in table \ref{table:binning}) for the results from the $12^\circ$ tracking telescopes. This 
analysis was performed by Brian Henderson, and details as well as an error analysis are documented
in his thesis \cite{henderson:thesis}.

\section{Systematic Uncertainty}

For this preliminary result I will consider systematic uncertainties from only a few sources, and leave conservative
estimates for the uncertainties introduced from all others. I choose to divide systematic effects into two categories:
effects uncorrelated from bin to bin, and those that are correlated over all of the bins. An example of an uncorrelated
systematic effect is a localized error in the simulated acceptance. This will affect the result in the bin corresponding
to the affected scattering angle (though this is only approximately true, since we are making a coincidence measurement). 
An example of a correlated systematic effect is the uncertainty in the species relative luminosity. An error in the
relative luminosity will introduce an equal asymmetry to all of the bins. I will break down my estimates in the following
sections.

\subsection{Uncorrelated Systematic Estimate}

There are several different sources of uncorrelated systematic uncertainty; species-dependent tracking inefficiency, 
event selection inefficiency, and uncertainty produced by the background subtraction procedure. For a final published
result, each of these sources needs to be studied in detail. For the preliminary result of this chapter, I will make
a conservative estimate of the systematic uncertainty by looking at the differences in the lepton-left and lepton-right
asymmetries, shown in figure \ref{fig:results_asym_LR_final}. I estimate the uncertainty to be the full-scale left-right
difference in each bin added in quadrature to an additional 0.5\% uncertainty, covering any effects which do not manifest
themselves in a left-right asymmetry difference. I choose 0.5\%, since that is approximately the scale of the largest
left-right difference. 

\subsection{Correlated Systematic Estimate}

The most correlated source of systematic uncertainty is the uncertainty on the species relative luminosity. 
Luminosity differences affect the asymmetry by the same amount for all bins. There are other sources of 
uncertainty that are correlated over the bins, although not to the same degree. One of those sources is
the uncertainty in the true shape of the radiative tails, i.e., whether the true radiative tail is
better described using the Method 1 approach (with exponentiation), or using the Method 2 approach 
(with a single radiated photon) in the radiative generator. As a conservative estimate, I will assign,
as a correlated uncertainty, the quadratic sum of the luminosity uncertainty (0.15\%) and the full difference 
between the Method 1 and Method 2 simulated asymmetries.

\section{Results}

\begin{table}
  \centering
  \begin{tabular}{| c c c c c c c |}
    \hline
    Bin & $Q^2$ & $A_{2\gamma}$ & $A_{2\gamma}$ & $\delta$ stat. & $\delta$ sys. & $\delta$ sys. \\
    & (GeV/$c$)$^2$ & soft from \cite{Maximon:2000hm} & soft from \cite{Mo:1968cg} & & uncorr. & corr. \\
    \hline \hline
    0 & 0.165 & -0.0013 & -0.0016 & 0.0005 & 0.0027 & 0.0015 \\
    \hline
    1 & 0.60 & -0.0022 & -0.0032 & 0.0005 & 0.0063 & 0.0017 \\
    2 & 0.80 & 0.0023 & 0.0014 & 0.0007 & 0.0063 & 0.0018 \\
    3 & 1.00 & 0.0034 & 0.0027 & 0.0010 & 0.0053 & 0.0020 \\
    4 & 1.20 & 0.0044 & 0.0041 & 0.0015 & 0.0060 & 0.0023 \\
    5 & 1.40 & 0.0069 & 0.0069 & 0.0021 & 0.0056 & 0.0027 \\
    6 & 1.60 & 0.0128 & 0.0134 & 0.0028 & 0.0061 & 0.0031 \\
    7 & 1.80 & 0.0058 & 0.0071 & 0.0038 & 0.0080 & 0.0035 \\
    8 & 2.075 & 0.0080 & 0.0105 & 0.0040 & 0.0072 & 0.0044 \\
    \hline
  \end{tabular}
  \caption[OLYMPUS results]{\label{table:results} This table shows the preliminary lepton sign asymmetry result from OLYMPUS 
    for two different definitions of soft two-photon exchange. }
\end{table}

The preliminary results of the OLYMPUS experiment are shown in \ref{table:results}. I present two different
asymmetries, each corresponding to a slightly different way of making of radiative corrections. The first
asymmetry (in column 3) uses the Maximon and Tjon definition of soft two-photon exchange \cite{Maximon:2000hm}. 
The second asymmetry (in column 4) uses the Mo and Tsai definition \cite{Mo:1968cg}. The two definitions of
where to draw the hard/soft boundary are arbitrary, so I present our result for both so that 
the relevant asymmetry can be used for comparisons. For example, the form factor fits of Bernauer et al.\
\cite{Bernauer:2013tpr} use data that have been corrected (in many cases, retroactively) using Maximon and Tjon.
For that reason any comparisons to Bernauer fits will be made with our Maximon and Tjon asymmetry.
In contrast, the other contemporary two-photon exchange experiments use the Mo and Tsai definition of soft 
two-photon exchange, and so I will only compare our Mo and Tsai asymmetry to their results. For clarification,
both asymmetries I present use the Method 1 radiative corrections. The difference between Method 1 and Method 2
is included in the correlated systematic uncertainty estimate.

\begin{figure}[htpb]
\centering
\includegraphics{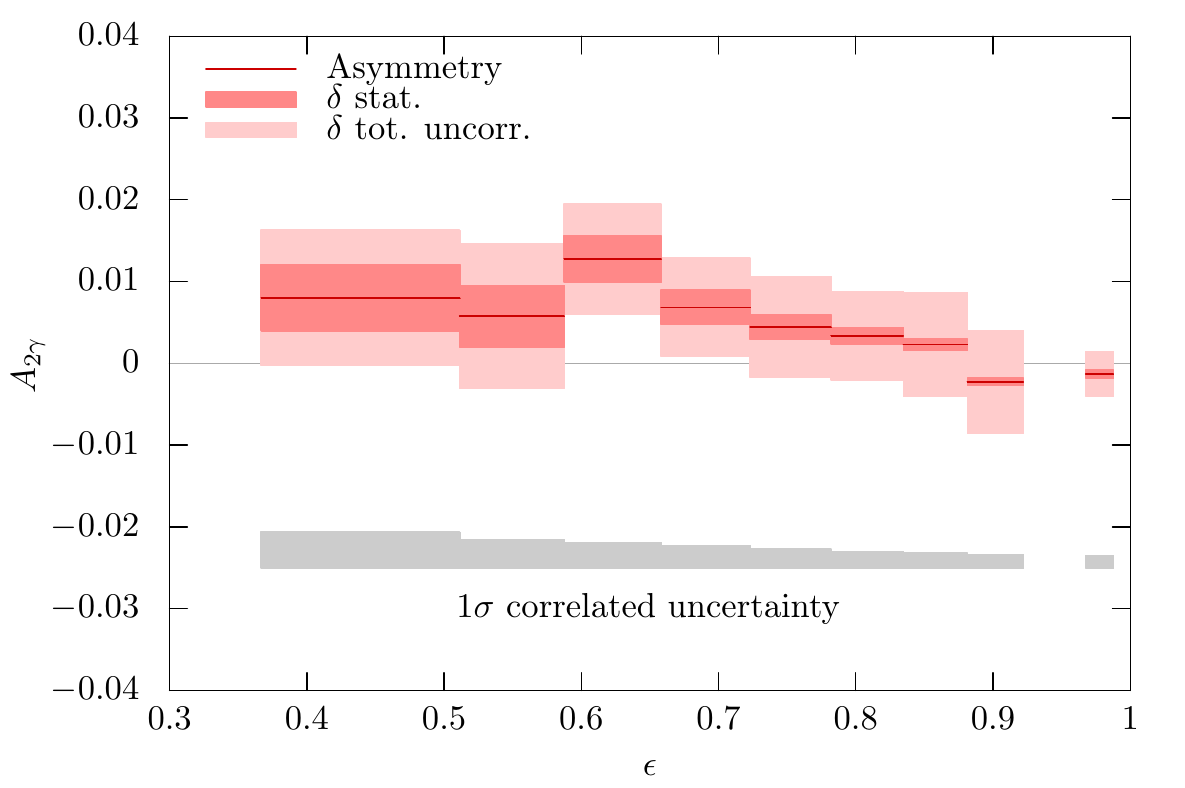}\\
\includegraphics{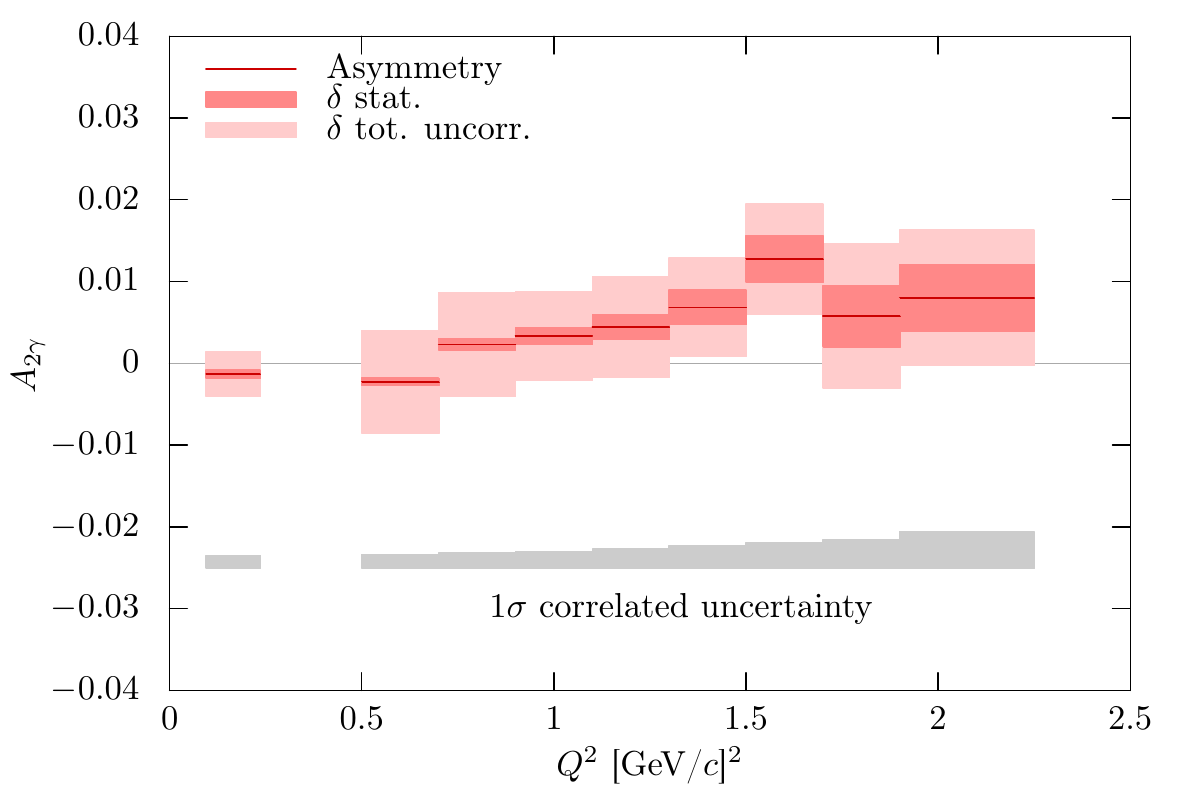}
\caption[Results with Maximon and Tjon soft TPE definition]
        {\label{fig:results_max} The asymmetry with the Maximon and Tjon definition of 
soft TPE is presented as a function of $\epsilon$ (top plot) and $Q^2$ (bottom plot). }
\end{figure}

\begin{figure}[htpb]
\centering
\includegraphics{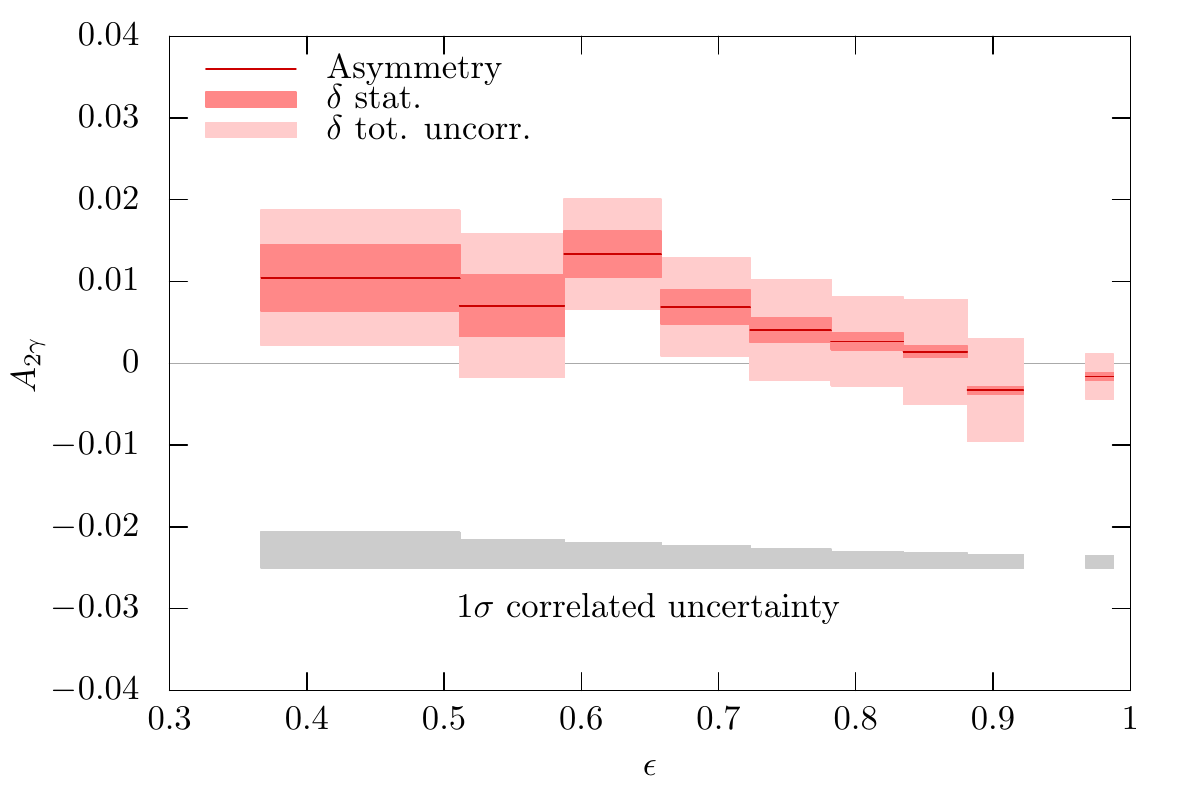}\\
\includegraphics{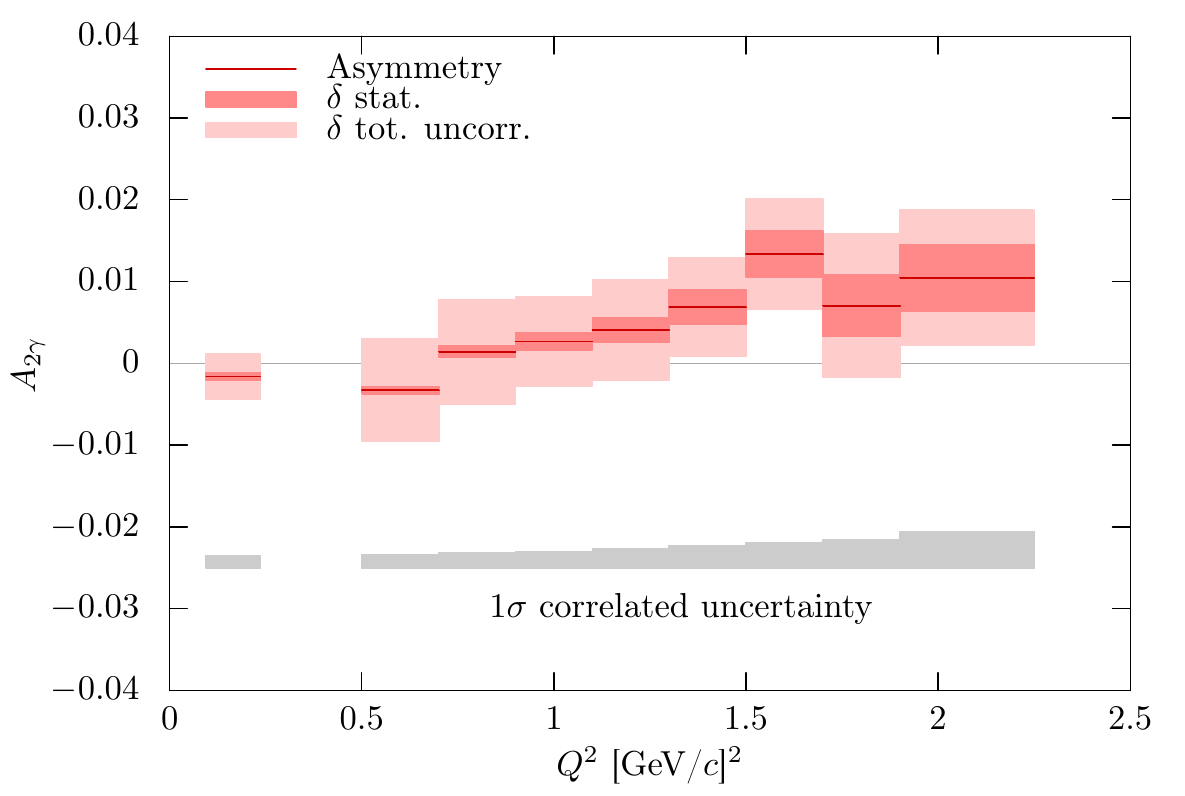}
\caption[Results with Mo and Tsai soft TPE definition]{\label{fig:results_mo} The asymmetry with the Mo and Tsai definition of 
soft TPE is presented as a function of $\epsilon$ (top plot) and $Q^2$ (bottom plot). }
\end{figure}

Figures \ref{fig:results_max} and \ref{fig:results_mo} show the asymmetries in table \ref{table:results}
plotted as functions of $\epsilon$ and $Q^2$. Figure \ref{fig:results_max} shows the results using the
Maximon and Tjon definition of soft TPE, while figure \ref{fig:results_mo} shows the results according
to the Mo and Tsai definition. The statistical and uncorrelated uncertainties are shown as colored bands around
the measured asymmetry. The correlated systematic uncertainties are shown in a band below. 

Qualitatively, the data show an asymmetry that varies over the kinematic range, increasing with $Q^2$ and
increasing as $\epsilon \longrightarrow 0$. This specific slope indicates that the two-photon exchange 
contribution reduces the size of the form factor discrepancy. However, the magnitude of the asymmetry is 
not large: only growing to about 1\% at the end of the OLYMPUS acceptance. Many of the points have an 
asymmetry that is within error of zero. That is partially due to my conservative estimate of the systematic
uncertainties. The goal is that by studying the various systematic effects in detail, we can, with 
confidence, reduce those uncertainties.

\section{Comparison with Novosibirsk and CLAS}

\begin{figure}[htpb]
\centering
\includegraphics{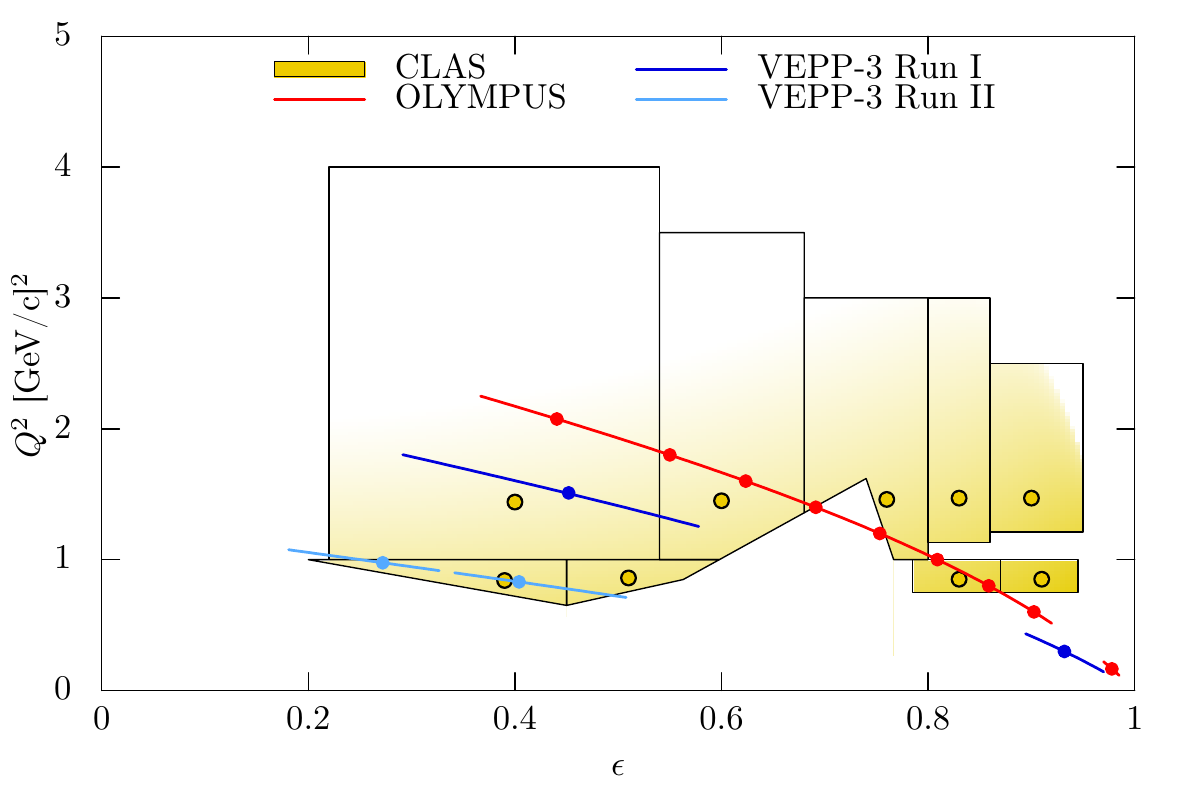}
\caption[Kinematic reach of published data points]
        {\label{fig:reach_wPoints} The kinematics of the published data points of the CLAS and Novosibirsk two-photon experiments
are shown in comparison to the OLYMPUS data points presented in this thesis.}
\end{figure}

OLYMPUS is not the only contemporary lepton sign asymmetry experiment with new results. Both the experiments
in Novosibirsk and at CLAS have recently published results. The kinematic reach of the data from all three
contemporary experiments is shown in figure \ref{fig:reach_wPoints}. In this section, I show the results
of the other two experiments, and make a comparison with the OLYMPUS results in this thesis.

\subsection{Novosibirsk Results}

\begin{figure}[htpb]
\centering
\includegraphics{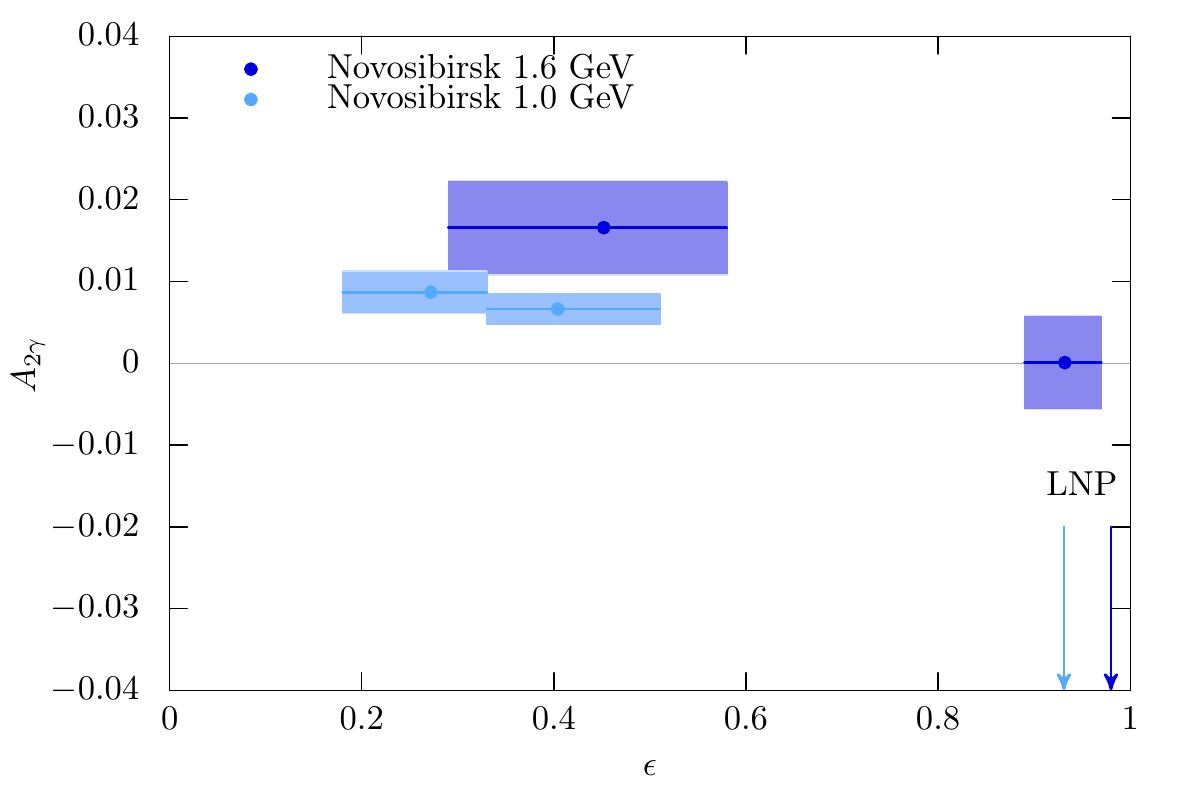}
\caption[Results of the Novosibirsk experiment]
        {\label{fig:results_novosibirsk} The Novosibirsk experiment published four data points, at two different
beam energies. The experiment was statistics limited; the systematic errors are nearly invisible in this plot.}
\end{figure}

The results of the Novosibirsk two-photon experiment were published in a letter in 2015 \cite{Rachek:2014fam}.
I have rendered the results from this paper in figure \ref{fig:results_novosibirsk}. The Novosibirsk experiment
reported four data points, at two different beam energies. The experiment was severely statistics limited; in fact,
the quadratic sum of the systematic and statistical error bars is practically invisible in figure \ref{fig:results_novosibirsk}
on top of the statistical-only errors. The Novosibirsk experiment monitored the relative $e^-p$ and $e^+p$ luminosities
through the rate of forward elastic $ep$ scattering. As a consequence, the results are relative to the asymmetry
at these luminosity normalization points (LNPs), shown with arrows in figure \ref{fig:results_novosibirsk}.

With only four data points, the Novosibirsk results are hardly conclusive, but are certainly suggestive of a
two-photon exchange effect which helps resolve the discrepancy. At both beam energies, the asymmetry rises
as $\epsilon$ decreases. Furthermore, at higher beam energy the asymmetry rise is greater, which is consistent
with a larger discrepancy at greater $Q^2$.

\subsection{CLAS Results}

\begin{figure}[htpb]
\centering
\includegraphics{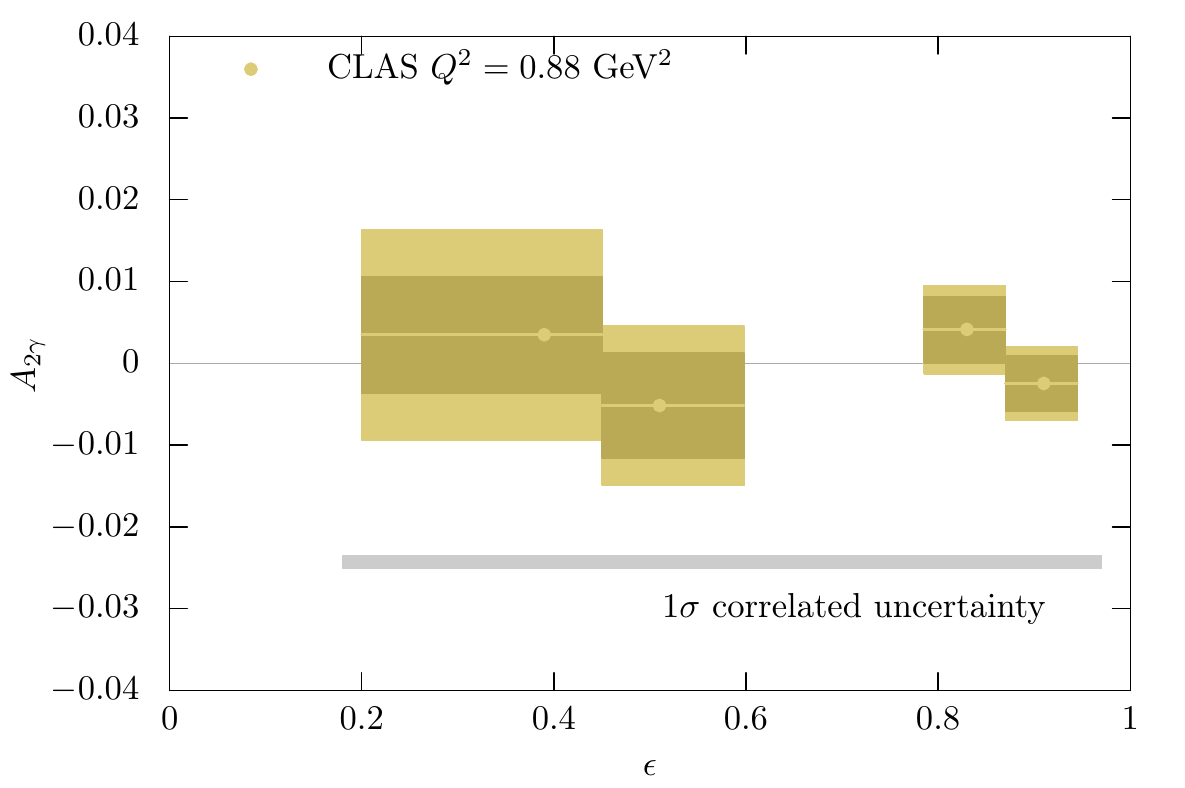}\\
\includegraphics{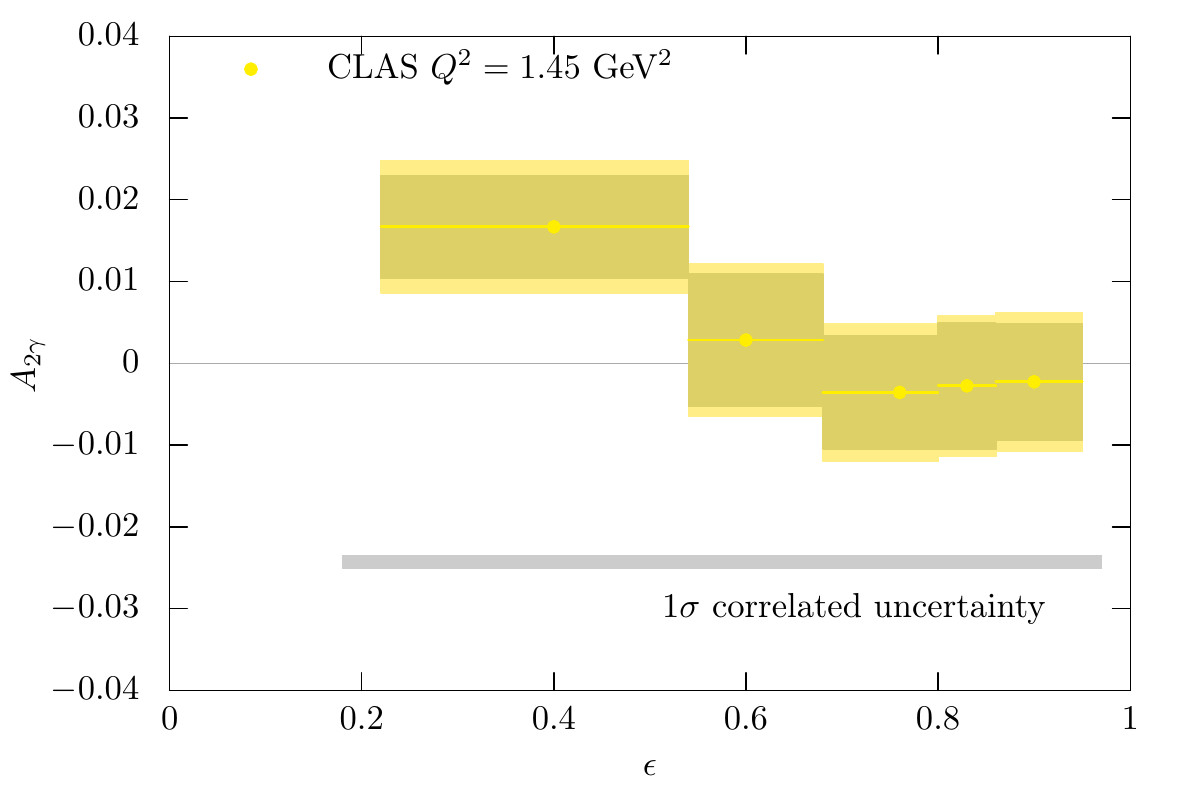}
\caption[Results of the CLAS experiment]
{\label{fig:results_clas} The CLAS two-photon experiment results for their fixed $Q^2$ binning scheme are
shown. The inner bands for each bin represent statistical error only, while the outer bands represent the quadratic
sum of point-to-point errors. The correlated luminosity uncertainty is shown by the gray band below the data.}
\end{figure}

The CLAS two-photon experiment has released results \cite{Rimal:2016toz}, which at the time of this writing are
in the midst of the peer-review process. The results were presented in two different binning schemes. In one
scheme the bin centers are positioned along lines of constant $\epsilon$, and in the other scheme the bin
centers are positioned along lines of constant $Q^2$. The bins in the two schemes overlap, and so are not
statistically independent. For that reason, in this work I will only show the constant $Q^2$ scheme, 
which allows me to plot variations in $\epsilon$, in figure \ref{fig:results_clas}.
For the constant $Q^2$ scheme, the CLAS experiment has released four points at $Q^2=0.88$~GeV$^2$, and five
points at $Q^2=1.45$~GeV$^2$. Since the experiment used a tertiary beam that had a range of energies, the
bins are not contours in the $\epsilon,Q^2$ plane, but polygons, and are shown in figure \ref{fig:reach_wPoints}.

The data at $Q^2=0.88$~GeV$^2$, shown in the top plot of figure \ref{fig:results_clas} are all consistent 
with zero asymmetry. This is not surprising, because the form factor discrepancy is not large at this
$Q^2$. The $Q^2=1.45$~GeV$^2$ data, shown in the bottom plot, have a slight rising trend, as $\epsilon \rightarrow 0$.
However, only the lowest $\epsilon$ point has an asymmetry greater than zero by more than the quoted uncertainty. 

\subsection{Comparison with OLYMPUS}

\begin{figure}[htpb]
\centering
\includegraphics{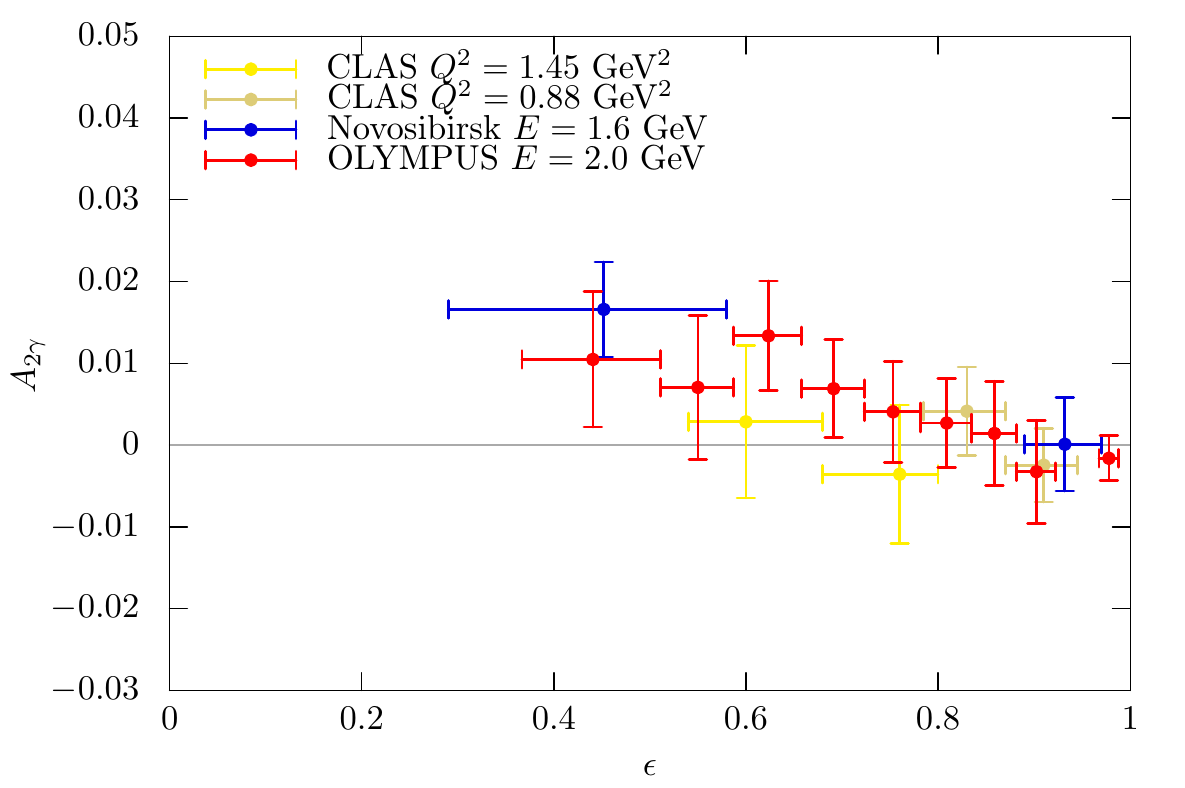}
\caption[OLYMPUS results compared with CLAS and Novosibirsk]
{\label{fig:results_all_3} Only the CLAS and Novosibirsk data close to the $E=2$~GeV contour are plotted
for comparison with the OLYMPUS results from this thesis.}
\end{figure}

Making a comparison between the CLAS and Novosibirsk data and OLYMPUS is difficult as the three
experiments measure slightly different kinematic points as seen in figure \ref{fig:reach_wPoints}.
The three experiments are not measuring the same quantities exactly, and need not agree. In the 
interest of making a comparison anyway, I observe that a few of the CLAS and Novosibirsk data
points fall close to the OLYMPUS points presented in this thesis, and that the results for the
three experiments should be similar for this limited set of points. A plot of the asymmetries
of these points is shown in figure \ref{fig:results_all_3}. The three experiments are consistent;
however, the uncertainties are so large that it would be difficult for that not to be the case.

\section{Two-Photon Exchange Hypothesis}

The OLYMPUS results quantify the contribution to the elastic $ep$ cross section from hard 
two-photon exchange. The motivation for this measurement, however, was addressing the proton
form factor discrepancy. The crucial question is whether or not the OLYMPUS result supports
or contradicts the hypothesis that hard two-photon exchange is the source of the discrepancy.
This is a difficult question to answer for several reasons. The size of the form factor discrepancy 
is only as well quantified as the form factors have been quantified via Rosenbluth separation. 
Measurements of the lepton sign asymmetry at a few kinematic points is not enough to fully
constrain the two-photon exchange correction's dependence on $\epsilon$ and $Q^2$. The measurements
themselves also have limited precision and accuracy. 

It is the purview of phenomenology to determine the asymmetry needed to fully resolve that 
discrepancy, and there have been several such calculations 
\cite{Chen:2007ac,Guttmann:2010au,Borisyuk:2010ep,Bernauer:2013tpr}. 
In appendix \ref{app:tpe}, I derive a very simple method for estimating the asymmetry from the size of 
the form factor discrepancy with the help of a few simplifying assumptions. To answer whether the OLYMPUS 
results are consistent with a resolution with the discrepancy, we should compare the results to these calculations.

\begin{figure}[htpb]
\centering
\includegraphics{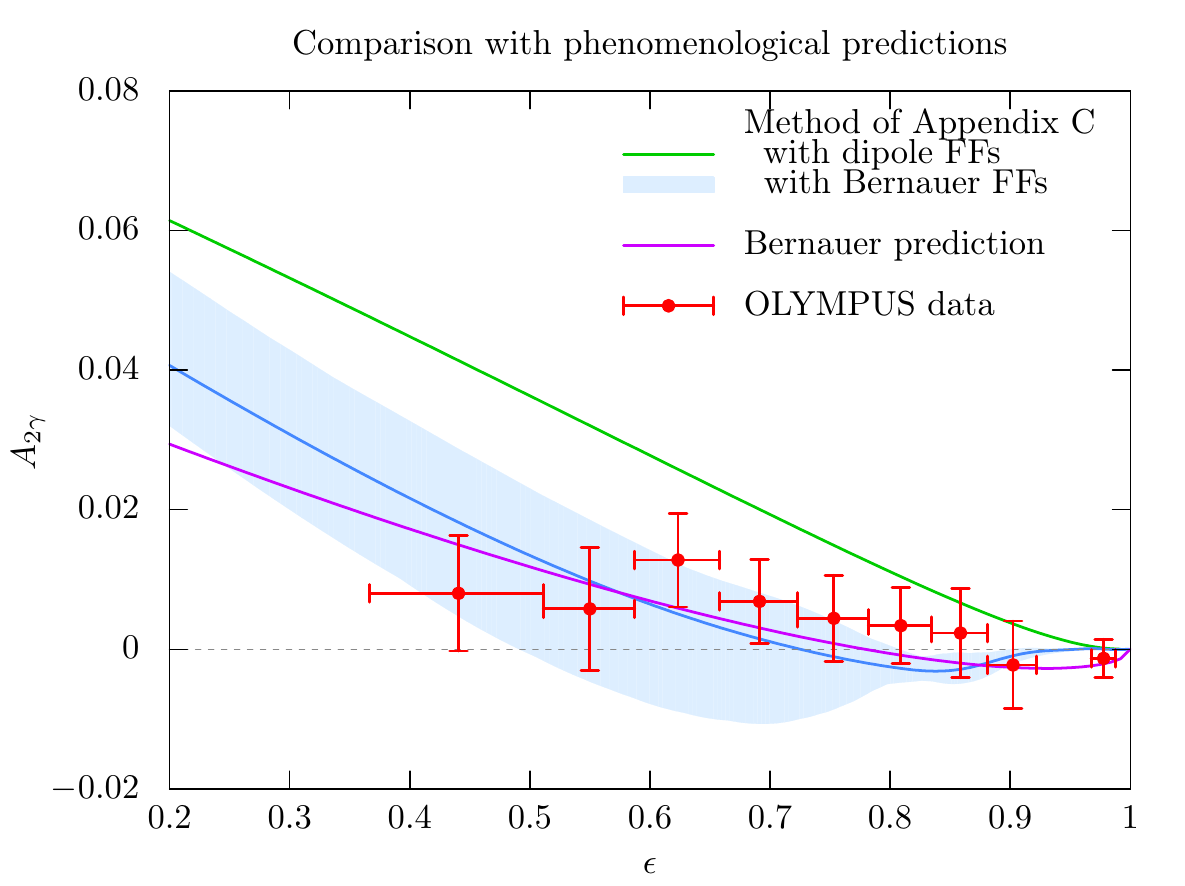}
\caption[OLYMPUS results compared with phenomenological predictions]
{\label{fig:olympus_v_phenom} The OLYMPUS results are compared with three phenomenological
extractions of how large of an asymmetry is needed to resolve the discrepancy. The light-blue error 
band is associated with my estimate using the Bernauer form factor fits, and is described in more
detail in appendix \ref{app:tpe}.} 
\end{figure}

Figure \ref{fig:olympus_v_phenom} shows the OLYMPUS results compared to my simple estimate, for two
different form factor models, as well as a more sophisticated estimate by Bernauer et al.\ \cite{Bernauer:2013tpr}.
My estimate using dipole form factors predicts an asymmetry that grows much faster than both the 
OLYMPUS data and the other predictions. This is to be expected, since under the dipole assumption,
$\mu_p G_E / G_M = 1$, while in many fits to Rosenbluth data, $\mu_p G_E / G_M$ is below 1. The dipole
model assumes that the discrepancy is larger than more sophisticated fits suggest. 

My simple estimate using the Bernauer form factor fits predicts roughly the same qualitative trends
as the Bernauer prediction, which should not be surprising since both estimates use the same underlying
form factors. The light blue error band is associated with my simple estimate, and is a way of gauging
the uncertainty in the magnitude of the form factor discrepancy. 
The slope of the OLYMPUS data does not seem to be as great as the phenomenological calculations predict.
However OLYMPUS is certainly consistent within the blue error band, and I do not believe that the OLYMPUS
results are inconsistent with the two-photon exchange hypothesis. 

\section{Conclusions}

OLYMPUS has measured the lepton sign asymmetry over a wide-range of angles, and the results in this
thesis are a determination of the hard TPE contribution to $ep$ scattering. The asymmetry has a 
clear slope, and the direction of that slope suggests a mitigation of the form factor discrepancy.
The uncertainty estimates in their current state are conservative, and therefore make it to difficult
to draw definite conclusions about whether or not two-photon exchange is the sole cause of the
form factor discrepancy, especially since the magnitude of the measured asymmetry is only about 1\%. 
Fortunately, the analysis of the systematic uncertainties will progress and as OLYMPUS nears a final
result, the goal is to reduce, with confidence and justification, the size of the uncertainty estimates.
With smaller systematic uncertainties, it may be possible for more decisive conclusions to be drawn. 
It will certainly be interesting once the OLYMPUS data can be combined with the data from Novosibirsk
and CLAS in order to make fits and extrapolations of the hard TPE effect.

\appendix
\chapter{Bremsstrahlung Cross Section}

\label{chap:brems}

The cross section for tree-level bremsstrahlung is an important component to the radiative generator.
In this appendix I will show some important derivations. I have a couple of notational preferences
which I will use in this appendix. I will use $\vec{v}$ to refer to a three-vector, but $\fv{v}$
to refer to four-vectors, whose individual components will be labeled as $v_\mu$. The expression 
$v^2$ will mean $v_\mu v^\mu$, or equivalently $\fv{v}\cdot\fv{v}$. The norm of a three vector,
defined as $\sqrt{\vec{v}\cdot\vec{v}}$, will be written as $|\vec{v}|$. 

I will use the Dirac gamma matrices $\gamma_\mu$. I will write contractions of the form
$\gamma_\mu v^\mu$ as $\slashed{v}$.

\section{Bremsstrahlung Matrix Element}

\begin{figure}[htpb]
  \centering
  \includegraphics{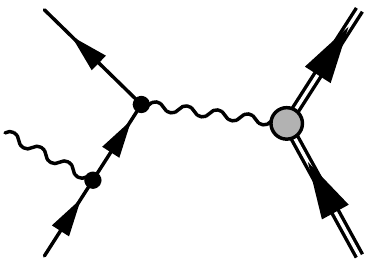}
  \includegraphics{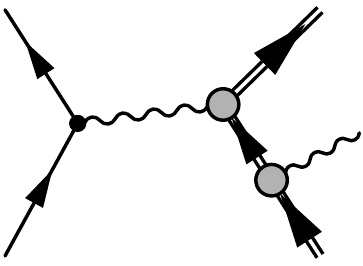}
  \includegraphics{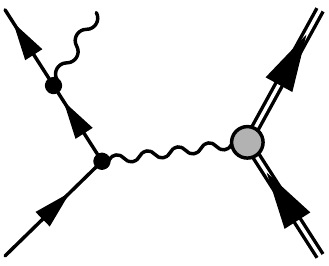}
  \includegraphics{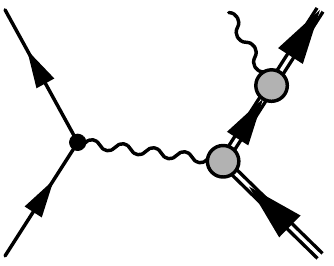}
\caption[The four tree-level bremsstrahlung diagrams]
{\label{fig:brems_diagrams} There are four tree-level bremsstrahlung diagrams.}
\end{figure}

There are four tree-level bremsstrahlung diagrams, shown in figure \ref{fig:brems_diagrams},
whose matrix elements must be summed together. I'll label these matrix elements as 
$\mathcal{M}_{li}$, $\mathcal{M}_{lf}$, $\mathcal{M}_{pi}$, and $\mathcal{M}_{pf}$ based on
if the photon emerges from a lepton or proton, and an initial or final leg. Let's call
the incoming lepton 4-momentum $\fv{p}_1$, the incoming proton 4-momentum $\fv{p}_2$,
the outgoing lepton 4-momentum $\fv{p}_3$, and the outgoing proton 4-momentum $\fv{p}_4$. 
The 4-momentum of the outgoing photon will be labeled with $\fv{k}$. Following Feynman
rules, the four matrices can be written as:
\begin{align}
\mathcal{M}_{li} =& -e^3 \epsilon^\ast_\lambda(\fv{k}) \bar{u}(\fv{p}_3) 
\gamma_\mu \frac{\slashed{p}_1 - \slashed{k} + m}{2\fv{p}_1 \cdot \fv{k}} \gamma^\lambda 
u(\fv{p}_1) \frac{1}{(\fv{p}_1 - \fv{k} - \fv{p}_3)^2} \bar{u}(\fv{p}_4) 
\Gamma^\mu(\fv{p}_4 - \fv{p}_2)u(\fv{p}_2) \\
\mathcal{M}_{lf} =& e^3 \epsilon^\ast_\lambda(\fv{k}) \bar{u}(\fv{p}_3) 
\gamma_\lambda \frac{\slashed{p}_3 + \slashed{k} + m}{2\fv{p}_3 \cdot \fv{k}} \gamma^\mu
u(\fv{p}_1) \frac{1}{(\fv{p}_1 - \fv{k} - \fv{p}_3)^2} \bar{u}(\fv{p}_4) 
\Gamma^\mu(\fv{p}_4 - \fv{p}_2)u(\fv{p}_2)\\
\mathcal{M}_{pi} =& e^3 \epsilon^\ast_\lambda(\fv{k}) \bar{u}(\fv{p}_3) \gamma_\mu u(\fv{p}_1)
\frac{1}{(\fv{p}_1 - \fv{p}_3)^2} \bar{u}(\fv{p}_4) \Gamma^\mu (\fv{p}_1 - \fv{p}_3 )
\frac{\slashed{p}_2 - \slashed{k} + M}{2\fv{p}_2\cdot\fv{k}} \Gamma^\lambda (-\fv{k}) u(\fv{p}_2)\\
\mathcal{M}_{pf} =& -e^3 \epsilon^\ast_\lambda(\fv{k}) \bar{u}(\fv{p}_3) \gamma_\mu u(\fv{p}_1)
\frac{1}{(\fv{p}_1 - \fv{p}_3)^2} \bar{u}(\fv{p}_4) \Gamma^\lambda (-\fv{k})
\frac{\slashed{p}_4 + \slashed{k} + M}{2\fv{p}_4\cdot\fv{k}} \Gamma^\mu (\fv{p}_1 - \fv{p}_3 ) u(\fv{p}_2).
\end{align}
Taking the sum of all four, we get:
\begin{equation}
  \begin{split}
  \mathcal{M}_\text{brems.} = e^3 \epsilon^\ast_\lambda(\fv{k}) \Big\{
  \bar{u}(\fv{p}_3) A_{\lambda\mu}(\fv{p}_1,\fv{p}_3,\fv{k}) u(\fv{p}_1) \frac{1}{(\fv{p}_4 - \fv{p}_2)^2}
\bar{u}(\fv{p}_4) \Gamma_\mu(\fv{p}_4 - \fv{p}_2) u(\fv{p}_2)\\
  - \bar{u}(\fv{p}_3) \gamma_\mu u(\fv{p}_1) \frac{1}{(\fv{p}_1 - \fv{p}_3)^2}\bar{u}(\fv{p}_4) B^{\lambda\mu} (\fv{p}_2,\fv{p}_4,\fv{k}) u(\fv{p}_2)
    \Big\},
\end{split}
\end{equation}

where the tensors $A_{\lambda\mu}(\fv{p}_1,\fv{p}_3,\fv{k})$ and $B^{\lambda\mu}(\fv{p}_2,\fv{p}_4,\fv{k})$ are given by:
\begin{align}
A_{\lambda\mu}(\fv{p}_1,\fv{p}_3,\fv{k})={}& \left[\gamma_\lambda \frac{\slashed{p}_3 + \slashed{k}+ m}{2\fv{p}_3\cdot\fv{k}} \gamma_\mu 
  - \gamma_\mu \frac{\slashed{p}_1 - \slashed{k}+ m}{2\fv{p}_1\cdot\fv{k}} \gamma_\lambda \right],\\
\begin{split}
B^{\lambda\mu}(\fv{p}_2,\fv{p}_4,\fv{k})={}& \Bigg[ 
\Gamma^\lambda(-\fv{k}) \frac{\slashed{p}_4 + \slashed{k} + M}{2\fv{p}_4\cdot\fv{k}}\Gamma^\mu(\fv{p}_4 + \fv{k} -\fv{p}_2) \\
{}&-\Gamma^\mu(\fv{p}_4 + \fv{k} -\fv{p}_2)\frac{\slashed{p}_2 - \slashed{k} + M}{2\fv{p}_2\cdot\fv{k}}\Gamma^\lambda(-\fv{k}) \Bigg].
\end{split}
\end{align}

To calculate a cross section, we'll need to square this matrix element. And since we want an unpolarized cross section, 
we must average the initial spin states and sum over the final spin-states:
\begin{align}
\left\langle |\mathcal{M}_\text{brems.}|^2 \right\rangle =& \frac{1}{4} \sum_\text{spins}
\left[ \mathcal{M}^\dagger_\text{brems.} \mathcal{M}_\text{brems.} \right]\\
\begin{split}
=& \frac{e^6}{4}\sum_\text{spins} \Bigg[ \epsilon^\ast_\lambda \epsilon_\kappa \\
  &\bigg\{ u^\dagger(\fv{p}_1) A^{\dagger\kappa\nu}(\fv{p}_1,\fv{p}_3,\fv{k}) \gamma_0 u(\fv{p}_3) \frac{1}{(\fv{p}_4 - \fv{p}_2)^2}
  u^\dagger(\fv{p}_2) \Gamma^\dagger_\nu(\fv{p}_4 - \fv{p}_2) \gamma_0 u(\fv{p}_4)\\
 & - \bar{u}(\fv{p}_1) \gamma_\mu u(\fv{p}_3) \frac{1}{(\fv{p}_1 - \fv{p}_3)^2}
u^\dagger(\fv{p}_2) B^{\dagger\kappa\nu} (\fv{p}_2,\fv{p}_4,\fv{k}) \gamma_0 u(\fv{p}_4) \bigg\} \\
 & \bigg\{ \bar{u}(\fv{p}_3) A_{\lambda\mu}(\fv{p}_1,\fv{p}_3,\fv{k}) u(\fv{p}_1) \frac{1}{(\fv{p}_4 - \fv{p}_2)^2}
\bar{u}(\fv{p}_4) \Gamma_\mu(\fv{p}_4 - \fv{p}_2) u(\fv{p}_2)\\
 & - \bar{u}(\fv{p}_3) \gamma_\mu u(\fv{p}_1) \frac{1}{(\fv{p}_1 - \fv{p}_3)^2}
\bar{u}(\fv{p}_4) B^{\lambda\mu} (\fv{p}_2,\fv{p}_4,\fv{k}) u(\fv{p}_2) \bigg\}
\Bigg].
\end{split}
\end{align}
We'll have to drag $\gamma_0$ accross the operators $\Gamma$, $A$, and $B$.

For the case of $\Gamma$:
\begin{align}
\Gamma_\mu^\dagger(\fv{q}) \gamma_0 
&= \left( F_1(Q^2)\gamma_\mu + \frac{i\kappa F_2(Q^2)}{2 m_p} q_\nu \sigma^{\mu\nu}\right)^\dagger \gamma_0\\
&= \left( F_1(Q^2)\gamma_\mu^\dagger - \frac{i\kappa F_2(Q^2)}{2M} q_\nu \sigma^{\dagger\mu\nu} \right) \gamma_0\\
&= \gamma_0 \left( F_1(Q^2)\gamma_\mu - \frac{i\kappa F_2(Q^2)}{2M} q_\nu \sigma^{\mu\nu} \right) \\
&= \gamma_0 \Gamma_\mu^\dagger(-\fv{q}).
\end{align}

For the case of $A$:
\begin{align}
A^\dagger_{\mu\nu}(\fv{p}_1,\fv{p}_3,\fv{k})\gamma_0 
&= \left[\gamma_\mu \frac{\slashed{p}_3 + \slashed{k}+ m}{2\fv{p}_3\cdot\fv{k}} \gamma_\nu 
 - \gamma_\nu \frac{\slashed{p}_1 - \slashed{k}+ m}{2\fv{p}_1\cdot\fv{k}} \gamma_\mu \right]^\dagger \gamma_0\\
&= \left[\gamma_\nu^\dagger \frac{\slashed{p}_3^\dagger + \slashed{k}^\dagger+ m}{2\fv{p}_3\cdot\fv{k}} \gamma_\mu^\dagger 
  - \gamma_\mu^\dagger \frac{\slashed{p}^\dagger_1 - \slashed{k}^\dagger+ m}{2\fv{p}_1\cdot\fv{k}} \gamma_\nu^\dagger \right] \gamma_0
\end{align}
\begin{align}
A^\dagger_{\mu\nu}(\fv{p}_1,\fv{p}_3,\fv{k})\gamma_0 
&= \gamma_0 \left[\gamma_\nu \frac{\slashed{p}_3 + \slashed{k}+ m}{2\fv{p}_3\cdot\fv{k}} \gamma_\mu 
  - \gamma_\mu \frac{\slashed{p}_1 - \slashed{k}+ m}{2\fv{p}_1\cdot\fv{k}} \gamma_\nu \right] \\
&= \gamma_0 A_{\nu\mu}(\fv{p}_1,\fv{p}_3,\fv{k}).
\end{align}
Note that the indices of the $A$ tensor have become reversed.

Lastly, for the case of $B$:
\begin{align}
  \begin{split}
    B^\dagger_{\mu\nu}(\fv{p}_2,\fv{p}_4,\fv{k})\gamma_0 
    &= \Bigg[ \Gamma_\mu(-\fv{k}) \frac{\slashed{p}_4 + \slashed{k} + M}{2\fv{p}_4\cdot\fv{k}}\Gamma_\nu(\fv{p}_4 + \fv{k} -\fv{p}_2) \\
{}&-\Gamma_\nu(\fv{p}_4 + \fv{k} -\fv{p}_2)\frac{\slashed{p}_2 - \slashed{k} + M}{2\fv{p}_2\cdot\fv{k}}\Gamma_\mu(-\fv{k}) \Bigg]^\dagger \gamma_0
\end{split}\\
  \begin{split}
    &= \Bigg[ \Gamma^\dagger_\nu(\fv{p}_4+\fv{k}-\fv{p}_2) \frac{\slashed{p}^\dagger_4 + \slashed{k}^\dagger + M}{2\fv{p}_4\cdot\fv{k}}
      \Gamma^\dagger_\mu(-\fv{k}) \\
{}&-\Gamma^\dagger_\mu(-\fv{k})\frac{\slashed{p}^\dagger_2 - \slashed{k}^\dagger + M}{2\fv{p}_2\cdot\fv{k}}
\Gamma^\dagger_\nu(\fv{p}_4+\fv{k}-\fv{p}_2) \Bigg] \gamma_0
\end{split}\\
  \begin{split}
    &= \gamma_0 \Bigg[ \Gamma_\nu(-\fv{p}_4-\fv{k}+\fv{p}_2) \frac{\slashed{p}_4 + \slashed{k} + M}{2\fv{p}_4\cdot\fv{k}}\Gamma_\mu(\fv{k}) \\
{}&-\Gamma_\mu(\fv{k})\frac{\slashed{p}_2 - \slashed{k} + M}{2\fv{p}_2\cdot\fv{k}} \Gamma_\nu(-\fv{p}_4-\fv{k}+\fv{p}_2) \Bigg]
\end{split}\\
  &\equiv \gamma_0 C_{\nu\mu}(\fv{p}_2,\fv{p}_4,\fv{k}).
\end{align}
The new tensor I've defined, $C$ will be useful in simplifying the notation.

Now, we can continue with our evaluation of the spin-summed matrix element:
\begin{equation}
  \begin{split}
    \left\langle |\mathcal{M}_\text{brems.}|^2 \right\rangle &= \frac{e^6}{4}\sum_\text{spins}
    \Bigg[ \epsilon^\ast_\lambda \epsilon_\kappa \bigg\{ 
      \bar{u}(\fv{p}_1) A^{\nu\kappa}(\fv{p}_1,\fv{p}_3,\fv{k}) u(\fv{p}_3) \frac{1}{(\fv{p}_4 - \fv{p}_2)^2}
      \bar{u}(\fv{p}_2) \Gamma_\nu(\fv{p}_2 - \fv{p}_4) u(\fv{p}_4) \\
      & - \bar{u}(\fv{p}_1) \gamma_\mu u(\fv{p}_3) 
      \frac{1}{(\fv{p}_1 - \fv{p}_3)^2}\bar{u}(\fv{p}_2) C^{\nu\kappa} (\fv{p}_2,\fv{p}_4,\fv{k}) u(\fv{p}_4) \bigg\} \\
        & \bigg\{ \bar{u}(\fv{p}_3) A_{\lambda\mu}(\fv{p}_1,\fv{p}_3,\fv{k}) u(\fv{p}_1) \frac{1}{(\fv{p}_4 - \fv{p}_2)^2}
      \bar{u}(\fv{p}_4) \Gamma_\mu(\fv{p}_4 - \fv{p}_2) u(\fv{p}_2)\\
        & - \bar{u}(\fv{p}_3) \gamma_\mu u(\fv{p}_1) \frac{1}{(\fv{p}_1 - \fv{p}_3)^2}
      \bar{u}(\fv{p}_4) B^{\lambda\mu} (\fv{p}_2,\fv{p}_4,\fv{k}) u(\fv{p}_2) \bigg\} \Bigg].
  \end{split}
\end{equation}
\begin{equation}
  \label{eq:brems_matsq}
  \begin{split}
    \left\langle |\mathcal{M}_\text{brems.}|^2 \right\rangle &= \frac{-e^6}{4} \Bigg[ \frac{1}{(\fv{p}_4 - \fv{p}_2)^4} 
      \Tr\left\{ (\slashed{p}_1 + m) {A^\nu}_\lambda(\fv{p}_1,\fv{p}_3,\fv{k}) (\slashed{p}_3 + m) A^{\lambda\mu}(\fv{p}_1,\fv{p}_3,\fv{k}) \right\} \\
      & \times \Tr\left\{ (\slashed{p}_2 + M) \Gamma_\nu (\fv{p}_2-\fv{p}_4) (\slashed{p}_4 + M) \Gamma_{\mu}(\fv{p}_4-\fv{p}_2) \right\}\\
      & - \frac{1}{(\fv{p}_4 - \fv{p}_2)^2(\fv{p}_1 - \fv{p}_3)^2} \Bigg(
      \Tr\left\{(\slashed{p}_1 + m)\gamma_\nu(\slashed{p}_3 + m)A^{\lambda\mu}(\fv{p}_1,\fv{p}_3,\fv{k})\right\} \\
      & \times \Tr\left\{ (\slashed{p}_2 + M) {C^\nu}_\lambda (\fv{p}_2,\fv{p}_4,\fv{k}) (\slashed{p}_4 + M) \Gamma_{\mu}(\fv{p}_4-\fv{p}_2) \right\}\\
      & + \Tr\left\{ (\slashed{p}_1 + m) {A^\nu}_\lambda(\fv{p}_1,\fv{p}_3,\fv{k}) (\slashed{p}_3 + m)\gamma_\mu \right\} \\
      & \times \Tr\left\{ (\slashed{p}_2 + M) \Gamma_\nu (\fv{p}_2-\fv{p}_4) (\slashed{p}_4 + M) B^{\lambda\mu}(\fv{p}_2,\fv{p}_4.\fv{k}) \right\}\Bigg)\\
      & + \frac{1}{(\fv{p}_1 - \fv{p}_3)^4}\Tr \left\{ (\slashed{p}_1 + m) \gamma_\nu (\slashed{p}_3 + m) \gamma_\mu \right\}\\
      & \times \Tr\left\{ (\slashed{p}_2 + M) {C^\nu}_\lambda (\fv{p}_2,\fv{p}_4,\fv{k}) (\slashed{p}_4 + M) B^{\lambda\mu}(\fv{p}_2,\fv{p}_4.\fv{k}) 
      \right\}\Bigg].
  \end{split}
\end{equation}
In equation \ref{eq:brems_matsq}, we can see that there are three terms. The term with the denominator of $(\fv{p}_4 - \fv{p}_2)^4$
represents the contribution from bremsstrahlung from the lepton. The term with the denominator of $(\fv{p}_1 - \fv{p}_3)^4$ represents
the contribution from bremsstrahlung from the proton. The mixed term, with denominator $(\fv{p}_4 - \fv{p}_2)^2(\fv{p}_1 - \fv{p}_3)^2$
is the interference between lepton and proton bremsstrahlung. This term changes sign when electrons are substituted for positrons. The
sign in equation \ref{eq:brems_matsq} is for electrons.

The denominators of the propagators contained in the tensors $A$, $B$, and $C$ are of the form $2\fv{p}\cdot\fv{k}$. This means
in the limit of the photon energy going to zero, the matrix element will diverge as $|\vec{k}|^2$. We can factor this divergence
out from the matrix element such that
\begin{equation}
\left\langle|\mathcal{M}'|^2\right\rangle \equiv \frac{4|\vec{k}|^2}{e^6} \left\langle|\mathcal{M}|^2\right\rangle
= \frac{4 |\vec{k}|^2}{(4\pi\alpha)^3} \left\langle|\mathcal{M}|^2\right\rangle.
\end{equation}
This notation will be valuable in discussion of cancelling divergences in the generator weights.

\section{Bremsstrahlung Phase Space}

From Fermi's Golden Rule, we know that:
\begin{equation}
d\sigma = \frac{ \left\langle|\mathcal{M}|^2\right\rangle }{2 E_1 2 m_p |v_1 - v_2|} \frac{d^3 \vec{p}_3}{(2\pi)^3 2E_3}
\frac{d^3 \vec{p}_4}{(2\pi)^3 2E_4} \frac{d^3 \vec{k}}{(2\pi)^3 2k} (2\pi)^4 \delta^4 (\fv{p}_1 + \fv{p}_2 - \fv{p}_3 - \fv{p}_4 \fv{k} ).
\end{equation}
Let's work in the lab frame, in which $|\vec{p}_2|=0$, and take the limit $m_e \rightarrow 0$. Then we have:
\begin{equation}
  \begin{split}
    d\sigma = & \frac{ \left\langle|\mathcal{M}|^2\right\rangle }{1024 \pi^5 E_1 m_p} \frac{E_3^2 dE^3 d\Omega_l}{E_3}
    \frac{|\vec{p}_4|^2 d|\vec{p}_4| d\Omega_4}{E_4} \times \\
    & \;\; |\vec{k}| d|\vec{k}| d\Omega_\gamma \delta( E_1 + m_p - E_3 - E_4 - k ) \delta^3 (\vec{p}_1 - \vec{p}_3 - \vec{p}_4 -\vec{k} ).
  \end{split}
\end{equation}
Let's use the momentum part of the delta function to remove the integral over $|\vec{p}_4|$:
\begin{equation}
d\sigma = \frac{ \left\langle|\mathcal{M}|^2\right\rangle }{1024 \pi^5 E_1 m_p} \frac{E_3 k}{E_4}
dE_3 d\Omega_l dk d\Omega_\gamma \delta( E_1 + m_p - E_3 - E_4 - k ).
\end{equation}
Let's use the energy delta function to eliminate the integral over $E_3$. To do this we'll need
to solve a Jacobian, because after fixing the proton momentum, $E_4$ has dependence on $E_3$. 
Recall that:
\[
\delta(f(x)) = \frac{\delta(x-x_0)}{|f'(x_0)|}.
\]
In our case:
\begin{align}
f(E_3) =& E_1 + m_p - E_3 - E_4(E_3) - k\\
f'(E_3) =& -1 + \frac{\partial}{\partial E_3}\sqrt{(\vec{p}_1 - \vec{p}_3 - \vec{k})^2 + m_p^2 }\\
=& -1 - \frac{1}{2E_4}\left( -2E_1\cos\theta_{l} + 2E_3 + 2k\cos\theta_{l \gamma } \right) \\
=& -\frac{E_4 + E_3 + k\cos\theta_{l\gamma} - E_1\cos\theta_l}{E_4}.
\end{align}

Plugging this back into our phase space calculation, we get:
\begin{equation}
d\sigma = \frac{ \left\langle|\mathcal{M}|^2\right\rangle }{1024 \pi^5 E_1 m_p} \frac{E_3 k}{E_4}
dE_3 d\Omega_l dk d\Omega_\gamma \frac{E_4}{\left|E_4 + E_3 + k\cos\theta_{l\gamma} - E_1\cos\theta_l \right|} \delta( E_3 - E_3' ).
\end{equation}
Integrating over the delta function and rearranging, we get an expression for the cross section:
\begin{equation}
\frac{d\sigma}{d\Omega_l d\Omega_\gamma dk} = \frac{ \left\langle|\mathcal{M}|^2\right\rangle }{1024 \pi^5 E_1 m_p}
 \frac{E_3 k}{\left|E_4 + E_3 + k\cos\theta_{l\gamma} - E_1\cos\theta_l \right|}.
\end{equation}
To use the notation of our previous section:
\begin{align}
\frac{d\sigma}{d\Omega_l d\Omega_\gamma dk} =&
 \frac{ (4\pi)^3 \alpha^3 \left\langle|\mathcal{M}'|^2\right\rangle }{1024 \pi^5 E_1 m_p k^2}
 \frac{E_3 k}{\left|E_4 + E_3 + k\cos\theta_{l\gamma} - E_1\cos\theta_l \right|}\\
=&  \frac{ \alpha^3 \left\langle|\mathcal{M}'|^2\right\rangle }{ 64 \pi^2 E_1 m_p k}
 \frac{E_3}{\left|E_4 + E_3 + k\cos\theta_{l\gamma} - E_1\cos\theta_l \right|}.
\end{align}
We can see that in the limit $k\rightarrow 0$, this cross section diverges as $k^{-1}$.

\section{Jacobian for $k \longrightarrow \Delta E_l$}

Many standard radiative correction prescriptions use $\Delta E_l$ as the variable for determining elasticity.
To work with these corrections, it would be useful to define the bremsstrahlung cross section
in the form
\[
\frac{d\sigma}{d\Omega_l d\Omega_\gamma d\Delta E_l}.
\]
To get this, we'll need the Jacobian to take $k$ to $\Delta E_l$:
\[
\mathcal{J}(k\rightarrow \Delta E_l) \equiv \left | \frac{\partial k}{\partial \Delta E_l}\right|.
\]
First, we'll need to define $k$ as a function of $\Delta E$. To get this, let's recall that
\begin{equation}
\fv{p}_2 - \fv{p}_4 = - (\fv{p}_1 - \fv{p}_3 -\fv{k} ).
\end{equation}
Squaring both sides, we get:
\begin{align}
\fv{p}_2 - \fv{p}_4 =& - (\fv{p}_1 - \fv{p}_3 -\fv{k} ) \\
m_p^2 - \fv{p}_2 \cdot \fv{p}_4 =& - \fv{p}_1 \cdot \fv{p}_3  - \fv{p}_1 \cdot \fv{k} + \fv{p}_3 \cdot \fv{k} \\
-E_4 =& \frac{ - \fv{p}_1 \cdot \fv{p}_3  - \fv{p}_1 \cdot \fv{k} + \fv{p}_3 \cdot \fv{k} + m_p^2 }{ m_p }.
\end{align}
Now let's use a statement of the conservation of energy, $k=E_1 - E_3 + m_p - E_4$ to write:
\begin{align}
k =& E_1 - E_3 + m_p + \frac{ - \fv{p}_1 \cdot \fv{p}_3  - \fv{p}_1 \cdot \fv{k} + \fv{p}_3 \cdot \fv{k} + m_p^2 }{ m_p }\\
=& E_1 - E_3 + \frac{ - \fv{p}_1 \cdot \fv{p}_3  - \fv{p}_1 \cdot \fv{k} + \fv{p}_3 \cdot \fv{k} }{ m_p }\\
=& E_1 - E_3 + \frac{ -E_1 E_3 (1-\cos\theta_l) - E_1 k (1 - \cos\theta_\gamma) + E_3 k ( 1 - \cos\theta_{l\gamma})}{ m_p }\\
=& \frac{m_p (E_1 - E_3) - E_1 E_3 (1-\cos\theta_l)}{m_p + E_1(1-\cos\theta_\gamma) + E_3(1-\cos\theta_{l\gamma})}.
\end{align}
From here, let's recall that $E_3 = E^\text{el.}_3 - \Delta E_l$:
\[
E_3 = \frac{E_1 m_p}{m_p + E_1(1-\cos\theta_l)} - \Delta E_l.
\]
That means:
\begin{align}
k=&\frac{m_p E_1 - \frac{E_1 m_p^2}{m_p + E_1(1-\cos\theta_l)} + m_p \Delta E_l - \frac{E_1^2 m_p}{m_p + E_1(1-\cos\theta_l)} (1-\cos\theta_l)
+ \Delta E_l E_1(1-\cos\theta_l)}
{m_p + E_1(1-\cos\theta_\gamma) + E_3^\text{el.} (1-\cos\theta_{l\gamma}) - \Delta E_l (1-\cos\theta_{l\gamma})}\\
=& \frac{ \Delta E_l (m_p + E_1 (1 - \cos\theta_l))}
{m_p + E_1(1-\cos\theta_\gamma) + E_3^\text{el.} (1-\cos\theta_{l\gamma}) - \Delta E_l (1-\cos\theta_{l\gamma})}.\label{eq:kOfE}
\end{align}

Now that we have an expression for $k$ as a function of $\Delta E_l$, we can differentiate it. To make this somewhat easier on
ourselves, let's notice that the expression in equation \ref{eq:kOfE} can be re-written with constants $\Sigma$, $\Psi$, $\Lambda$:
\[
k = \frac{\Psi \Delta E_l}{\Sigma + \Lambda \Delta E_l}
\]
so that
\[
\frac{\partial k}{\partial \Delta E_l} = \frac{\Psi}{\Sigma + \Lambda\Delta E_l}\frac{\Sigma}{\Sigma + \Lambda \Delta E_l} = 
\frac{k}{\Delta E_l} \frac{\Sigma}{\Sigma + \Lambda \Delta E_l}.
\]
Substituting for the constants gives us:
\begin{equation}
\frac{\partial k}{\partial \Delta E_l} = \frac{k}{\Delta E_l} \frac{m_p + E_1(1-\cos\theta_l) - E_3^\text{el.}(1-\cos\theta_{l\gamma})}
     {m_p + E_1(1 - \cos\theta_\gamma) - E_3(1-\cos\theta_{l\gamma})},
\end{equation}
and our Jacobian is 
\begin{align}
\mathcal{J}(k\rightarrow \Delta E_l) =&\frac{k}{\Delta E_l} \left|\frac{m_p + E_1(1-\cos\theta_l) - E_3^\text{el.}(1-\cos\theta_{l\gamma})}
     {m_p + E_1(1 - \cos\theta_\gamma) - E_3(1-\cos\theta_{l\gamma})}\right|\\
     \equiv & \frac{k}{\Delta E_l} \tilde{\mathcal{J}}.
\end{align}
We can now express the bremsstrahlung cross section as:
\begin{equation}
  \frac{d^5\sigma}{d\Omega_l d\Omega_\gamma d\Delta E}
  = \frac{ \alpha^3 \left\langle|\mathcal{M}'|^2\right\rangle }{ 64 \pi^2 E_1 m_p}
  \frac{E_3}{\left|E_4 + E_3 + k\cos\theta_{l\gamma} - E_1\cos\theta_l \right|} \frac{\tilde{\mathcal{J}}}{\Delta E_l} .
\label{eq:brems_tree_level}
\end{equation}
There is a divergence of the form $(\Delta E_l)^{-1}$.

\clearpage
\newpage

\chapter{Magnet Coil Displacement}

\label{app:magnet}

\section{Overview}

One of the arguments about why a measurement of the magnetic field was needed, as opposed to a calculation, was 
the observation during the BLAST experiment that the magnet flexed under magnetic forces when current was passed through it.
Even though the coils may sit close to their nominal positions at rest, when the magnet is turned on they 
can move by several millimeters. A displacement of 7~mm towards the beamline at 6730~A was quoted frequently. 

During the magnetic field measurements, we had a chance to test this phenomenon. We used the total
station to monitor a collection of survey targets that we glued to the surface of one of the coils. We
confirmed that the magnet coils do move by several millimeters. These results influenced our choice of
free parameters in our magnet coil model in our analysis of the magnetic field measurements.

\section{Measurements}

\begin{figure}[htb]
\centering
\includegraphics[width=12cm]{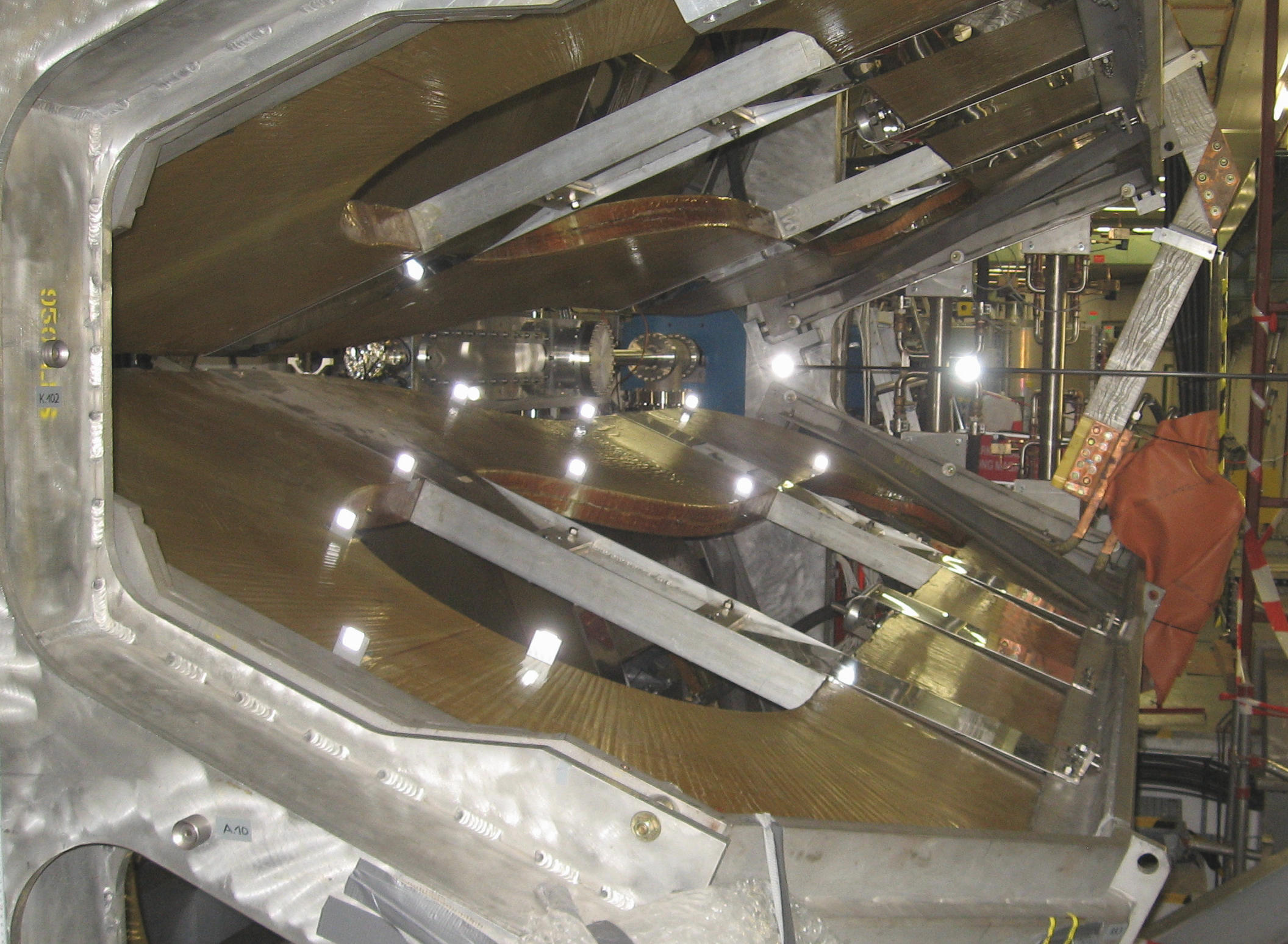}
\caption[Photograph of the magnetic coil displacement study]
{\label{fig:magnet_disp_photo} We estimated the coil displacement by glueing survey targets to
the surface of one of the coils. The survey targets are extremely reflective and appear as bright spots
due to the camera flash. Also visible are two survey targets still attached to one of the measurement rods.}
\end{figure}

The total station was a really valuable survey tool, since, with its laser tracker, it could determine the
coordinates in three dimensions of a target. During the magnetic field measurements, we glued several targets
to surfaces of one of the coils, as can been seen in figure \ref{fig:magnet_disp_photo}. We chose a coil surface
that was completely visible to the total station and we arranged the survey targets around the edges of the coil
face to bound as much of the coil as possible. We surveyed these targets while the magnet had no current. Then
we turned on 5000~A of current and surveyed the targets again. We repeated this procedure several times and tested
both magnet polarities. Using the total station's telescope, it was clear the magnet coil moved from changes in 
current because the survey targets would leave the cross hairs as the current turned on. Some analysis was needed
to determine how much displacement there was.

\section{Analysis}

\begin{figure}[htpb]
\centering
\includegraphics{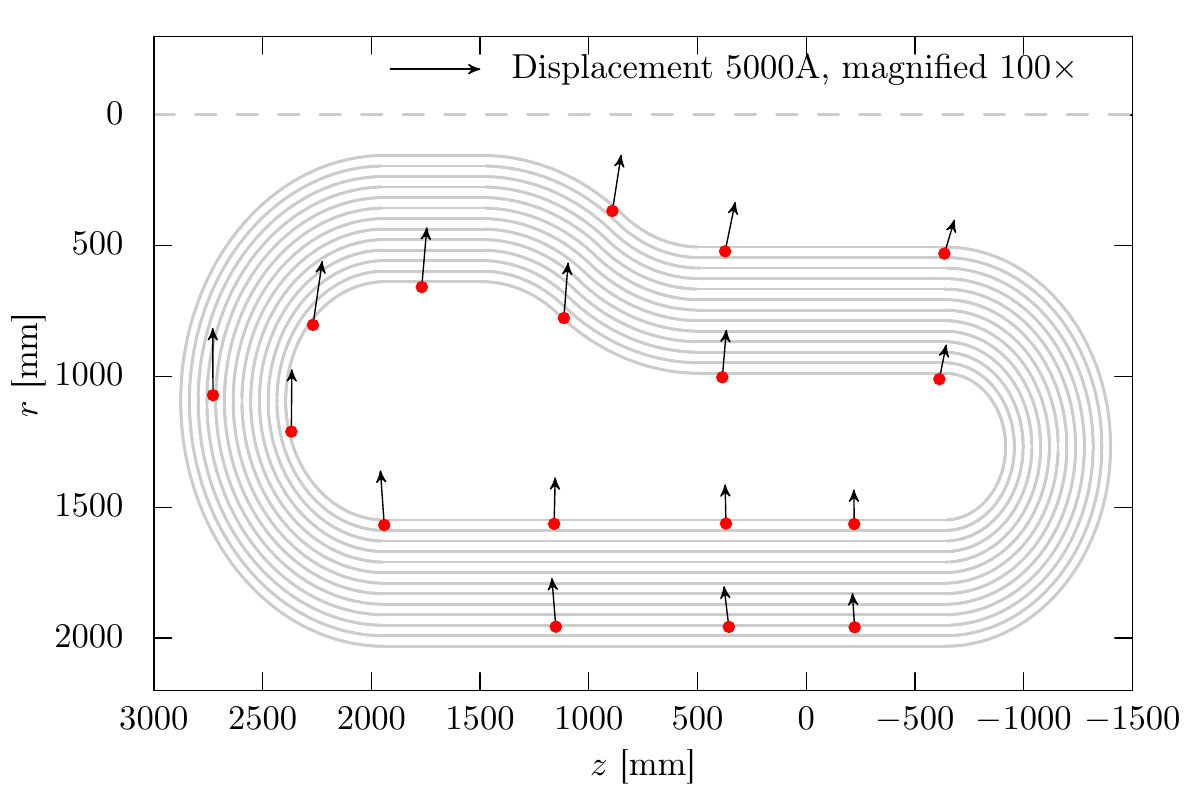}
\caption[Results of the magnetic coil displacement study]
{\label{fig:magnet_disp} Turning on the current caused the coil to move in plane. Most of the
motion was towards the beamline, but there was also a slight rotation. The arrows show the displacement
magnified by a factor of 100. The largest measured displacement was 2.5~mm.}
\end{figure}

Using the daily calibration data from the total station, we transformed the total station angles and 
distances into three dimensional positions in the OLYMPUS coordinate system. The results were very clear.
When current passed through the magnet, the coils moved by several millimeters. The movement was consistent
(to within the $\approx 200~\mu$m resolution of the total station) and was the same magnitude and direction
regardless of magnet polarity. The motion is almost entirely in the plane of the coil. The in-plane displacements
are shown in figure \ref{fig:magnet_disp}. 

\section{Discussion}

Most of the motion appears to be toward the beamline. This was consistent with our expectation. By moving
towards the beamline, the coils get closer together, reducing the length of the magnetic field lines. However,
it was clear that there was slight rotation of the coils. The wide part of the coil appears displaced farther
than the narrow part. This observation lead us to try an in-plane rotation as one of the free parameters 
when fitting our magnetic field model. In the fit we decided to use (parameters in table \ref{tab:params}),
the in-plane rotation of the upper and lower pair of coils was fit to $0.05^\circ$. 

\chapter{A Simple Parameterization for Time-to-Distance Functions}

\label{app:axelttd}

\section{Overview}

\begin{figure}[htpb]
\centering
\includegraphics[width=12cm]{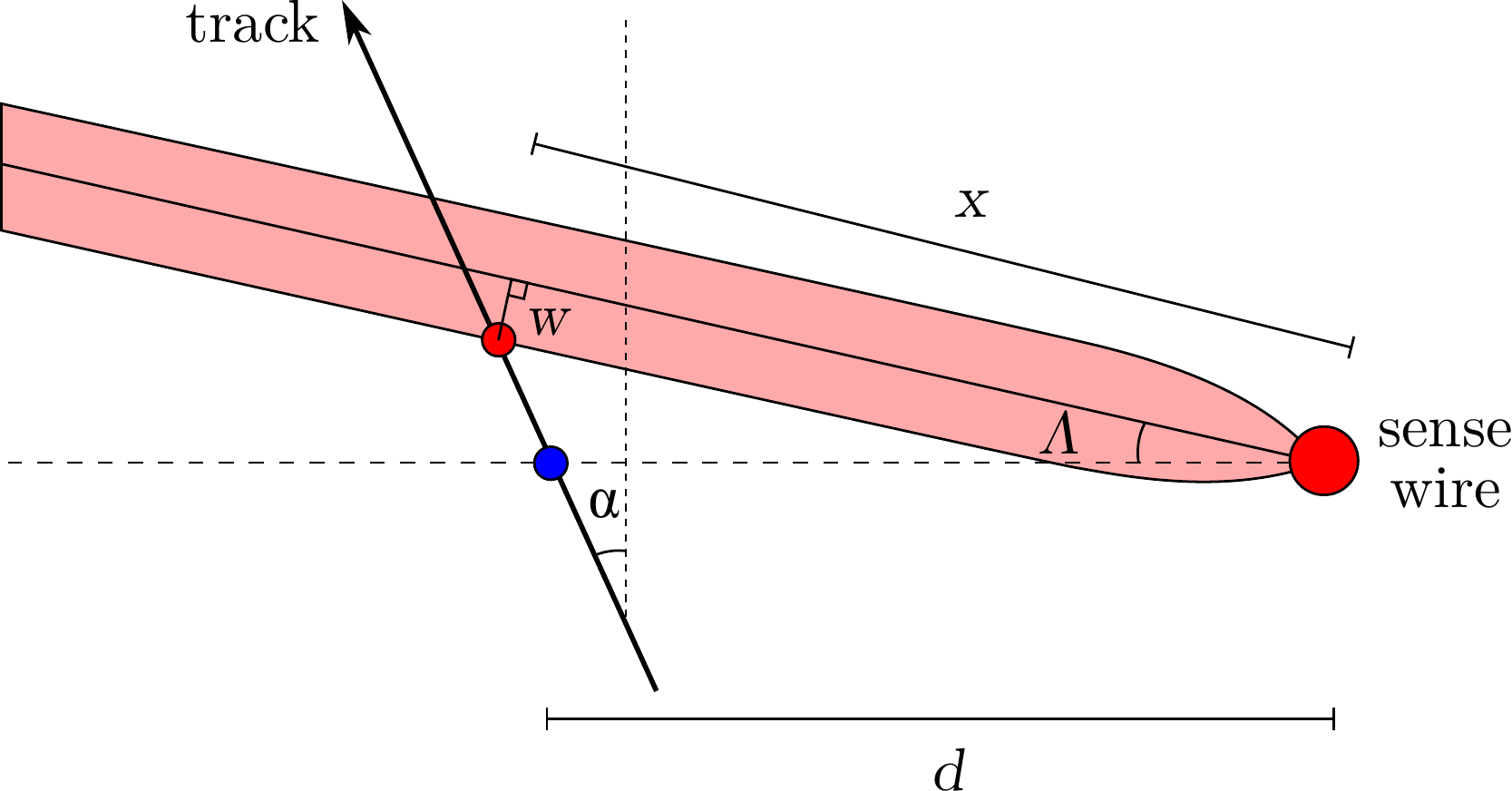}
\caption[Geometric approach to TTD]{\label{fig:axel_ttd_deriv} We used a simple geometric model
to determine the functional dependence on $\alpha$.} 
\end{figure}

In this appendix, I present the derivation of the simple parameterization for the TTD functions
discussed in section \ref{ssec:ttd_axel}. The derivation makes use of a simple two-dimensional
geometric model, illustrated in figure \ref{fig:axel_ttd_deriv}. First, I will use the model
to derive the approximate dependence on $\alpha$ in the region where the function is linear in 
$t$. Then I will consider the regions in which the TTD function is no longer linear, when the
drift times are very short or very long. Lastly, I will add parameters to allow the function
to vary with $\phi$. 

In my derivation, I will use the following conventions. The Lorentz angle parameter $\Lambda$ is
taken always to be positive. The incidence angle $\alpha$ is 0 when the track is perpendicular
to the sense wire plane, positive when the track points further upstream, and negative when
the track points further downstream. A track with $\alpha=0$ has a scattering angle of approximately 
$74^\circ$; for most tracks entering the drift chambers $\alpha < 0$. I use $\phi$ here to refer to the 
azimuthal angle from the horizontal plane; that is, a track traveling horizontally left will have
$\phi=0^\circ$, and a track traveling horizontally right will also have $\phi=0^\circ$. This is different
from how I use $\phi$ in other chapters, in which a track traveling horizontally to the 
right has $\phi=180^\circ$. 

\section{Linear Region}

In the simple geometric model, I assume that the lines of drift form a jet of constant width far 
from the sense wire, that the jet emerges from the sense wire with an angle $\Lambda$ from the 
sense wire plane, and that the point at which the first-arriving electrons are ionized 
is the intersection point of the jet and the track, which sits at approximately distance $x$ from the sense
wire. For tracks that intersect the jet in the linear region: 
\begin{equation}
d = x \left( \cos\Lambda - \sin\Lambda \tan\alpha \right) + w \left| \sin\Lambda + \cos\Lambda \tan\alpha\right|.
\end{equation}
To simplify things slightly, I assume $\Lambda$ is always small enough to justify replacing
$\sin\Lambda$ with $\Lambda$ and $\cos\Lambda$ with $1$. This leads to:
\begin{equation}
d = x \left( 1 - \Lambda \tan\alpha \right) + w \left| \Lambda + \tan\alpha\right|.
\end{equation}
Next, I assume a constant drift velocity $v$ so that $x$ can be replaced with $vt$:
\begin{equation}
d = vt \left( 1 - \Lambda \tan\alpha \right) + w \left| \Lambda + \tan\alpha\right|.
\end{equation}
I choose to add an extra parameter, a constant $r$, to allow $d(t)$ to have a non-zero
intercept, even when $\Lambda = -\tan\alpha$:
\begin{equation}
d = vt \left( 1 - \Lambda \tan\alpha \right) + w \left| \Lambda + \tan\alpha\right| + r.
\label{eq:axelttd_prelim}
\end{equation}
The parameter $r$ can be interpreted as a radius around the sense wire in which the
ionization electrons cease to drift at a constant velocity and instead accelerate toward
the wire. 

\section{Non-linear Regions}

Equation \ref{eq:axelttd_prelim} describes a function that is linear in $t$, but captures
the non-trivial dependence of $d$ on $\alpha$. This parameterization is suitable for the
linear part of the TTD function, between approximately t=0.2~$\mu$s and t=1.1~$\mu$s for
most wires (example data are shown in figure \ref{fig:ttd_data}). Special modifications are needed for very 
small times, where the higher electric fields near the wire increase the drift velocity,
and for very large times, where the TTD function plateaus. 

For very small times, I choose to approximate the function with a cubic polynomial in $t$.
I join the cubic and linear regions at the cut-off point $\psi$ by keeping the first two derivatives continuous:
\begin{equation}
  d(t,\alpha) = 
  \begin{dcases}
    t \left[ v\left( 1 - \Lambda \tan\alpha \right) + \frac{w \left| \Lambda + \tan \alpha \right| + r}{\psi}
    \left( 3 - \frac{3t}{\psi} + \frac{t^2}{\psi^2} \right) \right] &  t < \psi , \\
    vt \left( 1 - \Lambda \tan\alpha \right) + w \left| \Lambda + \tan\alpha\right| + r &  t \geq \psi.    
  \end{dcases}
\end{equation}
We typically used a cut-off value $\psi=0.2$~$\mu$s.

For very long times, the TTD function plateaus at a constant value, $d_\text{max}$, corresponding
to the largest possible drift distance, beyond which the track no longer passes through  the active 
region of the cell. I use the geometric model to determine $d_\text{max}$. Given $c$, the distance
between the ground plane and the sense wire, then:
\begin{equation}
d_\text{max} = c - y \tan\alpha,
\end{equation}
\begin{equation}
  y = 
  \begin{dcases}
    y_\text{min} & \\
    \pm \left[ c\Lambda - \text{sign}(\alpha) w \right] & \\
    y_\text{max} &
  \end{dcases}.
  \label{eq:ttd_plateau_y}
\end{equation}
The variable $y$ represents the distance between the sense plane and the intersection point of
the jet and the ground plane. It has a positive sign for upstream sides of wires on the left
sector and downstream sides of wires on the right sector, and otherwise a negative sign. $y$ must
be restricted to be within the bounds $[y_\text{min},y_\text{max}]$, set by the vertical width of 
the cell: [$-10$~mm,~$30$~mm] for inner sense wires, [$-20$~mm,~$20$~mm] for middle sense wires,
and [$-30$~mm,~$10$~mm] for outer sense wires. 

By setting $d_\text{max}$, I have also set a maximum time:
\begin{equation}
t_\text{max} = \frac{d_\text{max} - w |\Lambda + \tan\alpha| - r}{v(1 - \Lambda \tan\alpha)}.
\end{equation}
Putting all three regions together, our parameterization becomes:
\begin{equation}
  d(t,\alpha) = 
  \begin{dcases}
    t \left[ v\left( 1 - \Lambda \tan\alpha \right) + \frac{w \left| \Lambda + \tan \alpha \right| + r}{\psi}
    \left( 3 - \frac{3t}{\psi} + \frac{t^2}{\psi^2} \right) \right] &  t < \psi , \\
    vt \left( 1 - \Lambda \tan\alpha \right) + w \left| \Lambda + \tan\alpha\right| + r &  \psi \leq t < t_\text{max},\\
    d_\text{max} & t_\text{max} \leq t.
  \end{dcases}
\end{equation}
This parameterization has only four free parameters: $\Lambda$, $w$, $v$, and $r$. 

\section{Adding dependence on $\phi$}

So far, this derivation has been entirely two-dimensional; I have not discussed any dependence
on $\phi$. Looking at reconstructed tracks, we observe that TTD functions have a weak dependence 
on $\phi$. In order to allow such a dependence, I make $\Lambda$, $w$, and $v$ quadratic functions
of $\phi$:
\begin{align}
\Lambda & \longrightarrow \Lambda_0 + \Lambda_1 \phi + \Lambda_2 \phi^2, \\
w & \longrightarrow w_0 + w_1 \phi + w_2 \phi^2, \\
v & \longrightarrow v_0 + v_1 \phi + v_2 \phi^2,
\end{align}
raising the total number of free parameters to 10. I choose not to allow $r$ to vary with $\phi$, thinking
that if $r$ is determined by the electric field very close to the sense wire, then it should be constant
along the length of the wire. 

\chapter{An Estimate of the Asymmetry Needed to Resolve the Form Factor Discrepancy}

\label{app:tpe}

Hard TPE is a possible explanation for the form factor discrepancy. Therefore, in interpreting
the results of an asymmetry measurement, one would like to know: is the asymmetry large enough
to resolve the discrepancy? In this appendix, I will present a simple way of estimating how much
of an asymmetry is needed to fully resolve the form factor discrepancy. My method is by no means 
the only method for making such an estimate. In fact, you could describe my method as the least
sophisticated estimate one can make. I think it is useful for establishing the size scale of the
asymmetry and its kinematic dependencies, in the scenario where hard TPE is fully responsible for the 
discrepancy.

My estimate requires a few assumptions. The first assumption is that the form factor ratio
$R\equiv \mu_p G_E / G_M$ as measured by polarization asymmetry experiments describes the
true form factor ratio, i.e., the polarization measurements are completely insenstive to 
hard TPE. Second, I assume that the TPE effect preserves the linearity of Rosenbluth plots. 
Third, I assume that the TPE effect is zero at $\epsilon=1$. Let's look at the consequences
of these assumptions. 

\begin{figure}[htpb]
\centering
\includegraphics{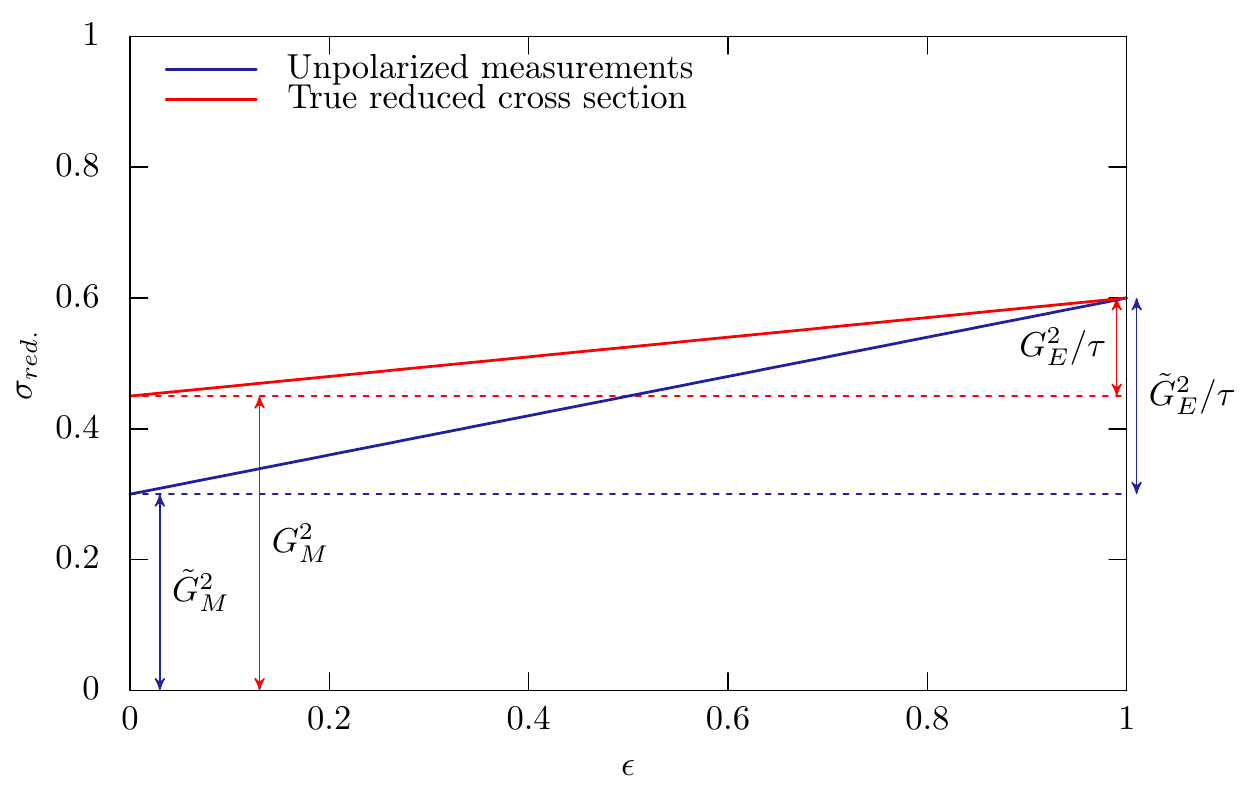}
\caption[Illustration of how TPE could modify Rosenbluth plots]
{\label{fig:rosenbluth_mod} I assume that hard TPE modifies the slope but preserves the linearity
of Rosenbluth plots.}
\end{figure}

To preserve the linearity of Rosenbluth plots, hard TPE must correct the reduced cross
section such that:
\begin{equation}
G_M^2(Q^2) + \frac{\epsilon}{\tau}G_E^2(Q^2) = \tilde{G}_M^2(Q^2) + \frac{\epsilon}{\tau}\tilde{G}_E^2(Q^2) + \delta(Q^2)(1-\epsilon),
\label{eq:tpe_mod}
\end{equation}
where $G_E$ and $G_M$ represent the true form factors, $\tilde{G}_E$ and $\tilde{G}_M$ represent 
the form factors extracted from Rosenbluth separation without accounting for hard TPE, and
$\delta(Q^2)$ represents a modification due to hard TPE. The effect of this correction is illustrated
in figure \ref{fig:rosenbluth_mod}. Given the $\epsilon$ dependence of
both sides of equation \ref{eq:tpe_mod}, we find two relationships that must hold at every value of $Q^2$:
\begin{align}
G_M^2 &= \tilde{G}_M^2 + \delta \\
G_E^2 &= \tilde{G}_E^2 - \tau \delta.
\end{align}
My first assumption was that the true form factor ratio $R$ can be taken from polarization asymmetry
measurements. This leads to:
\begin{align}
R^2 &= \frac{\mu_p^2 (\tilde{G}_E^2 - \delta \tau )}{\tilde{G}_M^2 + \delta} \\
\delta(R^2 + \tau \mu_p^2) &= \mu_p^2\tilde{G}_E^2 - R^2 \tilde{G}_M^2 \\
\delta &= \frac{\mu_p^2 \tilde{G}_E^2 - R^2 \tilde{G}_M^2}{R^2 + \mu_p^2 \tau}.
\end{align}
With an expression for $\delta$ in hand, it is easy to find the size of the asymmetry:
\begin{equation}
A = \frac{ \delta (1-\epsilon)}{ \tilde{G}_M^2 + \frac{\epsilon}{\tau} \tilde{G}^2_E + \delta(1-\epsilon)}.
\end{equation}
To use this method, one must supply $\tilde{G}_E$ and $\tilde{G}_M$ as taken from Rosenbluth measurements, 
as well as $R$ as taken from polarization asymmetry measurements, then calculate $\delta$ and $A_{2\gamma}$.

\begin{figure}[htpb]
\centering
\includegraphics{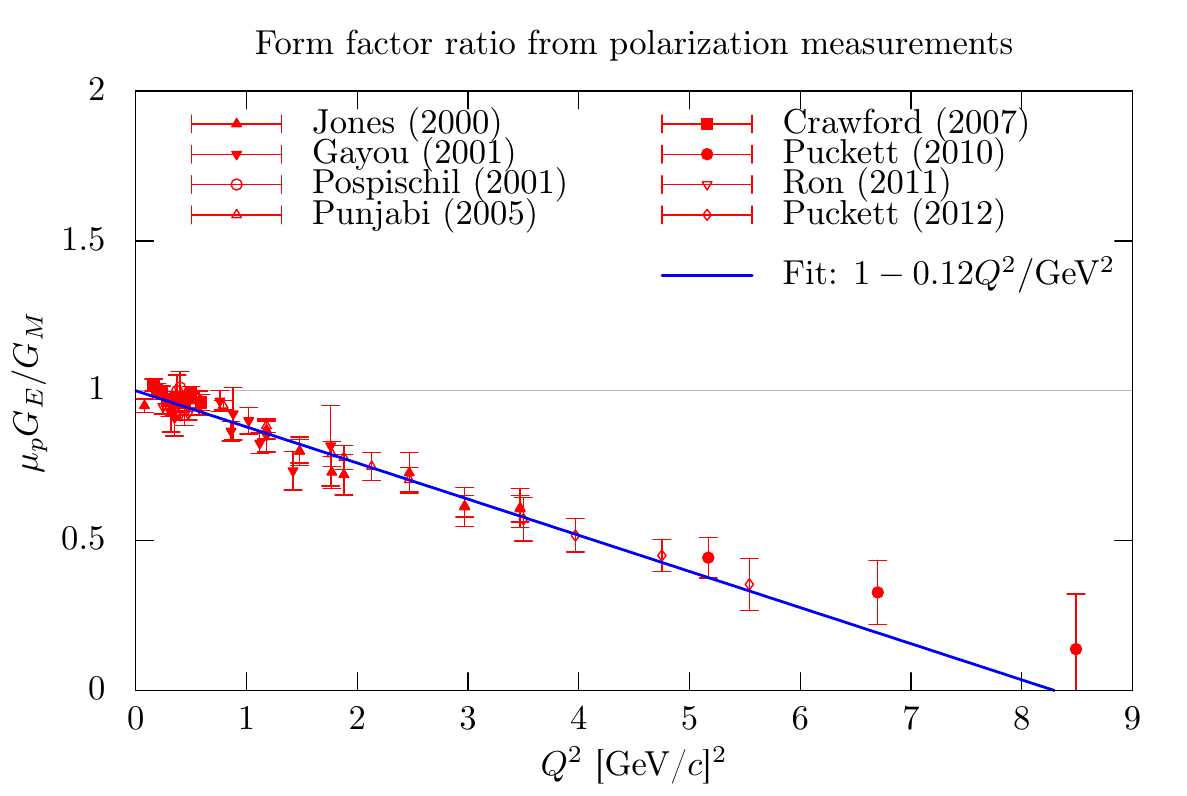}
\caption[Fits to polarization data]{\label{fig:fit_to_pol_data} Polarization data 
\cite{Jones:1999rz,Gayou:2001qt,Pospischil:2001pp,Punjabi:2005wq,Crawford:2006rz,Puckett:2010ac,Ron:2011rd,Puckett:2011xg}
are well fit by a line with slope 0.12~GeV$^{-2}$.}
\end{figure}

The values of $\tilde{G}_E$, $\tilde{G}_M$, and $R$ have been measured at several dozen discrete points in
$Q^2$. To make a prediction as a function of $Q^2$, or $\epsilon$, or scattering angle, one must interpolate
between the measured data points, or fit some functional form to the data. I will examine the effect of using
two such functional forms, one simple, and one sophisticated. For a simple model, I will approximate
$\tilde{G}_E$, $\tilde{G}_M$ as having the standard dipole form. For the sophisticated model, I will use the
2013 Bernauer et al.\ variable-knot spline fits to cross section data only \cite{Bernauer:2013tpr}. As a model 
for $R$, in both cases, I will use the model:
\[
R(Q^2) = 1 - (0.12 \text{~GeV}^{-2}) Q^2,
\]
which fits the existing polarization transfer data well enough for our purposes, as can be seen in figure
\ref{fig:fit_to_pol_data}

\begin{figure}[htpb]
\centering
\includegraphics{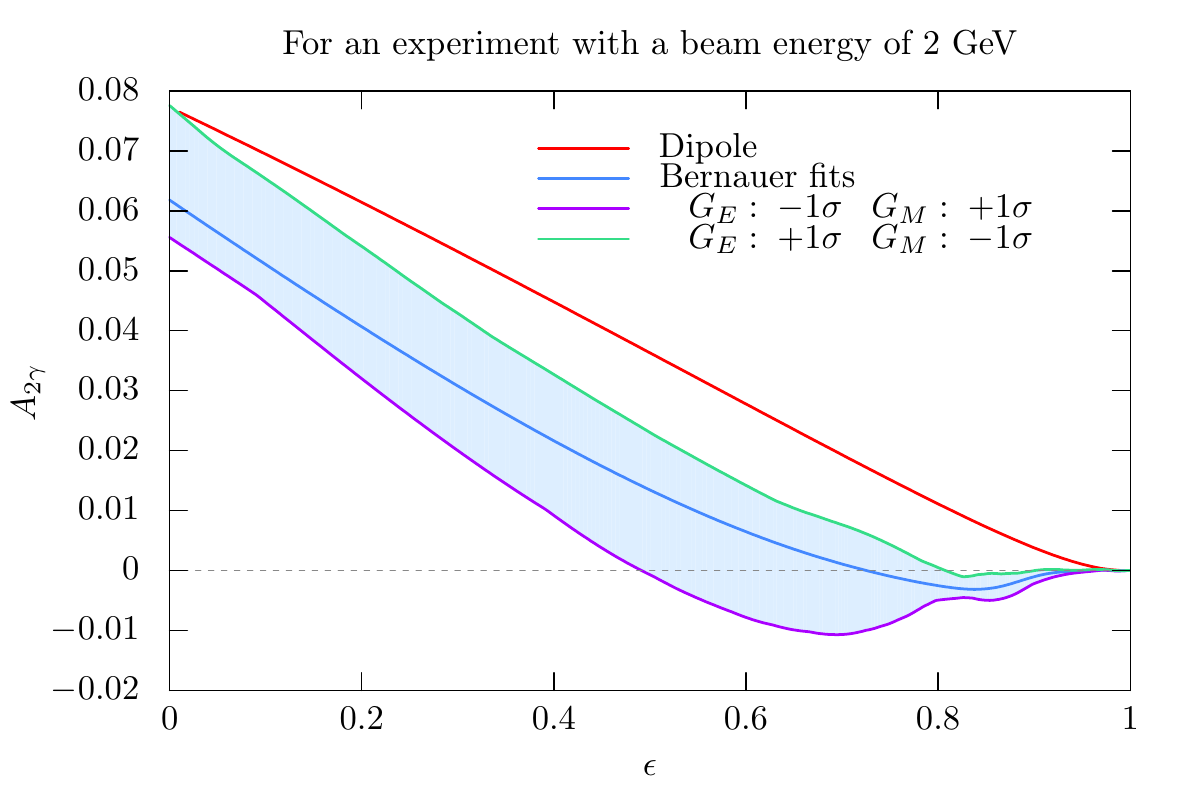}
\caption[Simple estimate of the asymmetry needed to resolve the form factor discrepancy]
{\label{fig:tpe_estimate_for_app} According to this simple estimate, the lepton sign asymmetry must rise by 
several percent with decreasing $\epsilon$ in order to resolve the form factor discrepancy. The particular form factors
used as input can have a big impact on how large of asymmetry is needed. The discrepancy is bigger if one assumes
dipole form factors. If one uses a sophisticated global fit to cross section data \cite{Bernauer:2013tpr}, the necessary
asymmetry is smaller. The uncertainty in the form factor fit introduces uncertainty in how much TPE is needed. Here, I 
define $1\sigma$ as the uncertainty from statistics and systematics added in quadrature, simply as a rough estimate.} 
\end{figure}

The results of this estimate procedure for both form factor models is shown in figure \ref{fig:tpe_estimate_for_app},
for the OLYMPUS beam energy of 2~GeV. These results suggest that the asymmetry should rise to several percent as
$\epsilon$ decreases. The choice of form factors can have a big effect on the magnitude of the estimate. Using dipole
form factors, the asymmetry needed to resolve the discrepancy rises by nearly 5\% over the OLYMPUS acceptance. If one
uses a global fit to cross section data, the asymmetry need not be so large. A rise of 3\% over the acceptance is sufficient.
I have also chosen to show a band corresponding to some measure of the fit uncertainty. This is essentially a measure of
how well the form factors are constrained by data, and thus, how well the form factor discrepancy is constrained by data.
If the lower error band represents the true form factors, then an asymmetry of less than 2\% is sufficient to resolve
the discrepancy. I interpret this spread in estimates as a call for caution against making a definite interpretation of
the data. An asymmetry that rises to 3\% seems suggestive that hard TPE is at play in the form factor discrepancy, but
is by no means conclusive.

\begin{singlespace}
\bibliography{references}
\bibliographystyle{hieeetr}
\end{singlespace}

\end{document}